\documentclass[twocolumn]{aastex63}
\usepackage{graphicx}
\usepackage[version=4]{mhchem}
\usepackage{soul} 
\usepackage{amssymb,amsfonts,indentfirst,times,mathtools,amsmath,amstext} 
\usepackage{comment,makecell} 
\newcounter{reactions} 

\usepackage{hyperref} 
\usepackage[caption=false]{subfig}
\usepackage{makecell} 
\usepackage{multirow}
\hypersetup{colorlinks=true,allcolors=blue,pdfpagemode=UseOutlines}
\allowdisplaybreaks 
\begin{document}
\title{A Comparative Study of Atmospheric Chemistry with VULCAN}

\author[0000-0002-8163-4608]{Shang-Min Tsai}
\affiliation{Atmospheric, Ocean, and Planetary Physics, Department of Physics, University of Oxford, UK}
\author[0000-0002-2110-6694]{Matej Malik}
\affiliation{Department of Astronomy, University of Maryland, College Park, MD 20742, USA }
\author[0000-0003-4269-3311]{Daniel Kitzmann}
\affiliation{University of Bern, Center for Space and Habitability}
\author{James R. Lyons}
\affiliation{Arizona State University, School of Earth and Space Exploration, Bateman Physical Sciences, Tempe, USA}
\author[0000-0003-2863-2707]{Alexander Fateev}
\affiliation{Technical University of Denmark, Department of Chemical and Biochemical, Denmark}
\author[0000-0002-3052-7116]{Elspeth Lee}

\affiliation{University of Bern, Center for Space and Habitability, Switzerland}
\author[0000-0003-1907-5910]{Kevin Heng}
\affiliation{University of Bern, Center for Space and Habitability, Switzerland}

\begin{abstract}
We present an update of the open-source photochemical kinetics code VULCAN (\cite{tsai17}; \url{https://github.com/exoclime/VULCAN}) to include C-H-N-O-S networks and photochemistry. Additional new features are advection transport, condensation, various boundary conditions, and temperature-dependent UV cross-sections. First, we validate our photochemical model for hot Jupiter atmospheres by performing an intercomparison of HD 189733b models between \cite{Moses11}, \cite{Venot12}, and VULCAN, to diagnose possible sources of discrepancy. Second, we set up a model of Jupiter extending from the deep troposphere to upper stratosphere to verify the kinetics for low temperature. Our model reproduces hydrocarbons consistent with observations, and the condensation scheme successfully predicts the locations of water and ammonia ice clouds. We show that vertical advection can regulate the local ammonia distribution in the deep atmosphere. Third, we validate the model for oxidizing atmospheres by simulating Earth and find agreement with observations. Last, VULCAN is applied to four representative cases of extrasolar giant planets: WASP-33b, HD 189733b, GJ 436b, and 51 Eridani b. We look into the effects of the C/O ratio and chemistry of titanium/vanadium species for WASP-33b; we revisit HD 189733b for the effects of sulfur and carbon condensation; the effects of internal heating and vertical mixing ($K_{\textrm{zz}}$) are explored for GJ 436b; we test updated planetary properties for 51 Eridani b with \ce{S8} condensates. We find sulfur can couple to carbon or nitrogen and impact other species such as hydrogen, methane, and ammonia. The observable features of the synthetic spectra and trends in the photochemical haze precursors are discussed for each case. 
\end{abstract}




\section{Introduction}
Understanding the chemical compositions has been a central aspect in atmospheric characterization for planets within and beyond the Solar System. Photochemical kinetics models establish the link between our knowledge of chemical reactions and various planetary processes (e.g., atmospheric dynamics, radiative transfer, outgassing process, etc.), providing a theoretical basis for interpreting observations and addressing habitability. 

Hot Jupiters are the first discovered and best characterized class of exoplanets. Transit and eclipse observations have made various initial detections of chemical species in their atmospheres such as Na, K, \ce{H2O}, \ce{CH4}, CO, \ce{CO2} \citep[e.g. see the review of ][]{Kreidberg2018}. An extreme class of exceedingly irradiated hot Jupiter around bright stars have equilibrium temperature higher than 2000 K. They are prime targets for emission observations, and recent high-resolution spectroscopic measurements reveal atomic and ionic features that make their atmospheres resemble low-mass stars \citep[e.g., ][]{Birkby2013,Brogi2014,Jens2018}. 

The majority of discovered exoplanets have sizes between Earth and Neptune. Their heavy elemental abundances (i.e. metallicity) can vary considerably, as often inferred by the water detection \citep[e.g., ][]{Wakeford2017,Chachan2019}. While \ce{CH4} is expected to be more abundant in cooler (T$_{\textrm{eq}}$ $\lesssim$ 1000 K) atmospheres, understanding how disequilibrium chemistry and other processes alter the \ce{CH4}/CO abundance ratio remains an ongoing task.

The direct imaging technique provides a complementary window to resolve young planets at a far orbit \citep[e.g., see the reviews of][]{Crossfield2015,Pueyo2018}). The new generation of instruments like GPI and SPHERE \citep{Chauvin2018} have identified a number of interesting young Jupiter analog. These young planets are self-luminous from their heat of formation and receive UV fluxes from the star at the same time, giving insights on the planet forming conditions outside the snow lines and the transition between planets and brown dwarfs.  

Across the various types of aforementioned planetary atmospheres, photochemical kinetics and atmospheric transport are the dominant mechanisms that control the major chemical abundances. Photodissociation occurs when molecules are split into reactive radicals by high-energy photons while atmospheric transport shapes the abundance distribution. Disequilibrium processes can drive abundances considerably away from the chemical equilibrium state and are best studied in chemical kinetics models. 

Kinetics models stem from simulating the atmospheric compositions in Solar System planets \citep[e.g., ][]{Kasting1979,Yung1984,Nair1994,Wilson2004,lavvas2008,Hu2012,Krasnopolsky2012}, which focus on photochemistry and radical reactions. The low temperature regime makes thermochemistry less relevant in most cases. \cite{Liang2003} first applied a photochemical kinetics model, Caltech/JPL KINETICS \citep{Allen1981}, to the hot Jupiter HD 209458b and identified the photochemical source of water for producing atomic H. However, some reaction rates in their study are extrapolated from measurements at low temperatures and not suitable for hot Jupiter conditions. \cite{Line2010} adopt the high-temperature rate coefficients for the major molecules and use the lower boundary to mimic mixing from the thermochemical equilibrium region. A new group of models incorporating kinetics data valid at high temperatures started to emerge since then. \cite{Zahnle09} reverse the reactions to ensure kinetics consistent with thermodynamic calculations and consider sulfur chemistry on hot Jupiters. \cite{Moses11} implement high-temperature reactions in KINETICS to model hot Jupiters HD 189733b and HD 209458b with detailed pathway analysis. \cite{Venot12} adopt the combustion mechanisms validated for industrial applications to model the same canonical hot Jupiters but find different quenching and photolysis profiles from \cite{Moses11}. \cite{Hobbs2021} recently extend \cite{Zahnle09} to include sulfur photochemistry and find the inclusion of sulfur can impact other non-sulfur species on HD 209458b and 51 Eridani b. As the discovery of diverse exoplanets progresses, more kinetics models have been applied to study a wide range of aspects, such as the compositional diversity within an atmospheric-grid framework \citep{Moses2013,Miguel2014,Karan2019}, atmospheric evolution with loss and/or outgassing processes \citep{Hu2015,Wordsworth2018,Lincowski2018}, prebiotic chemistry driven by high-energy radiation \citep{rimmer16,Rimmer2019}, and detectability of habitable planets \citep{Arney2019,bio_review}.


A number of recent attempts of atmospheric composition measurements are hindered by aerosol layers \citep{Kreidberg2014,wasp80b}. Aerosol particles are possibly ubiquitous, with diverse compositions \citep{Gao2020} including cloud particles formed from condensation or produced by photolysis at high altitudes. Microphysics models \citep{Helling2006,Lavvas2017,Yui2018,Gao2018b,Ohno2020} have investigated trends and properties of aerosols for various environments. One particularly interesting candidate of aerosols is the sulfur family, such as sulfuric clouds \citep{Hu2013,Misra2015,Loftus2019} in an oxidizing atmosphere or elemental sulfur in a reducing atmosphere \citep{Hu2013,Gao2017}. Photochemistry generally sets off the initial steps in the gas phase, then the condensable species can form particles when saturated in a broad range of altitudes \citep{Gao2017}. The relatively simple sulfur particles in \ce{H2}-dominated atmospheres allow a consistent photochemical-aerosol kinetics modeling, which we will conduct in this work. Although the formation pathways of organic haze particles are highly complex, we will focus on a group of haze precursors and investigate their photochemical stability in the hope of providing complementary insights on the haze-forming conditions.

The exclusive access to often proprietary chemical models motivates us to develop an open-source, chemical kinetics code VULCAN \citep{tsai17}. The initial version of VULCAN includes a reduced-size C-H-O thermochemical network and treats eddy diffusion. In \cite{tsai17}, VULCAN is validated by comparing the quench behavior with A{\footnotesize RGO} \citep{rimmer16} and \cite{Moses11}. Since then, VULCAN has been continuously updated and applied to several studies such as \cite{Zilinskas2020} who identify key molecules of hot super-Earths with nitrogen-dominated atmospheres, and \cite{Shulyak2020} who explore the effects of XUV for different stellar types. 

In this work, we present the new version of 1-D photochemical model VULCAN, with embedded chemical networks now including hydrogen, oxygen, carbon, nitrogen, and sulfur. The chemical network is customizable and does not require separating fast and slow species. The major updates of VULCAN from \cite{tsai17} are:
\renewcommand\labelitemi{\tiny$\bullet$}
\begin{itemize}

\item C-H-N-O-S chemical networks with about 100 species, including a simplified benzene forming mechanism

\item Photochemistry with options for temperature-dependent UV cross sections input

\item Condensation and particle settling included

\item Advection, eddy diffusion, and molecular diffusion included for the transport processes
 
\item Choice of various boundary conditions

\end{itemize}

In Section \ref{model}, we describe model details that have been updated since \cite{tsai17}. In Section \ref{validation}, we validate photochemistry and various new features of VULCAN with simulations of HD 189733b, Jupiter, and Earth. A comprehensive model comparison for HD 189733b between \cite{Moses11}, \cite{Venot12}, and VULCAN is given. In Section \ref{case}, we perform case studies with focus on the effects of sulfur chemistry and haze precursors. We discuss caveats, implications and opportunities for future work in Section \ref{discussion} and summarize the highlights in Section \ref{sec:summary}.

\section{Kinetics model}\label{model}
\subsection{Basic Equations and Numerics}
The 1D photochemical kinetics model solves a set of Eulerian continuity equations,
\begin{equation}
\frac{\partial n_i}{\partial t} = {\cal P}_i - {\cal L}_i - \frac{\partial \phi_i}{\partial z},
\label{eq:master}
\end{equation}
where $n_i$ is the number density (cm$^{-3}$)  of species $i$ and $t$ denotes the time.  ${\cal P}_i$ and ${\cal L}_i$ are the production and loss rates (cm$^{-3}$ s$^{-1}$) of species $i$, from both thermochemical and photochemical reactions. The system of (\ref{eq:master}) has the same form as that in \cite{tsai17}, except only eddy diffusion is considered for the transport flux $\phi_i$ in \cite{tsai17}. The transport flux including advection, eddy diffusion, molecular and thermal diffusion while assuming hydrostatic balance is now written as \citep[e.g., ][]{topa87}
\begin{equation}
\phi_i = n_i \, v -K_{\rm zz} n_{\rm tot} \frac{\partial X_i}{\partial z} -D_i[\frac{\partial n_i}{\partial z} + n_i(\frac{1}{H_i} + \frac{1+\alpha_T}{T}\frac{dT}{dz})],
\label{eq:flux}
\end{equation}
where $v$ is the vertical wind velocity, $K_{\rm zz}$ and $D_i$ are the eddy diffusion and molecular diffusion coefficient, respectively, $H_i$ is the molecular scale height for species $i$ with molecular mass $m_i$ , i.e. $H_i$ = $\frac{m_i g}{k_BT}$ ($g$: gravity; $T$: temperature; $k_B$: the Boltzmann constant ), and $\alpha_T$ is the thermal diffusion factor. While advection is commonly ignored in 1-D models, we keep the advection term and distinguish it from eddy diffusion with respect to their intrinsic differences. For example, a plume of smoke transports the initial abundance along the direction of wind until diffusion becomes important and dissipates the smoke to the surrounding air.

Physically, the first term of the transport flux (\ref{eq:flux}) describes advection in the direction of the wind. The second term is eddy diffusion that acts to smear out the compositional gradient. The molecular diffusion in the third term becomes important at low pressure and drives each constituent toward diffusive equilibrium, which is different for each species based on its individual scale height. The direction of thermal diffusion depends on the sign of the thermal diffusion factor. Positive sign means the component will diffuse toward colder region and vice versa. Thermal diffusion is often a secondary effect compared to eddy diffusion or molecular diffusion, except for the light species in the thermosphere with large temperature gradients \citep{Nicolet1968}. The molecular diffusion coefficient has the expression of $b/N$ from the gas kinetic, where $b$ is a parameter for binary gas mixtures. The binary parameter $b$ and the thermal diffusion factor $\alpha_T$ are ideally determined experimentally for each binary mixture. In practice, we simplify the atmosphere to a binary system with the dominant gas as the main constituent and the rest in turn as the minor constituent. Specifically, we adopt the molecular diffusion coefficient of a binary mixture that is available from the experimental data and scale that of other mixtures based on the fact that $b$ is proportional to the mean relative speed of two gases, i.e. given $D_{1-2}$ for the dominant gas 1 and minor gas 2, the molecular diffusion coefficient for gas 1 and any other minor gas $i$ can be scaled as    
\begin{equation}\label{eq:D_scale}
D_{1-i} = D_{1-2} \sqrt{m_2/m_i ((m_1 + m_i)/(m_1 + m_2))}.
\end{equation}
The molecular diffusion coefficient and the thermal diffusion factor for atmospheres dominated by \ce{H2}, \ce{N2}, and \ce{CO2} are listed in Appendix \ref{app:Dzz}. 

A second-ordered central difference is used to discretize the spatial derivative of diffusion flux, as in \cite{tsai17}, except a first-order upwind scheme \citep{Jacob2017} is applied for advection. The finite difference form for the derivative of the transport flux of layer $j$ is
\begin{equation}
\frac{\phi_{i,j+1/2} - \phi_{i,j-1/2}}{\Delta z_j}
\label{diff_flux}
\end{equation}
, with the upper and lower interfaces of layer $j$ labeled as $j+1/2$ and $j-1/2$, respectively, in the staggered structure. The full expression for the transport flux in Equation (\ref{eq:flux}) at the upper and lower interfaces is then
\begin{equation}
\begin{split}
&\phi_{i,j+1/2} = \phi^{adv}_{i,j+1/2}   - (K_{{\rm zz},j+1/2} + D_{i,j+1/2}) n_{{\rm tot},j+1/2}\\
&\times \frac{X_{i,j+1} - X_{i,j}}{\Delta z_{j+1/2}} - D_{j+1/2}X_{i,j+1/2}(\frac{1}{H_i} - \frac{1}{H_0} + \\
&\frac{\alpha_T}{T_{j+1/2}} \frac{T_{j+1} - T_j}{\Delta z_{j+1/2}})\\
&\phi_{i,j-1/2} = \phi^{adv}_{i,j-1/2}   - (K_{{\rm zz},j-1/2} + D_{i,j-1/2}) n_{{\rm tot},j-1/2}\\
&\times \frac{X_{i,j} - X_{i,j-1}}{\Delta z_{j-1/2}} - D_{j-1/2}X_{i,j-1/2}(\frac{1}{H_i} - \frac{1}{H_0} + \\
&\frac{\alpha_T}{T_{j-1/2}} \frac{T_{j} - T_{j-1}}{\Delta z_{j-1/2}})\\
&\phi^{adv}_{i,j+1/2} = \begin{cases}
      v_{j+1/2} n_{i,j}, & \text{for } v_{j+1/2} > 0 \\
      v_{j+1/2} n_{i,j+1}, & \text{for } v_{j+1/2} < 0
\end{cases}\\
&\phi^{adv}_{i,j-1/2} = \begin{cases}
      v_{j-1/2} n_{i,j-1}, & \text{for } v_{j-1/2} > 0 \\
      v_{j-1/2} n_{i,j}, & \text{for } v_{j-1/2} < 0
\end{cases}
\end{split}
\label{eq:flux2}
\end{equation}
where $H_0$ is the atmospheric scale height with altitude dependent gravity and we have approximated the physical quantities at the interface by the average of two adjacent layers
$n_{{\rm tot},j \pm1/2} =  \frac{n_{{\rm tot},j} + n_{{\rm tot},j \pm1}}{2}$, $X_{i,j\pm1/2}$ =  $\frac{X_{i,j} + X_{i,j\pm1}}{2}$, and $T_{{\rm tot},j \pm1/2} =  \frac{T_{{\rm tot},j} + T_{{\rm tot},j \pm1}}{2}$. The advection flux $\phi^{adv}$ in Equation (\ref{eq:flux2}) only depends on the property of the upstream layer in the upwind scheme. Equation (\ref{eq:master}) can be reduced to a system of ordinary differential equations (ODEs) after replacing the spatial derivative of transport flux in Equation (\ref{eq:master}) with (\ref{diff_flux}) and (\ref{eq:flux2}) and assigning proper boundary conditions. The numerical scheme using the Rosenbrock method to integrate the ``stiff'' system (\ref{eq:master}) forward in time until steady state is achieved is described in detail in \cite{tsai17}. 

\subsection{Boundary Conditions}
The solutions to the system of ODEs derived from Equation (\ref{eq:master}) need to satisfy the given boundary conditions. The boundary conditions encompass various planetary processes that are crucial in regulating the atmosphere. There are three basic quantities commonly used to describe the boundary conditions \citep[e.g.][]{Hu2012}: flux, velocity, and mixing ratio. We will elucidate their corresponding implications for the lower and upper boundaries. 

The flux term in Equation (\ref{eq:flux2}) depends on the layers above and below. Hence the fluxes at the top and bottom are unspecified. Assigning constant fluxes is common to represent surface emission at the lower boundary for rocky planets and inflow/outflow at the upper boundary.
For example, CO and \ce{CH4} surface sources play a key role to Earth's troposphere; meteoritic inflow or hydrodynamic escape outflow can be prescribed as constant flux at the upper boundary \citep[e.g., ][]{Wordsworth2018}. Alternatively, diffusion-limited flux can be assigned at the upper boundary, which assumes the escape flux is limited by the diffusion transport into exosphere. The diffusion-limited flux reads 
\begin{equation}
\phi_{i,\textrm{top}} = - D_{i,\textrm{top}} n_i (\frac{1}{H_i} - \frac{1}{H_0})
\end{equation}
and can be applied to any set of light species in the code. Without additional constraints, we often simply assume the flux to be zero, which means no net material exchange. This zero-flux boundary condition is generally suited for the lower boundary conditions while placed at a sufficient depth of most gas giants \citep{Moses11,rimmer16,tsai17}. While not specifying the boundary condition, zero flux is implied as default in VULCAN.   

In addition to the flux, velocity is useful to represent sources and sinks that scale with the species abundance. For example,  (dry/wet) deposition velocity is conventionally used to parametrize removal processes such as gas absorption or uptake into the surface \citep{Hu2012,Seinfeld2016}. At the upper boundary, upward velocity can be assigned to account for escape velocity or for any process producing inflow/outflow \citep{Krasnopolsky2012}. The flux and velocity can also be assigned together to describe the final boundary condition of a single species.

Constant mixing ratios are prescribed for the boundary condition when the detail exchange is complex but the knowledge of precise abundance is available. For example, the water vapor at the surface is expected to be set by saturation according to relative humidity on an ocean planet with a substantial reservoir of water. Assigning constant mixing ratios is also practical for regional models, such as the composition around the cloud layers for the Venus model with lower boundary placed at the cloud layer \citep{Krasnopolsky2012}. Since constant mixing ratio does not allow changes of the composition at the boundary, this boundary condition should not be used in conjunction with flux or velocity boundary conditions.  

\subsection{Chemical Networks}\label{sec:network}
We have extended the previous C-H-O network in \citep{tsai17} to include nitrogen and sulfur in a hierarchical manner, e.g.,  C-H-O\footnote{We have updated C-H-O network from \citep{tsai17} by adding \ce{HO2} and \ce{H2O2}.}, C-H-N-O, C-H-N-O-S networks. Each network is provided with a reduced version and a full version, where ``reduced" is referred to both oxidation state and network size. The reduced version has species and mechanisms (e.g., the ozone cycle) that are only important in oxidizing conditions stripped off, which are more computationally efficient and suited for the general hydrogen-dominated atmospheres. The full version of networks are designed for a wide range of main atmospheric constituents, from reducing to oxidizing. Hydrocarbon species are truncated at two carbons, while some higher-order hydrocarbons are present as necessary sinks for the two-carbon species or hazy precursors. The chemical network files with rate coefficients for the forward reactions can be found at \url{https://github.com/exoclime/VULCAN/tree/master/atm}.  

The full version of C-H-N-O-S network includes 96 species: \ce{H}, \ce{H2}, \ce{O}, \ce{^1O}, \ce{O2}, \ce{O3}, \ce{OH}, \ce{H2O}, \ce{HO2}, \ce{H2O2}, \ce{CH}, \ce{C}, \ce{CH2}, \ce{^1CH2}, \ce{CH3}, \ce{CH4}, \ce{C2}, \ce{C2H2}, \ce{C2H}, \ce{C2H3}, \ce{C2H4}, \ce{C2H5}, \ce{C2H6}, \ce{C4H2}, \ce{C3H3}, \ce{C3H2}, \ce{C3H4}, \ce{C6H5}, \ce{C6H6}, \ce{C4H3}, \ce{C4H5}, \ce{CO}, \ce{CO2}, \ce{CH2OH}, \ce{HCO}, \ce{H2CO}, \ce{CH3O}, \ce{CH3OH}, \ce{CH3CO}, \ce{H2CCO}, \ce{HCCO}, \ce{CH3O2},  \ce{CH3OOH}, \ce{N}, \ce{N(^2D)}, \ce{N2}, \ce{NH}, \ce{CN}, \ce{HCN}, \ce{NH2},  \ce{NH3}, \ce{NO}, \ce{N2H2}, \ce{N2H}, \ce{N2H3}, \ce{N2H4}, \ce{HNO}, \ce{H2CN}, \ce{HC3N}, \ce{CH3CN}, \ce{CH2CN}, \ce{C2H3CN}, \ce{HNCO}, \ce{NO2}, \ce{N2O}, \ce{CH2NH2}, \ce{CH2NH}, \ce{CH3NH2}, \ce{CH3CHO},  \ce{NO3}, \ce{HNO3}, \ce{HNO2}, \ce{NCO}, \ce{N2O5}, \ce{S}, \ce{S2}, \ce{S3}, \ce{S4}, \ce{S8}, \ce{SH}, \ce{H2S}, \ce{HS2}, \ce{SO}, \ce{SO2}, \ce{SO3}, \ce{CS}, \ce{OCS}, \ce{CS2}, \ce{NS}, \ce{HCS}, \ce{HSO}, \ce{HSO3}, \ce{H2SO4}, \ce{CH3S}, \ce{CH3SH},  \ce{S2O} and about 570 forward thermochemical reactions and 69 photodissociation branches. All thermochemical reactions are reversed using the equilibrium constant derived from the NASA polynomials as described in \cite{tsai17} to ensure chemical equilibrium can be kinetically achieved\footnote{We report a significant discrepancy in the new NASA 9-polynomials of \ce{CH2NH} (\url{http://garfield.chem.elte.hu/Burcat/NEWNASA.TXT}) compared to the early NASA 7-polynomials and other sources, which can lead to several orders of magnitude errors. We use the fit from the NASA 7-polynomials for \ce{CH2NH} instead.}. We also provide an option for customizing modular networks. A subgroup of species can be freely picked and only reactions that involve the selected species will form a new modular chemical network. Unlike minimizing Gibbs free energy for equilibrium chemistry, caution is required in this process to incorporate trace species that are important intermediates to set up a sensible network.

We have incorporated a simplified benzene mechanism into the generally two-carbon based kinetics, with the motivation of considering it in the context of haze precursors, as will be discussed in Section \ref{sec:haze}. The intention is to capture the main formation pathways at minimum cost in terms of the size of the network. We adopt one of the possible benzene forming pathways through propargyl (\ce{C3H3}) recombination \ce{C3H3 + C3H3 ->[\textrm{M}] C6H6} \citep{Frenklach2002}, whereas \ce{C3H3} is produced by \ce{CH3 + C2H -> C3H3 + H}. We then add hydrocarbons such as \ce{C3H2}, \ce{C3H4}, and \ce{C6H5} for the hydrogen abstraction reactions of \ce{C3H3} and \ce{C6H6} to complete the mechanism.

The rate coefficients of the reactions are broadly drawn from the following: (1) NIST database\footnote{\url{https://kinetics.nist.gov}}  (2) KIDA database\footnote{\url{http://kida.obs.u-bordeaux1.fr/}} (3) literature sources including \cite{Moses2005,lavvas2008,Moses11,Zahnle2016}. Although most rate coefficients are chosen to be validated for a wide range as possible (300 - 2500 K), some of the rate coefficients are still only measured at limited temperature ranges, which has been a long standing issue in kinetics. The kinetics becomes even more uncertain while sulfur is involved. For example, elemental sulfur in the gas phase exists in many allotropic forms but the chain-forming reactions between the allotropes were poorly constrained. The recombination rates of S that form the first sulfur bond \ce{S + S ->[\textrm{M}] S2} from two early measurements \cite{Fair1969} and \cite{Nicholas1979} are differed by four orders of magnitude. A recent calculation by \cite{Du2008} confirms the value by \cite{Fair1969} and we adopt the rate coefficient from \cite{Du2008} in our network. To address the uncertainties in sulfur kinetics, we perform sensitivity tests for selective key reactions in Section \ref{case}.  

\subsection{Computing Photochemistry}
Stars are the ultimate energy source of disequilibrium chemistry. The stellar radiation interacting with the atmosphere can be converted into internal energy or initiates chemical reactions. Photodissociation describes the process in which energetic photons break molecules apart, schematically written as an unimolecular reaction with photons (h$\nu$)
\begin{equation}\label{re:photolysis}
\ce{A ->[h$\nu$] B + C}.
\end{equation}
Photodissociation typically produces active free radicals and initiate a chain of reactions that are essential to atmospheric chemistry (e.g., the ozone cycle on Earth or the organic haze formation on Titan).

The radiative flux that drives photolysis is conventionally defined by the number of photons {\it from all directions} per unit time per unit area per unit wavelength and referred as the actinic flux, $J(z,\lambda)$, with $z$ being altitude $\lambda$ being wavelength.

$J(z,\lambda)$ consists of two components, direct beam and diffuse radiation:  
\begin{equation}
J(z,\lambda) =  J(\infty,\lambda)e^{-\tau(z,\lambda)/\mu} + J_{\textrm{diff}}(z,\lambda).
\label{eq:sflux}
\end{equation}
where $\tau$ is the optical depth and $\mu$ = cos$\theta$ with $\theta$ being the zenith angle of the incident beam. The first term of Equation(\ref{eq:sflux}) describes the attenuated actinic flux reaching the plane perpendicular to the direction of beam (there is no cosine pre-factor as for radiative heating since the number of intercepted molecules is randomly oriented and independent of the direction of the stellar beam). 

The optical depth $\tau$ accounts for the extinction from both absorption and scattering is calculated as 
\begin{equation}
\tau =  \int [\Sigma_i (\sigma_{a,i} + \sigma_{s,i}) n_i] dz
\label{eq:tau}
\end{equation}
where $\sigma_{a,i}$ and $\sigma_{s,i}$ are the cross section of absorption and scattering, respectively. The absorption cross section $\sigma_{a,i}$ can be different from the photodissociation cross section because absorption is not necessarily followed by dissociation. The diffusive flux $J_{\textrm{diff}}$ is the scattered radiation defined by integrating the diffuse specific intensity over all directions.
We use the two-stream approximation in \cite{Malik2019} to first solve for the diffuse flux and convert it to total intensity using the first Eddington coefficient \citep{Heng2018}: 
\begin{equation}
J_{\textrm{diff}}(z,\lambda) = F_{\textrm{diff}} / \epsilon 
\label{eq:}
\end{equation}
where $F_{\textrm{diff}}$ is the total diffuse flux given by $F_{\textrm{diff}}$ $\equiv$ $F_{\uparrow}^{\textrm{diff}} + F_{\downarrow}^{\textrm{diff}}$ and $\epsilon$ is the first Eddington coefficient with value 0.5 for isotropic flux. Although multiple scattering is not explicitly included in the expression in \cite{Malik2019}, the process can be approached through iteration and we find the equilibrium state of multiple scattering can normally be achieved within 200 iterations for a strongly irradiated hot Jupiter. In the code, we have the option to update the actinic flux periodically to save computing time.  


Once the actinic flux has been obtained, the photolysis rate coefficient can be determined from integrating the actinic flux and the absorption cross section over the wavelength
\begin{equation}
k =  \int_{\lambda} q(\lambda) \sigma_a(\lambda) J(z,\lambda) d\lambda.
\label{eq:photo_rate}
\end{equation}
and the photolysis rate of Reaction (\ref{re:photolysis}) is
\begin{equation}
\frac{d n_{\textrm{A}} }{dt} = - k n_{\textrm{A}} 
\end{equation}
, where $q(\lambda)$ is the quantum yield (photons$^{-1}$), describing the probability of triggering a photolysis branch for each absorbed photon. In VULCAN, we adopt the cross sections from the Leiden Observatory database\footnote{\url{http://home.strw.leidenuniv.nl/~ewine/photo}} \citep{Heays2017} whenever possible, which provides tabulated data of photoabsorption, photodissociation, and photoionisation cross sections with uncertainty ranking. The data has been benchmarked against other established databases such as the PHIDRATES database\footnote{\url{http://phidrates.space.swri.edu}} \citep{Huebner1992,Huebner2015} which is detailed in \cite{Heays2017}. The full list of photolysis reactions and references are listed in Table \ref{tab:photo_rates}.

The spectral resolution with respect to the stellar flux and cross sections can be important while computing Equation (\ref{eq:photo_rate}) numerically. The minimum resolution used in the model should be capable of resolving the line structures in the stellar spectra and cross sections. We discuss the errors from under-resolving in Appendix \ref{app:resolution}.

\subsection{Temperature-Dependent UV Cross Sections}\label{sec:Tcross}
Most laboratory measurements of UV cross sections are conducted at room temperature or lower, which might raise reliability issues with application to high-temperature atmospheres.  \cite{Heays2017} suggested that as temperatures increased by a few hundred K, the excitation of vibrational and rotational levels (limited to $v \leq 2$) in many cases only cause minor broadening of the cross sections and does not alter its wavelength integration. However, for molecules with prominent transition between excited vibrational states (e.g. \ce{CO2}), the temperature dependence on the cross section and photolysis rate can be important. 

Recent work has started to investigate the high-temperatures UV cross sections of a few molecules
 \citep{Venot2013,Venot2018}. Given the available data, we have included temperature-dependent photoabsorption cross sections of \ce{H2O} (EXOMOL\footnote{\url{http://www.exomol.com/data/data-types/xsec_VUV/}}), \ce{CO2} \citep{Venot2018} (with 1160 K from EXOMOL), \ce{NH3}(EXOMOL), \ce{O2} \citep{O2-lowT,O2-highT}, SH \citep{SH_cross}, \ce{H2S} \citep{SH_cross},  \ce{COS} \citep{SH_cross}, \ce{CS2} \citep{SH_cross} in the current version of VULCAN. The temperature dependence of the UV cross sections of these molecules can be found in Figure \ref{fig:cross_T}. It is evident that both the absorption threshold and cross sections of \ce{CO2} exhibit strong temperature dependence. For \ce{H2O}, we have incorporated the recent measurement for the cross section above 200 nm \citep{Ranjan20}. We follow \cite{Ranjan20} taking a log-linear fit for the noisy data above 216 nm. In addition, we have included measured data from \cite{Schulz2002} for temperature above 1500 K, .

A layer-by-layer interpolation for the temperature-dependent cross sections is implemented in the model, i.e. the cross section of one single species is allowed to vary across the atmosphere due to the temperature variation. The interpolation is linear in the temperature space and logarithmic in the cross-section space. With limited data, we find the linear interpolation in temperature generally underestimates the cross sections and therefore our implementation is considered as a conservative estimate for how photolysis increases with temperature.

\subsection{Condensation and rainout}
VULCAN handles condensation and evaporation using the growth rate of particles, assuming sufficient activated nuclei. For a schematic condensation/evaporation reaction
\begin{equation}
\ce{A_{(gas)} <-> A_{(particle)}},
\end{equation}
the reaction rate is given by the mass balance equation \citep{Seinfeld2016}
\begin{equation}
\frac{d n_{\textrm{A}} }{dt} = - \frac{D_{\ce{A}} m_{\ce{A}}}{\rho_p r_p^2} (n_{\ce{A}} - n^{\textrm{sat}}_{\ce{A}}) n_{\ce{A}}
\label{eq:conden-rate}  
\end{equation}
where $D_{\ce{A}}$ and $m_{\ce{A}}$ are the molecular diffusion coefficient and molecular mass of gas A, $\rho_p$ and $r_p$ are the density and radius of the particle, $n_{\ce{A}}$ and $n^{\textrm{sat}}_{\ce{A}}$ are the number density and saturation number density of gas A, respectively. Equation (\ref{eq:conden-rate}) describes the growth rate by diffusion for particles with size $r_p$ in the continuum regime (particles larger than the mean free path i.e. Knudsen number ($K_n$) smaller than 1). The negative value of Equation (\ref{eq:conden-rate}) corresponds to condensation when $n_{\ce{A}} > n^{\textrm{sat}}_{\ce{A}}$ and the positive value corresponds to evaporation when $n_{\ce{A}} < n^{\textrm{sat}}_{\ce{A}}$. Our condensation expression takes the same form as \cite{Hu2012,rimmer16}, except that the growth rate of particles in the kinetic regime (particles smaller than their mean free path i.e. Knudsen number ($K_n$) greater than 1) is used in \cite{Hu2012,rimmer16}. When applying $K_n = \frac{\pi \mu v_{th}}{4 P}$ where $\mu$ is the dynamic viscosity, $v_{th}$ the thermal velocity, and $P$ the pressure, a \ce{H2} atmosphere enters the kinetics regime with $K_n > 10$ above 1mbar for a temperature of 400 K and above 0.1 $\mu$bar for a temperature of 1000 K. We find that for most of the application, condensation occurs in the lower atmosphere with micron-size or larger particle and the continuum regime is more suitable. Since condensation typically operates in a relatively short timescale, we implement an option to switch off condensation and fix the abundances of condensing species and the condensates after the dynamic equilibrium has reached. The approach is similar to the quasi-steady-state assumption (QSSA) method, which decouples the fast and slow reactions to ease the computational load.

After the gas condenses to particles, they fall following the terminal settling velocity ($v_s$) derived from the Stoke's law \citep{Seinfeld2016} as
\begin{equation}
v_s = \frac{2}{9} \frac{\rho_p r_p^2 g}{\mu}
\label{eq:vs}
\end{equation} 
where $\mu$ is the atmospheric dynamic viscosity with value taken from \cite{Cloutman2000} for the corresponding background gas. We have again assumed large particle size to simplify the slip correction factor (the correction for non-continuum) to unity in Equation (\ref{eq:vs}). In this work, we have implemented and will demonstrate the condensation of \ce{H2O}, \ce{NH3}, \ce{S2}, and \ce{S8} in the following sections. 
\subsection{Chemistry of Ti and V Compounds}\label{sec:Ti}
TiO (titanium oxide) and VO (vanadium oxide) are present in the gas phase in cool stars and brown dwarfs where temperature exceeds 2000 K. The highly irradiated hot Jupiters have been suggested to 
manifest inverted temperature structures due to the strong optical absorption of TiO and VO vapor \citep{Hubeny03} in the stratosphere. The pioneering work proposing the role of TiO and VO in irradiated atmosphere \citep{Fortney08} is based on equilibrium chemistry, where the authors argue that the conversion between TiO and \ce{TiO2} is fast enough for TiO to remain in chemical equilibrium. However, it is not clear for conversion reactions with Ti or other titanium species. For example, the interconversion of \ce{CO <-> CO2} is relatively fast but the ultimate CO abundance is still controlled by the slower \ce{CO <-> CH4} interconversion. In addition to TiO, titanium hydride (TiH), has been suggested important in brown dwarfs by \cite{Burrows2005}. As the thermodynamics data of TiH is not available in the literature or standard databases, \cite{Burrows2005} perform ab initio calculations of the Gibbs free energy of TiH (based on the partition function obtained from the spectroscopic constants). To explore the kinetics of titanium and vanadium, we expand the species list to include Ti, TiO, \ce{TiO2}, TiH, TiC, TiN, V, VO. As only Ti, TiO, and TiO2 are available for titanium compounds in the NASA polynomials, we adopt the thermodynamics data of TiH from \cite{Burrows2005}, TiC from \cite{Woitke2018}, and the rest from \cite{tsuji73}.

While there are a few measurements for the reactions of titanium/vanadium species with laser vaporization at low temperature, the kinetics data at high temperature is nearly non-existent. 
As a first step, we perform simple estimates on the unknown rate constants of titanium/vanadium species. First, we look for kinetics data of analogous transition metals, such as Fe. We assume the same rate coefficient as the analogous reaction if it is measured at high temperature. When high-temperature data are not available, we estimate the temperature dependence based on transition state theory. For an endothermic reaction, we approximate the activation energy (the exponential term in the Arrhenius expression) by the enthalpy difference between the products and reactants, assuming the energy increase of the transition state is small compared to the enthalpy difference for reactions involving radicals \footnote{To verify our approach, we compared the activation energy estimated from the enthalpy difference to that of well measured reactions. e.g., endothermic reactions \ce{H2O + H -> OH + H2} and \ce{CO2 + H -> CO + OH} have activation energy 10800 K \citep{Davidson1989} and 13300 K \citep{Tsang1986}, respectively, whereas our estimate yields 7200 K and 10300 K, respectively.}. 
Once the activation energy is obtained, the pre-exponential factor is adjusted to fit the reference value at low temperature. The titanium/vanadium kinetics we adopted is listed in Table \ref{tab:tio}. For photolysis, we include photodissociation of TiO, \ce{TiO2}, TiH, TiC, and VO. We estimate their UV cross sections from FeO \citep{FeO2005} at 252.39 nm and scale the photolysis threshold according to their bond dissociation energy.

\subsection{Photochemical Hazy Precursors}\label{sec:haze} 
Observations have informed us that clouds or photochemical hazes are ubiquitous in a diverse range of planetary atmospheres. Microphysics models that  include processes such as nucleation, coagulation, condensation, and evaporation of particles \citep[e.g., ][]{Gao2018b,Yui2019,Lavvas2017} simulate the formation and distribution of various-size aerosol particles. Given the complexity and uncertainty of the polymerizing pathways, one common approach is to select precursor species as a proxy and assume they will further grow into complex hydrocarbons \citep{Morley2013,Yui2018}. Typical choices of haze precursors include \ce{C2H}$_x$ and HCN, which is also limited by our kinetics knowledge and computing capacity.

In this work, we preferentially consider precursors that are more closely related to forming polycyclic aromatic hydrocarbon (PAH) or nitriles. PAH is a group of complex hydrocarbon made of multiple aromatic rings, which has been commonly found in the smog pollution on Earth and expected to be associated with the organic haze on Titan \citep{Zhao2018}. In the polar region of Jupiter where charged particles are the main energy source, ionchemistry has also been suggested to promote the formation of PAHs and organic haze \citep{Wong2003}. Once the first aromatic ring, benzene, has formed, the thermodynamics state (enthalpy and entropy) does not vary much with the processes of attaching and arranging the rings. From the kinetics point of view, the classic mechanism of making complex hydrocarbons, H-Abstraction-Carbon-Addition (HACA), requires aromatic hydrocarbon and acetylene in the primary abstraction and addition steps \citep[e.g.][]{Frenklach2020}. It is conceivable that benzene formation is the rate-limiting step in forming complex hydrocarbons as the growth rate increases downstream from benzene. In practice, while the fundamental pathways leading to PAH remain elusive \citep{Wang2011,Zhao2018}, the combustion study can provide a good handle on the formation of benzene to a certain degree. Therefore, we suggest considering benzene as an important haze precursor.

One important caveat about modeling benzene is that its photodissociation branches are poorly quantified across various branches \citep[see e.g.][]{Lebonnois2005}. The main photolysis products are possibly phenyl radical (\ce{C6H5}) and benzyne radical (\ce{C6H4}) \citep{Suto1992}. If they further absorb photons again, they could fragment into smaller, linear molecules like \ce{C4H3} and \ce{C3H3}. We adopt the cross sections of \ce{C6H6} from \cite{Boechat04} and \cite{Capalbo16}. For simplicity, we assume the main dissociation of benzene primarily goes into phenyl radical (\ce{C6H5}) with a small fraction leading to \ce{C3H3} ($\sim$ 15$\%$ based on \citep{Kislov2004}). 

Although HCN is the basic molecule for nitrile chemistry, it is unlikely that most of HCN will convert into complex nitriles. The nitrile formation is more likely to be limited by the less abundant \ce{H2CN}, \ce{CH2NH}, or \ce{CH3CN}.  Hence we include these species along with \ce{HC3N} to represent the nitrile family precursor. For sulfur gases, in addition to the condensation of sulfur allotropes (S$_x$), we also consider \ce{CS2} according to the laboratory experiments by \cite{He2020}. Overall, we compose the following species as photochemical haze precursors: \ce{C2H2}, \ce{C2H6}, \ce{C4H2}, \ce{C6H6}, HCN, \ce{HC3N}, \ce{CH2NH}, \ce{CH3CN}, \ce{CS2}.

\section{Model Validation}\label{validation}

\begin{table*}
\begin{center}
\caption{Model Validation Setup}
\begin{tabular}{lllllll}
\hline
\hline
Planet & P-T profile & Network\footnote{files available in supplementary material} & stellar UV & Gravity\footnote{at the surface for Earth and defined at 1 bar for gaseous planet}& Upper Boundary & Lower Boundary\\
&&&&(cm$^2$/s)&\\
\hline
\hline
HD 189733b & \cite{Moses11} & N-C-H-O & Eps Eri\footnote{from the StarCat database (\url{https://casa.colorado.edu/~ayres/StarCAT}) \citep{Ayres2010} and following the same scaling adjustment as \cite{Moses11}} & 2140 & H escape\footnote{Assuming diffusion-limited escape rate} & zero-flux\\
\hline
\multirow{2}*{Jupiter} & \cite{Moses2005} & N-C-H-O-lowT & \cite{Gueymard2018} & 2479 & \ce{H2O}, CO, \ce{CO2} & zero-flux\\
 &+ dry adiabat& & & & inflow & \\
 \hline
Earth & COSPAR & S-N-C-H-O-full & \cite{Gueymard2018}& 980 & H, \ce{H2} escape & Table \ref{tab:BC_Earth}\\ 
\hline
\hline
\end{tabular}
\end{center}
\label{tab:validation}
\end{table*}

\subsection{HD 189733b}\label{sec:hd189}
We have benchmarked our thermochemical kinetics results using a C-H-O network with vertical transport against \cite{Moses11} for HD 189733b and HD 209458b in \cite{tsai17}. In this work, we compare our results including N-C-H-O photochemistry to \cite{Moses11} and \cite{Venot12} (M11 and V12 hereafter). V12 use a chemical kinetics scheme derived from combustion application and find different disequilibrium abundances of \ce{CH4} and \ce{NH3} from those in M11. Since then, a size-reduced network based on V11 has been developed \citep{Venot2019}, with the motivation to support computationally heavy simulations. In particular, the controversial methanol mechanism, which has been identified to cause the differences in \ce{CH4}-CO conversion \citep{Moses11,Moses2014}, is further updated and analyzed in \citep{Venot2020}. Therefore, aiming to consolidate the model discrepancy, we run an additional model with VULCAN but implemented with the updated reduced network from \cite{Venot2020}. The planetary parameters and model setting are listed in Table \ref{tab:validation}. Before diving into the detailed comparison, we provide an overview of the chemical profiles and absorption properties for HD 189733b and HD 209458b in Figure \ref{fig:HD189-209}.

\subsubsection{Disequilibrium Effects}
\begin{figure*}[tph]
\begin{center}
\includegraphics[width=\columnwidth]{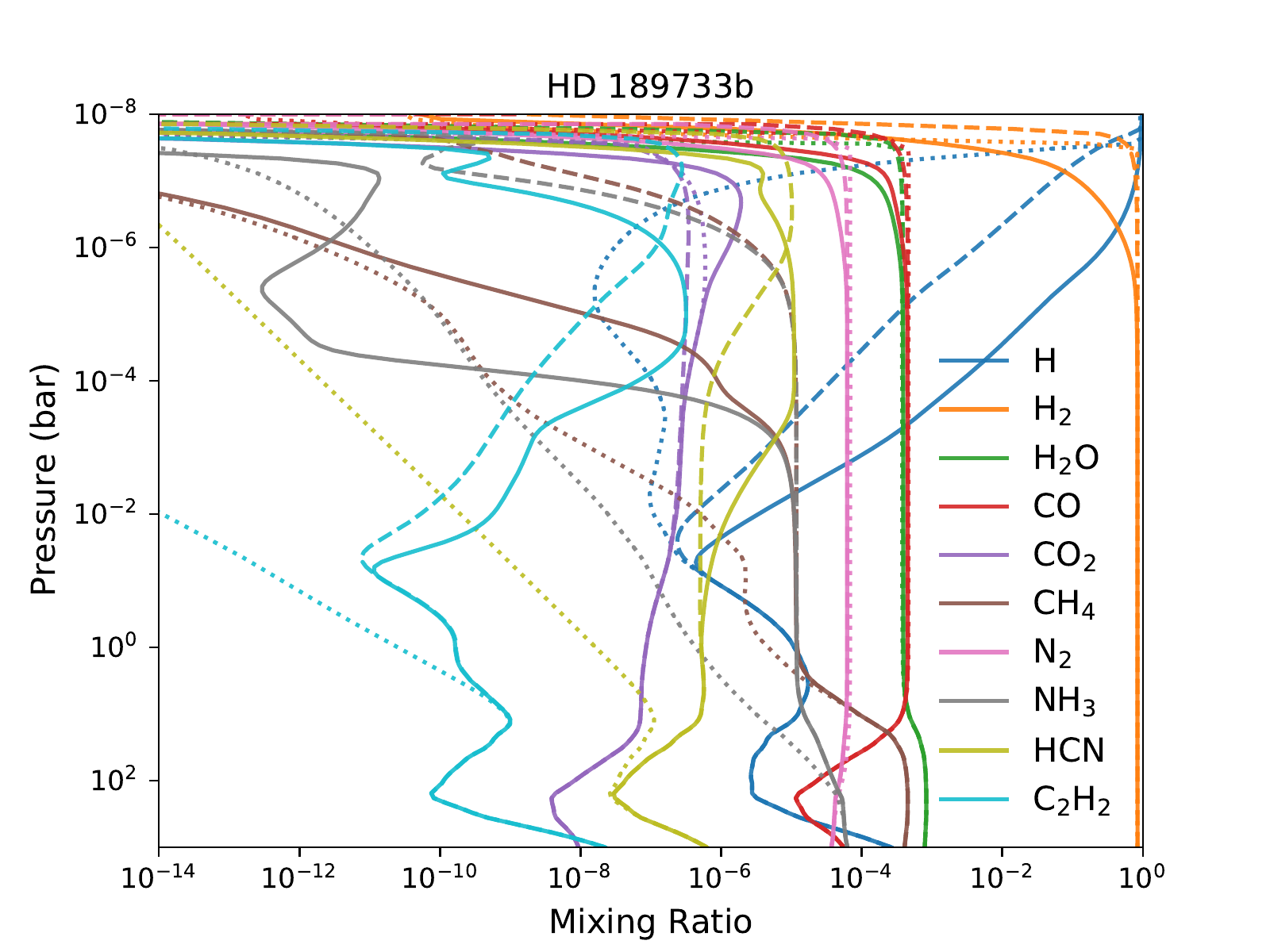}
\includegraphics[width=\columnwidth]{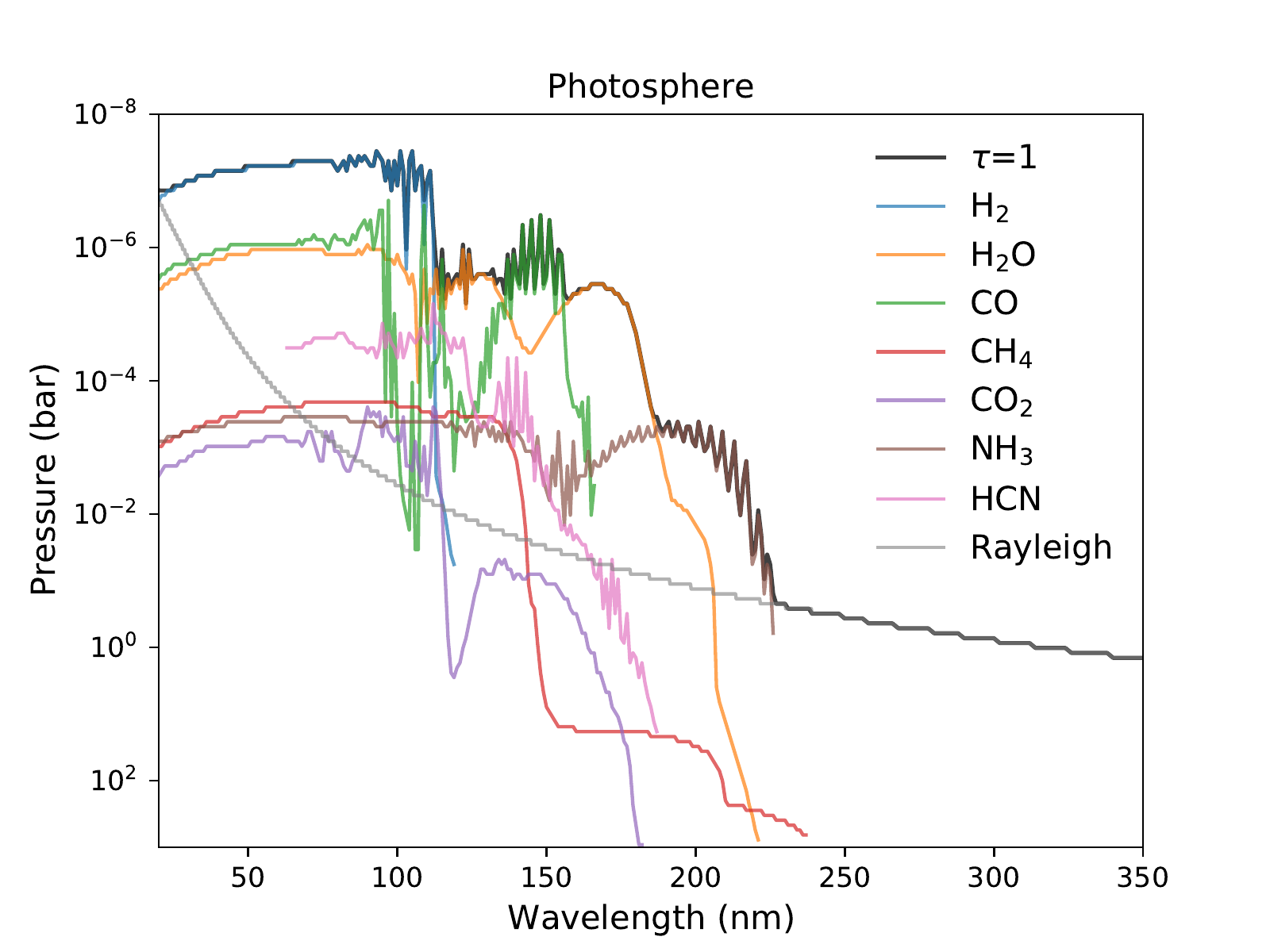}
\includegraphics[width=\columnwidth]{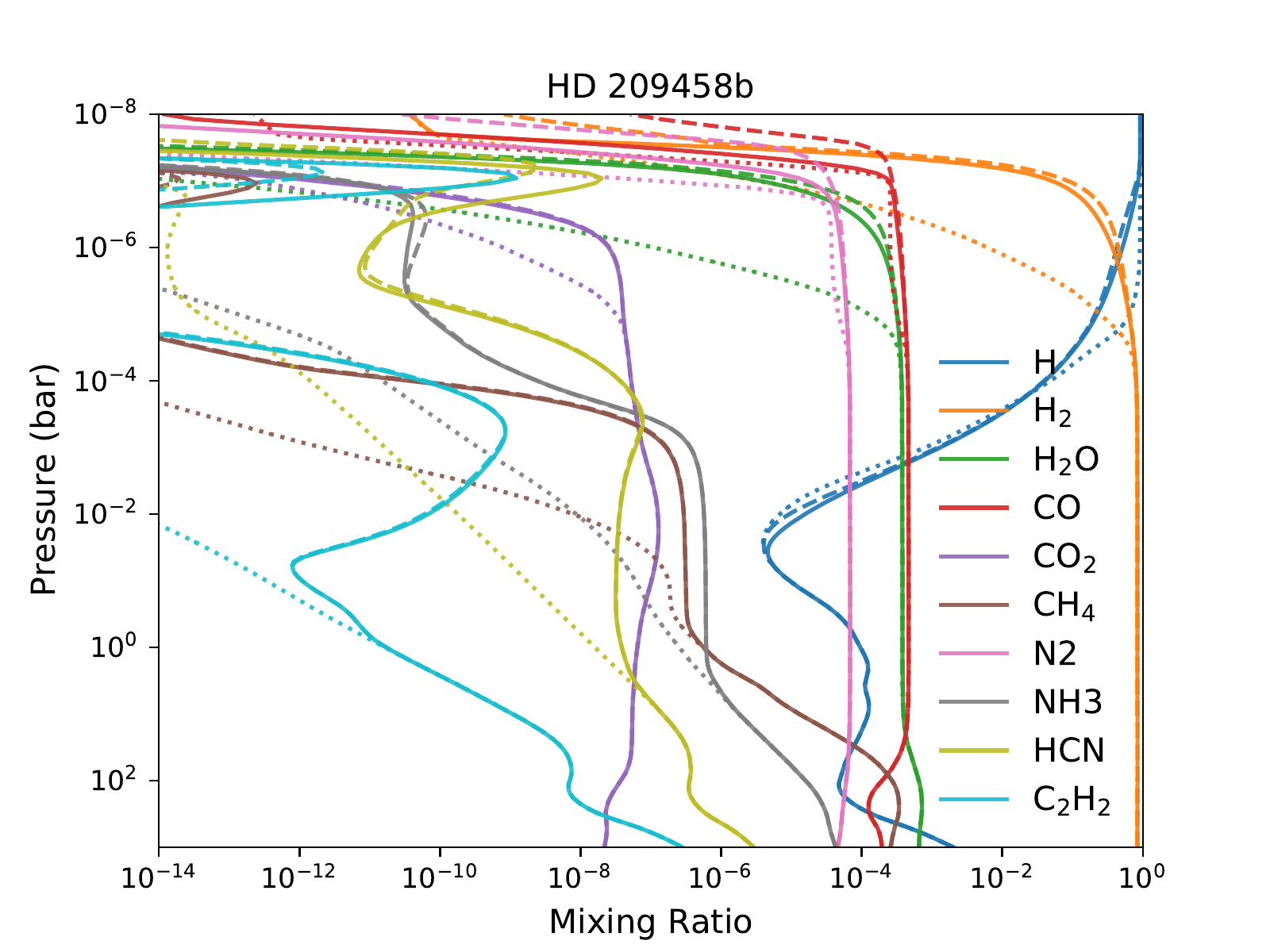}
\includegraphics[width=\columnwidth]{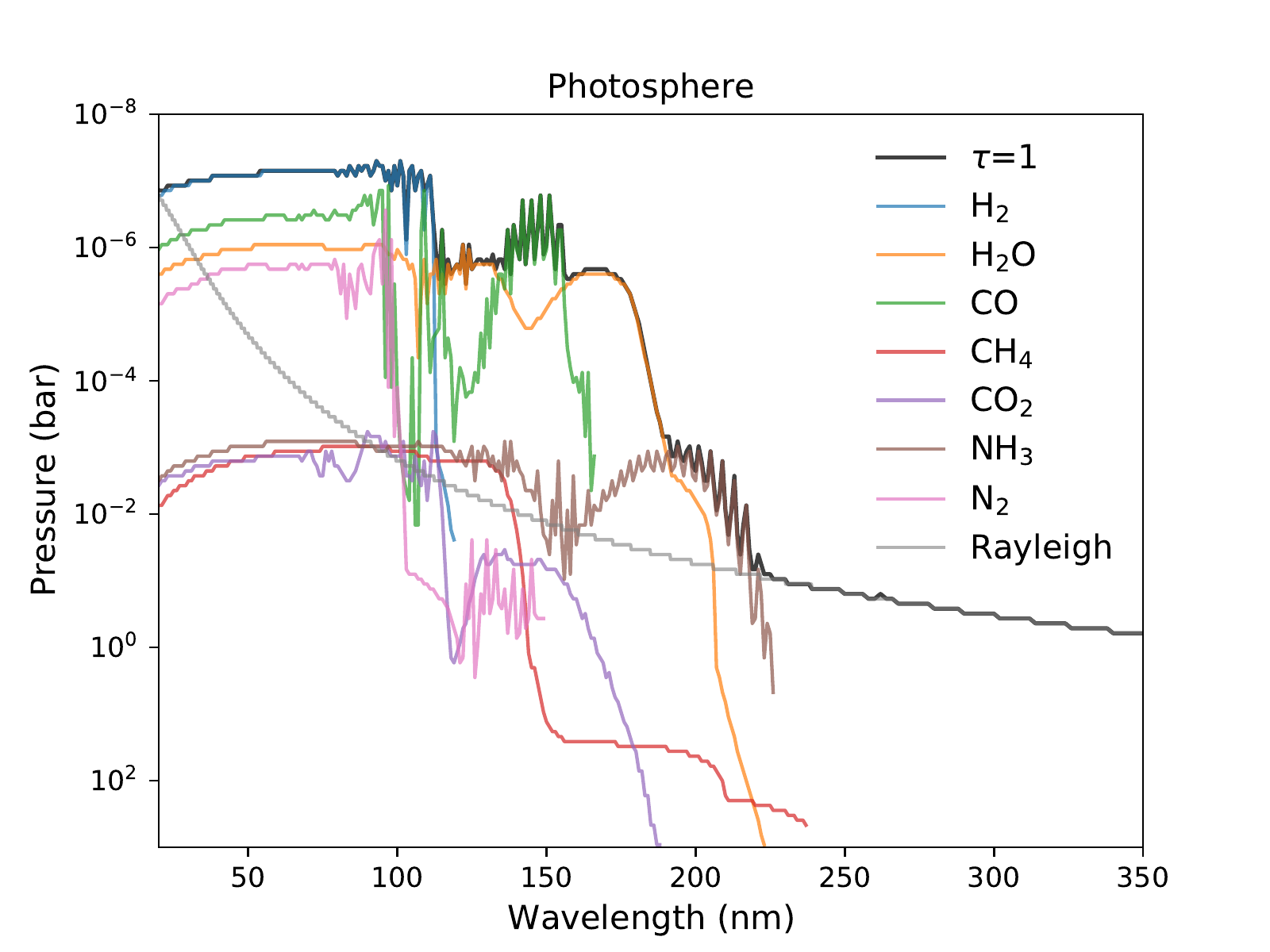}
\end{center}
\caption{C-H-N-O Photochemical kinetics results (top-left) of HD 189733b (solid), compared with including vertical mixing but no photochemistry (dashed), and thermochemical equilibrium (dotted). The temperature-pressure structure and eddy diffusion (K$_{zz}$) profile are taken from the dayside-average profile in \cite{Moses11} (their Figure 1 and 2). On the top right, we show the pressure level where energetic photons are mostly absorbed, i.e. optical depth $\tau$ = 1 (black), and decomposed into the main absorbers. The bottom panels show same as the top panels except for HD 209458 b.} 
\label{fig:HD189-209}
\end{figure*}

The left panels of Figure \ref{fig:HD189-209} depict how vertical mixing and photochemistry drives the compositions out of equilibrium on HD 189733b, by isolating the two effects. The underlying processes can be understood as a general property of hot Jupiters, as discussed in \citep{Moses11,Venot12,Moses2014,Hobbs19,Karan2019}. Equilibrium chemistry prevails in the deep, hot region whereas energetic photons dissociate molecules and produce reactive radicals in the upper atmosphere. Between the two regions, the composition distribution is controlled by vertical transport, viz., species in equilibrium at depth are transported upward and become quenched when vertical mixing predominates chemical reactions; photochemical products are also mixed downward and initiate a sequence of reactions.  

The right panels of Figure \ref{fig:HD189-209} show the UV photosphere where the optical depth equals one, with decomposition of contribution from the main molecules. Our photochemical model captures several general transmission properties of irradiated \ce{H2}-atmospheres: \ce{H2} provides the dominant absorption in EUV (10--120 nm) whereas \ce{H2O} and \ce{CO} are the dominant absorbers in FUV (120--200 nm). The window around 160--200 nm is particularly important for water dissociation, which makes a catalytic cycle turning \ce{H2} into atomic H \citep{Liang2003,Moses11}. In the NUV (300--400nm), radiation can penetrate deep down to $\sim$ 1 bar until being scattered. The photospheres in Figure \ref{fig:HD189-209} descend from about 1 $\mu$bar to 10 mbar (from the end of \ce{H2}-shielding to the tail of ammonia absorption) which denote the photochemically active region in the atmosphere.

HD 209458b shares qualitatively similar results with HD 189733b. Owning to its higher temperature and the inverted thermal structure (see Figure 1. in \cite{Moses11}), the quench level is lifted higher and the photolysis has little influence, as can be seen in Figure \ref{fig:HD189-209}. The composition distribution on HD 209458b can be described by a lower equilibrium region and an upper quenched region. We will now only focus on HD 189733b for the model comparison as disequilibrium processes contribute more compared to the hotter HD 209458b (see \cite{Hobbs19} for model comparison of HD 209458b).

\subsubsection{Model Comparison with \cite{Moses11} and \cite{Venot12}}
The HD 189733b model comparisons between VULCAN, M11, and V12 are showcased in Figure \ref{fig:HD189-VM}, where the top row highlights the major species and the following rows are grouped into carbon, oxygen, and nitrogen species. For the major species, VULCAN produces profiles more consistent with M11, while there are notable differences with V12 in H, \ce{CH4}, \ce{NH3}, and HCN. \ce{CH4} and \ce{NH3} are quenched from below 1 bar level until being attacked by H around 1 mbar. Hence the differences with V12 in the photospheric region ($\sim$ 1 bar -- 1 mbar) are due to thermochemical kinetics, rather than photochemical sources. Nitrogen species generally manifest higher variances, reflecting the kinetics uncertainties.   

\paragraph{Quenching of \ce{CH4} and \ce{NH3}}\mbox{}

The sharp gradients of the equilibrium distribution of \ce{CH4} and \ce{NH3} (Figure \ref{fig:HD189-209}) imply the abundances are sensitive to the quench levels, viz. small differences in the quench levels can lead to considerable differences. The key reactions responsible for the conversion at quench levels deserve a closer look.  

The match of quenched \ce{CH4} abundance between VULCAN and M11 has been discussed in \cite{tsai17}, in which we identify a similar pathway of \ce{CH4} destruction as M11. The inclusion of nitrogen does not change the fact since nitrogen does not participate in the \ce{CH4}-CO conversion. It can be seen that \ce{CH4} is quenched at a higher level with lower mixing ratio in V12, as a result of faster \ce{CH4}-CO conversion. \cite{Moses2014} identified the faster methanol decomposition \ce{H + CH3OH -> CH3 + H2O} measured by \cite{Hidaka1989} adopted in V12 as the key reaction that \ce{CH4} exhibits a shorter timescale in V12. \cite{Moses2014} suggested the rate is overestimated by \cite{Hidaka1989} based on the high energy barrier of the reaction. In response, \cite{Venot2019} removed the controversial reaction by \cite{Hidaka1989} and updated their chemical scheme with a newly validated \ce{CH3OH} combustion work \citep{Burke2016}, given the importance of methanol  as an intermediate species for \ce{CH4}-CO conversion. Intriguingly, \cite{Venot2019} still find a methane abundance rather close to that in V12. 

Attempting to resolve this mystery, we further run our model with the \cite{Venot2020} reduced scheme\footnote{The reduced scheme captures the key reactions at work from V12 and has been benchmarked against V12 \citep{Venot2019}. The two schemes are approximately equivalent regarding the quenching of main species.} integrated with new \ce{CH3OH} mechanism. We did not incorporate the same photolysis scheme from V12 but here photolysis has no effects on the quenching comparison below 1 bar. Contrary to the findings in \cite{Venot2020}, the new scheme of \cite{Venot2020} implemented in our model indeed shows a slower \ce{CH4}-CO conversion and brings the \ce{CH4} profile closer to VULCAN and M11 (dotted line in Figure \ref{fig:HD189-VM}-(b)). Our model implemented with the \cite{Venot2020} scheme predicts a quenched methane mixing ratio 1.13 $\times 10^{-5}$, close to 1.51 $\times 10^{-5}$ in M11 and 1.26 $\times 10^{-5}$ in our nominal model, whereas V12 with the faster methanol decomposition from \cite{Hidaka1989} predicts 5.20 $\times 10^{-6}$. We conclude that the methanol decomposition indeed results in faster \ce{CH4}-CO conversion and subsequently lowers the \ce{CH4} abundance in V12.  

For nitrogen chemistry, the high-temperature kinetics is more uncertain and many reducing  reactions relevant for \ce{H2}-atmospheres are not available on the NIST database. We drew data from the combustion literature (\cite{Dean2000}, same as M11) and the KIDA database. In particular, there are considerable uncertainties regarding the rates for the reactions that control the \ce{NH3}-\ce{N2} conversion, as extensively discussed in \cite{Moses2014}. We follow the suggestions in \cite{Moses2014} and adopt the rate coefficient of \ce{ NH3 + NH2 -> N2H3 + H2} from \cite{Dean1984} and that of \ce{ NH2 + NH2 -> N2H2 + H2} from \cite{Klippenstein2009}, since \cite{Konnov2001} used in V12 is measured at low temperatures.

As \ce{NH3} progressively become fully quenched in the region between a few hundreds bar and 1 bar, there are more than a single pathway and rate-limiting step for \ce{NH3}-\ce{N2} conversion that effectively control the \ce{NH3} abundance. For pressure greater than $\sim$ 30 bar, we identify the pathway
\begin{eqnarray}
\begin{aligned}
\ce{ NH3 + H &-> NH2 + H2}\\
\ce{ NH3 + NH2 &-> N2H3 + H2} \; (\textrm{i})\\
\ce{ N2H3 &->[\textrm{M}] N2H2 + H} \; (\textrm{ii})\\
\ce{ N2H2 + H &-> N2H + H2}\\
\ce{ N2H &->[\textrm{M}] N2 + H}\\
\noalign{\vglue 5pt} 
\hline
\noalign{\vglue 5pt}  
\mbox{net} : \ce{2NH3 &-> N2 + 3H2}.
\end{aligned}
\label{path-nh3-1}
\end{eqnarray}
where the rate-limiting step switches from (\ref{path-nh3-1})-(i) to (\ref{path-nh3-1})-(ii) with increasing pressure. In the region with pressure between 30 and 1 bar, we find two pathways with close contribution: 
\begin{eqnarray}
\begin{aligned} 
2(\ce{ NH3 + H &-> NH2 + H2})\\
\ce{ NH2 + H &-> NH + H2}\\
\ce{ NH + NH2 &-> N2H2 + H} \; (\textrm{iii})\\
\ce{ N2H2 + H &-> N2H + H2}\\
\ce{ N2H &->[\textrm{M}] N2 + H}\\
\ce{ H2 &->[\textrm{M}] 2H}\\
\noalign{\vglue 5pt} 
\hline %
\noalign{\vglue 5pt} 
\mbox{net} : \ce{2NH3 &-> N2 + 3H2}.
\end{aligned}
\label{path-nh3-2}
\end{eqnarray}
and
\begin{eqnarray}
\begin{aligned} 
\ce{ NH3 + H &-> NH2 + H2}\\
\ce{ NH2 + H &-> NH + H2}\\
\ce{ NH + H &-> N + H2}\\
\ce{ NH3 + N &-> N2H + H2}  \; (\textrm{iv})\\
\ce{ N2H &->[\textrm{M}] N2 + H}\\
\ce{ H2 &->[\textrm{M}] 2H}\\
\noalign{\vglue 5pt} 
\hline %
\noalign{\vglue 5pt} 
\mbox{net} : \ce{2NH3 &-> N2 + 3H2}.
\end{aligned}
\label{path-nh3-3}
\end{eqnarray}
where (\ref{path-nh3-2})-(iii) and (\ref{path-nh3-3})-(iv) are the rate-limiting steps. 

Our pathways (\ref{path-nh3-1}) and (\ref{path-nh3-2}) are identical to those in M11 ((5) and (6) in \cite{Moses11}), although we find (\ref{path-nh3-1})-(i) still play a role for controlling \ce{NH3} quenching, even with the high energy barrier given by \cite{Dean1984}. As we adopt the same rates for several key reaction relevant for \ce{NH3}-\ce{N2} conversion, our model reproduces \ce{NH3} very close to M11, whereas V12 with a faster \ce{NH3}-\ce{N2} conversion predicts a higher quench level and lower abundance for \ce{NH3} (Figure \ref{fig:HD189-VM}-(b)). In all, we reiterate that further investigation for the key reactions (e.g., (\ref{path-nh3-1})-(i), (\ref{path-nh3-1})-(ii), (\ref{path-nh3-2})-(iii), (\ref{path-nh3-3})-(iv)) at high temperatures are required to improve our ability to accurately model the \ce{NH3}-\ce{N2} system.


\begin{figure*}[!ht] 
\begin{center}
\includegraphics[width=\columnwidth]{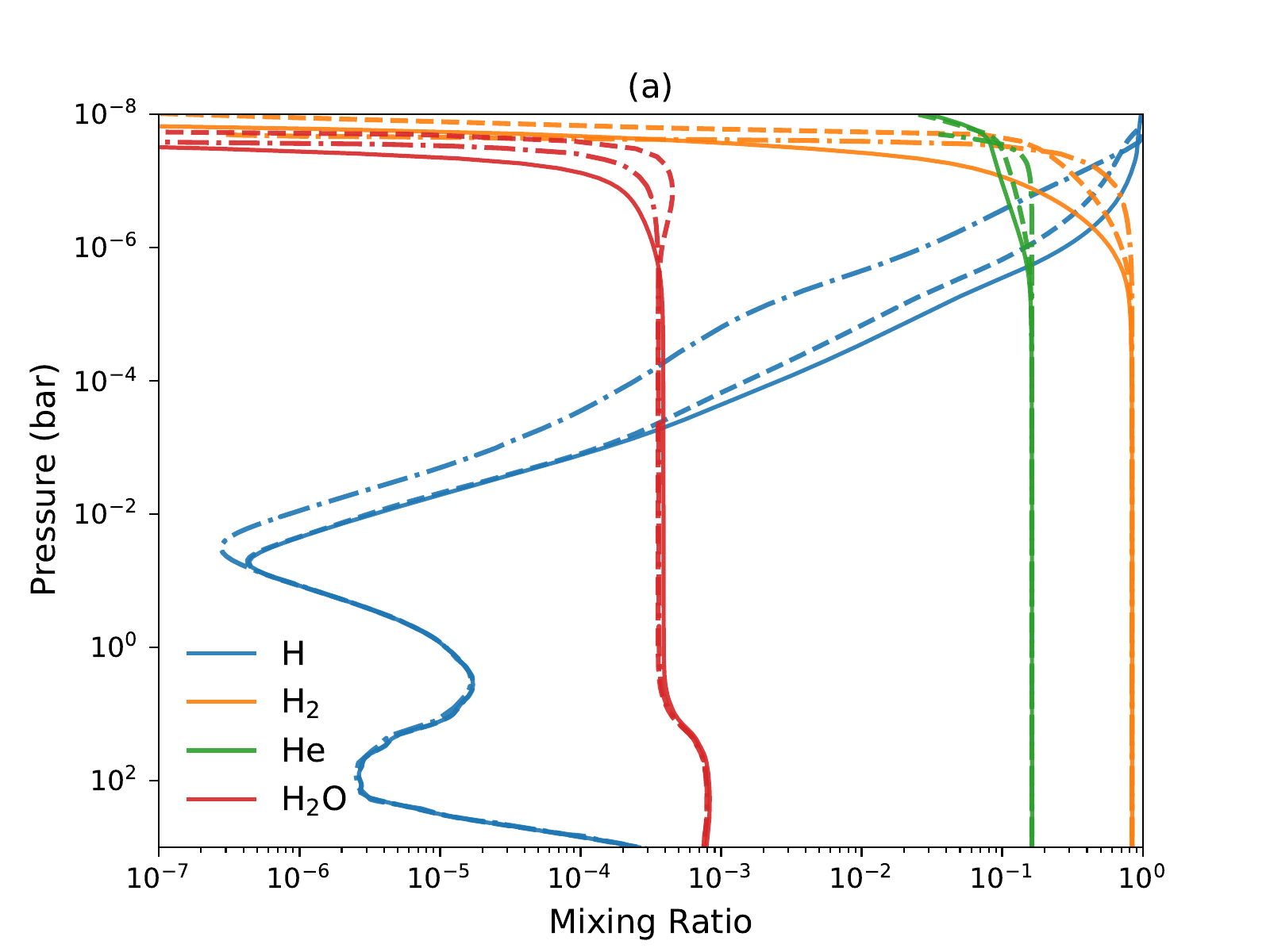}
\includegraphics[width=\columnwidth]{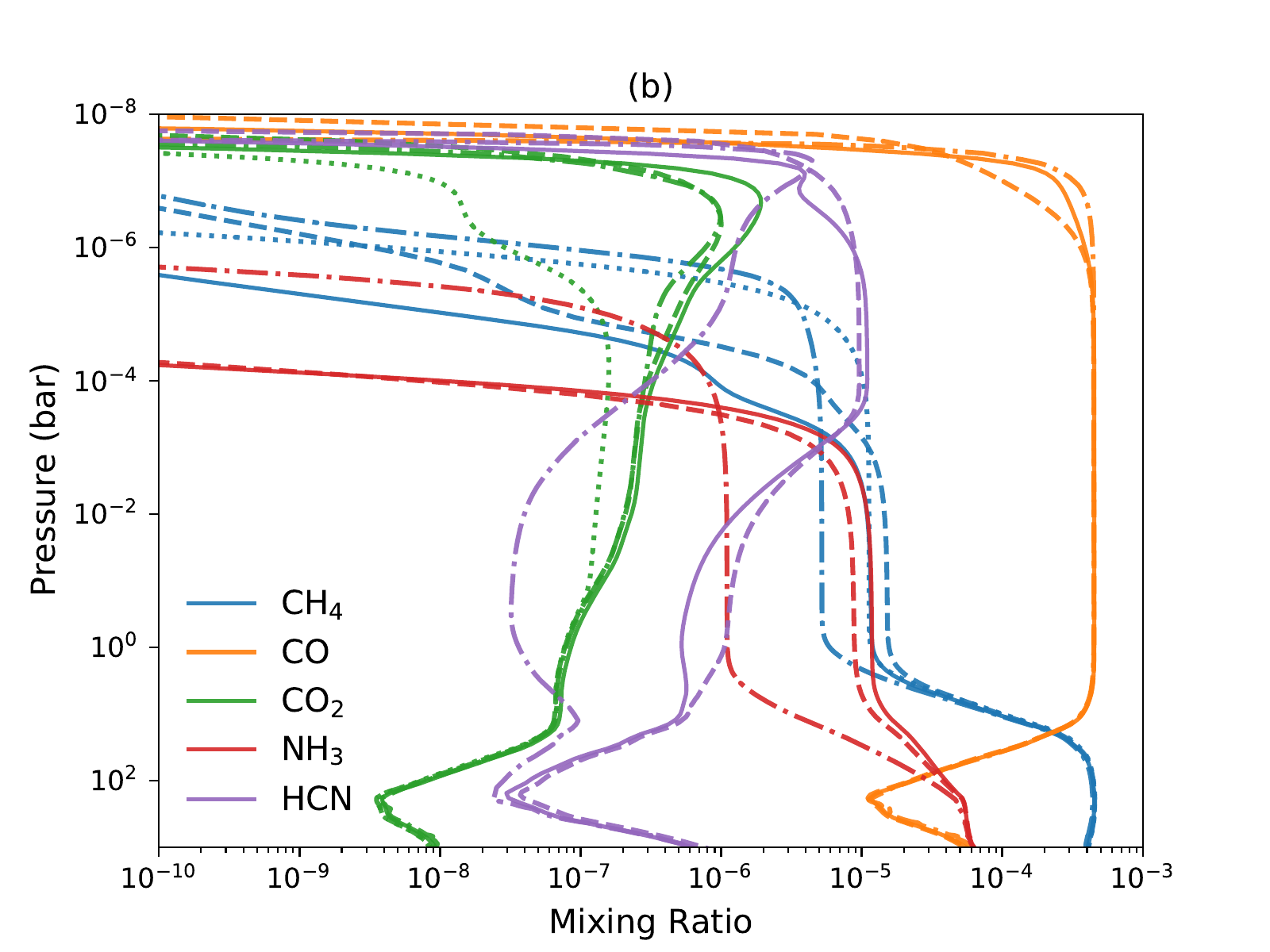}
\includegraphics[width=\columnwidth]{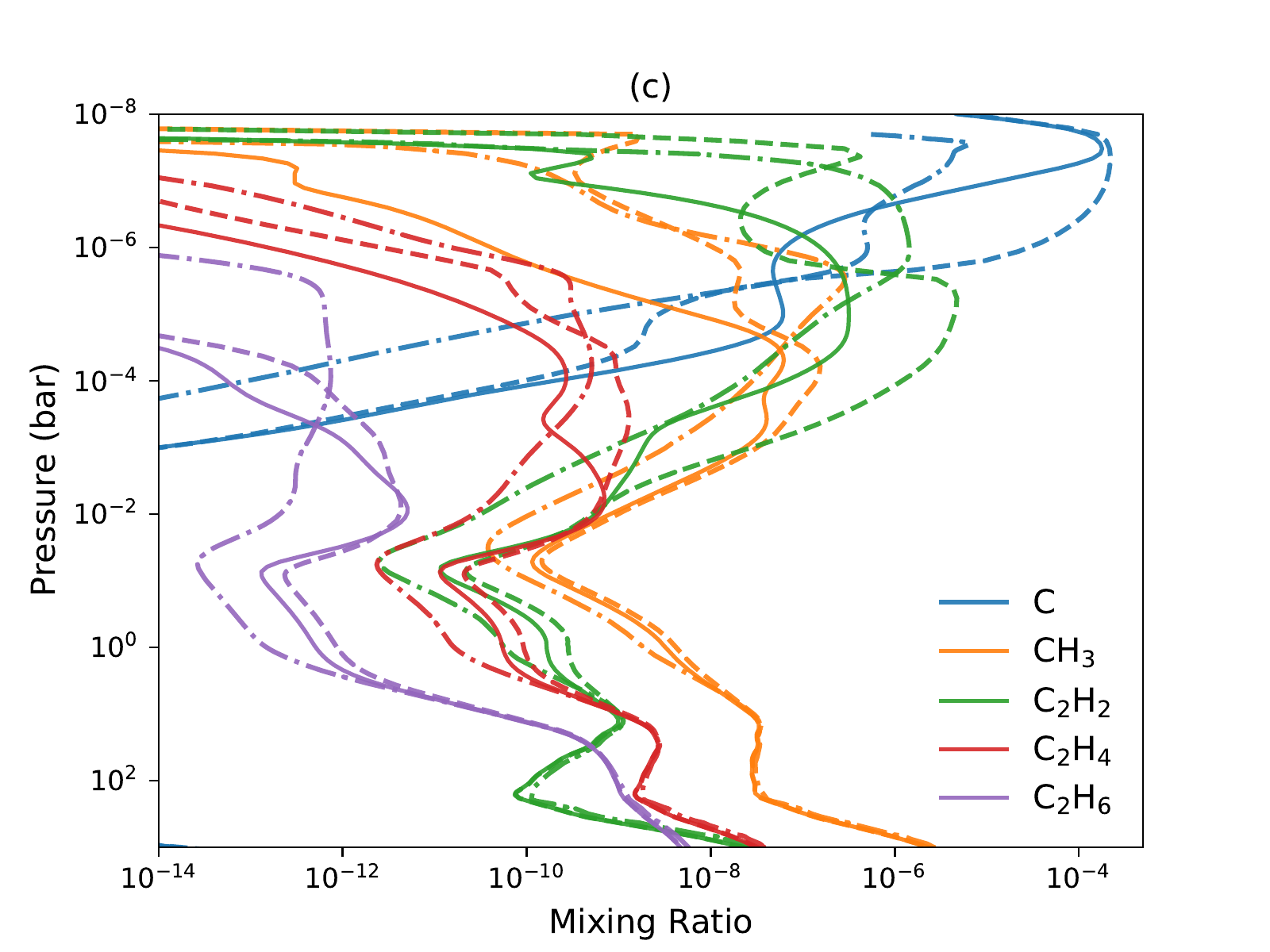}
\includegraphics[width=\columnwidth]{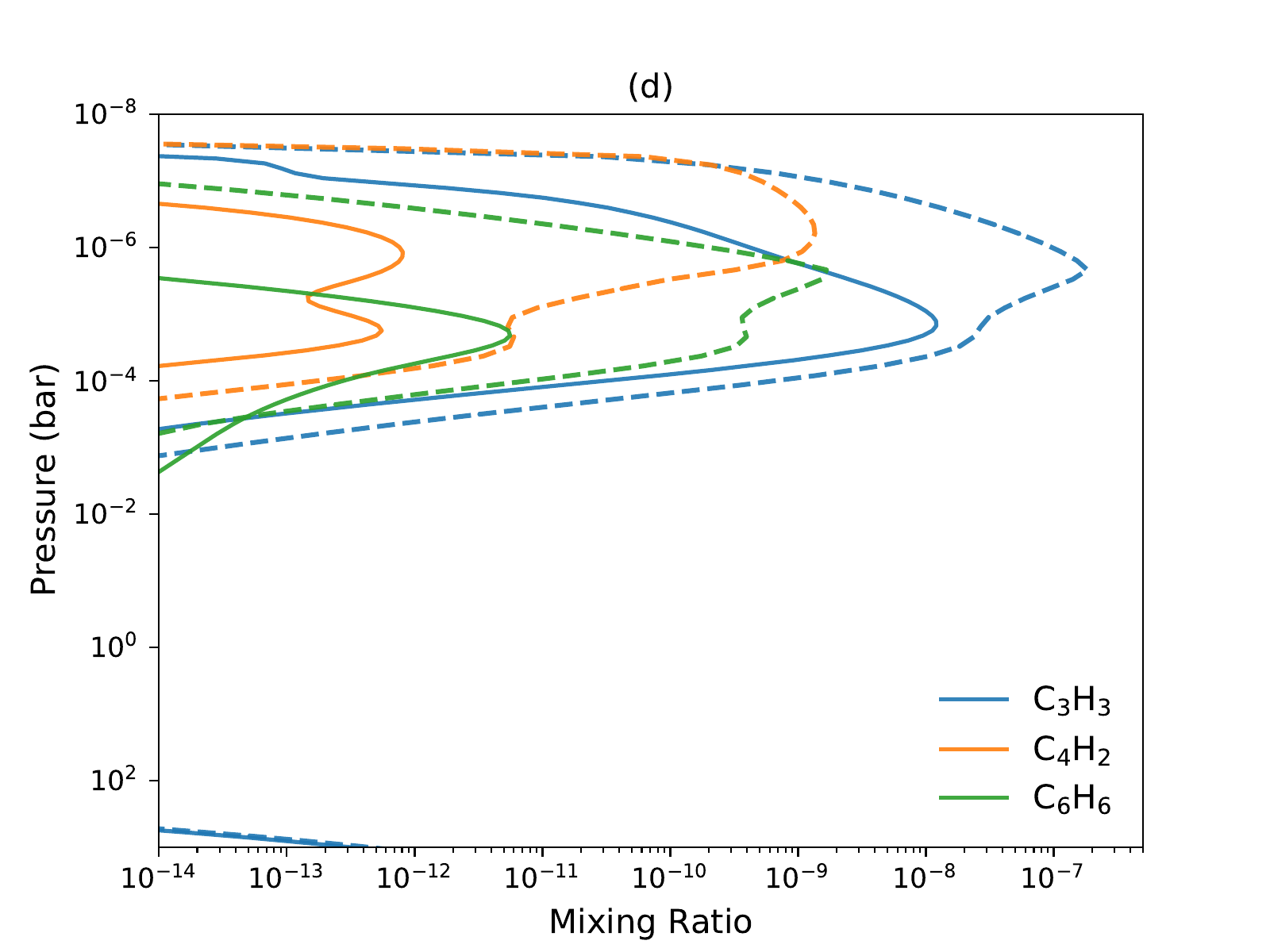}
\includegraphics[width=\columnwidth]{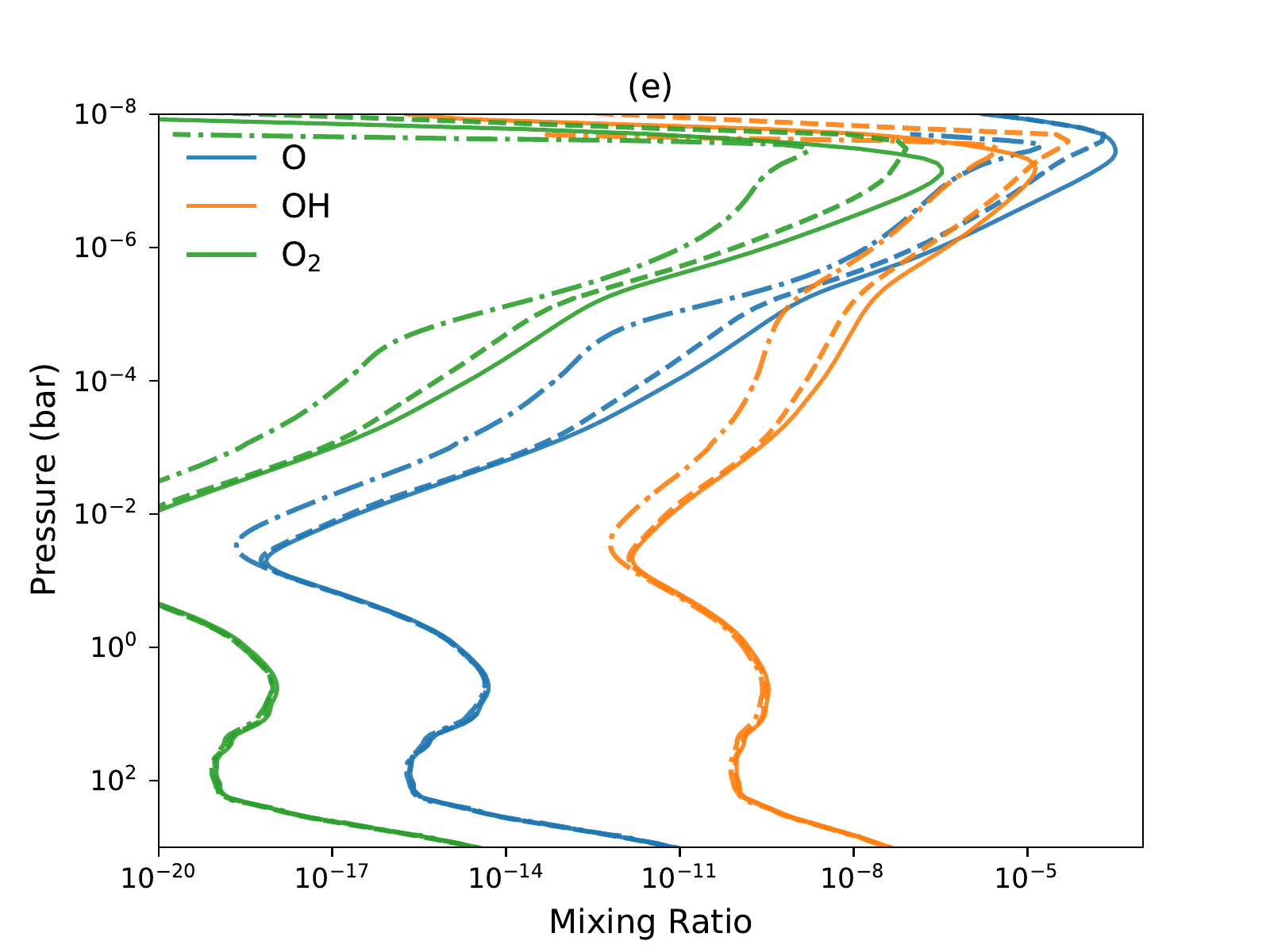}
\includegraphics[width=\columnwidth]{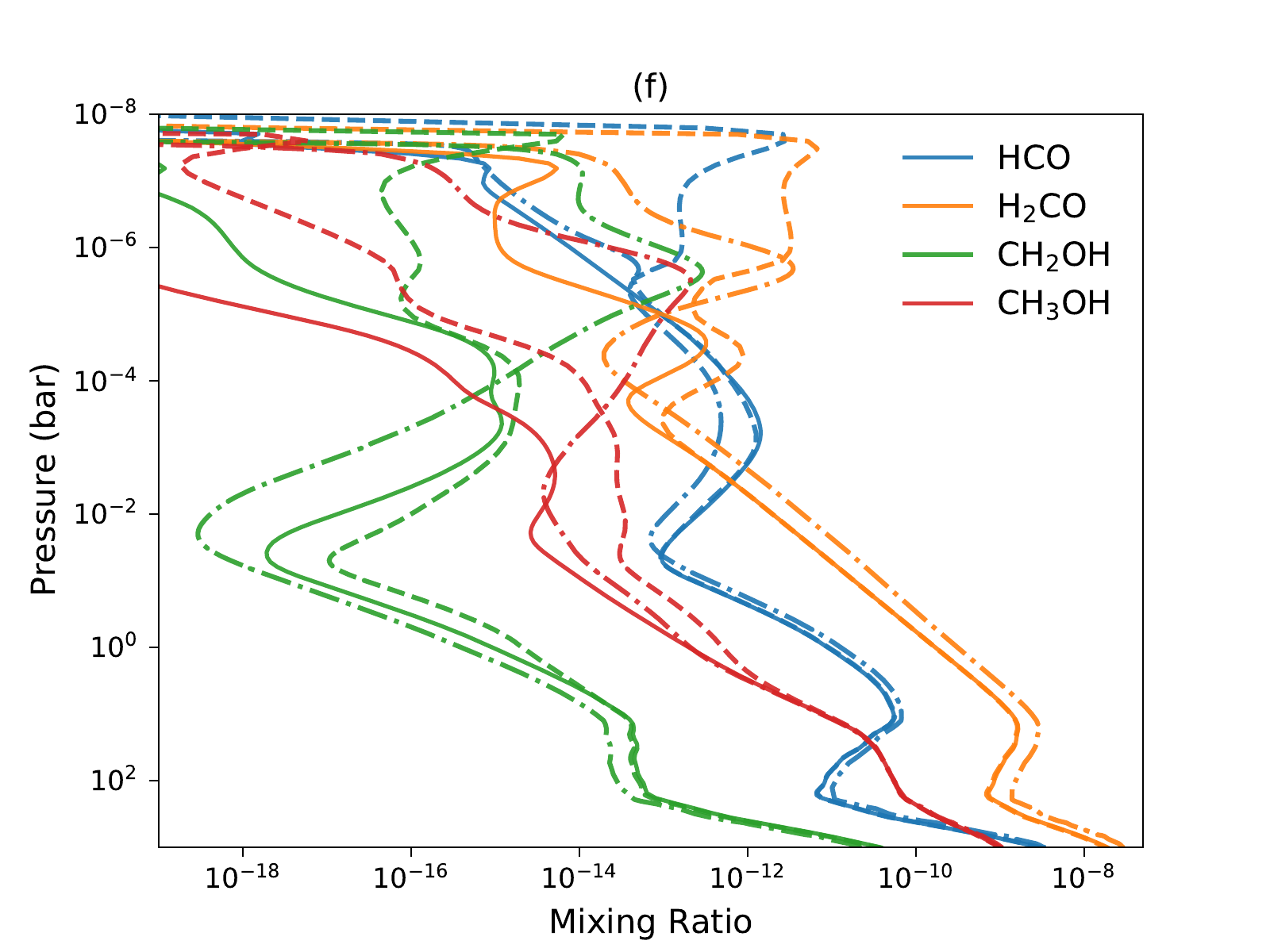}
\end{center}
\caption{Comparison of atmospheric compositions on HD 189733b computed by VULCAN (solid), \cite{Moses11} (dashed), and \cite{Venot12} (dashed-dotted), showing volume mixing ratios of main species ((a), (b)), carbon species ((c), (d)), oxygen species ((e), (f)), and nitrogen species ((g),(h)) (some species not included in V12). Additionally, dotted lines for \ce{CH4} and \ce{CO2} are from running VULCAN with the updated methanol scheme from \cite{Venot2020}.}
\label{fig:HD189-VM}
\end{figure*}
\begin{figure*}[!ht] 
\ContinuedFloat
\begin{center}
\includegraphics[width=\columnwidth]{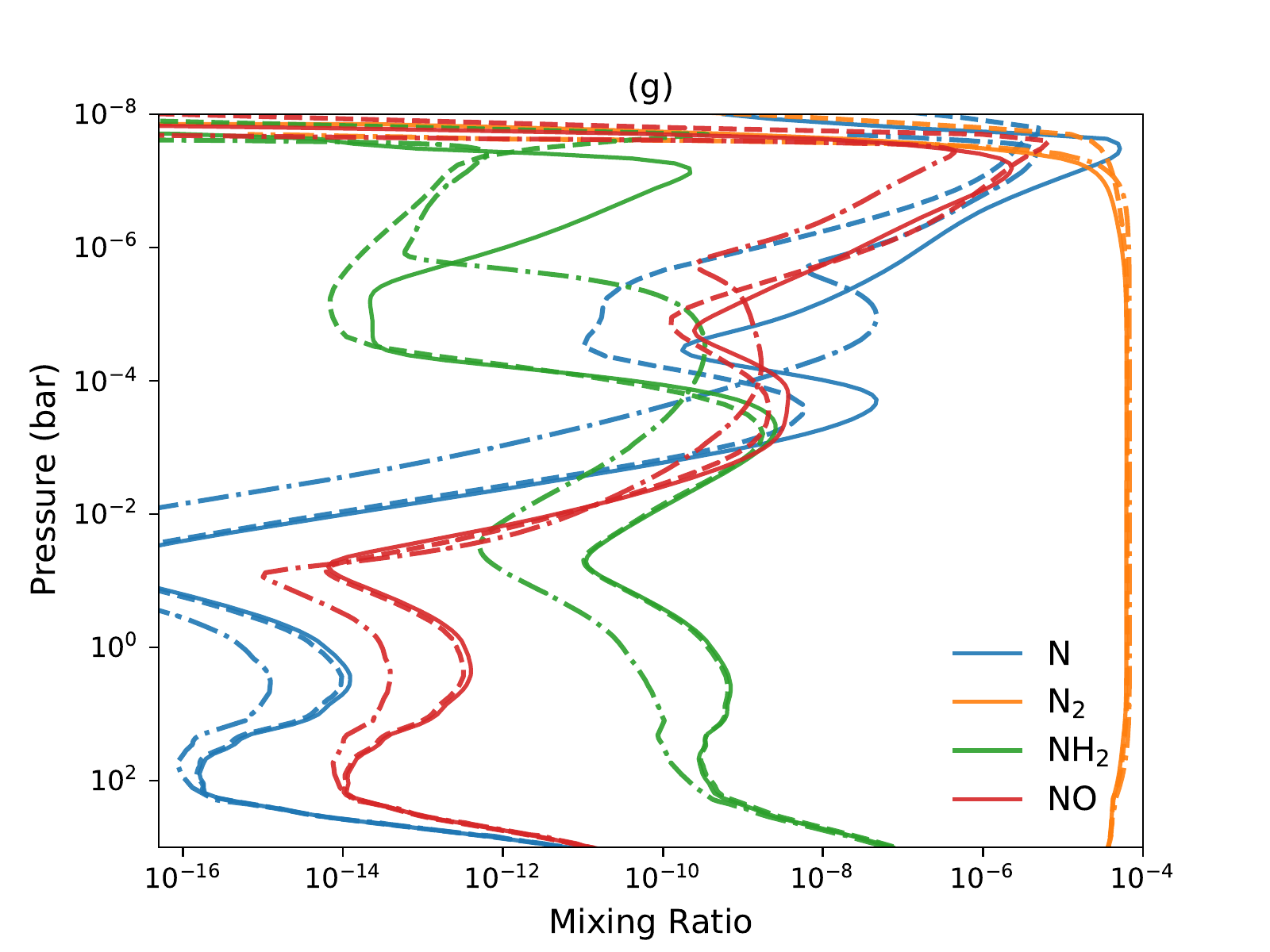}
\includegraphics[width=\columnwidth]{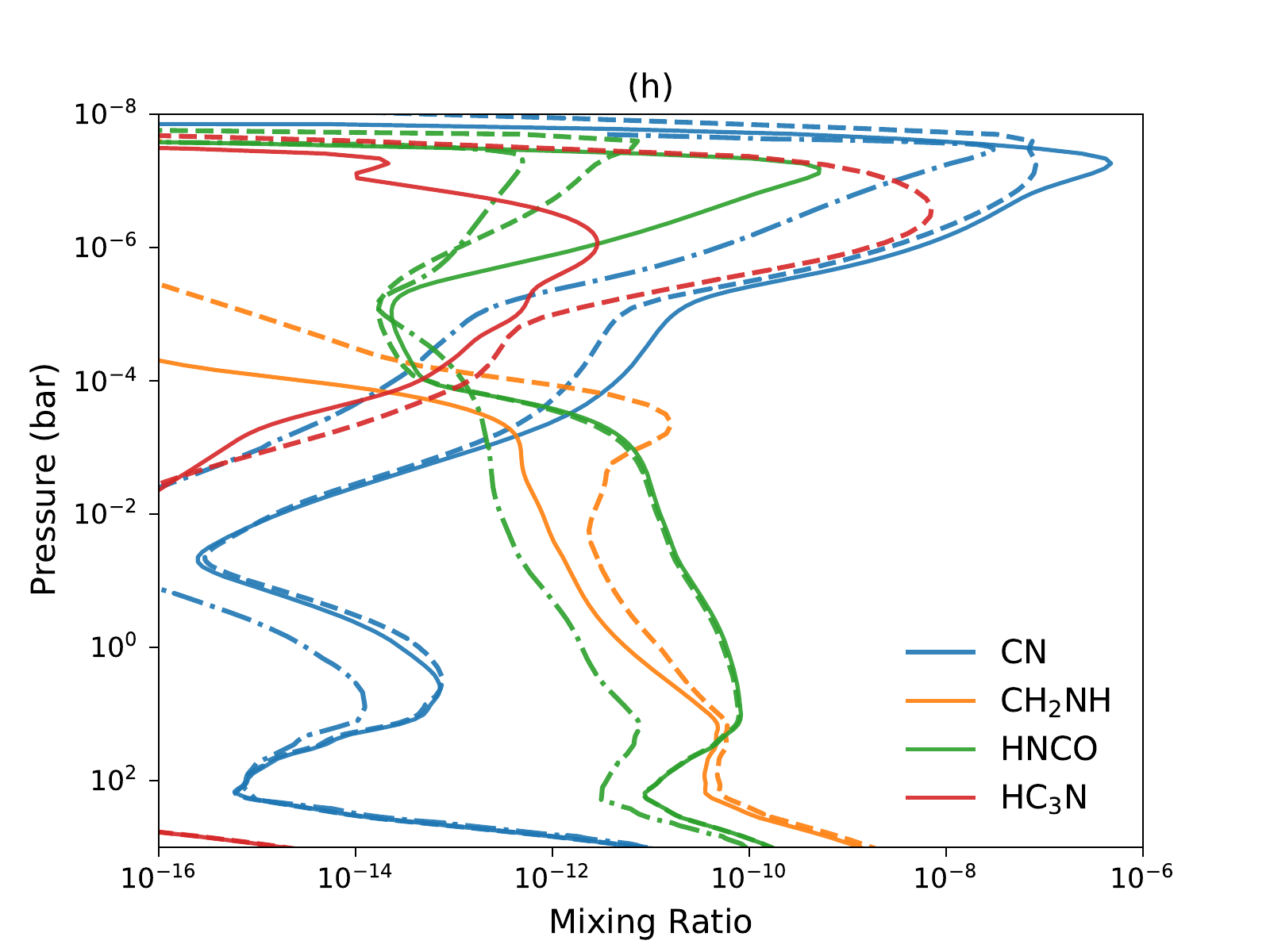}
\end{center}
\caption{(cont.)}
\end{figure*}

\paragraph{Production of \ce{CO2} and HCN}\mbox{}

Another unexpected change in \cite{Venot2020} is that \ce{CO2} remains in chemical equilibrium across the atmosphere. Our model with the implementation of \cite{Venot2020} scheme confirmed the same result. This is remarkably differed from all other models, including V12, where \ce{CO2} is enhanced by photochemically produced OH: 
\begin{equation}
\vspace{-0.2cm}
\ce{CO + OH -> CO2 + H}
\vspace{+0.03cm}
\label{re:CO}
\end{equation}
This reaction with the OH radical is expected to rapidly convert CO into \ce{CO2} while the reaction rate is well studied owning to its importance in the terrestrial atmosphere as well as combustion kinetics. The rate coefficient of Reaction (\ref{re:CO}) adopted in \cite{Venot2020}: 2.589 $\times$ 10$^{-16}$ ($T$/300)$^{1.5}$ exp(251.4 / $T$), has a pre-exponential factor about two orders of magnitude smaller than the typical values listed on NIST, as compared in Figure \ref{fig:rate_CO2}. The slow CO oxidation shuts off the \ce{CO2} production and makes \ce{CO2} retain chemical equilibrium in \cite{Venot2020}. We are not sure if this rate constant is part of the updated methanol scheme from \cite{Burke2016} at this point, as to our knowledge, the base network in \cite{Burke2016} takes the rate coefficient of Reaction (\ref{re:CO}) from \cite{joshi06}, which is consistent with the literature and faster than that in \cite{Venot2020}.

The dissociation of \ce{CH4} and \ce{NH3} leads to the formation of HCN, the primary photochemical product that coupled carbon and nitrogen on HD 189733b. HCN becomes the most abundant carbon-bearing molecule next to CO in the upper atmosphere. We identify the pathway in the HCN-dominated region between 1 mbar and 1 $\mu$bar as
\begin{eqnarray}
\begin{aligned} 
\ce{ CH4 + H &-> CH3 + H2}\\
\ce{ NH3 + H &-> NH2 + H2}\\
\ce{ NH2 + H &-> NH + H2}\\
\ce{ NH + H &-> N + H2}\\
\ce{ CH3 + N &-> H2CN + H}\\
\ce{ H2CN + H &-> HCN + H2}\\
2(\ce{ H2O &->[h\nu] OH + H})\\
2(\ce{ OH + H2 &-> H2O + H})\\
\noalign{\vglue 5pt} 
\hline %
\noalign{\vglue 5pt} 
\mbox{net} : \ce{CH4 + NH3 &-> HCN + 3H2}
\end{aligned}
\label{path-hcn}
\end{eqnarray}
, which is identical to (14) of \cite{Moses11}. HCN in V12 naturally follows the more scarce \ce{CH4} and \ce{NH3} and presents in a lower abundance. We note that \cite{Pearce2019} have run simulations and discovered previous unknown rate coefficients, e.g., the destruction of HCN by reacting with the excited \ce{N(^2D)} could be an important sink of HCN. 




\paragraph{Photolysis Effects}\label{sec:photo}\mbox{}

In the upper stratosphere above 1 mbar, the model differences most likely come from photochemical sources. However, it is less straightforward to compare model discrepancy originated from photochemistry as each step in converting photon fluxes into photolysis rates can give rise to deviation, including stellar fluxes, cross sections, branching ratios, and radiative transfer, etc. For simplicity, we will directly inspect the computed photolysis rates from M11, V12, and VULCAN. We limit our comparison to water photolysis, owing to its importance of producing H radicals and the frontline role of H in reacting with molecules such as \ce{CH4} and \ce{NH3} \citep{Liang2003,Moses11}.  

\begin{figure}[tph]
\begin{center}
\includegraphics[width=\columnwidth]{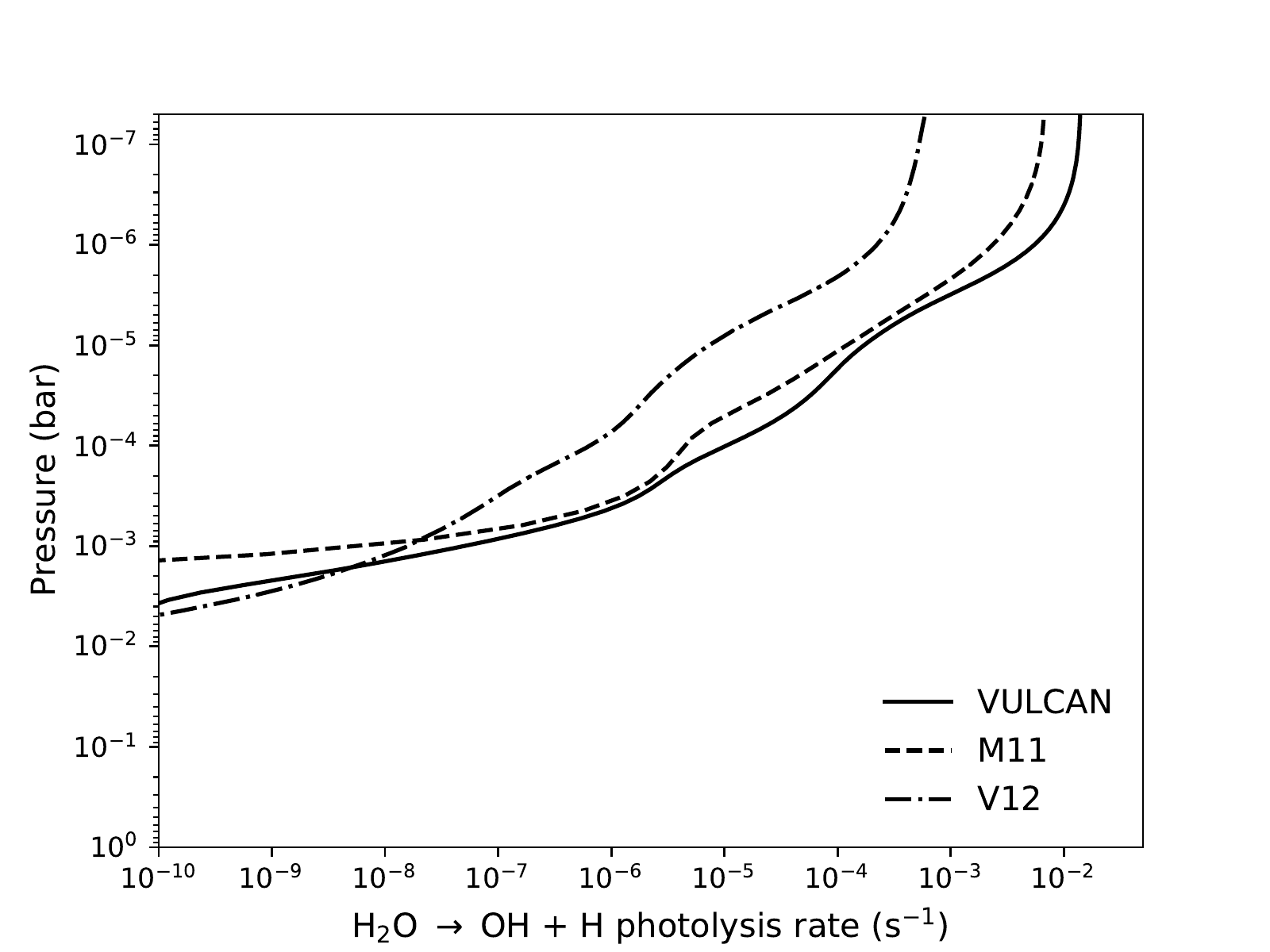}
\end{center}
\caption{Comparison of the photodissociation rates (s$^{-1}$) of the main branch of \ce{H2O} in HD 189733b computed by VULCAN, M11 \citep{Moses11}, and V12 \citep{Venot12}.}
\label{fig:hd189-H2OJ}
\end{figure}

Figure \ref{fig:hd189-H2OJ} compares the photodissociation rates of the main branch \ce{H2O ->[h\nu] OH + H} computed by three models. The water photodissociation rate in VULCAN is about twice as large as that in M11 and around one order of magnitude larger than that in V12. The \ce{H2O} photolysis rates evidently correlate with the H and OH profiles in Figure \ref{fig:HD189-VM}-(a), -(e) and molecules in V12 (e.g. \ce{CH4}) generally tend to survive toward higher altitude. The disagreement started even from the top of the models, with the same deviation also found across other photolytic species, such as \ce{CH4} and \ce{NH3}. This implies the model implementation of stellar fluxes is the first-order contribution to photochemical differences. Although according to \cite{Venot12}, they found no differences in switching to the same stellar flux from M11 and suggested that Rayleigh scattering could be the source of disagreement. We have tested switching off Rayleigh scattering and found negligible changes, since Rayleigh scattering only dominates in the deep region where photochemistry has ceased (see Figure \ref{fig:HD189-209}). We note that potential errors with insufficient spectral resolution can contribute to the photolysis rates as well (see Appendix \ref{app:resolution}). Overall, more attention should be paid to calibrating the stellar irradiation for future photochemical model benchmarks and we suggest using \ce{H2O} photolysis as a baseline.

\vspace{-0.1cm}

\paragraph{Carbon Species Comparison}\mbox{}

Panels (c) and (d) in Figure \ref{fig:HD189-VM} show the same comparison for other important carbon-bearing species. Atomic carbon is liberated from CO photodissociation near the top atmosphere. CO photolysis appears to be stronger in M11 and generates more atomic carbon around $\mu$bar level. The carbon vapor exceeds saturation and can potentially condense in the upper atmosphere. We will examine the implication of C condensation in Section \ref{HD189-C-conden}. In the lower stratosphere, various hydrocarbon production is initiated by methane abstraction, i.e. H being successively stripped from \ce{CH4} to form more reactive unsaturated hydrocarbons. The hydrocarbon profiles predicted by M11, V12, and our model are consistent with the divergence of parent \ce{CH4}, except that acetylene (\ce{C2H2}) is also governed by atomic C in the upper atmosphere.

\ce{C2H2} is the most favoured unsaturated hydrocarbon on HD 189733b. In the CO-photolysis region, atomic C can couple with nitrogen into CN and eventually produce \ce{C2H2} by dissociation of \ce{HC3N}. Yet we find \ce{CH4} to still be the dominant source for producing \ce{C2H2} below 1 $\mu$bar via a pathway such as\vspace{-0.3cm}
\begin{eqnarray}
\begin{aligned}
2(\ce{ CH4 + H &-> CH3 + H2})\\
\ce{ CH3 + H &-> CH2 + H2}\\
\ce{ CH2 + H &-> CH + H2}\\
\ce{ CH + H &-> C + H2}\\
\ce{ CH3 + C &-> C2H2 + H}\\
2(\ce{ H2O &->[h\nu] H + OH})\\
2(\ce{ OH + H2 &-> H2O + H})\\
\noalign{\vglue 5pt} 
\hline %
\noalign{\vglue 5pt} 
\mbox{net} : \ce{2CH4 &-> C2H2 + 3H2}.
\end{aligned}
\end{eqnarray}
Our scheme predicts \ce{C2H2} with the maximum abundance a few factor smaller than V12 and about an order-of-magnitude smaller than M11.

Ethylene (\ce{C2H4}) is the next most abundant hydrocarbon after acetylene and peaks around 10 mbar. \ce{C2H4} and other \ce{C2H_x} production stems from \ce{CH3} association reaction via the pathway
\begin{eqnarray}
\vspace{-2cm}
\begin{aligned}
\ce{ H2O &->[h\nu] H + OH}\\
\ce{ OH + H2 &-> H2O + H}\\
2(\ce{ CH4 + H &-> CH3 + H2})\\
\ce{ CH3 + CH3 &->[M] C2H6}\\
\ce{ C2H6 + H &-> C2H5 + H2}\\
\ce{ C2H5 &->[M] C2H4 + H}\\
\noalign{\vglue 5pt} 
\hline %
\noalign{\vglue 5pt} 
\mbox{net} : \ce{2CH4 &-> C2H4 + 2H2}
\end{aligned}
\vspace{-0.5cm}
\end{eqnarray}
, where forming \ce{C2H6} is usually the rate-limiting step. The abundances of \ce{C2H4} and \ce{C2H6} in our model are in agreement with M11 within an order of magnitude.

The kinetics beyond C$_2$ hydrocarbons becomes less constrained \citep{Moses11,Venot2013}. As discussed in Section \ref{sec:network}, we intended to capture the major pathways of producing \ce{C6H6} as a proxy for haze precursors without invoking an exhaustive suite of hydrocarbons. In our model, \ce{C6H6} is formed by the pathway
\begin{eqnarray}
\label{re:path-c6h6}
\begin{aligned}
9(\ce{ H2O + h$\nu$ -> H + OH})\\
9(\ce{ OH + H2 -> H2O + H})\\ 
6(\ce{ CH4 + H -> CH3 + H2})\\
4(\ce{ CH3 + H -> CH2 + H2})\\
4(\ce{ CH2 + H -> CH + H2})\\
4(\ce{ CH + H -> C + H2})\\
2(\ce{ CH3 + C -> C2H2 + H})\\
2(\ce{ C + C2H2 -> C3H2})\\
2(\ce{ C3H2 + H + M -> C3H3 + M})\\
\ce{ C3H3 + C3H3 + M -> C6H6 + M}\\
\hline \nonumber
\mbox{net} : \ce{6CH4 -> C6H6 + 9H2} .
\end{aligned}
\end{eqnarray}
where the recombination of \ce{C3H3} is the rate-limiting step (akin to the cooler atmosphere of Jupiter \citep{Moses2005}). Figure \ref{fig:HD189-VM}-(d) shows that \ce{C4H2} and \ce{C6H6} predicted by our reduced scheme have considerably lower abundances than those in M11. Given the agreement of \ce{C3H3} up until 10$^{-5}$ bar, we suspect that the differences of \ce{C6H6} between VULCAN and M11 are due to photodissociation effects from \ce{C6H6} as well as other species such as CO. Given all the uncertainties and complexity as we mentioned in Section \ref{sec:haze}, we do not consider the predicted abundances of \ce{C4H2} and \ce{C6H6} to be accurate, but it should rather serve the purpose for accessing the haze precursors.

\paragraph{Oxygen Species Comparison}\mbox{}

Panels (e) and (f) of Figure \ref{fig:HD189-VM} compare oxygen-bearing species. The deviation of O, OH, and \ce{O2} again follows the discrepency in \ce{H2O} photolysis, similar to H. There is a minor shift of the equilibrium abundance of \ce{H2CO} in V12, possibly from the thermodynamic data difference between JANAF and the NASA polynomial, as pointed out in \cite{tsai17}. All three models exhibit somewhat different quench levels and profiles for \ce{CH2OH} and \ce{CH3OH}, which are generally important intermediates for \ce{CH4}-CO interconversion \cite{Moses11,tsai18,Venot2020}. Nevertheless, this does not reflect in the \ce{CH4} abundance since \ce{CH4} has already quenched in the deeper region. The updated methanol scheme in \cite{Venot2020} also provides more consistent \ce{CH2OH} and \ce{CH3OH} distributions with M11 and VULCAN. Since VULCAN adopted the same rate coefficients from the ab initio calculation from M11 for the three methanol reactions, the difference between VULCAN and M11 is more likely associated with reactions involving \ce{CH2OH}.

\paragraph{Nitrogen Species Comparison}\mbox{}

Panel (g) and (h) of Figure \ref{fig:HD189-VM} compare nitrogen-bearing species. N and \ce{NH2} follow the same quench level as \ce{NH3} (panel (b)), since they are part of the \ce{NH3}-\ce{N2} conversion.
Considerable amount of atomic N is produced above mbar level by hydrogen abstraction of ammonia, similar to that of methane. Atomic N is oxidized by OH into NO in the upper atmosphere. NO reacts rapidly with atomic C into CN, as C-N bond is stronger than N-O bond. CN is an important source of nitrile production, e.g., CN reacts with \ce{C2H2} to form \ce{HC3N}. Our model shows a slower \ce{HC3N} production and predicts \ce{HC3N} with a peak value about two orders of magnitude lower than M11.

The carbon-nitrogen-bearing species are grouped in Figure \ref{fig:HD189-VM}-(h). Since \ce{NH3} quenched first in the deeper layers than \ce{CH4}, the quench levels of general carbon-nitrogen-bearing species also follow \ce{NH3}. Despite in trace abundance, \ce{CH2NH} and HNCO participate in HCN forming mechanism and become important at high pressures. We find HCN formed around 10 mbar via \ce{CH2NH} and \ce{CH3NH2} in a pathway identical to (7) in \cite{Moses11}.\\  

We conclude that we validate our model of HD 189733b by thoroughly reproducing composition distribution within the uncertainty range enclosed by M11 and V12. The kinetics data we employed generally yields quenching behavior close to M11, while our model appeared to predict lower \ce{C2H2}, \ce{C4H2}, \ce{C6H6}, and \ce{HC3N} than M11 in the upper atmosphere. Contrary to what have been reported in \cite{Venot2020}, we find that the update methanol scheme in fact increases the quenched \ce{CH4} abundance and more consistent with that in M11 and this work. The photochemical part of the atmosphere is more complex to diagnose but we suggest the implementation of stellar fluxes is the main factor in the discrepancy between M11, V12, and VULCAN.

\subsection{Jupiter}
The modeling work for Jovian chemistry broadly falls into two categories addressing two main regions: the stratosphere and the deep troposphere. The stratospheric compositions are governed by photochemical kinetics with the main focus on understanding the formation of various hydrocarbons. For stratospheric models, fixed mixing ratios or fluxes at the lower boundary need to be specified \citep{Yung1980,Moses2005}. As for the deep tropospheric compositions below the clouds with sparse observational constraints, kinetics models attempt to infer the interior water content based on other quenched species \citep{Visscher2010,Wang2016}. Since chemical equilibrium is expected to hold in the deep interior, the elemental ratio essentially controls the reservoir of gases and vertical mixing determines the quenched compositions in the upper troposphere. 

In this validation, our objective is to validate the chemical scheme at low temperatures with observed hydrocarbons and verify the condensation scheme. We take a general approach by connecting the deep troposphere to the stratosphere and solve the continuity equations consistently. Our lower boundary at 5 kbar is far down in the region ruled by equilibrium chemistry and zero flux can be applied to the lower boundary. In this setup, fixed-abundance lower boundary conditions are not required as in the stratosphere models \citep[e.g., ][]{Moses2005,Hue2018}. The compositions at the lower stratosphere are physically determined by condensation and transport from the troposphere in the model.   

\subsubsection{Model Setup}
\begin{figure}[htp]
\begin{center}
\includegraphics[width=\columnwidth]{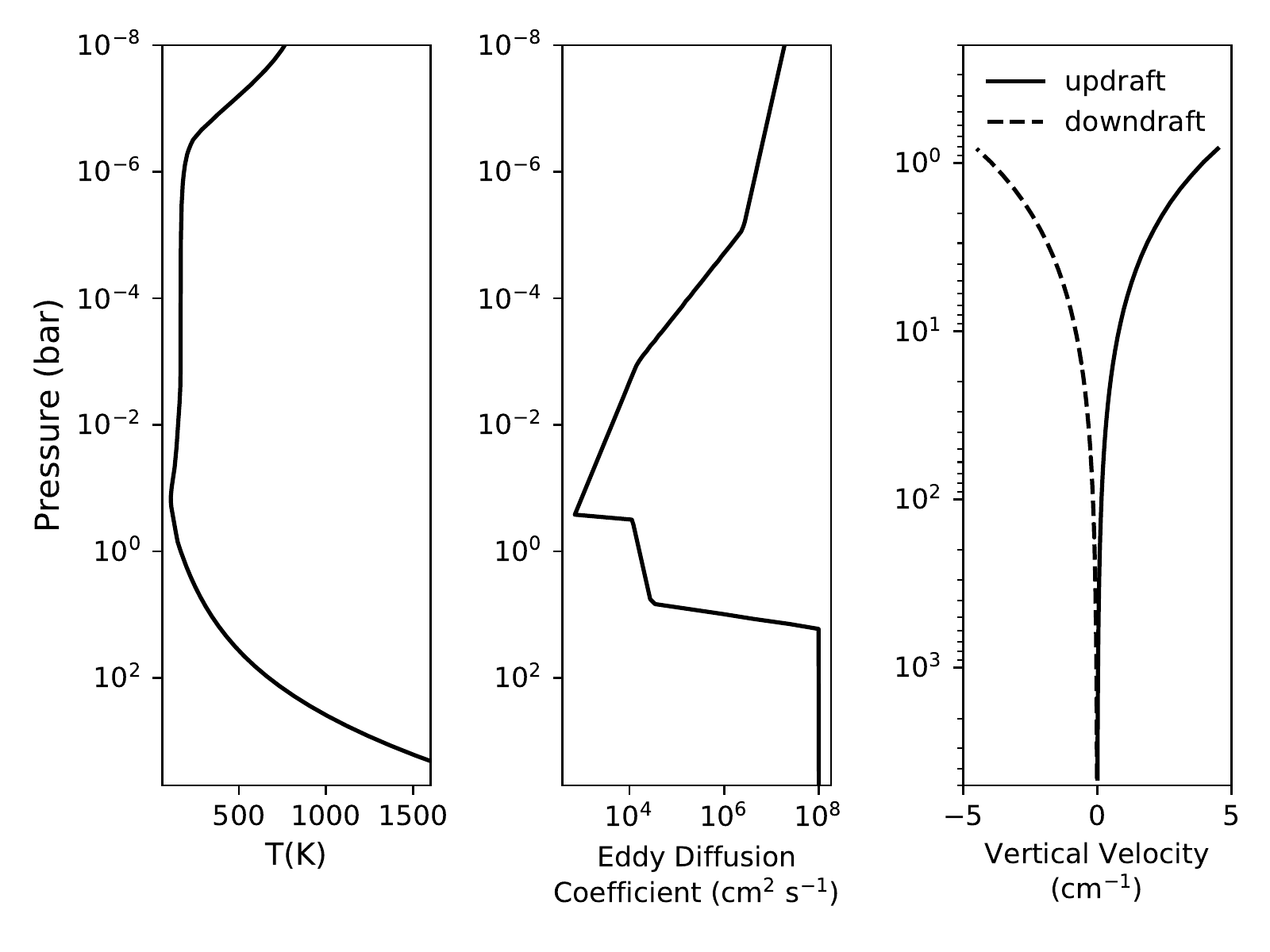}
\end{center}
\caption{The temperature, eddy diffusion, and deep vertical velocities used for our Jupiter model. The temperature and eddy diffusion in the stratosphere are taken from \cite{Moses2005}, while a dry adiabat and uniform eddy diffusion with $K_\textrm{zz}$ = 10$^8$ (cm$^2$/s) are assumed for the troposphere. The upward (positive) and downward (negative) vertical velocities are prescribed by Equation (\ref{vz}) with the maximum speed of 5 cm/s at 0.7 bar.}
\label{fig:Jupiter-TP}
\end{figure}

The temperature profile in the stratosphere and top of the troposphere (above 6.7 bar) is taken from \cite{Moses2005} and extended to 5000 bar following the dry adiabat, with T = 427.71 K at 22 bar measured by the Galileo probe as the reference point. We use the same eddy diffusion profile for the stratosphere as Model A of \cite{Moses2005}, which is derived from multiple observations. The eddy diffusion is assumed to be constant with 10$^8$ (cm$^2$/s) in the convective region below 6.7 bar. The temperature and eddy-diffusion profiles adopted for our Jupiter model are shown in Figure \ref{fig:Jupiter-TP}. 

Heavy elements in Jupiter are enhanced compared to solar metallicity, except the oxygen abundance is still unclear. We assign the elemental abundances for the Jupiter model as He/H = 0.0785 \citep{Atreya2020}, C/H = 1.19$\times$10$^{-3}$ \citep{Atreya2020}, O/H = 3.03$\times$10$^{-4}$ (0.5 times solar), and N/H = 2.28$\times$10$^{-4}$ \citep{Li2017}. Sulfur is not included in our Jupiter validation for simplicity. We include condensation of \ce{H2O} and \ce{NH3}, assuming a single particle size with average radius equal to 0.5 $\mu$m for the cloud condensates. Oxygen sources from micrometeoroids are prescribed at the upper boundary at 10$^{-8}$ bar following \cite{Moses2005}, with influx (molecules cm$^{-2}$ s$^{-1}$) of \ce{H2O} = 4 $\times$ 10$^4$, \ce{CO} = 4 $\times$ 10$^6$, and \ce{CO2} = 1 $\times$ 10$^4$.

\subsubsection{Comparing to Stratospheric Observations and \cite{Moses2005}}
\begin{figure}[htp]
\begin{center}
\includegraphics[width=\columnwidth]{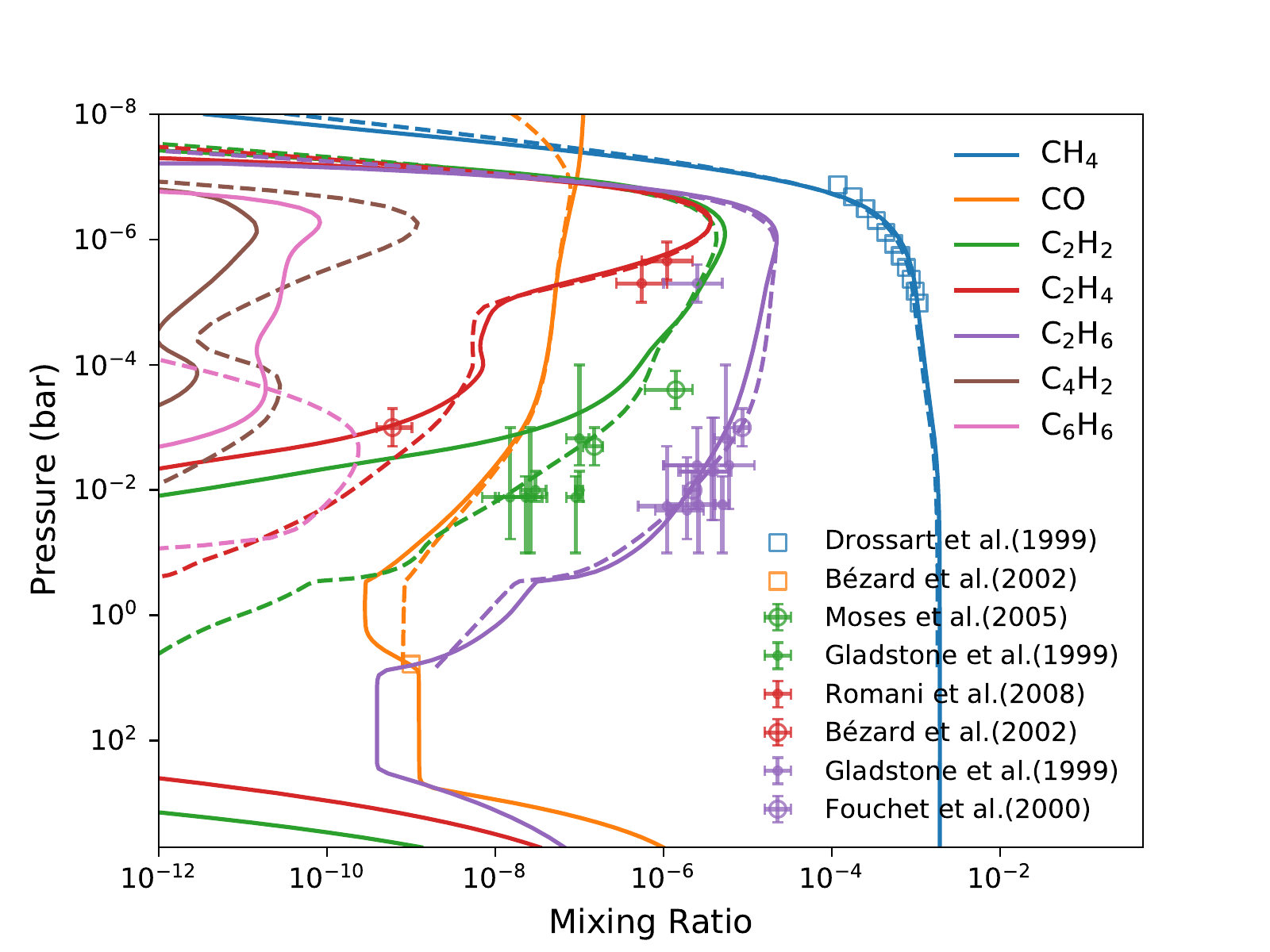}
\includegraphics[width=\columnwidth]{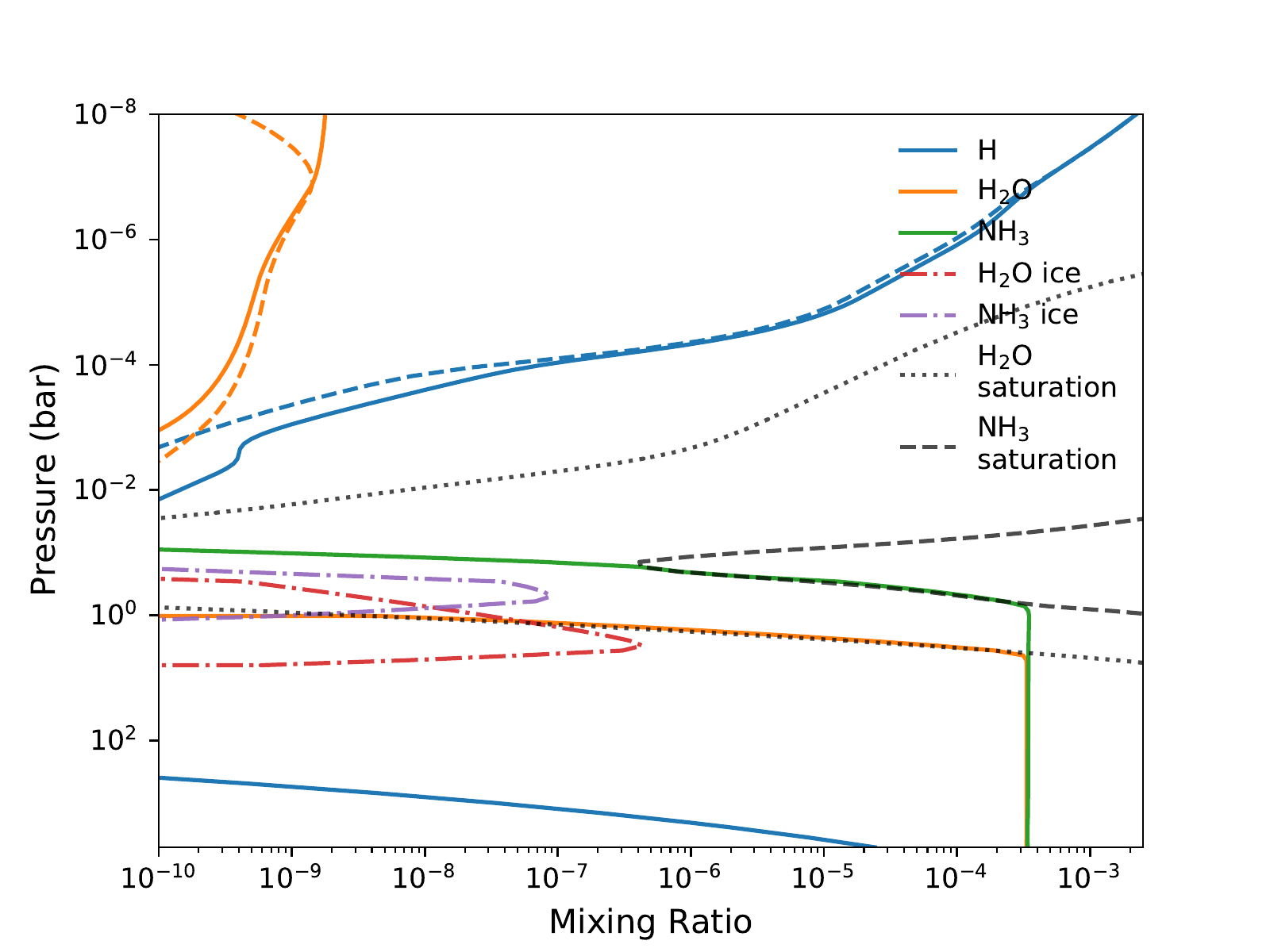}
\includegraphics[width=\columnwidth]{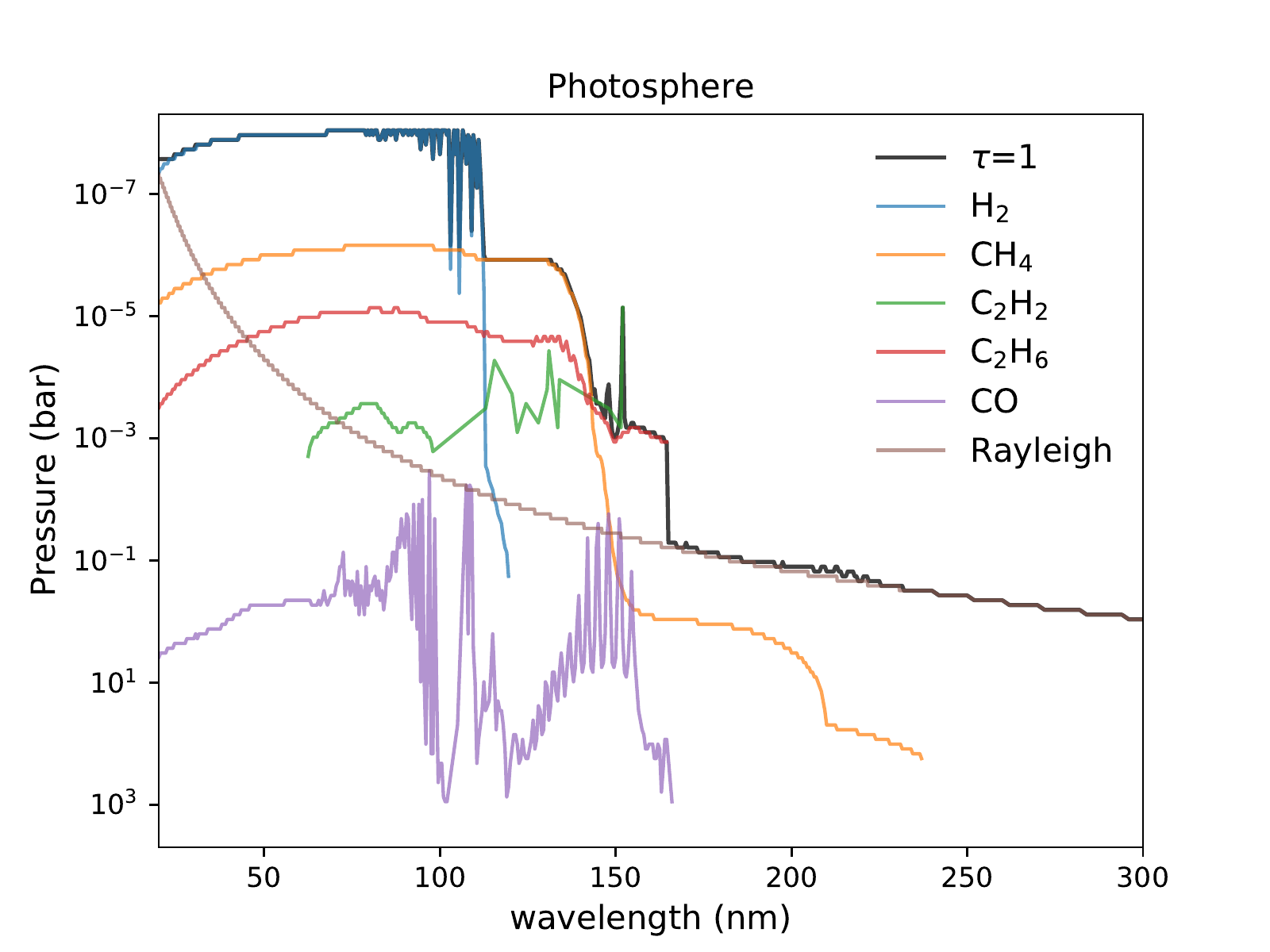}
\end{center}
\caption{The top panel shows the vertical mixing profiles of important chemical species in our Jupiter model (solid), compared with various observations (data points) of hydrocarbons and the stratospheric distributions from Model A of \cite{Moses2005} (dashed). We follow \cite{rimmer16} placing a factor of two error bars in pressure when they are not given in the observational data. The vapor mixing ratios and cloud densities (g/cm$^3$) of the condensible \ce{H2O} and \ce{NH3} are displayed in the middle panel. The bottom panel illustrates the UV photosphere where $\tau$ = 1 with decomposition of main absorbers.}\label{fig:Jupiter-mix}
\end{figure}

The top panel of Figure \ref{fig:Jupiter-mix} displays the vertical distribution of key species computed by our model, compared to \cite{Moses2005} and various observations. First, \ce{CH4} is the major carbon-bearing species across the atmosphere. It is well-mixed until photolysis and separation by molecular diffusion take in place at low pressure. The \ce{CH4} distribution in our model matches well with the observation \citep{Drossart1999}. We verify that our treatment of molecular diffusion accurately reproduces the decrease of \ce{CH4} due to molecular diffusion above the homopause. 

Second, our model successfully predicts the major \ce{C2} hydrocarbons, which stem from \ce{CH4} photolysis in the stratosphere. Our model tends to predict lower abundances for the unsaturated hydrocarbons \ce{C2H2} and \ce{C2H4} than \cite{Moses2005} in the lower stratosphere, but both profiles are within the observational constraints. The UV photosphere in Figure \ref{fig:Jupiter-mix} indicates that \ce{CH4} predominates the absorption from Ly-$\alpha$ to about 150 nm. We find the main scheme of converting \ce{CH4} to \ce{C2H6} in the upper atmosphere is
\begin{eqnarray}
\begin{aligned}
2(\ce{ CH4 &->[h\nu] CH3 + H})\\
2(\ce{ CH3 + CH3 &->[M] C2H6})\\
\ce{ 2H &->[M] H2}\\
\noalign{\vglue 5pt} 
\hline %
\noalign{\vglue 5pt} 
\mbox{net} : \ce{2CH4 &-> C2H6 + H2}
\end{aligned}
\end{eqnarray}
and the photodissociation branch of methane is replaced by \ce{ CH4 ->[h\nu] ^1CH2 + H2} followed by \ce{ ^1CH2 + H2 ->CH3 + H} at higher pressures. We confirm that hydrogen abstraction and three-body association reactions are sensitive to the formation of hydrocarbons on Jupiter as discussed in detail in \cite{Moses2005}. Particularly in the lower stratosphere where temperature drops below 200 K, the rate constants fall out of the valid temperature range or are not well constrained. We find it is particularly important to adopt the low-temperature rate constants for \ce{CH4} and \ce{C2Hx} recombination reactions, i.e. \ce{CH3 + H ->[M] CH4 }, \ce{H + C2H2 ->[M] C2H3}, \ce{H + C2H3 ->[M] C2H4}, \ce{H + C2H4 ->[M] C2H5}, and \ce{H + C2H5 ->[M] C2H6}. We also adopt the limit of rate constants below certain threshold temperatures derived by \cite{Moses2005}. 

Third, our condensation scheme predicts the location of water-ice clouds starts at 3.6 bar and ammonia clouds at 0.7 bar as shown in Figure \ref{fig:Jupiter-mix}, consistent with the thermodynamics prediction with 0.5 solar O/H \citep{Atreya2005,Weiden1973}. The ammonium hydrosulfide (\ce{NH4SH}) is not considered since sulfur is not included. Last, our model produces lower abundances of \ce{C4H2} and \ce{C6H6} is produced at higher altitude compared to those in \cite{Moses2005}, which reflects the uncertainties in high-order hydrocarbons and the photolysis branches of \ce{C6H6}.

\begin{figure}[htp]
\begin{center}
\includegraphics[width=\columnwidth]{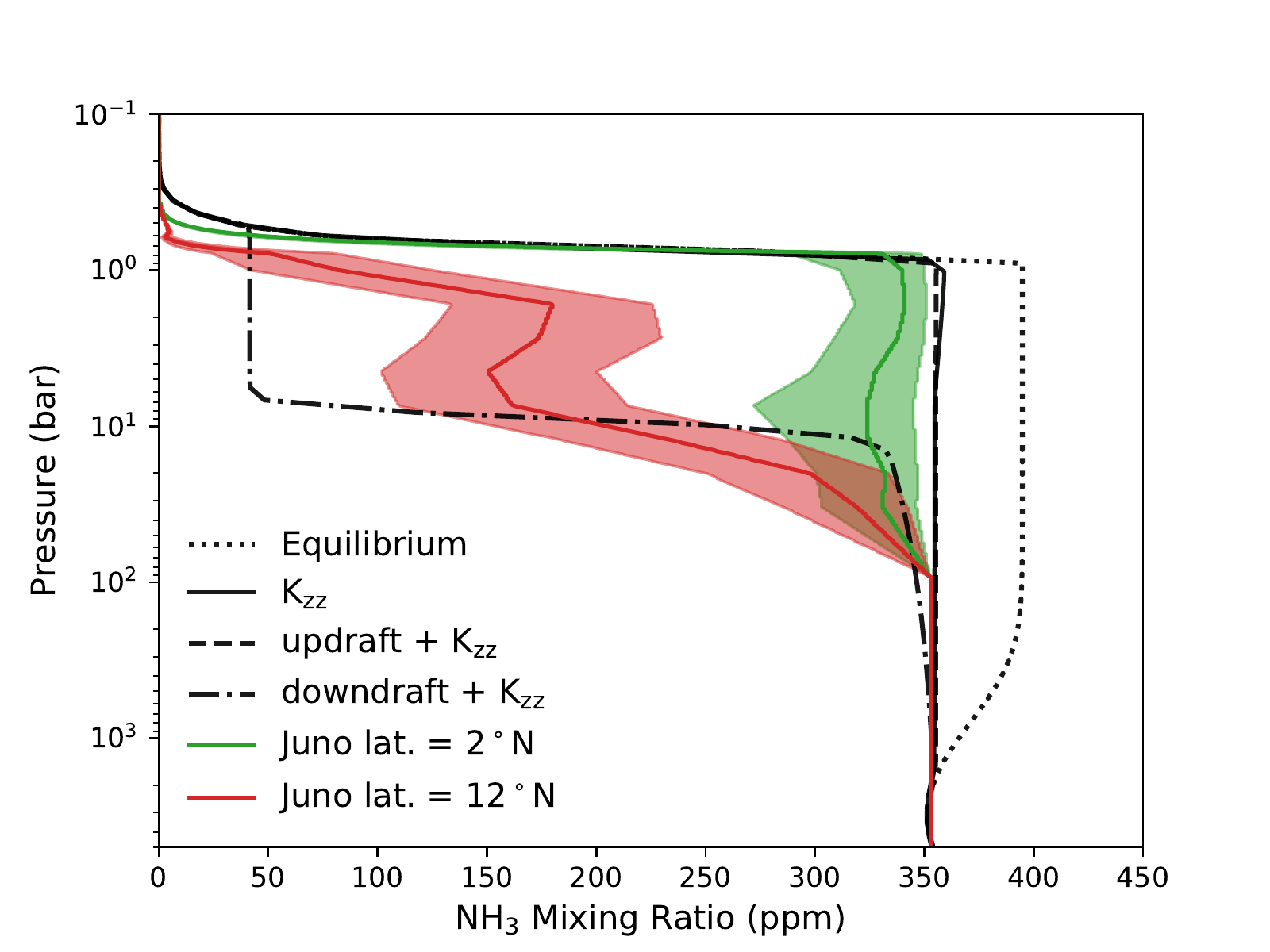}
\end{center}
\caption{The deep ammonia distribution in parts per million (ppm) computed by our Jupiter model (black), while assuming chemical equilibrium, with eddy diffusion only, and including upward/downward advection for the updraft/downdraft branch (Figure \ref{fig:Jupiter-TP}), respectively.  The red and green profiles show the inferred ammonia distribution at 2$^\circ$ N latitude and 12$^\circ$ N latitude based on Juno microwave measurement by \cite{Li2017}, where the shaded areas enclose the 16th and 84th percentile of the samples in their MCMC runs.}
\label{fig:nh3-Jupiter}
\end{figure}

\subsubsection{Spatial Variation of Ammonia Due to Vertical Advection}

During the Juno spacecraft's first flyby in 2016, the microwave radiometer (MWR) on Juno measured the thermal emission below the clouds, which was inverted to global distribution of ammonia from the cloud level down to a few hundreds bar level. A plume-like feature was curiously seen associated with latitudinal variation of ammonia \citep{juno17}. To explore the local impact of advection, we test how the upward and downward motion in a plume can shape the deep ammonia distribution in Jupiter. 

Although Galileo probe has provided constraints on the structure of Jupiter’s deep zonal wind \citep{Atkinson1997} and Juno also sheds light on the vertical extension of the zonal wind \citep{Stevenson2020}, we do not have observational constraints on the deep vertical wind. Hence we consider a simple but physically motivated (mass-conserving) vertical wind structure without tuning to fit the data. We assume an updraft and downdraft plumes starting from the bottom of \ce{NH3}-ice clouds at 0.7 bar, in addition to eddy diffusion, as depicted in the right panel of Figure \ref{fig:Jupiter-TP}. For the non-divergent advection to conserve mass in a 1-D column, the vertical velocity at layer $j$ with number density n$_j$ follows 
\begin{equation}
v_j n_j = v_\textrm{top} n_\textrm{top} = \textrm{constant}
\label{vz}
\end{equation}
such that the net flux remains zero at each layer. For this test, we arbitrarily choose the maximum wind velocity at the top to be 5 cm/s. This choice has advection timescale shorter than diffusion timescale in the lower pressure region, i.e. $v_j$ $\gtrsim$ K$_\textrm{zz} / H$, which allows us to see the influence of advection. Figure \ref{fig:nh3-Jupiter} compares the computed distribution of ammonia to that retrieved from Juno measurements (\cite{Li2017}; also see updates in \cite{Li2020}) at two different latitudes. First, the ammonia distribution predicted by chemical equilibrium is rather uniform with depth, only slightly increasing from 350 ppm to 400 ppm. Next, vertical mixing by eddy diffusion alone makes ammonia quenched from the deep interior below 1000 bar and thus brings ammonia to a slightly lower but uniform concentration of 300 ppm. There is almost no visible difference while including the upward advection since ammonia has already been quenched by eddy diffusion from the deep region. Last, the uniform distribution of ammonia is altered in the downdraft, where the downward motion transports the lower concentration of \ce{NH3} from the condensing region. Our \ce{NH3} distribution is qualitatively consistent with the \ce{NH3}-depleted branch at 12$^{\circ}$ N from \cite{Li2017}, where \ce{NH3} reaches a local minimum around 7 bar. We emphisize that this shape cannot be reproduced by eddy diffusion alone.

Although eddy diffusion is probably still essential in practice for parametrizing a range of mixing processes, we demonstrate that including vertical advection can be useful. The advection processes can especially play a bigger part in 2-D systems \citep{Zhang2013,Hue2015}.

\subsection{Present-Day Earth}
Our chemical network has only been applied to hydrogen-dominated, reducing atmospheres so far. In this section, we validate our full S-N-C-H-O network with the oxidizing atmosphere of present-day Earth. The interaction with the surface is particularly crucial in regulating the composition for 
the terrestrial atmosphere. Surface emission and deposition via biological and geological activities have to be taken into account. Our implementation of the top boundary fluxes and condensation scheme has been validated for Jupiter in the previous section. We will proceed to verify the lower boundary with surface emission and deposition in the Earth model.

\begin{table}[h]
\begin{center}
\caption{Lower Boundary Conditions for the Earth Validation}\label{tab:BC_Earth}
\begin{tabular}{|l|c|c|}
\hline
Species & Surface Emission\footnote{Global emission typically measured in mass budget (Tg/yr), which is converted to molar flux with the surface area of the Earth = 5.1 $\times$ 10$^{18}$ cm$^2$ for our 1-D photochemical model.} & V$_{\textrm{dep}}$ \footnote{Adopted from \cite{Hauglustaine1994}}\\
 & (molecules cm$^{-2}$ s$^{-1}$) & (cm s$^{-1}$)\\
\hline
CO\footnote{\cite{IPCC2001}} & 3.7 $\times$ 10$^{11}$ & 0.03\\
\ce{CH4}\footnote{\cite{Seinfeld2016}\label{S16}} & 1.6 $\times$ 10$^{11}$ & 0\\
NO\textsuperscript{\ref{S16}} & 1.3 $\times$ 10$^{10}$ & 0.001\\
\ce{N2O}\textsuperscript{\ref{S16}} & 2.3 $\times$ 10$^{9}$ & 0.0001\\
\ce{NH3}\textsuperscript{\ref{S16}} & 1.5 $\times$ 10$^{9}$ & 1\\
\ce{NO2} & 0 & 0.01\\
\ce{NO3} & 0 & 0.1\\
\ce{SO2}\textsuperscript{\ref{S16}} & 9 $\times$ 10$^{9}$ & 1\\
\ce{H2S}\textsuperscript{\ref{S16}} & 2 $\times$ 10$^{8}$ & 0.015\\
\ce{COS}\textsuperscript{\ref{S16}} & 5.4 $\times$ 10$^7$ & 0.003\\
\ce{H2SO4}\textsuperscript{\ref{S16}} & 7 $\times$ 10$^8$ & 1\\
\ce{HCN}\footnote{\cite{Li2003}\label{L03}} & 1.7 $\times$ 10$^8$ & 0.13\\
\ce{CH3CN}\textsuperscript{\ref{L03}} & 1.3 $\times$ 10$^8$ & 0.13\\
\ce{HNO3} & 0 & 4\\
\ce{H2SO4} & 0 & 1\\
\hline
\end{tabular}
\end{center}
\end{table}

\begin{figure}[h]
\begin{center}
\includegraphics[width=\columnwidth]{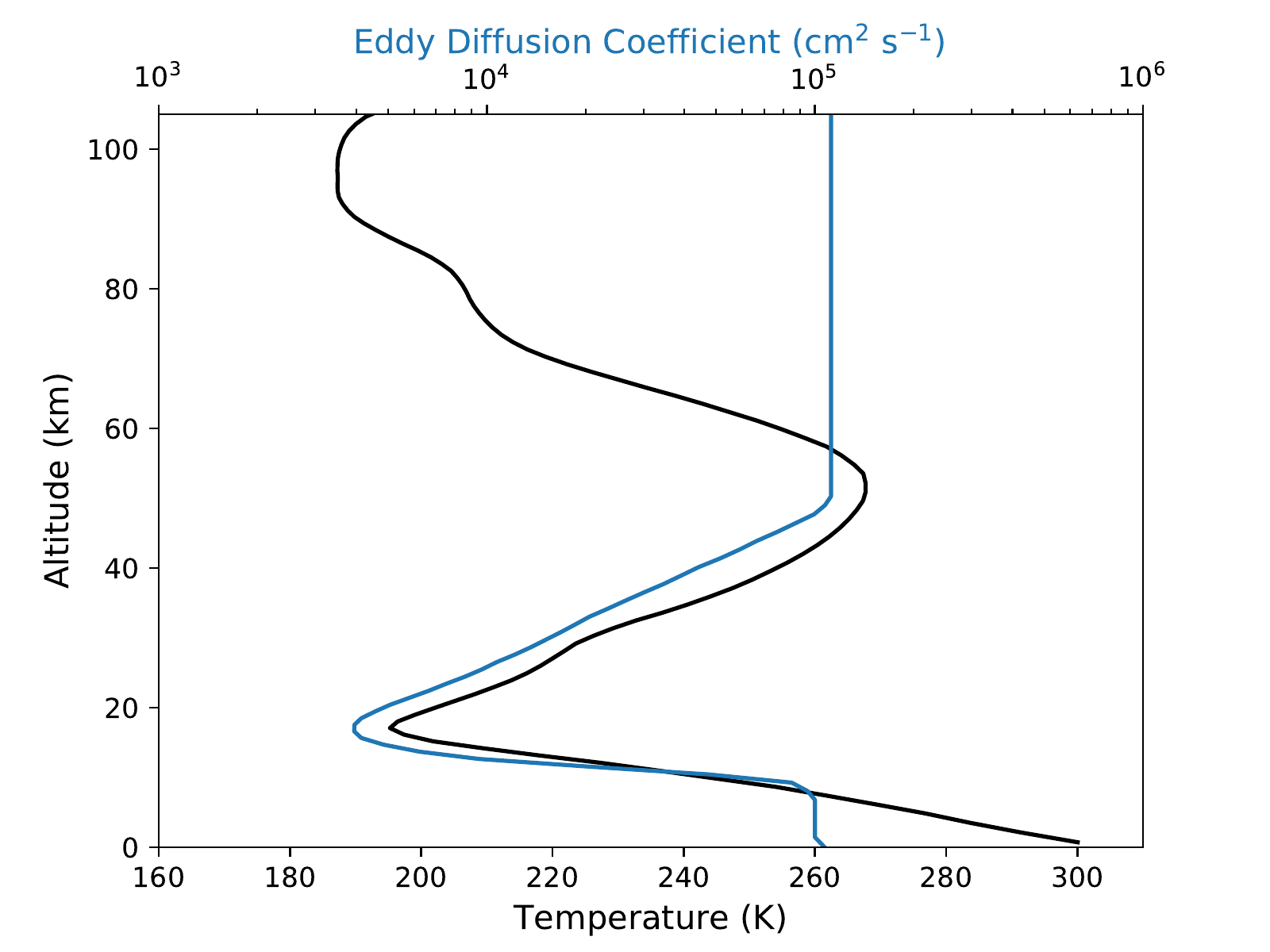}
\caption{The temperature (at the equator in January from CIRA-86 with references described in the text) and eddy diffusion (K$_\textrm{zz}$) profiles \citep{Massie1981} for the Earth model.}\label{fig:Earth-TP} 
\end{center}
\end{figure}

\subsubsection{Model Setup}
We follow \cite{Hu2012}, taking the monthly mean temperature at the equator in January 1986 (CIRA-86) from the empirical model COSPAR International Reference Atmosphere\footnote{\url{https://ccmc.gsfc.nasa.gov/pub/modelweb/atmospheric/cira/cira86ascii}} \citep{COSPARI,COSPARII}  as the background temperature profile and the eddy diffusion coefficients from \cite{Massie1981}, as shown in Figure \ref{fig:Earth-TP}. The winter atmosphere has a colder and hence drier tropopause and better represents the global averaged water vapor content (see \cite{Chiou1997} and the discussion in \cite{Hu2012}). 

Unlike gas giants, terrestrial atmospheres typically do not extend to a thermochemical equilibrium region. Instead, biochemical (e.g.,  plants and anthropogenic pollution) and geological (e.g. volcanic outgassing) fluxes provide surface sources and sinks that are key to regulate the atmosphere. For the lower boundary condition, global emission budget provides estimates for the surface fluxes, which is conventionally recorded in the unit of mass rate (Tg yr$^{-1}$) and needed to convert to flux (molecules cm$^{-2}$ s$^{-1}$) in our 1-D model. 

For Earth and any ocean worlds with large bodies of surface water reservoir, the standard setup is to fix the surface-water mixing ratio \citep{Kasting1980,Hu2012,Lincowski2018}. We set the surface mixing ratio of water to 0.00894, corresponding to 25$\%$ relative humidity. Surface \ce{CO2} is also fixed at 400 ppm for simplicity, since we did not consider several major sources and sinks of \ce{CO2} at the surface, such as respiration, photosynthesis, ocean uptake, weathering, etc. The specific emission fluxes and deposition velocities for the lower boundary are listed in Table \ref{tab:BC_Earth}, while zero-flux boundary is assumed for all remaining species. We initialize the atmospheric composition with well-mixed (constant with altitude) 78$\%$ \ce{N2}, 20$\%$ \ce{O2}, 400 ppm \ce{CO2}, 934 ppm Ar, and 0.2 ppb \ce{SO2}.

For the solar flux, we adopt a recently revised high-resolution spectrum \citep{Gueymard2018}, which is derived from various observations and models (see Table 1. of \cite{Gueymard2018}). The solar radiation was cut from 100 nm in \cite{Hu2012} for the missing absorption from the thermosphere. We do not find it necessary as we set the top layer to the lower thermosphere around 110 km and the EUV absorption is naturally accounted for. We have also tried including the absorption from atomic oxygen and nitrogen and found no differences regarding the neutral chemistry in the lower atmosphere, since \ce{N2} and \ce{O2} have already screened out the bulk EUV. The chlorine chemistry and lightning sources for odd nitrogen are not included in this validation. 

\subsubsection{Results}

\begin{figure}
\begin{center}
\includegraphics[width=\columnwidth]{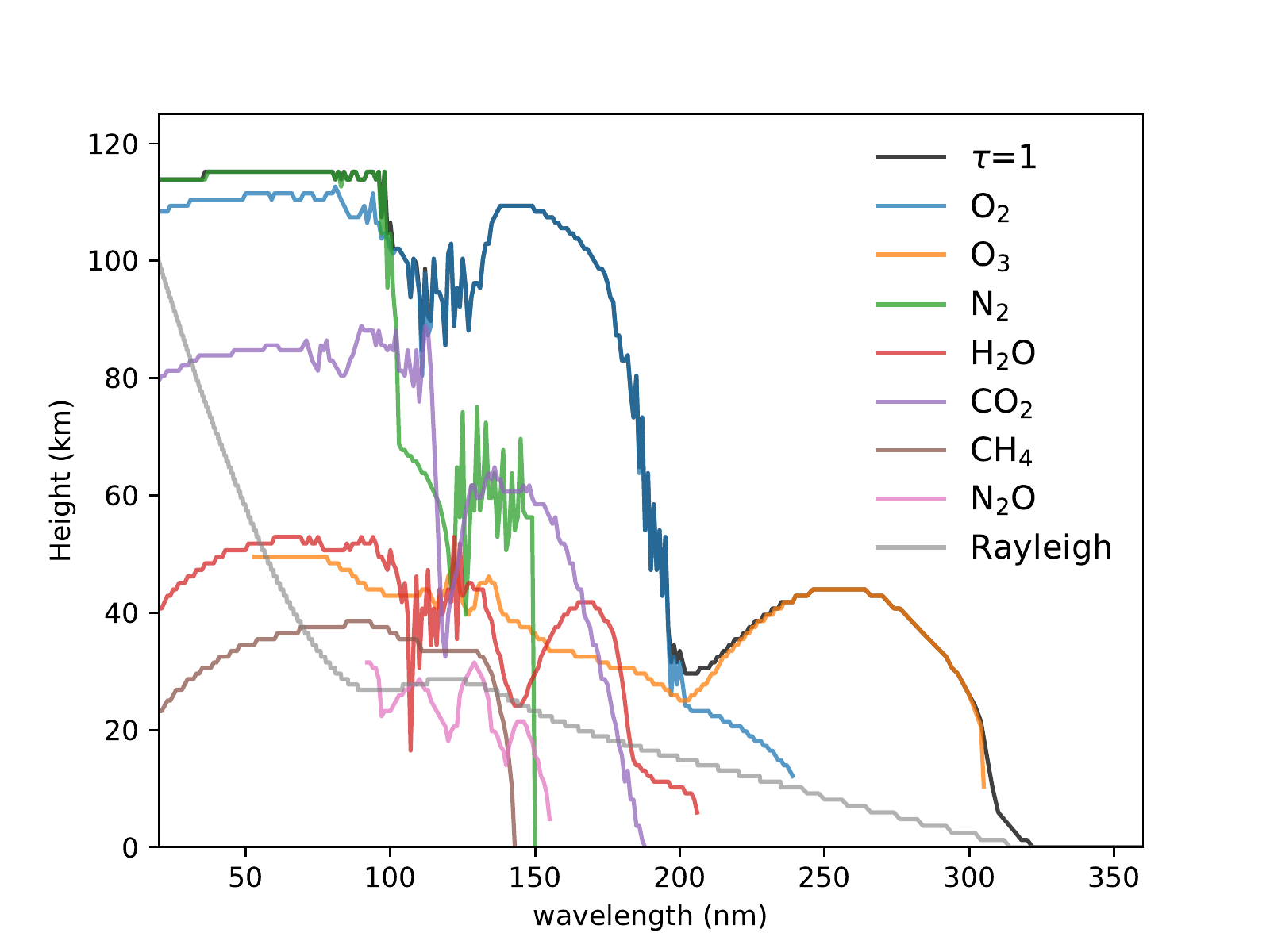}
\end{center}
\caption{The UV photosphere i.e. optical depth $\tau$ = 1 (black) in our Earth model, overlaid with the composition-decomposed photosphere for several key molecules.}
\label{fig:tau_Earth}
\end{figure}

\begin{figure}[htp]
\begin{center}
\includegraphics[width=\columnwidth]{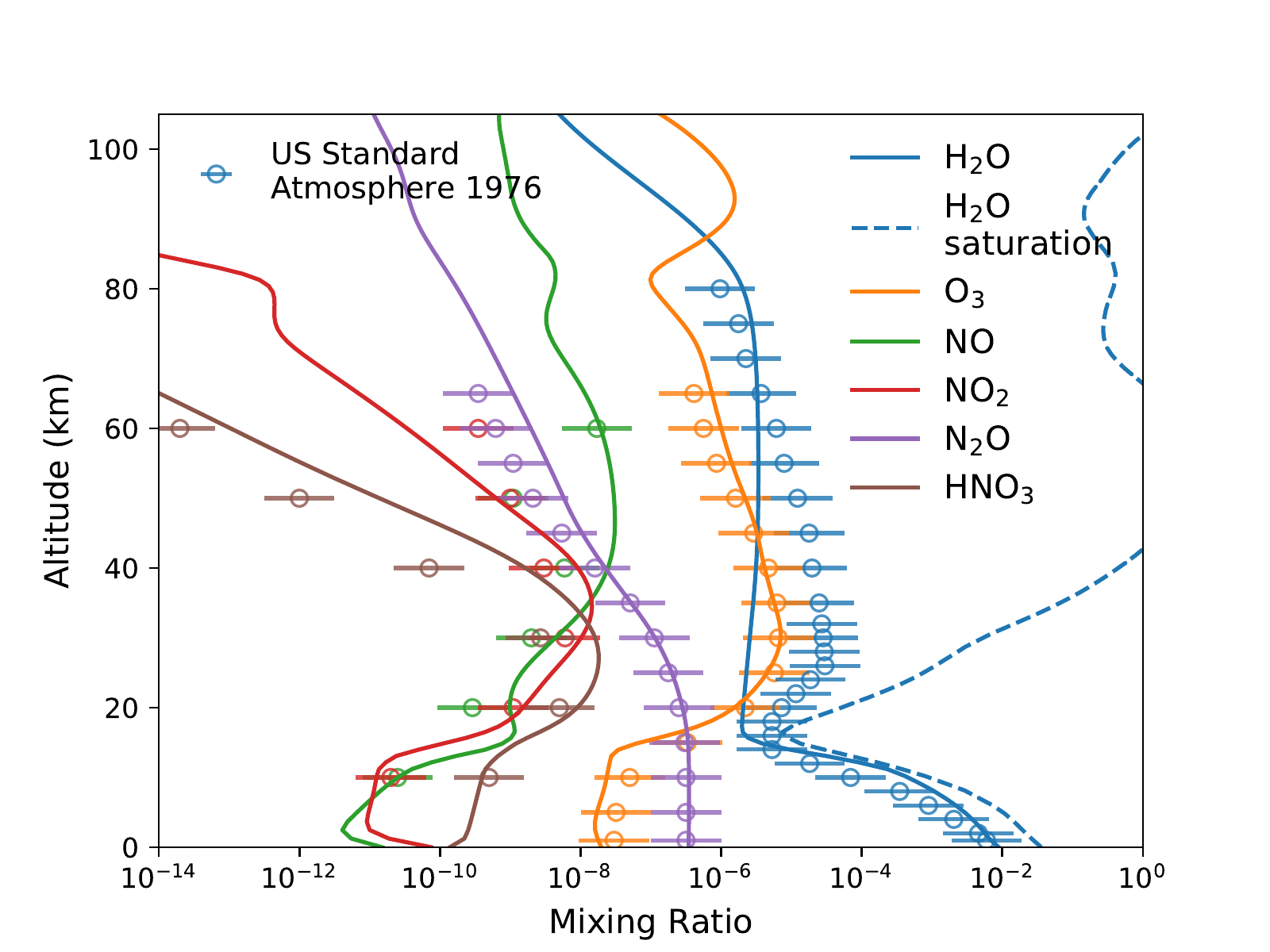}
\includegraphics[width=\columnwidth]{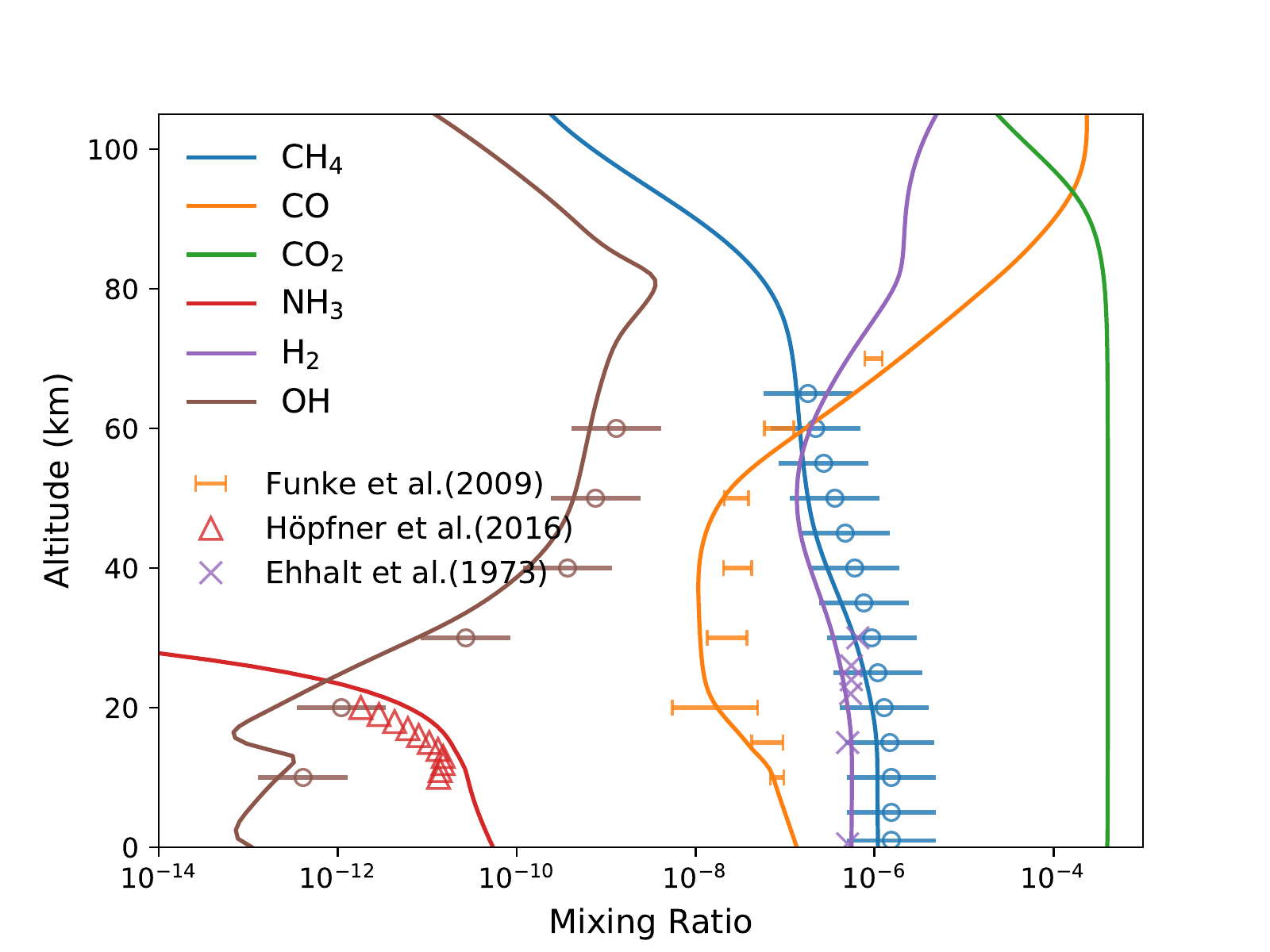}
\includegraphics[width=\columnwidth]{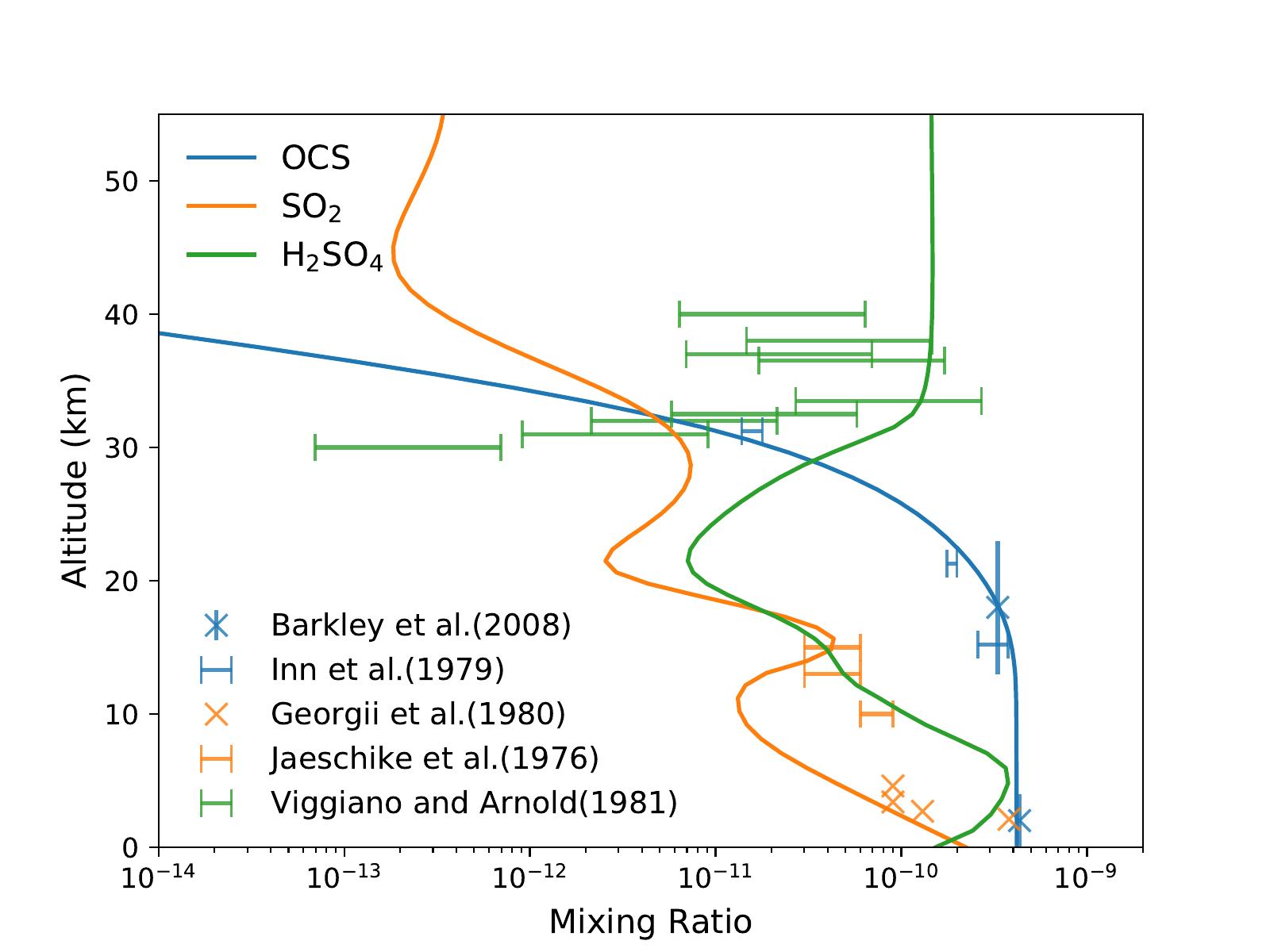}
\end{center}
\caption{The global average vertical distribution of key compositions in present-day Earth's atmosphere compared to observations. The \ce{H2O} mixing ratio is from the US Standard Atmosphere 1976\textsuperscript{*}. Satellite observations of CO in the tropics and \ce{NH3} within 30$^\circ$-- 40$^\circ$ N and 70$^\circ$-- 80$^\circ$ E in 2003 are measured by the Michelson Interferometer for Passive Atmospheric Sounding (MIPAS) \citep{Fischer2008}. The rest unlabelled observational data are from \cite{Massie1981,Hudson1979}. When errors are not included in the published observations, we follow \cite{Hu2012} placing one order of magnitude error bars for the diurnal and spatial variations.}
\small\textsuperscript{*}e.g. \url{https://www.digitaldutch.com/atmoscalc/help.htm}
\label{fig:earth-mix}
\end{figure}

Molecular oxygen (\ce{O2}) and ozone (\ce{O3}) are the main players in Earth's photochemistry. \ce{O2} absorbs VUV below 200 nm and \ce{O3} takes up the radiation longward than about 200 nm, which blocks the harmful UV from life on the surface. The penetration level of solar UV flux shown in Figure \ref{fig:tau_Earth} indicates that ozone absorbs predominately between 20 km and 50 km. The basics of oxygen--ozone cycle are described by the Chapman mechanism \citep[e.g., ][]{Yung1999,Jacob2011}. Our full chemical network encompasses the catalytic cycles involving hydrogen oxide and nitrogen oxide radicals that are responsible for the ozone sinks in the stratosphere. Although the catalytic cycle of chlorine which accounts for additional ozone loss is not included, we are able to reproduce the observed global average ozone distribution in Figure \ref{fig:earth-mix}. 

Our condensation scheme captures the cold trap of water in the troposphere, i.e. the water vapor entering the stratosphere is set by the tropopause temperature. Above the tropopause, water is supplied by diffusion transport from the troposphere and oxidation of \ce{CH4}. We find the conversion in the stratosphere go through the steps
\begin{eqnarray}
\begin{aligned}
\ce{CH4 + OH &-> H2O + CH3}\\
2(\ce{O3 &->[h\nu] O2 + O(^1D)})\\
2(\ce{O(^1D) + N2 &-> O + N2})\\
\ce{CH3 + O &-> H2CO + H}\\
\ce{O + H2CO &-> HCO + OH}\\
\ce{HCO + O2 &-> CO + HO2}\\
\ce{CO + OH &-> CO2 + H}\\
\ce{HO2 + OH &-> H2O + O2}\\
2(\ce{H + O2 &->[M] HO2})\\
\noalign{\vglue 5pt} 
\hline
\noalign{\vglue 5pt} 
\textrm{net} : \ce{CH4 + 2OH + 2O3 &-> CO2 + 2H2O + 2HO2}
\end{aligned}
\label{re:CH4-H2O}
\end{eqnarray}
, effectively turning one \ce{CH4} molecule into two \ce{H2O} molecules \citep{Noel2018}. \ce{H2O} eventually photodissociated in the mesosphere and produced \ce{H2}, as indicated by the profiles in Figure \ref{fig:earth-mix}. Overall, our model produces water distribution consistent with observations considering the diurnal and spatial variations. 

The two oxides of nitrogen, NO and \ce{NO2}, cycle rapidly in the presence of ozone: 
\begin{subequations}
\label{re:NOx}
\begin{align}
\begin{split} 
\ce{NO + O3 &-> NO2 + O2}\\
\ce{NO2 + O &-> NO + O2}\\
\hline \nonumber
\mbox{net} : \ce{O3 + O &-> 2O2} 
\end{split} 
\end{align}
\end{subequations}
Thus NO and \ce{NO2} are conventionally grouped as \ce{NO_x}. The burning of fossil fuel accounts for about half of the global tropospheric emission (e.g. Table 2.6 of \cite{Seinfeld2016}). \ce{NO_x} is mainly lost by oxidation into nitric acid (\ce{HNO3}): \ce{NO2 + OH ->[\textrm{M}] HNO3}. Our model reproduces distribution of \ce{NO_x}, whereas our higher \ce{HNO3} in the upper stratosphere is seemingly attributed to missing the hydration removal in the actual atmosphere. 

Nitrous oxide (\ce{N2O}) is mainly emitted by soil bacteria, prescribed by the surface emission at the lower boundary. There is no efficient \ce{N2O} removal reactions in the troposphere
and \ce{N2O} remains well-mixed as one of the important greenhouse gases. \ce{N2O} is predominantly removed by photodissociation in the stratosphere. Our calculated \ce{N2O} is in agreement with the observations for the troposphere and stratosphere. Although similar to \cite{Hu2012}, our model slightly overpredicts its abundance above 50 km, which is likely due to missing the photolysis branch of \ce{N2O} that produces excited oxygen \ce{O(^1S)}.  

\ce{CH4} is the most abundant hydrocarbon in Earth's atmosphere, with the surface emission largely comes from human activities (e.g. agriculture) as well as natural sources (e.g. wetlands). \ce{CH4} is oxidized into CO and eventually \ce{CO2} by OH through multiple steps similar to  (\ref{re:CH4-H2O}) in the stratosphere. CO is produced by combustion activities with about 0.1 ppm concentration near the surface \cite{Seinfeld2016}, as a result of the balance among the emission flux, OH oxdization, and dry deposition. CO is continuously removed by OH through the troposphere and generated by photodissociation of \ce{CO2} in the thermosphere and mesosphere, as depicted by their distributions in Figure \ref{fig:earth-mix}. As the major oxidizing agent, OH is an important diagnostic species for Earth's photochemical model. It is mainly produced in the stratosphere during daytime initiated by ozone photolysis and regenerated in the troposphere by \ce{NO_x} \citep[see e.g., ][]{Jacob2011}. The OH distribution in our model is consistent with that in \cite{Massie1981}. We will further discuss using calculated OH concentration to estimate the chemical timescale of long-lived species against oxidation in the next section. 

Carbonyl sulfide (OCS) is the main sulfur species in the troposphere, emitted by direct outgassing or oxidation of carbon disulfide (\ce{CS2}) and dimethyl sulfide (DMS) released by the ocean \citep{Seinfeld2016,Barkley2008}. OCS is rather stable in the troposphere until entering the stratosphere where it is photodissociated or oxidized by OH and ultimately turned into sulfuric acid. Sulfur dioxide (\ce{SO2}) is another important sulfur containing pollutant from fossil fuel combustion. \ce{SO2} oxidation begins from the troposphere with 
\begin{equation*}
\ce{SO2 + OH ->[\textrm{M}] HSO3}
\end{equation*}
\ce{HSO3} radical rapidly reacts with oxygen to form \ce{SO3}
\begin{equation*}
\ce{HSO3 + O2 -> SO3 + HO2}
\end{equation*}
followed by sulfuric acid formation
\begin{equation*}
\ce{SO3 + H2O -> H2SO4}
\end{equation*}
The sulfur-containing gases in our model generally agree with the global distribution, while the mismatch of \ce{H2SO4} is expected as our model does not include 
\ce{H2SO4} photodissociation and heterogeneous reactions that efficiently remove \ce{H2SO4} from the gas phase. 

\begin{figure}[htp]
\begin{center}
\includegraphics[width=\columnwidth]{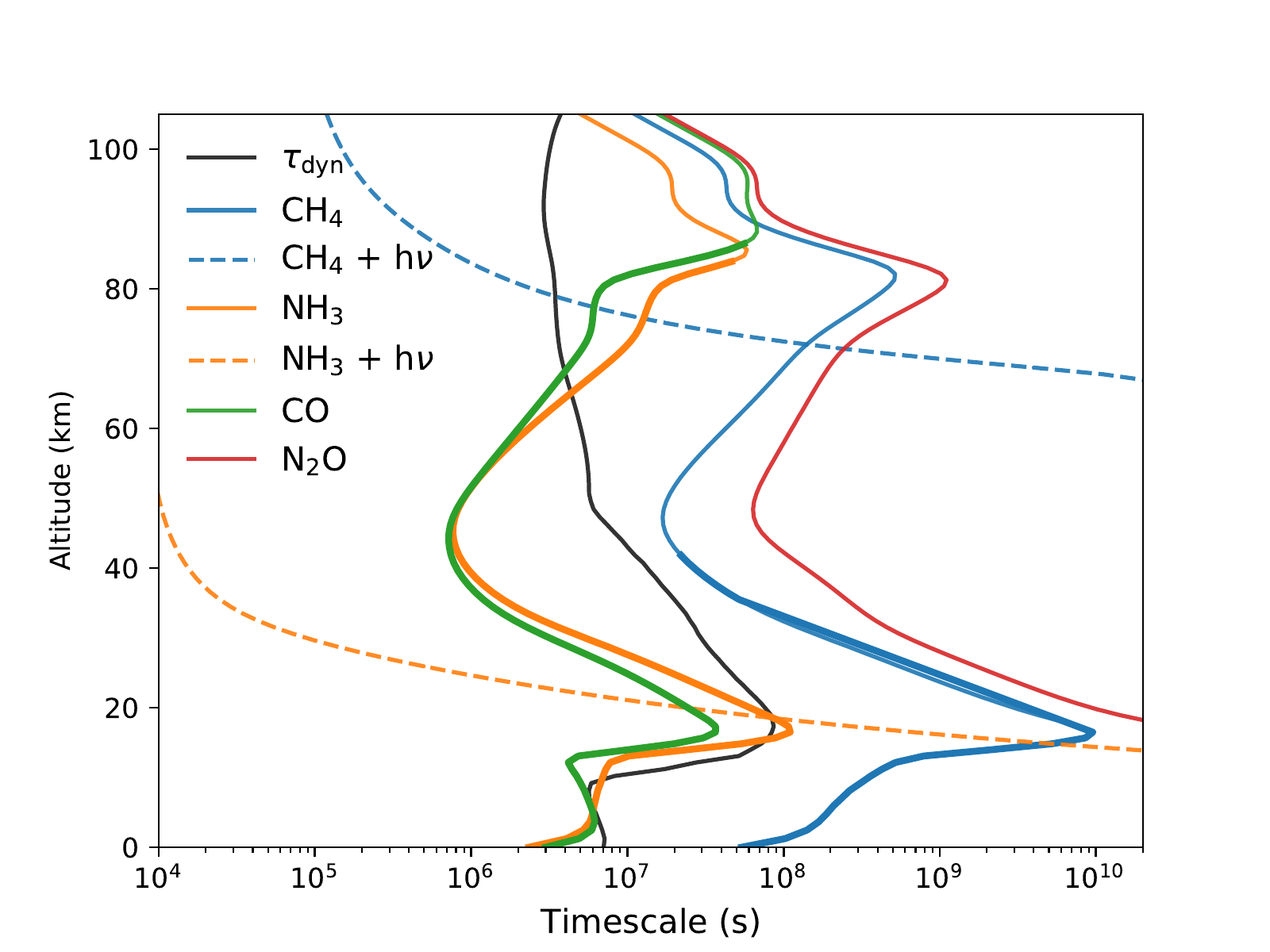}
\end{center}
\caption{Calculated chemical timescales of some environmentally important gases compared to the dynamical timescale of eddy diffusion in the Earth validation model. The thick lines indicate the region where the oxidation is dominated by OH (i.e. $\tau_{\ce{OH}}$ $\simeq$ $\tau_{\ce{chem}}$).}\label{fig:earth-tau}
\end{figure}

\subsubsection{Chemical Lifetime}
The oxidizing capacity of Earth's atmosphere is important for decontaminating toxic and greenhouse gases, such as CO, \ce{CH4}, and various volatile organic compounds . The oxidizing power is not only essential for regulating habitable conditions but also key to address the stability of biosignature gases for other terrestrial planets. Here we present a brief overview of the key timescales for some important trace gases from our Earth model.  

OH radical is the primary daytime oxidizing agent in our biosphere. The chemical timescale of species A against oxidization ($\tau^{\ce{A}}_{\ce{OH}}$) can be estimated by the computed OH concentration as 
\begin{equation}
\tau^{\ce{A}}_{\ce{OH}} = \frac{[\ce{A}]}{k_{\ce{A}-\ce{OH}}[\ce{A}][\ce{OH}]} = \frac{1}{k_{\ce{A}-\ce{OH}}[\ce{OH}]}
\end{equation}
where $k_{\ce{A-OH}}$ is the rate coefficient of the oxidizing reaction of \ce{A + OH}. In the upper atmosphere where molecular collision is less frequent, the excited \ce{O(^1D)} produced by ozone photolysis is not immediately stablized and becomes the main oxidant. We consider the two major oxidizing paths across the atmosphere and write the chemical timescale against oxidation as
\begin{equation}
\tau^{\ce{A}}_{\textrm{chem}} = \frac{1}{k_{\ce{A}-\ce{OH}}[\ce{OH}] + k_{\ce{A}-\ce{O(^1D)}}[\ce{O(^1D)}]}
\end{equation}
Figure \ref{fig:earth-tau} illustrates the chemical timescales ($\tau_{\textrm{chem}}$) along with photolysis timescales (1/$k_{\textrm{photo}}$) for several trace gases, where $\tau_{\ce{OH}}$ (thick lines) inversely correlates with temperature in general. We can gain some insights by comparing ($\tau_{\textrm{chem}}$) to the dynamical timescale of vertical mixing ($\tau_{\textrm{dyn}}$ = H$^2$/K$_\textrm{zz}$): In the troposphere, \ce{CH4} and \ce{N2O} display rather well-mixed abundances due to their longer chemical lifetime. CO and \ce{NH3} have comparable $\tau_{\textrm{chem}}$ with $\tau_{\textrm{dyn}}$ and and exhibit negative gradient with altitude from oxidation removal. In the stratosphere, \ce{NH3} is rapidly photodissociated while \ce{CH4} is transported from the troposphere and oxidized into CO. In the thermosphere above $\sim$ 80 km, the oxidation by \ce{O(^1D)} takes over for most species, but mixing processes with a shorter timescale here controls the chemical distribution. For example, CO abundance starts to increase with altitude from about 60 km as a result of downward transport of CO produced by \ce{CO2} photodissociation in the upper atmosphere.  

In summary, we validate our photochemical model with HD 189733b, Jupiter, and Earth, for a wide range of temperatures and oxidizing states. The inclusion of nitrogen and sulfur chemistry, along with the implementation of advection, condensation, and boundary conditions are verified by comparing with models and/or observations. The discrepancies in previous models of HD 189733b are identified for future investigation.

\begin{table}[t]
\setlength\tabcolsep{1pt} 
\caption{Parameters of the planetary systems.}
\begin{tabular}{p{1.8cm}p{1.6cm}p{1.75cm}p{1.25cm}p{1.25cm}} 
\hline
Parameter & WASP-33b & HD 189733b & GJ 436b & 51 Eri b\\
a\footnote{orbital distance} (AU) & 0.02558 & 0.03142 & 0.02887 & 11.1 \\
T$_\textrm{int}$ (K) & 200 & --- & 100/400 & 760\\
R$_\textrm{s}$ (R$_\odot$) & 1.51 & 0.805 & 0.464 & 1.45\\
R$_\textrm{p}$ (R$_\textrm{J}$) & 1.603  & 1.138 & 0.38 & 1.11\\
g\footnote{gravity at 1 bar level} (cm$^2$/s) & 2700  & 2140 & 1156 & 18197 \\
$\overline{\theta}$\footnote{mean stellar zenith angle} & 58 & 48 & 58 & 67\\
stellar type & A5 & K1-K2 & M2.5 & F0\\
\hline
\end{tabular}
\label{tab:planet_para}
\end{table}

\section{Case study}\label{case}
In this section, we select WASP-33b (ultra-hot Jupiter), HD 189733b (hot Jupiter), GJ 436b (warm Neptune), and 51 Eridani b (young Jupiter) to perform case studies. Each case represents a distinctive class among gas giants with \ce{H2}-dominated atmospheres. The effective temperatures of these objects span across 700--3000 K while having host stars of various stellar types. We investigate how disequilibrium processes play a part for these cases with additional attention on the effects of sulfur chemistry and photochemical haze precursors. 

\begin{figure}[!ht]
\begin{center}
\includegraphics[width=\columnwidth]{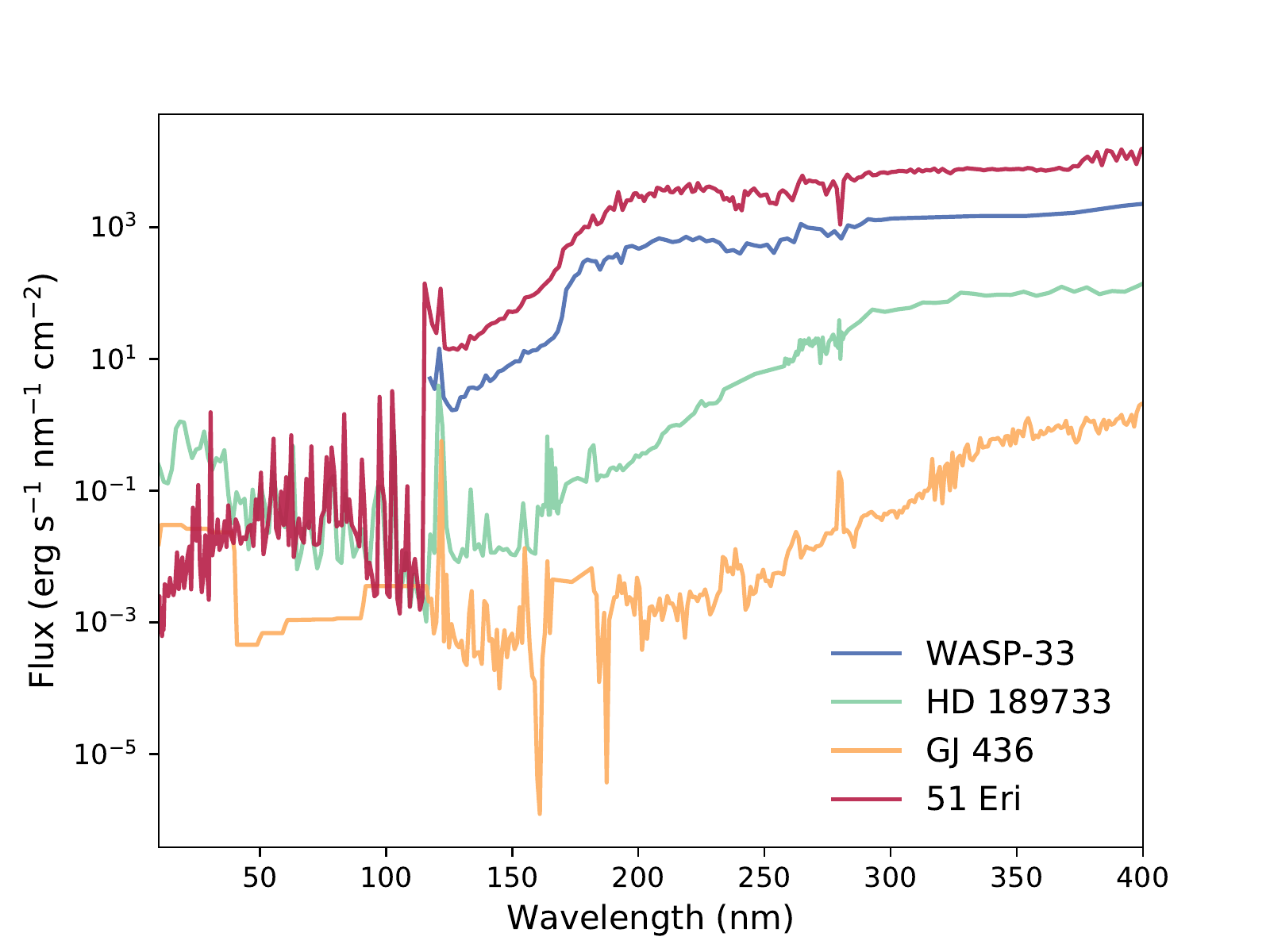}
\includegraphics[width=\columnwidth]{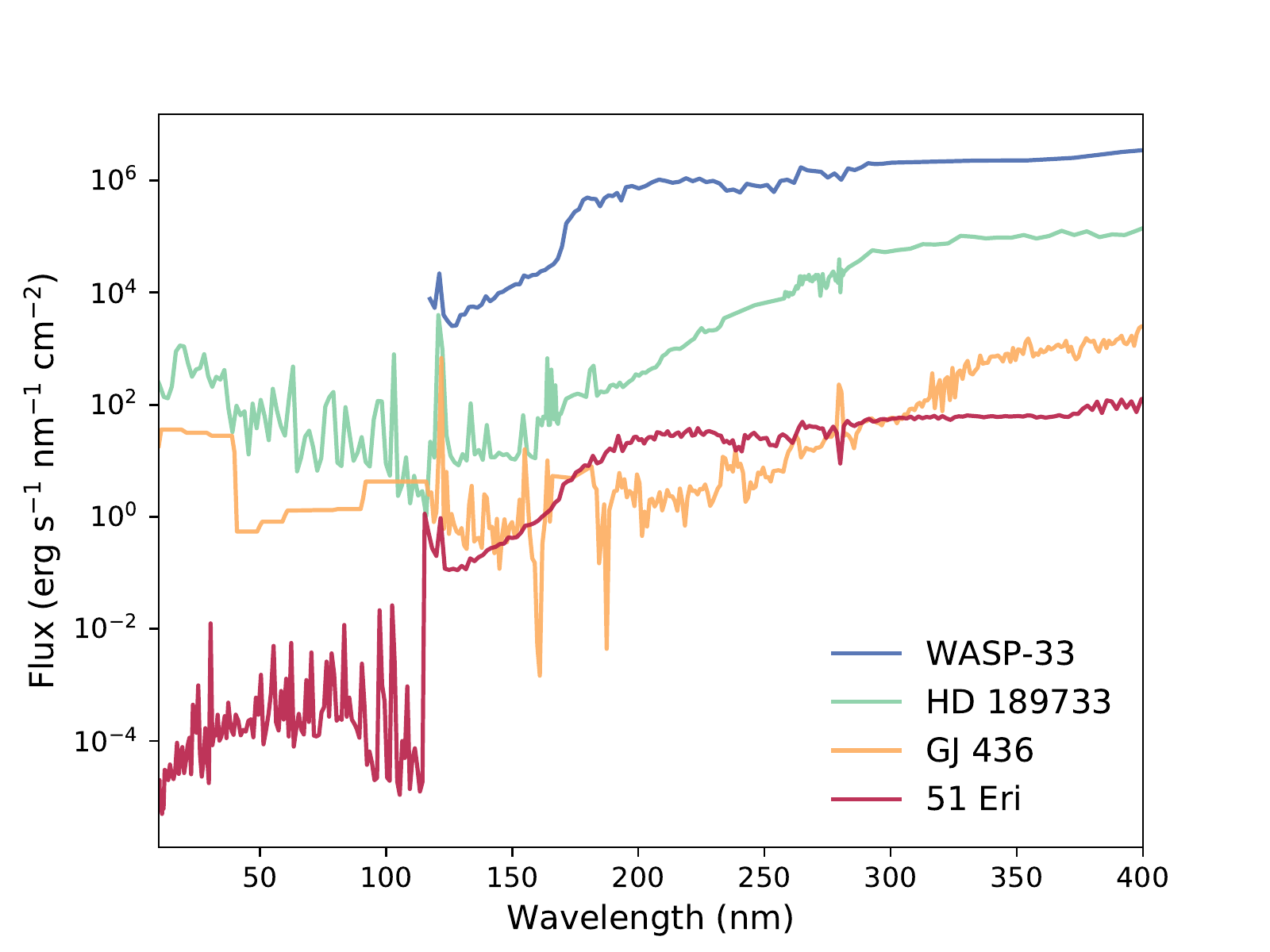}
\end{center}
\caption{Stellar UV fluxes normalized at 1 AU (top) and at the top of the planet's atmosphere (bottom) adopted in our case study models.}
\label{fig:case-sflux}
\end{figure}

All the P-T profiles in this section are generated using the open-source radiative-transfer model, HELIOS, except we keep the same P-T profile of HD 189733b as in Section \ref{sec:hd189} for comparative purposes. HELIOS employs two-stream approximation and correlated-k method to solve for the radiative-convective equilibrium temperature consistent with thermochemical equilibrium abundances. The gaseous opacities include \ce{H2O}, \ce{CH4}, CO, \ce{CO2}, \ce{NH3}, \ce{HCN}, \ce{C2H2}, NO, SH, \ce{H2S} \ce{SO2}, \ce{SO3}, SiH, CaH, MgH, NaH, AlH, CrH, AlO, SiO, CaO, CIA$_{\ce{H2-H2}}$, CIA$_{\ce{H2-He}}$, and additionally TiO, VO, Na, K, H- for WASP-33b. The P-T profiles are fixed without taking into account of the radiative feedback from disequilibrium chemistry (but see \cite{Drummond2016} for the effects on HD 189733b). The astronomical parameters  used are listed in Table \ref{tab:planet_para}. The dayside-average stellar zenith angle is used for WASP-33b and GJ436b and the global-average stellar zenith angle is used for 51 Eri b (see Appendix \ref{app:mu}), except that we keep the same value for HD 189733b to compare with the results in Section \ref{sec:hd189}. The stellar UV fluxes adopted for each system are compared in Figure \ref{fig:case-sflux}, with detailed description in each section.

For the eddy diffusion ($K_{\textrm{zz}}$) profiles in our case studies (except that we again retain the same profile for HD 189733b from \cite{Moses11}), we assume $K_{\textrm{zz}}$ to be constant in the convective region and increasing roughly with inverse square root of pressure in the stratosphere \citep{Lindzen1981,Vivien2013}. The expression as a function of pressure in bar ($P_{\textrm{bar}}$) takes a similar form as \cite{Charnay2015} or \cite{Moses2016}:
\begin{equation}
K_{\textrm{zz}} = K_{\textrm{deep}} (\frac{P_{\textrm{tran}}}{P_{\textrm{bar}}})^{0.4},
\label{eq:Kzz}
\end{equation}
where $P_{\textrm{tran}}$ is the transition pressure level informed by the radiative transfer calculation. The more irradiated atmospheres have deeper radiative-convective transition levels and greater P$_{\textrm{tran}}$. A common way of estimating $K_{\textrm{deep}}$ in the convective region is applying the mixing length theory with the knowledge of convective heat flux. For WASP-33b, most of the modeled atmosphere is in the radiative region. We choose $K_{\textrm{deep}}$ such that the overall pressure-dependent $K_{\textrm{zz}}$ profile matches that derived from the vertical wind in the general circulation model (GCM). For GJ 436b, $K_{\textrm{deep}}$ is treated as a loosely constrained free parameter along with the internal heating we explored (Section \ref{sec:GJ436b_input}). $K_{\textrm{deep}}$ is likely more important in controlling the quenched species for cooler planets, such as 51 Eri b, where we adopted a value of $K_{\textrm{deep}}$ that can produce quenched \ce{CH4} consistent with the observations.

We run nominal models for all planets in this section with the S-N-C-H-O chemical network\footnote{included in the supplementary material}. We recognize there are considerable uncertainties in sulfur kinetics, as discussed in Section \ref{sec:network}. In order to gauge the uncertainty effects of our sulfur scheme, we explore the sensitivity to sulfur chain-forming reactions for GJ 436b and OCS recombination for 51 Eridani b. After chemical abundances are obtained, we use the open-source tool PLATON \citep{Zhang2019,Zhang2020} to generate transmission spectra and HELIOS for the emission spectra.

\subsection{WASP-33b}
WASP-33b is among the hottest gas giants discovered with dayside temperature around 3000 K \citep{Essen2020}. To date it remains the only case showing evidence of both temperature inversion and \ce{TiO} features \citep{Serindag2021}, which makes WASP-33b an interesting target for testing the stability of TiO/VO along with other molecules. Previous work on ultra-hot Jupiters are limited by the assumption of chemical equilibrium chemistry \citep{Kitzmann2018,parmentier18,Zhang2018}. Here, we will verify the equilibrium assumption by exploring how disequilibrium processes affect the titanium and vanadium compounds with different C/O ratios.

\subsubsection{Stellar UV-flux and Eddy Diffusion}
The host star WASP-33 is an A5 type star with effective temperature about 7400 K. We use the 
UV spectrum of HD 40136 (F0 type) merged with a 7000 K atlas spectrum from \cite{Sarah2013} as an analogue. The star is fast rotating and exhibits pulsations, which might add more uncertainties to the UV flux. Nevertheless, as we will see in Section \ref{sec:wasp33b-NEQ}, photodissociation solely converts more molecules to atoms at this high temperature and the results should be qualitatively robust.   

\begin{figure}[pth]
\begin{center}
\includegraphics[width=\columnwidth]{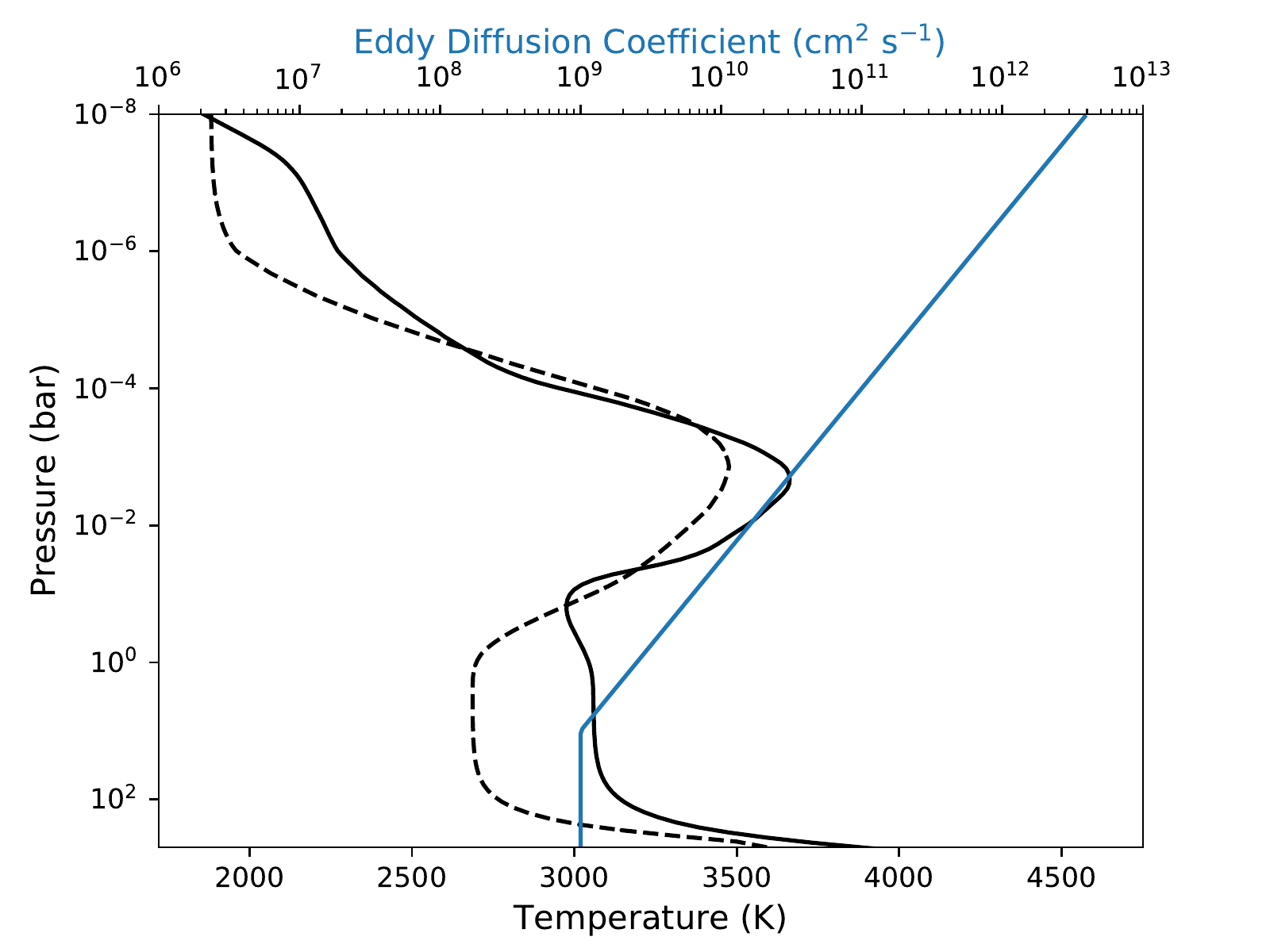}
\end{center}
\caption{The temperature-pressure and eddy diffusion ($K_\textrm{zz}$) profiles for WASP-33b. Solar elemental abundance (solid) and C/O = 1.1 (dashed) are assumed for calculating the temperature structure.} 
\label{fig:TP-wasp33b}
\end{figure}

Vertical wind generally correlates with the planet's effective temperature \citep{Tan2019,Komacek2019,Baxter2021}. We assume the value of $K_{\textrm{zz}}$ based on the simulations in \cite{Tan2019}, where the global RMS vertical wind increases with decreasing pressure and reaches about 100 m/s at 1 mbar (personal communication). The vertical wind translates to K$_\textrm{zz}$ $\sim$ 10$^{11}$ cm$^2$s$^{-1}$ around 1 mbar. The temperature and eddy diffusion profiles for WASP-33b are shown in Figure \ref{fig:TP-wasp33b}.

\subsubsection{Chemical Equilibrium} 
\begin{figure}[tph]
\begin{center}
\includegraphics[width=\columnwidth]{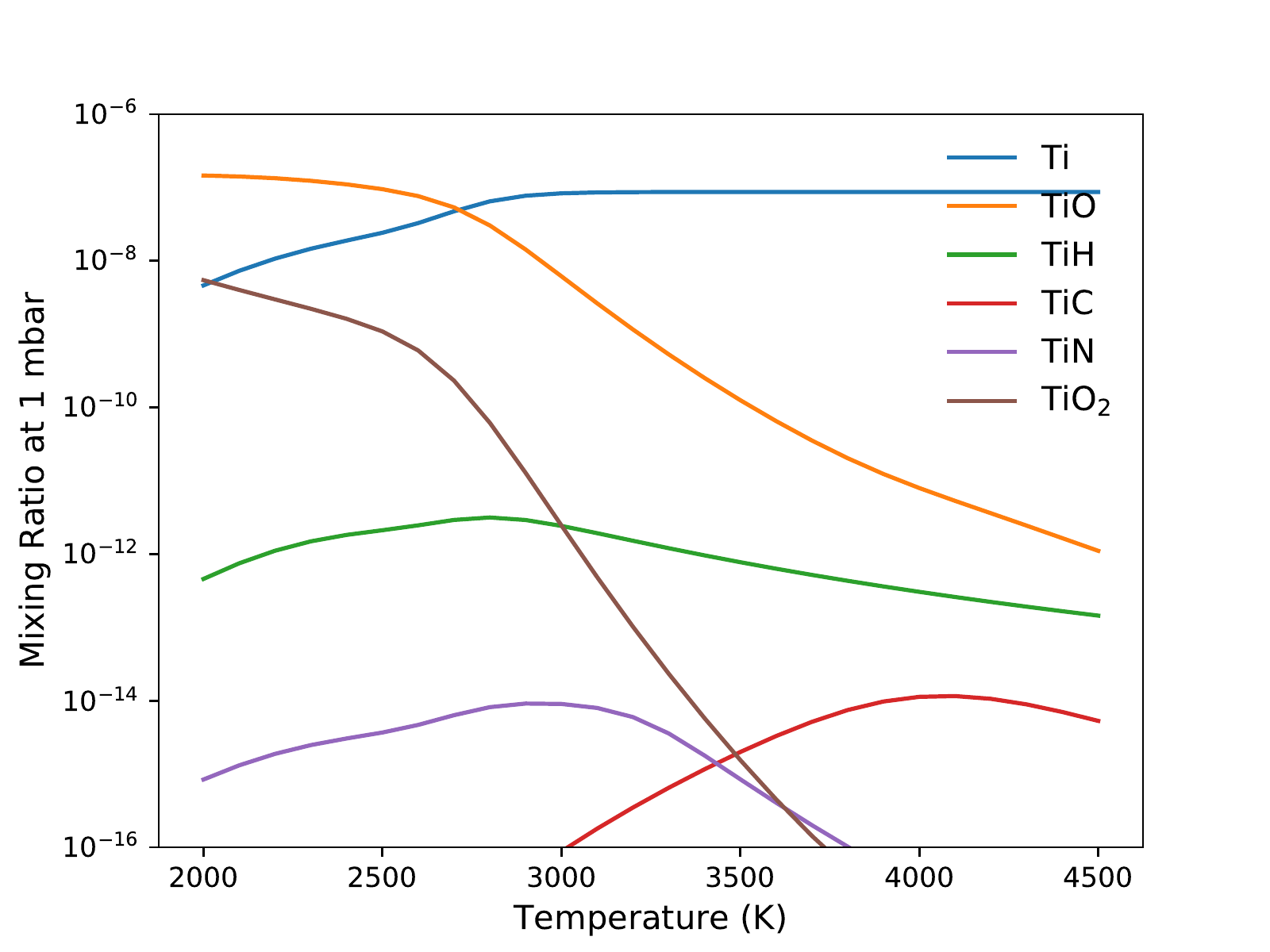}
\includegraphics[width=\columnwidth]{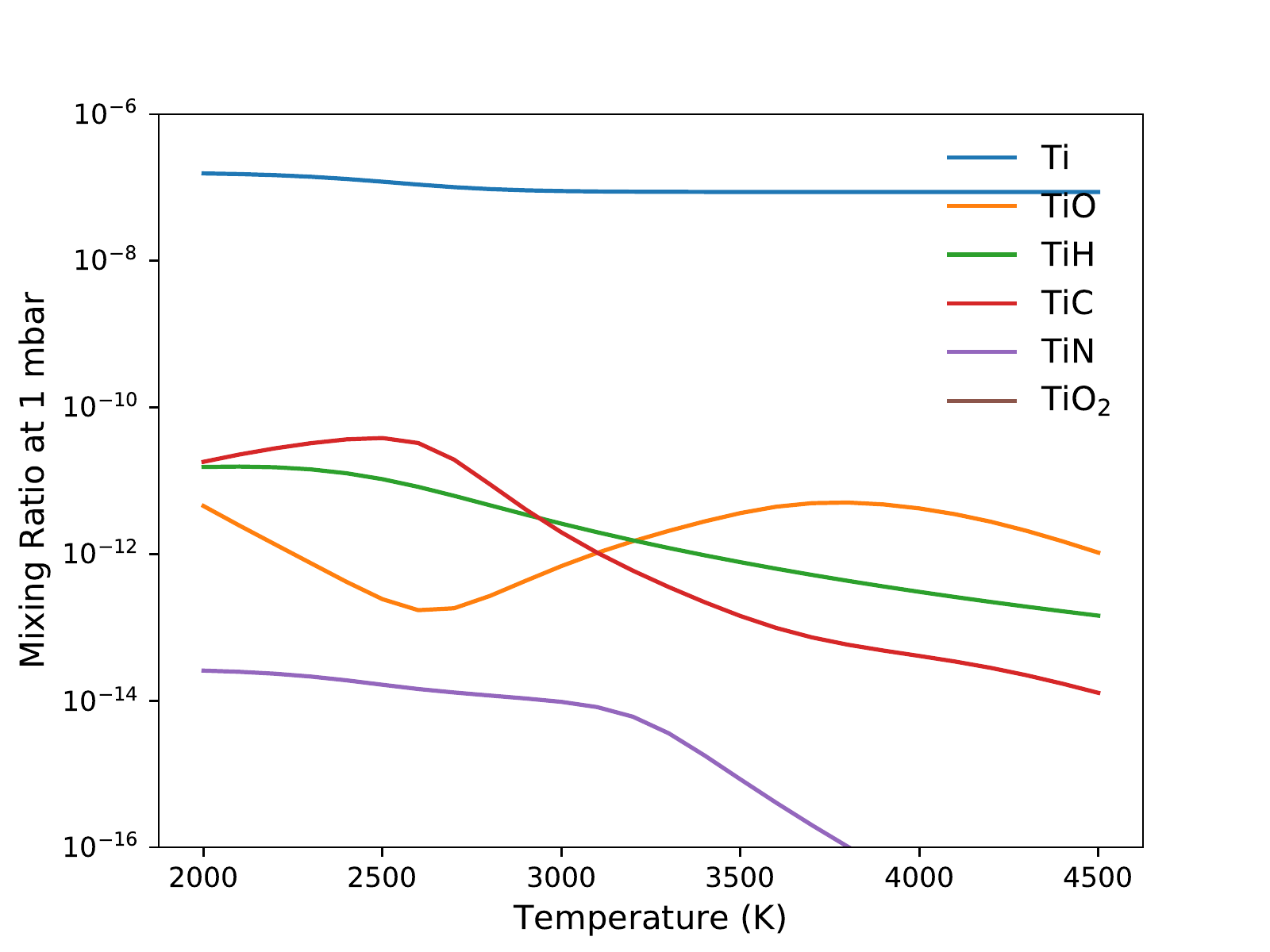}
\end{center}
\caption{The equilibrium mixing ratios of several gas phase titanium species at 1 mbar as a function of temperature for solar elemental abundance (top) and C/O = 1.1 (bottom).}
\label{fig:Ti-EQ}
\end{figure}

We first look at the trends associated with thermal dissociation governed by thermochemical equilibrium under carbon-poor and carbon-rich conditions, for which we assume a solar C/O and C/O = 1.1, respectively. Figure \ref{fig:Ti-EQ} illustrates how titanium compounds vary with temperature in equilibrium at 1 mbar. For solar C/O, titanium mainly exists in the form of Ti and TiO. As temperature exceeds about 2500 K, TiO becomes unstable against thermal dissociation and its abundance falls with temperature. For C/O = 1.1, TiO is depleted due to the scarcity of oxygen, as oxygen preferably combines with the excess carbon to form CO \citep{Madhu12}. Atomic titanium is the major species across this temperature range and TiC, TiH, and TiO have close abundances. 
 
The effects of thermal dissociation on WASP-33b are clearly visible in the equilibrium profiles in Figure \ref{fig:wasp33b-mix}. The blistering heat of WASP-33b makes all elements predominantly exist in the atomic form above 0.1 bar, where temperature starts to increase with altitude and exceeds 3000 K, while CO with the strong C--O bond is the only molecule that survives the high temperature. For solar C/O ratio, as the majority of C is locked in CO,  atomic C tracks the temperature structure whereas oxides such as \ce{H2O}, \ce{TiO}, VO, and \ce{TiO2} display inverse trends with temperature. For C/O = 1.1, atomic O swaps place with C and \ce{TiO} and VO are significantly depleted.

\subsubsection{The effects of Disequilibrium Chemistry}\label{sec:wasp33b-NEQ}
\begin{figure*}[tph]
\begin{center}
\includegraphics[width=\columnwidth]{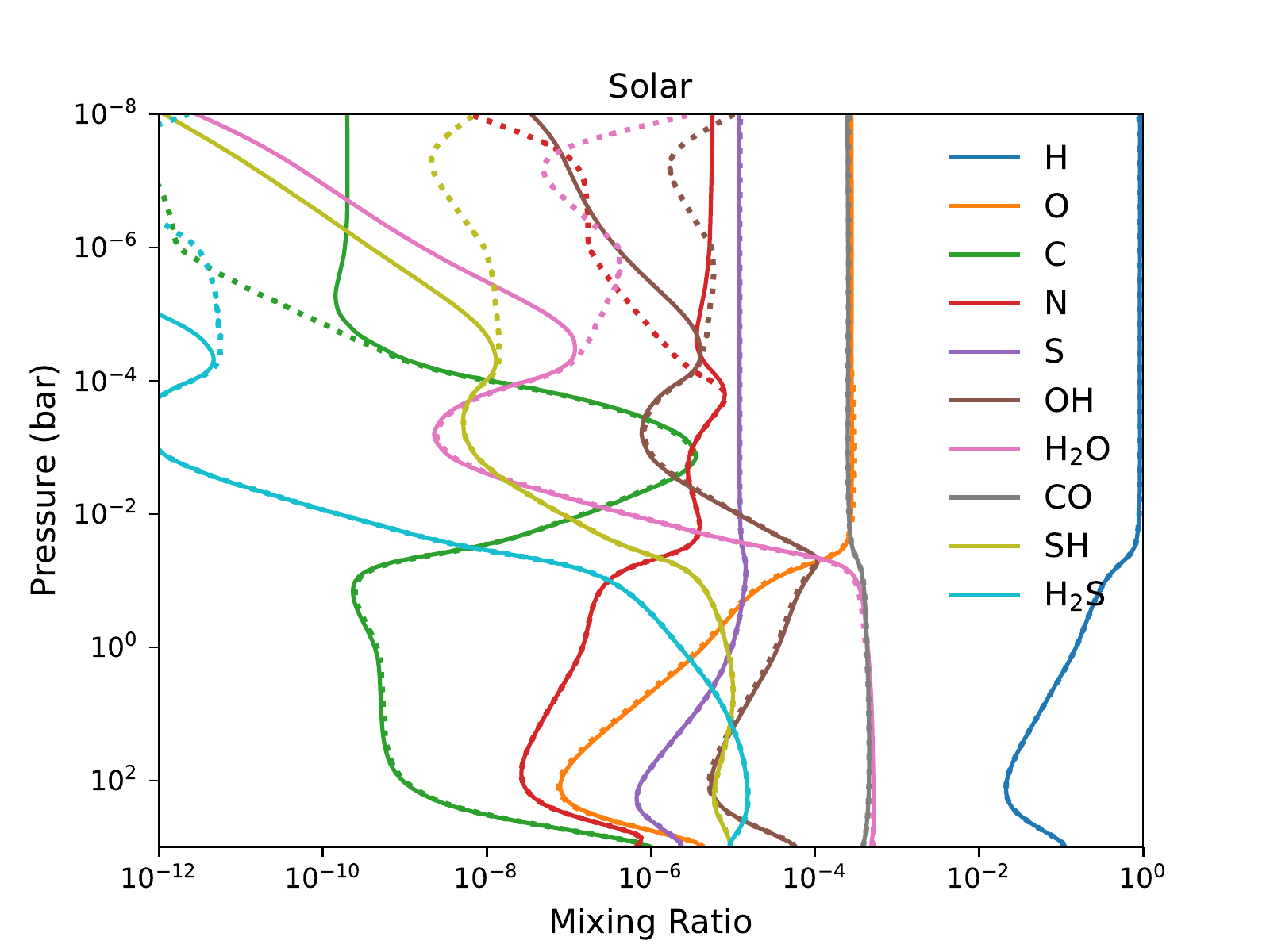}
\includegraphics[width=\columnwidth]{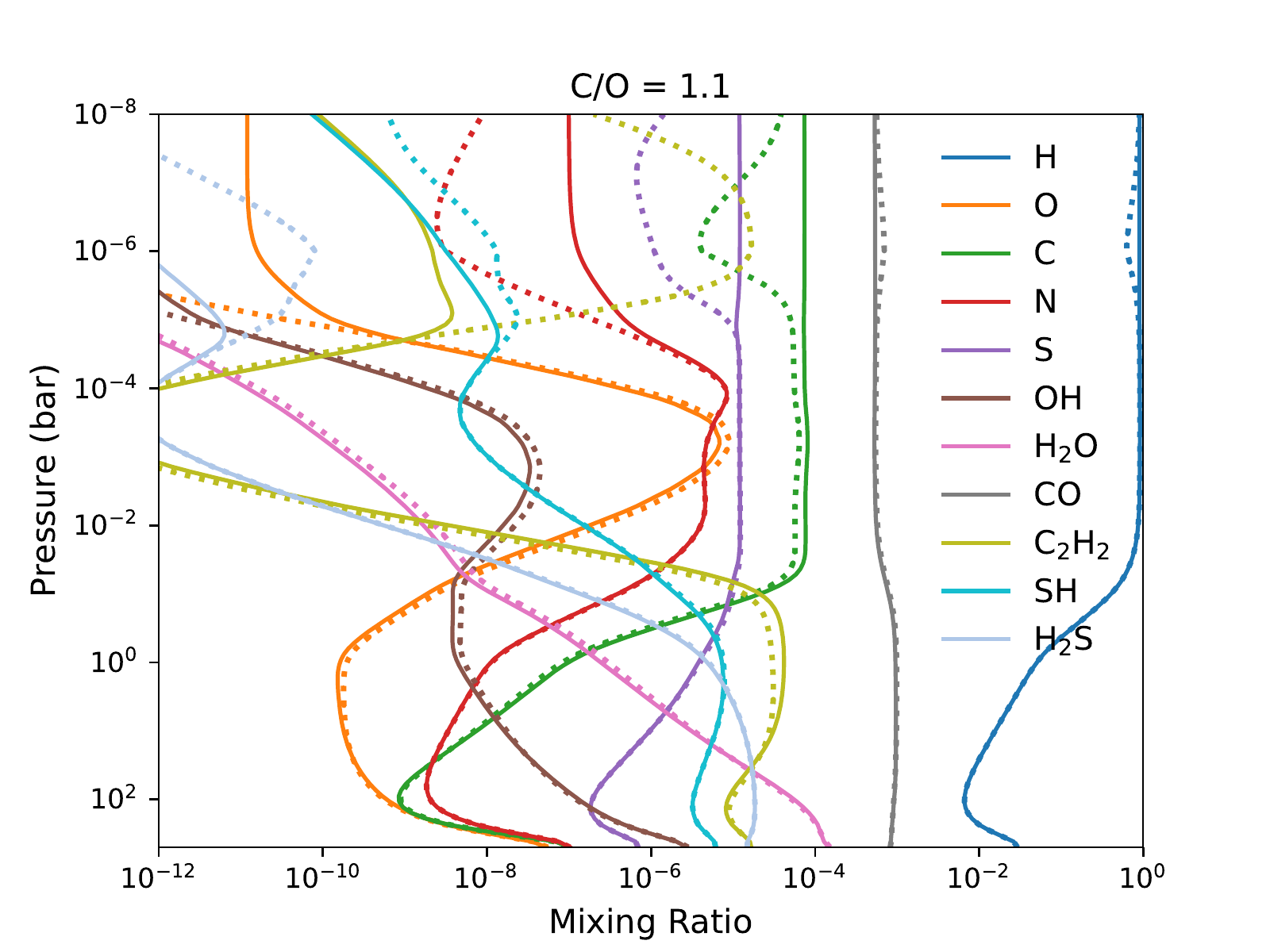}
\includegraphics[width=\columnwidth]{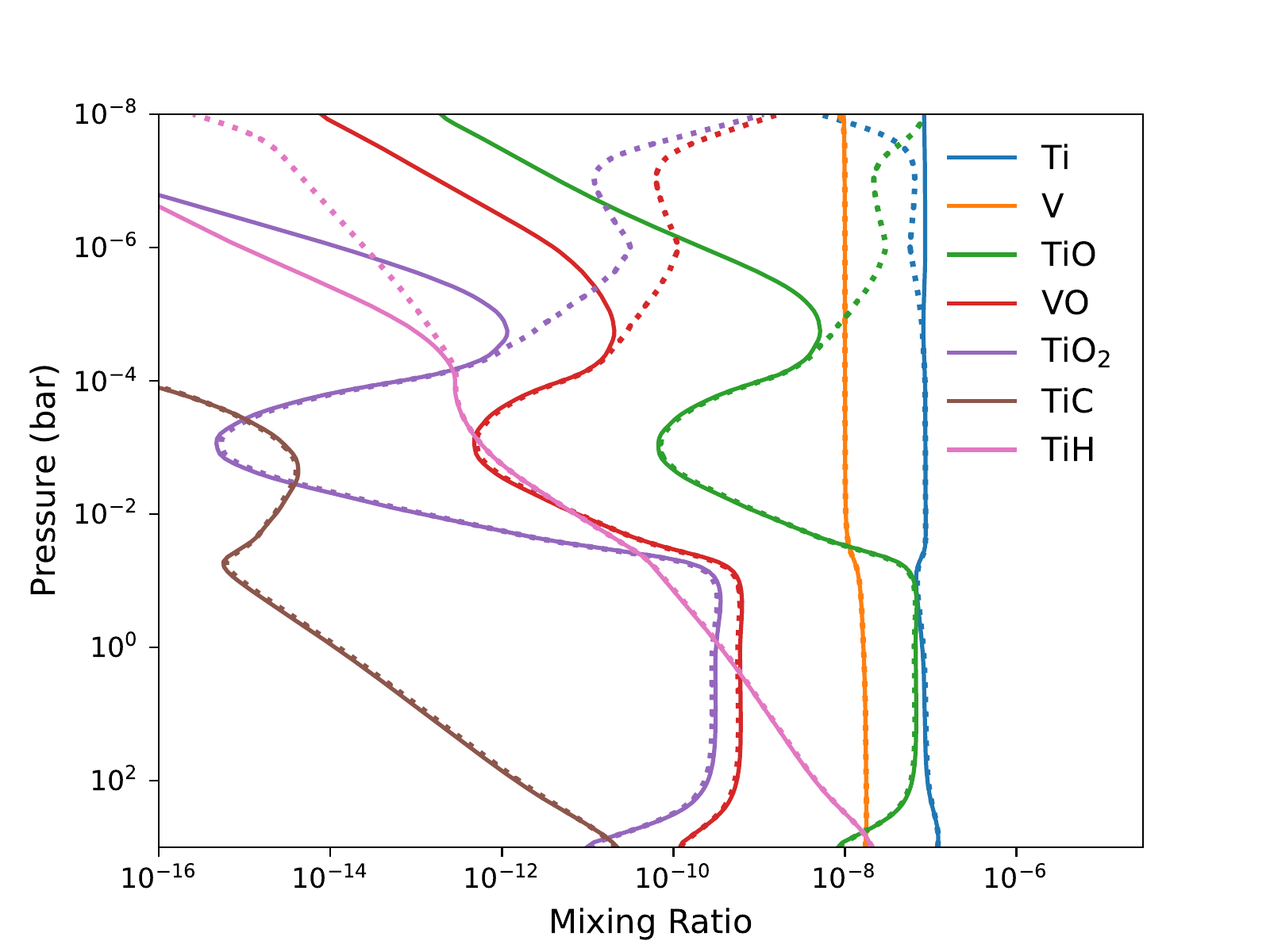}
\includegraphics[width=\columnwidth]{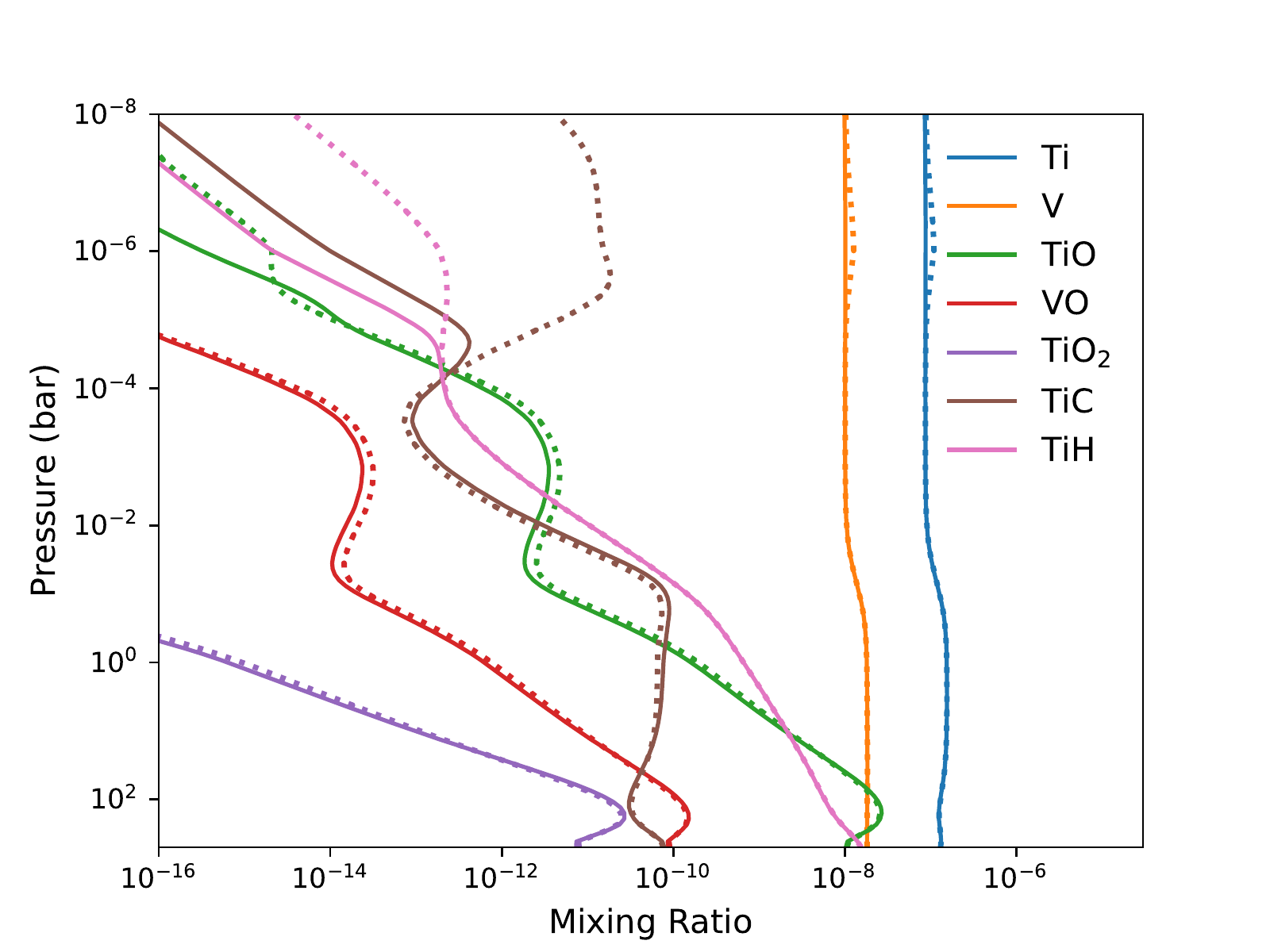}
\end{center}
\caption{The composition profiles for the main species of interest for WASP-33b, assuming solar C/O (left) and C/O = 1.1 (right). The equilibrium abundances are plotted in dotted lines.}
\label{fig:wasp33b-mix}
\end{figure*}
For a typical hot Jupiter (e.g. HD 189733b), vertical mixing plays a major role controlling the chemical distribution in the photosphere. However, it is not the case for WASP-33b as we compare the equilibrium and disequilibrium mixing ratio profiles in Figure \ref{fig:wasp33b-mix}. Although the strength of eddy diffusion also increases with temperature, faster thermochemical reactions still prevail upon vertical mixing. The deviation of disequilibrium profiles above the temperature-inverted region ( $\sim$ 10$^{-4}$ bar) is due to photodissociation, which reduces molecular species and produces more atoms. In the absence of vertical quenching, the depleted TiO in a carbon rich condition is unable to be replenished by vertical transport from the deep region, as seen in Figure \ref{fig:wasp33b-mix}. In the photodissociation region, in principle, stronger vertical mixing can transport more molecules upward against photodissociation.
We have perform additional tests with eddy diffusion profile varied by a factor of 10. Yet we found the change is minor and our results are not too sensitive to $K_\textrm{zz}$. 

For sulfur species, sulfur atom S is also the favored form followed by hydrogen sulfide (SH). The formation of \ce{S2} and other polysulfur (\ce{S_x}) is entirely shut down at this extremely high temperature. Sulfur does not couple to oxygen, carbon, or nitrogen since it mostly remains in the atomic form. Lastly, because the adopted stellar spectrum is truncated around Lyman-$\alpha$, we have further extended the stellar spectrum to include the EUV flux shorter than Lyman-$\alpha$ using the synthetic spectra by PHOENIX \footnote{\url{http://phoenix.astro.physik.uni-goettingen.de/}}. Apart from more C atoms from CO photodissociation above 10$^{-5}$ bar, we find no notable differences in all other species. Overall, the composition distribution of WASP-33b resembles that of a hot Jupiter except without a vertical quench region. The atmosphere of WASP-33b can be divided into a photochemically influenced region and a thermochemical equilibrium region, with the transition at the top of temperature-inverted layer around 10$^{-4}$ bar.

\begin{figure}[tph]
\begin{center}
\includegraphics[width=\columnwidth]{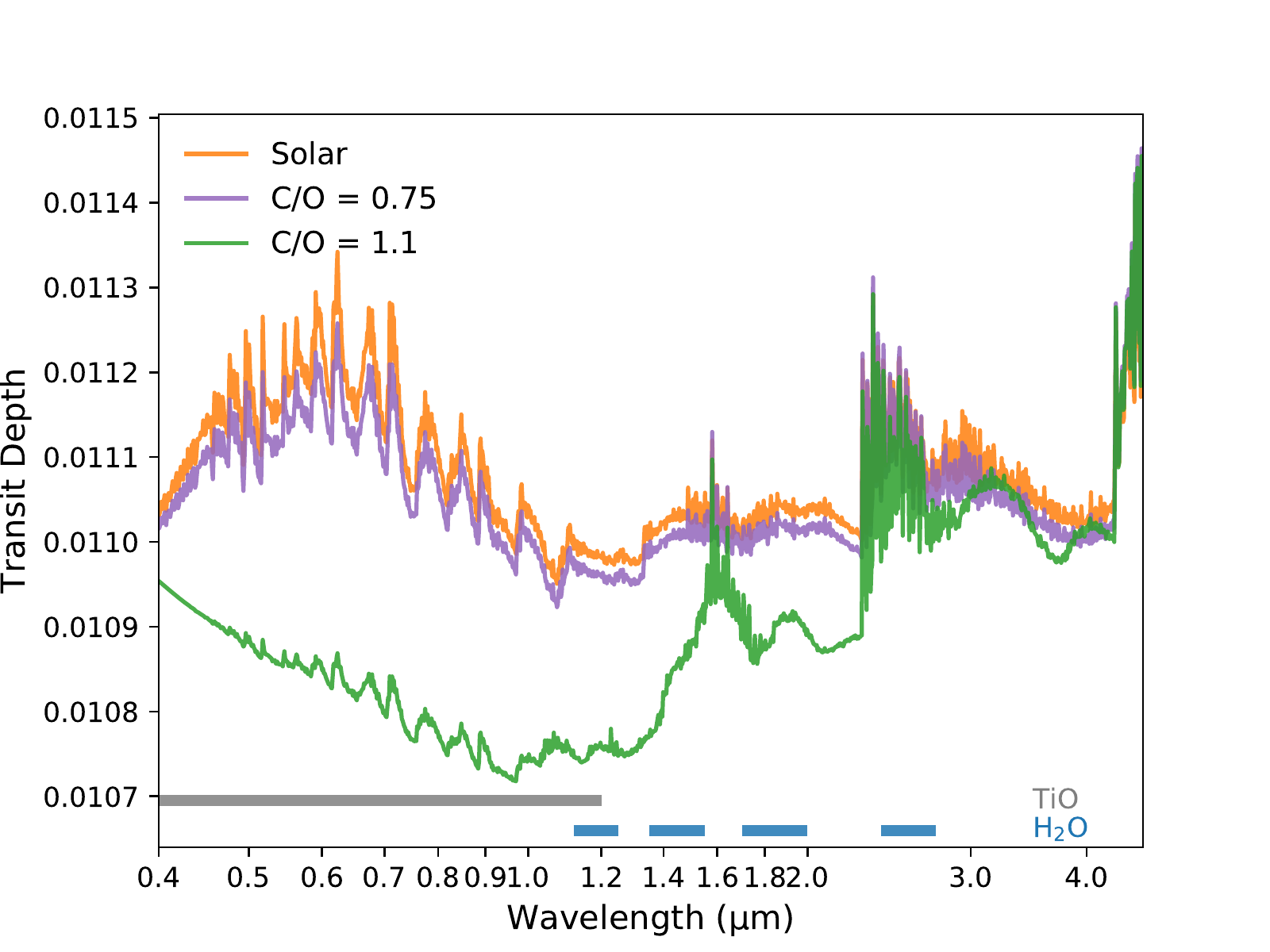}
\end{center}
\caption{Synthetic transmission spectra for WASP-33b computed from modeled compositions assuming solar elemental abundance, C/O = 0.75, and C/O = 1.1. The absorption features of TiO and \ce{H2O} are indicated by the color bands.}
\label{fig:wasp33b-transit}
\end{figure}
\subsubsection{Transmission Spectra}  
We have computed the synthetic spectra from equilibrium and disequilibrium abundances and found no observable differences in both transmission and emission spectra. The photochemical region above the temperature-inverted region is too optically thin, even when molecules like \ce{H2O} and TiO are strongly photodisscoiated here. High-resolution spectroscopy might be more sensitive to probe the atomic species in this region.

Alternatively, the equilibrium abundances of TiO/VO are sensitive to the change of elemental abundance. Figure \ref{fig:wasp33b-transit} demonstrates that the opacity in the optical is most sensitive to the change of TiO/VO as C/O is close to unity, which shows even greater variation than \ce{H2O} absorption between 1.2 and 2 $\mu$m. 

In conclusion, we find photodissociation only impacts the upper atmosphere of WASP-33b where P $<$ 0.1 mbar, chemical equilibrium is generally a valid assumption, as has been found for KELT9-b \citep{Kitzmann2018} and ultra hot Jupiters with dayside temperatures above 3000 K. Atmospheric mixing might still play an important role in an atmosphere with temperature lower than WASP-33b. Our first attempt to solve the kinetics of titanium species can provide an interesting avenue for investigating other transition metals such as Fe and Ca for future study of ultra hot Jupiters.

\subsection{HD189733b}
  
We have benchmarked our model of HD 189733b in Section \ref{sec:hd189}, where we attempt to keep the astronomical and chemical setup as close to \cite{Moses11,Venot12} as possible for comparison. In this section, we include the following updates and aspects that have not been considered in previous work:\\
{\tiny$\bullet$} Recently observed stellar UV-flux of HD 189733 \citep{Bourrier20}\\
{\tiny$\bullet$} Sulfur chemistry\\ 
{\tiny$\bullet$} Condensation of carbon vapor\\

\subsubsection{Stellar UV-flux}
\begin{figure}
\begin{center}
\includegraphics[width=\columnwidth]{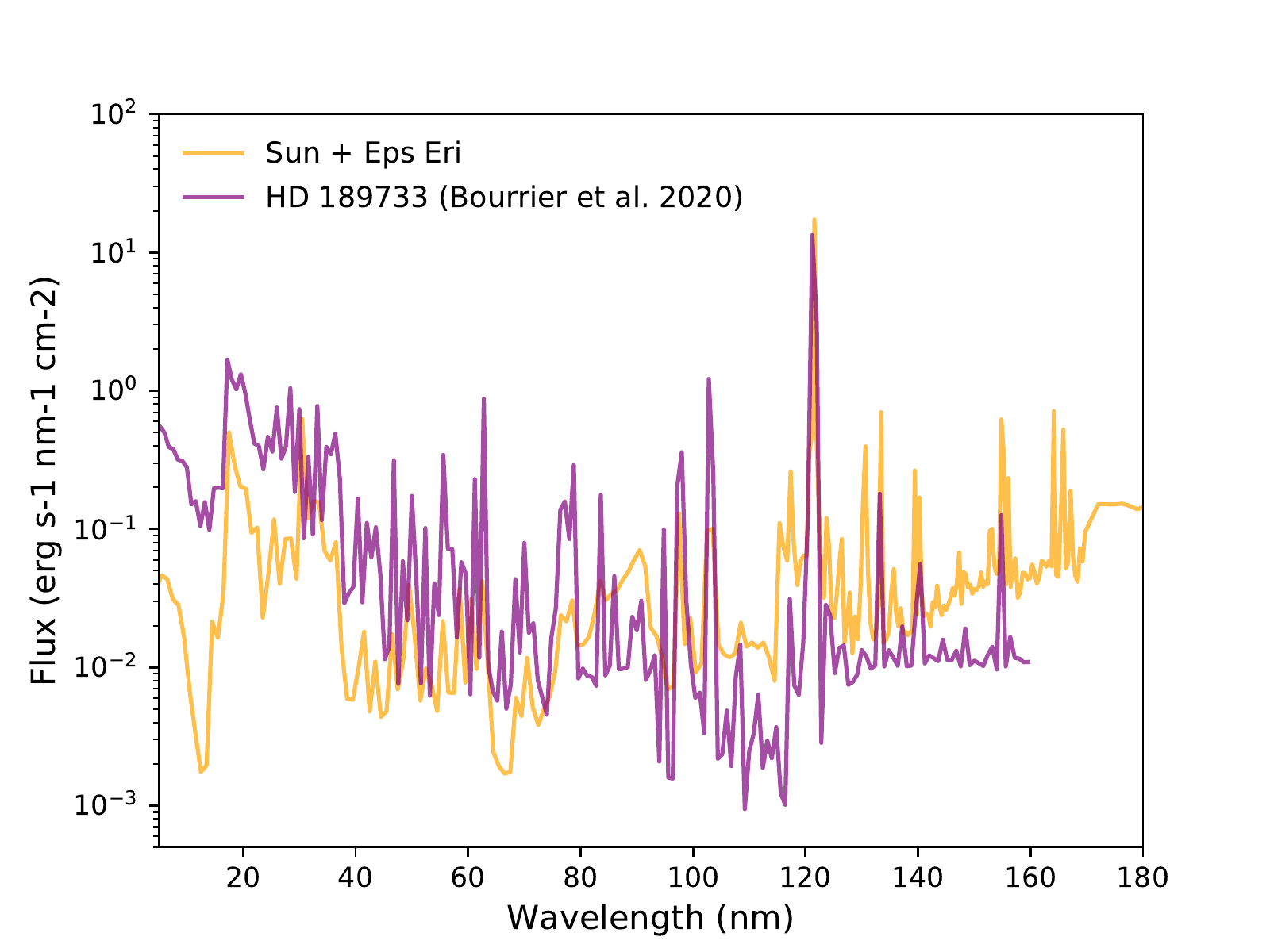}
\end{center}
\caption{The UV flux of HD 189733 received at 1 A.U. derived from recent observations \cite{Bourrier20} compared to the previously adopted spectrum, which consists of solar EUV and epsilon Eridani following the same approach as \cite{Moses11} and used in Section \ref{sec:hd189}. The spectra are binned for clarity.}
\label{fig:HD189-flux}
\end{figure}

\cite{Bourrier20} combine HST and XMM-Newton observations and derive semi-synthetic UV spectra up to 160 nm. For our model benchmark in Section \ref{sec:hd189}, solar flux is used for wavelengths below 115 nm and the observed spectrum of epsilon Eridani (a K2-type analogue) is adopted for 115 - 283 nm. The previously adopted and newly observed stellar fluxes are compared in Figure \ref{fig:HD189-flux}. The EUV flux of HD 189733 is modestly higher than that of the Sun, but the photochemically important FUV ($\lambda >$ 122 nm) flux appears to be weaker. Nevertheless, the change in the UV flux turns out to only slightly decrease the atomic H (by about 20$\%$). The overall impact of the updated EUV flux on neutral chemistry is in fact insignificant. 

\subsubsection{Sulfur Chemistry}\label{sec:HD189-S}
\begin{figure*}[ht]
\begin{center}
\includegraphics[width=\columnwidth]{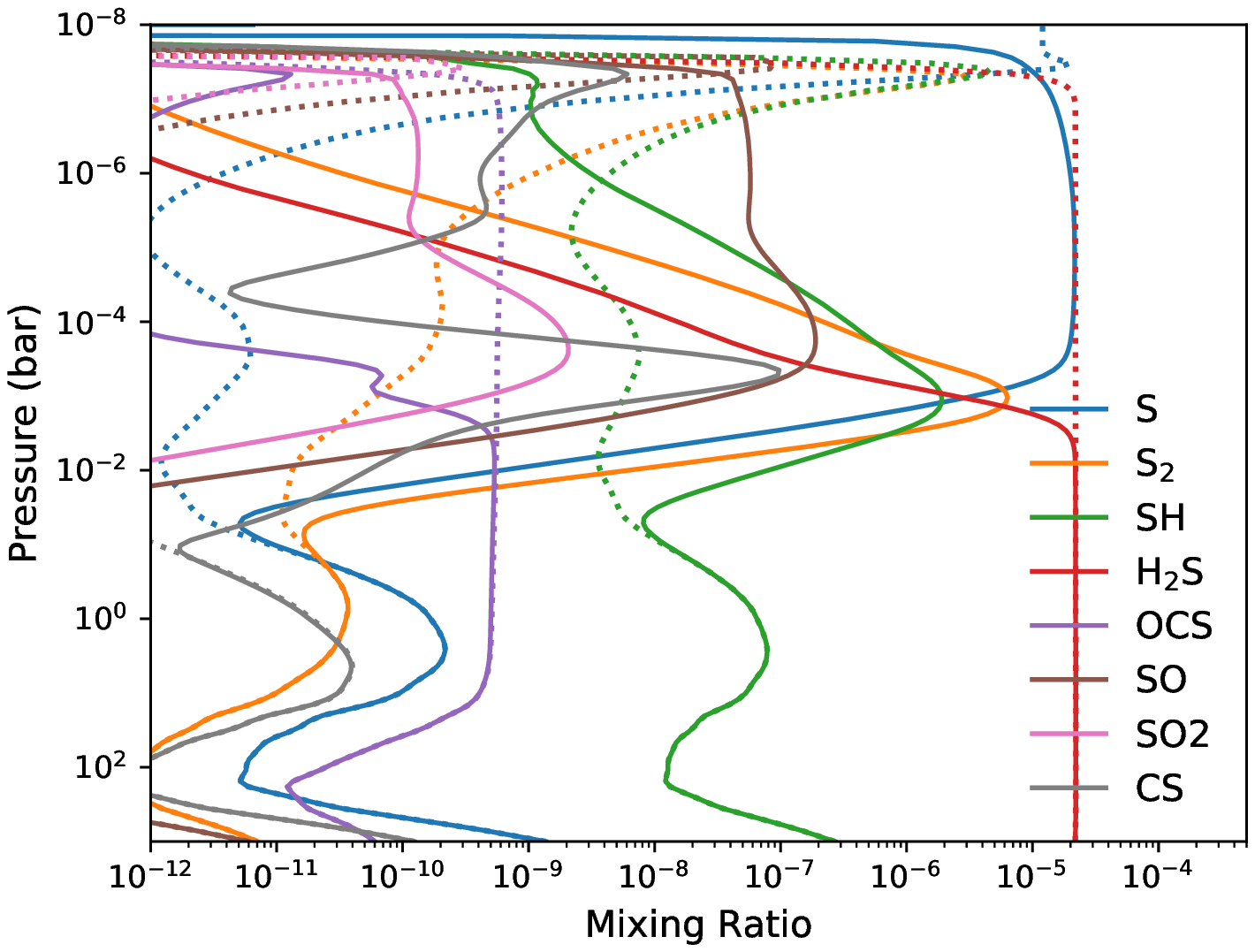}
\includegraphics[width=\columnwidth]{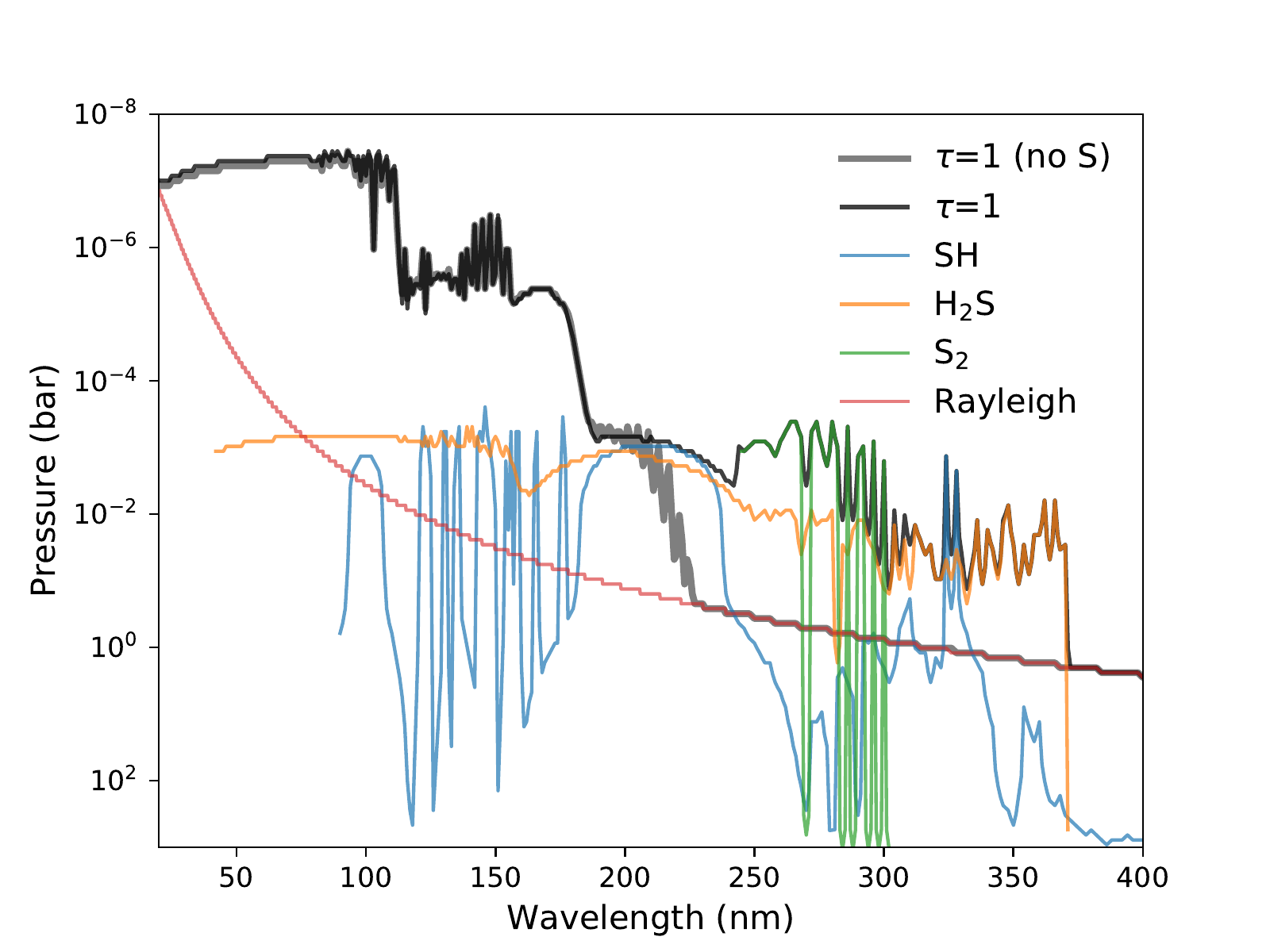}
\end{center}
\caption{Left: Mixing ratios of the major sulfur species computed in the model of HD 189733b. The photochemical kinetics results are in solid lines and equilibrium abundances in dotted lines. Right: The pressure level of optical depth $\tau$ = 1 as a function of wavelength while including (black) and excluding (grey) sulfur chemistry, along with the main individual contribution from sulfur species.}
\label{fig:HD189-S}
\end{figure*}

We next run the same model except including sulfur kinetics. The sulfur species from our photochemical calculation are illustrated in Figure \ref{fig:HD189-S} and are broadly consistent with previous work \citep{Zahnle09}. Hydrogen sulfide (\ce{H2S}) is the thermodynamically stable form of sulfur in a hydrogen-dominated atmosphere. \ce{H2S} is mostly destroyed by hydrogen abstraction 
\begin{equation}
\vspace{-0.1cm}
\ce{H2S + H -> SH + H2}
\label{re:H2S}
\vspace{-0.1cm}
\end{equation}
and restored by the reverse reaction of (\ref{re:H2S}). The forward and backward reactions of (\ref{re:H2S}) essentially dictates the level where \ce{H2S} starts to loss its stability. On HD 189733b, \ce{H2S} is dissociated above 1 mbar and predominantly turned into S. SH and \ce{S2} also reach maximum values at the level where \ce{H2S} dissociation kicks off. The implication is both \ce{SH} and \ce{S2} absorb shortwave radiation and could potentially provide stratospheric heating, especially with super-solar metallicity condition as discussed in \cite{Zahnle09}. 

We find SO accumulated in the upper atmosphere from the oxidation of \ce{S + OH} $\rightarrow$ \ce{SO + H}. The highly reactive SO is known to self-react into SO dimer (\ce{(SO)2}) and may facilitate formation of \ce{S2O} and \ce{S2} \citep{Pinto2021} or back into S and \ce{SO2}. What actually happened in our model is SO either photodissociated or reacted with atomic H back to S in the low pressure region. The elemental S might be subject to photoionization, as we will discuss in Section \ref{discussion}. 

One notable effect of photochemistry with sulfur is several sulfur species absorb in the MUV/NUV.
As illustrated in Figure  \ref{fig:HD189-S}, sulfur species raised the UV photosphere above $\sim$ 230 nm, compared to that without sulfur where no efficient absorption beyond the ammonia bands. We find \ce{H2S} responsible for the dominant absorbption in the NUV (300--400 nm), rather than SH as reported in \cite{Zahnle09}, which might be caused by the isothermal atmosphere at 1400 K used in \cite{Zahnle09}. The absorption of \ce{S2} between 250 and 300 nm and the SH peaks around 325 nm can make prospective observable features. 

\begin{figure}[htp]
\begin{center}
\includegraphics[width=\columnwidth]{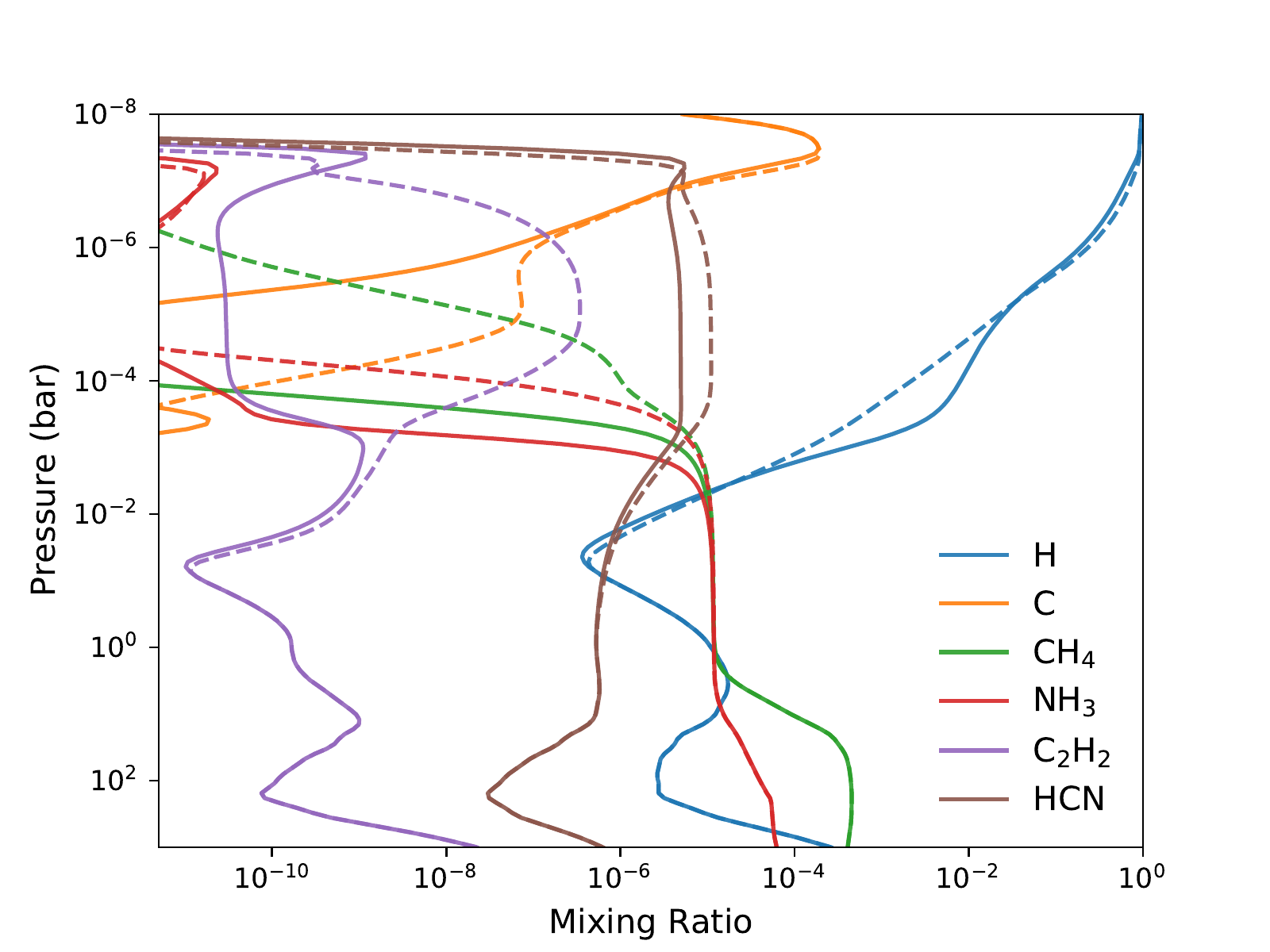}
\end{center}
\caption{Mixing ratio profiles of main species on HD 189733b that exhibit differences from models including sulfur kinetics (solid) and without sulfur kinetics (dashed).}
\label{fig:HD189-S-noS}
\end{figure}

Figure \ref{fig:HD189-S-noS} highlights the compositional differences when sulfur is present. Sulfur species can play an interesting role in catalyzing conversion schemes that take multiple steps. In particular, \ce{CH4} is more diminished down to about 1 mbar.  We find sulfur provide a catalytic pathway for \ce{CH4}-CO conversion. As \ce{CH4} and \ce{H2S} react with atomic H to liberate carbon and sulfur, they couple to form carbon monosulfide (CS). Carbon in CS is further oxidized into OCS and eventually ends up in CO through H-abstraction, via a pathway such as
\begin{eqnarray}
\begin{aligned} 
\ce{ CH4 + H &-> CH3 + H2}\\
\ce{ CH3 + H &-> CH2 + H2}\\
\ce{ CH2 + S &-> HCS + H}\\
\ce{ HCS + H &-> CS + H2}\\
\ce{ SH + H &-> S + H2}\\
\ce{ S + OH &-> SO + H}\\
\ce{ CS + SO &-> OCS + S}\\
\ce{ OCS + H &-> CO + SH}\\
\ce{H2O &->[h\nu] OH + H}\\
\ce{SH &->[h\nu] S + H}\\
\ce{S + H2 &-> SH + H}\\
\noalign{\vglue 5pt} 
\hline %
\noalign{\vglue 5pt} 
\mbox{net} : \ce{CH4 + H2O &-> CO + 3H2}.
\end{aligned}
\label{re:path-ch4-co-S}
\end{eqnarray}
Note there is no net change of sulfur species in the cycle. The rate-limiting reaction in pathway (\ref{re:path-ch4-co-S}) is the carbon-sulfur step \ce{ CH2 + S -> HCS + H}, which is about three orders of magnitude faster than the pathway without sulfur around 1mbar. Interestingly, we find SH play an analogous role to \ce{H2O} in catalytically converting \ce{H2} to H, causing the minor H increase in Figure \ref{fig:HD189-S-noS}. 

The presence of sulfur species enhances the destruction of methane and might partly contribute to 
the scarcity of methane detection on hot Jupiters \citep[e.g., ][and references within]{Baxter2021}. \ce{H2S} has also been reported to speed up the oxidation of methane in combustion experiment \citep{Gersen2017}, in the oxidizing and high-pressure conditions of gas engines. The decreasing of \ce{CH4} naturally reduces its offspring products to a great extent. The column density shown in Figure \ref{fig:haze-bar} reflects the reduction of haze precursors with the participation of sulfur. Based on our fiducial analysis on HD 189733 b, we suggest that organic haze formation is likely to be partly suppressed by sulfur kinetics on a hot Jupiter, as opposed to enhanced by sulfur kinetics in a \ce{CO2}-rich condition suggested by experimental simulations \citep{He2020}.

\subsubsection{Condensation of Carbon Vapor}\label{HD189-C-conden}
\begin{figure}[htp]
\begin{center}
\includegraphics[width=\columnwidth]{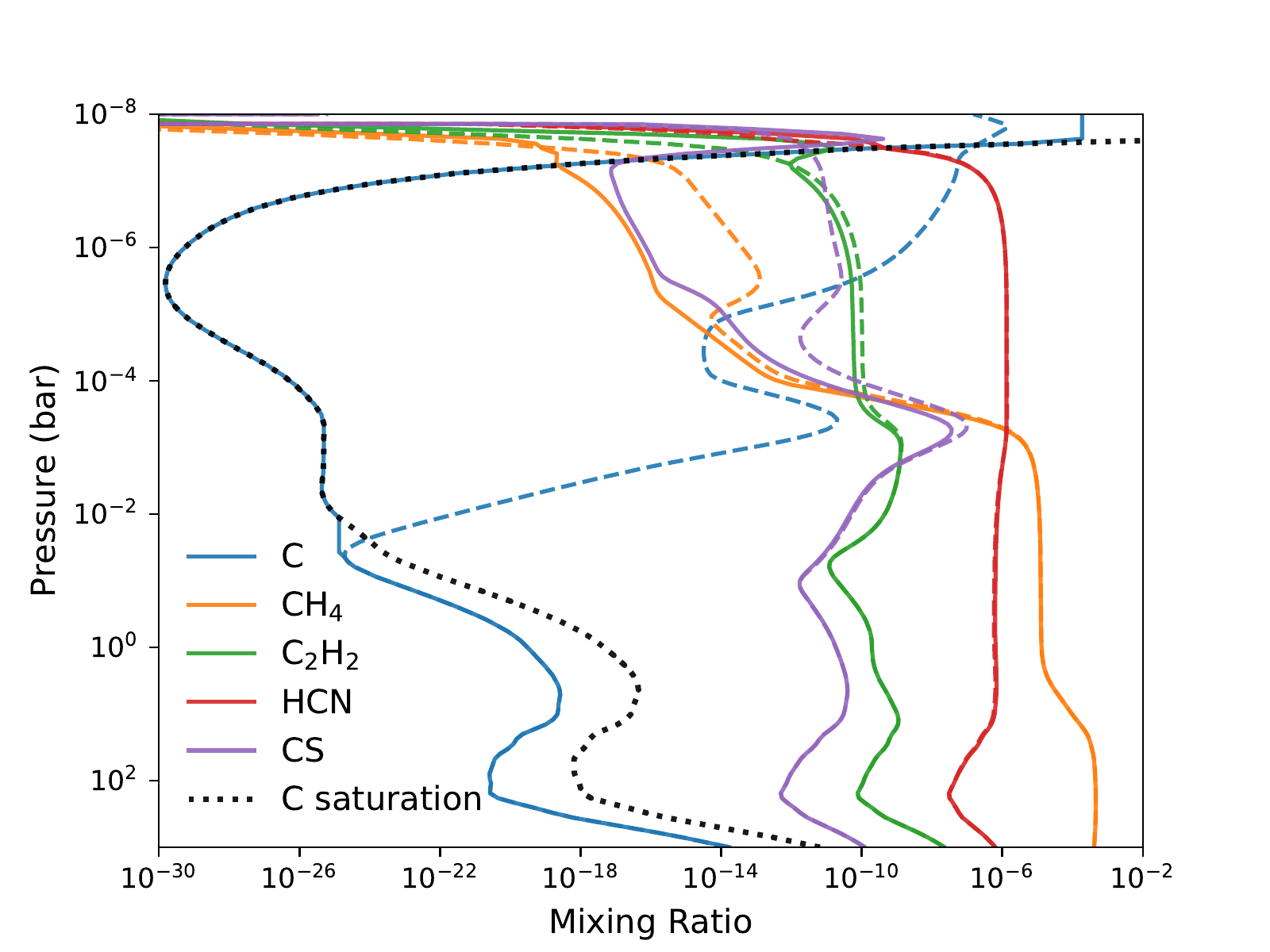}
\end{center}
\caption{Several carbon-containing species from the nominal model (dashed) compared to those from the model with limited C due to instantaneous condensation. The saturation mixing ratio of C is plotted in dotted curve.}
\label{fig:HD189-C-conden}
\end{figure}

Atomic carbon vapor (C) is produced by CO dissociation (including both photodissociation and thermal dissociation in the thermosphere) or the reaction \ce{N + CN -> C + N2} above $\sim$ 0.1 mbar and also by H abstraction with \ce{CHx} species in the lower region. The saturation vapor of C falls off rapidly with decreasing temperature in the upper atmosphere, as shown in Figure \ref{fig:HD189-C-conden}. In fact, the disequilibrium abundance of C starts to exceed the saturation concentration above 10 mbar. The realistic timescale for graphite growth by condensation involves detailed microphysics and is beyond the scope of this study. As a simple test, we explore the kinetic effects after carbon vapor is fully condensed. We run our HD 189733b model including sulfur chemistry and do not allow C to become supersaturated but simply fix the abundance of C in the gas phase to its saturation mixing ratio. This is physically equivalent to assuming instantaneous condensation and unlimited condensation nuclei. 
 
Figure \ref{fig:HD189-C-conden} demonstrates the consequences when C is instantaneously condensed, which mainly impacts the region above 0.1 mbar. Without the condensation of C, \ce{CH4} can be replenished by the hydrogenation sequence of C (i.e. C $\rightarrow$ CH $\rightarrow$ \ce{CH2} $\rightarrow$ \ce{CH3} $\rightarrow$ \ce{CH4}). This channel is closed as C condensed out and \ce{CH4} is further depleted in the upper atmosphere. CS is reduced in the same way but \ce{C2H2} and HCN remain almost unaffected (they are already reduced compared to the model without sulfur). In the end, we find that the condensation of C has limited effects to other gas compositions in the upper atmosphere. 

\subsubsection{Transmission Spectra}
\begin{figure}[htp]
\begin{center}
\includegraphics[width=\columnwidth]{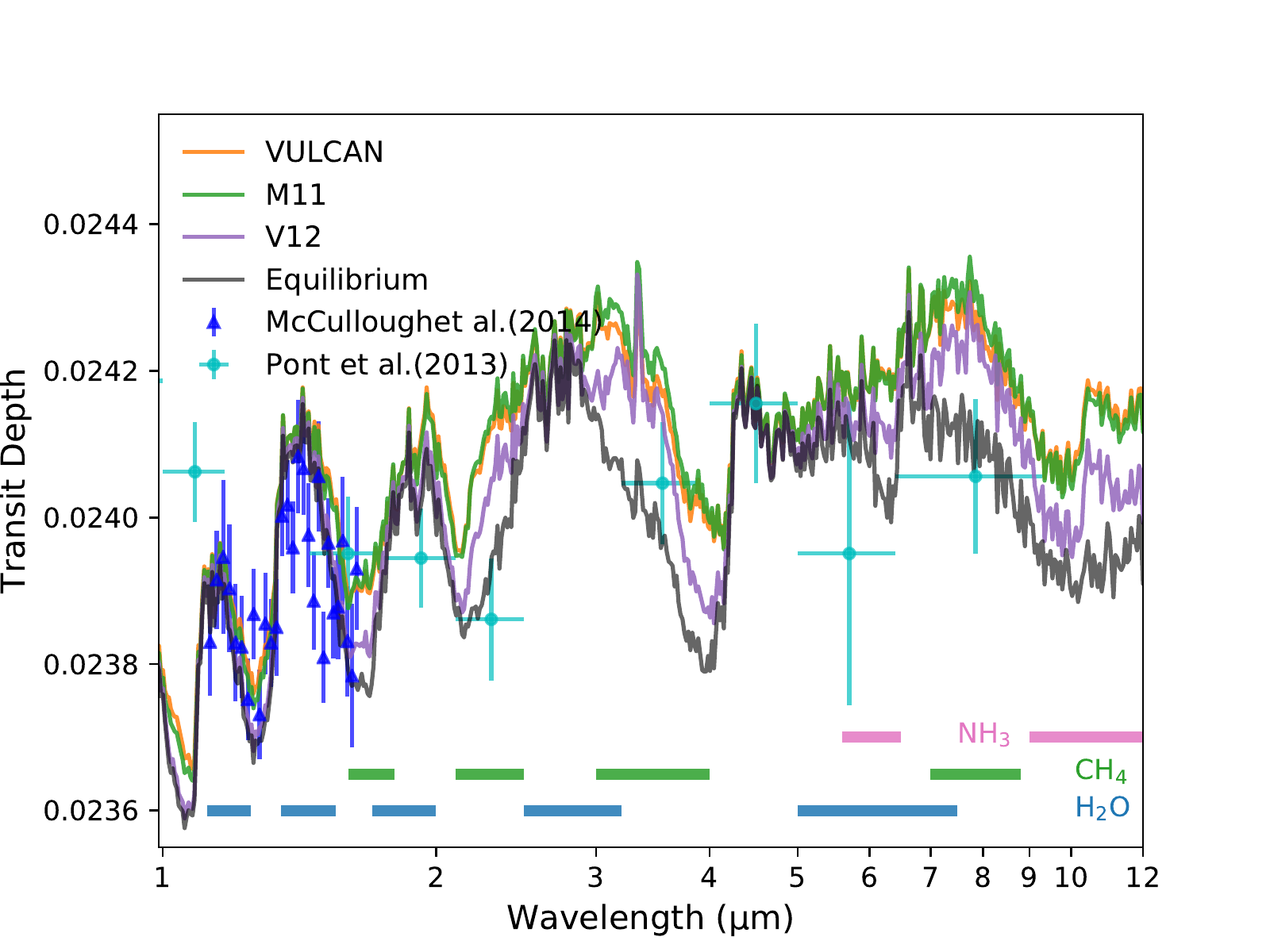}
\end{center}
\caption{Synthetic transmission spectra for HD 189733b generated from chemical abundances computed by VULCAN, \cite{Moses11}, \cite{Venot12}, and when assuming chemical equilibrium. Transit observations from \cite{Pont2013} and \cite{Mc2014} are shown as data points with error bars. The absorption features for the main molecules are indicated by the color bands.}
\label{fig:HD189-transit-VM11V12}
\end{figure}

Here we first take a look at the observational consequences due to model uncertainties among \cite{Moses11}, \cite{Venot12}, and VULCAN which we examined in Section \ref{sec:hd189}. Figure \ref{fig:HD189-transit-VM11V12} showcases the transmission spectra of HD 189733b generated from the compositions computed by VULCAN, \cite{Moses11}, and \cite{Venot12}. The lower quenched abundances of \ce{CH4} and \ce{NH3} in \cite{Venot12} are responsible for the primary spectral differences while the spectra from VULCAN and \cite{Moses11} are fairly close. The ammonia absorption around 8--12 $\mu$m could be a useful diagnosis for the quenching mechanism of nitrogen chemistry. Overall, we find the model uncertainties lead to about half of the spectral deviation caused by disequilibrium chemistry.  

\begin{figure}[htp]
\begin{center}
\includegraphics[width=\columnwidth]{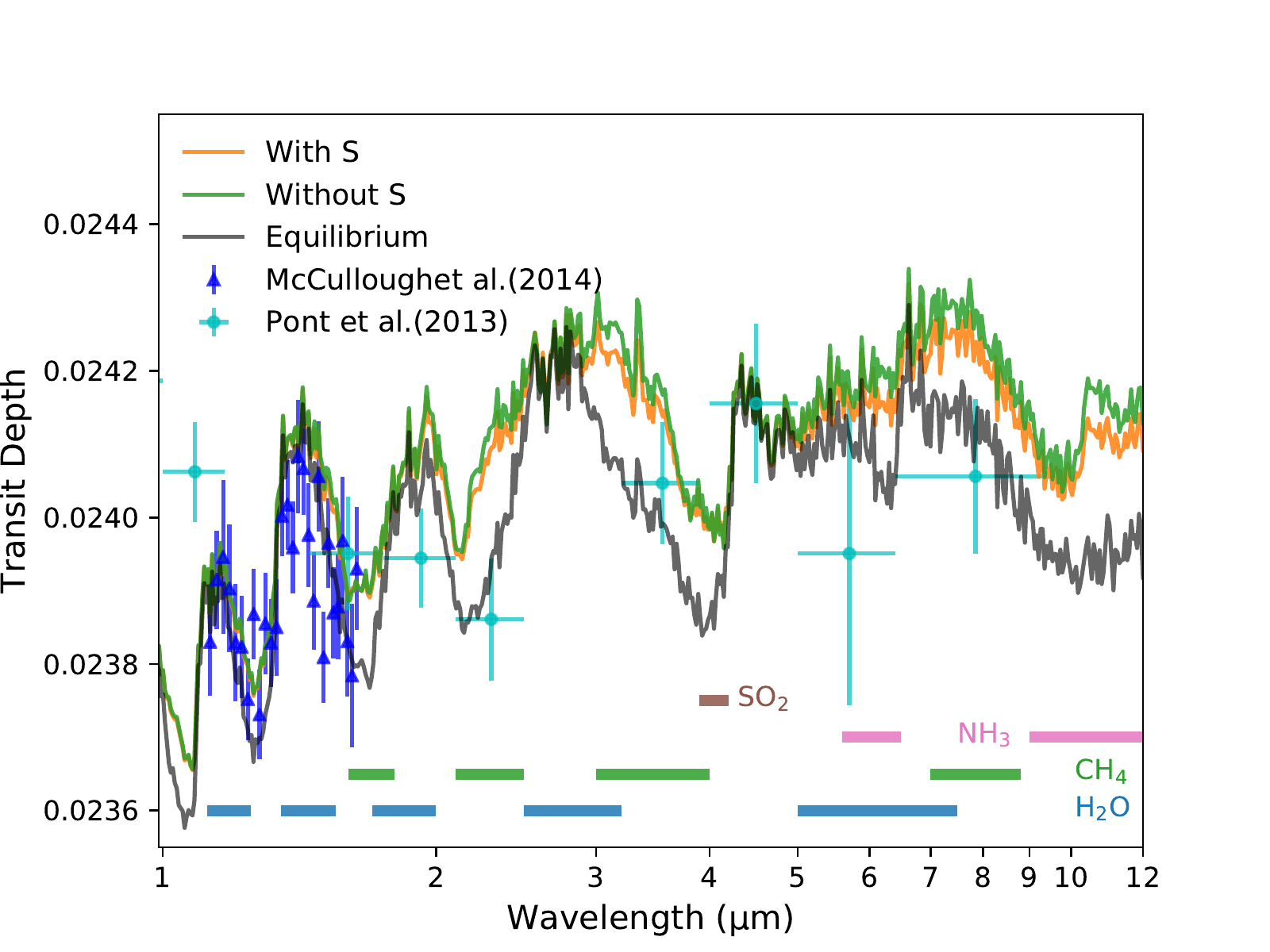}
\end{center}
\caption{Same as Figure \ref{fig:HD189-transit-VM11V12} but with abundances computed from our model while including or excluding sulfur species.}
\label{fig:HD189-transit-S}
\end{figure}

We then examine the effects of including sulfur chemistry to the transmission spectra in Figure \ref{fig:HD189-transit-S}. While the features from sulfur-containing species are almost obscured by other molecules such as \ce{H2O} and \ce{CH4} in the near-IR, there are still visible differences due to sulfur's impact on \ce{CH4} and \ce{NH3}. Since the coupling to sulfur reduces the abundances of \ce{CH4} and \ce{NH3}, the transit depth is smaller in the presence of sulfur.
The differences caused by sulfur chemistry is smaller than that between equilibrium and disequilibrium \ce{CH4} and \ce{NH3} abundances but not trivial. Current observations are not capable of placing conclusive constraints and we need to rely on future facilities with higher resolving power (e.g., JWST, ARIEL). 



\subsection{GJ 436b}\label{sec:GJ436b}
GJ 436b is a Neptune-sized planet in a close orbit around a M dwarf star. This warm Neptune has received tremendous attention since its first discovery \citep{Butler2004}, including multiple primary transit and secondary eclipse observed with {\it Spitzer}
 (\cite{Stevenson2010,Madhu2011,Morley2017} and references within), as well as transmission spectroscopic study with HST WFC3 \citep{Demory2007,Knutson2014}. {\it Spitzer} 3.6 $\mu$m and 4.5 $\mu$m emission indicates the atmosphere is rich in CO/\ce{CO2} and poor in \ce{CH4}. Yet precise constraint and the mechanism on CO/\ce{CH4} ratio still remain inconclusive. Forward models have suggested high metallicity \citep{Moses2013,Morley2017} and hot interior from tidal heating \citep{Agundez2014} can explain the observed CO/\ce{CH4} but inconsistent with the low water content (less than 10$^{-4}$) obtained by the retrieval model \cite{Madhu2011}. \cite{Hu2015} propose that a remnant helium-dominated atmosphere as a result of hydrogen escape can naturally deplete \ce{CH4} and \ce{H2O}. However, the Ly-alpha absorption still appears to indicate a hydrogen-dominated atmosphere for GJ 436b \citep{Khodachenko2019}. For this work, we restrict ourself to 100 times solar metallicity (Neptune-like) and explore the effects of vertical mixing and internal heat with the presence of sulfur.  



\begin{figure}[htp]
\begin{center}
\includegraphics[width=\columnwidth]{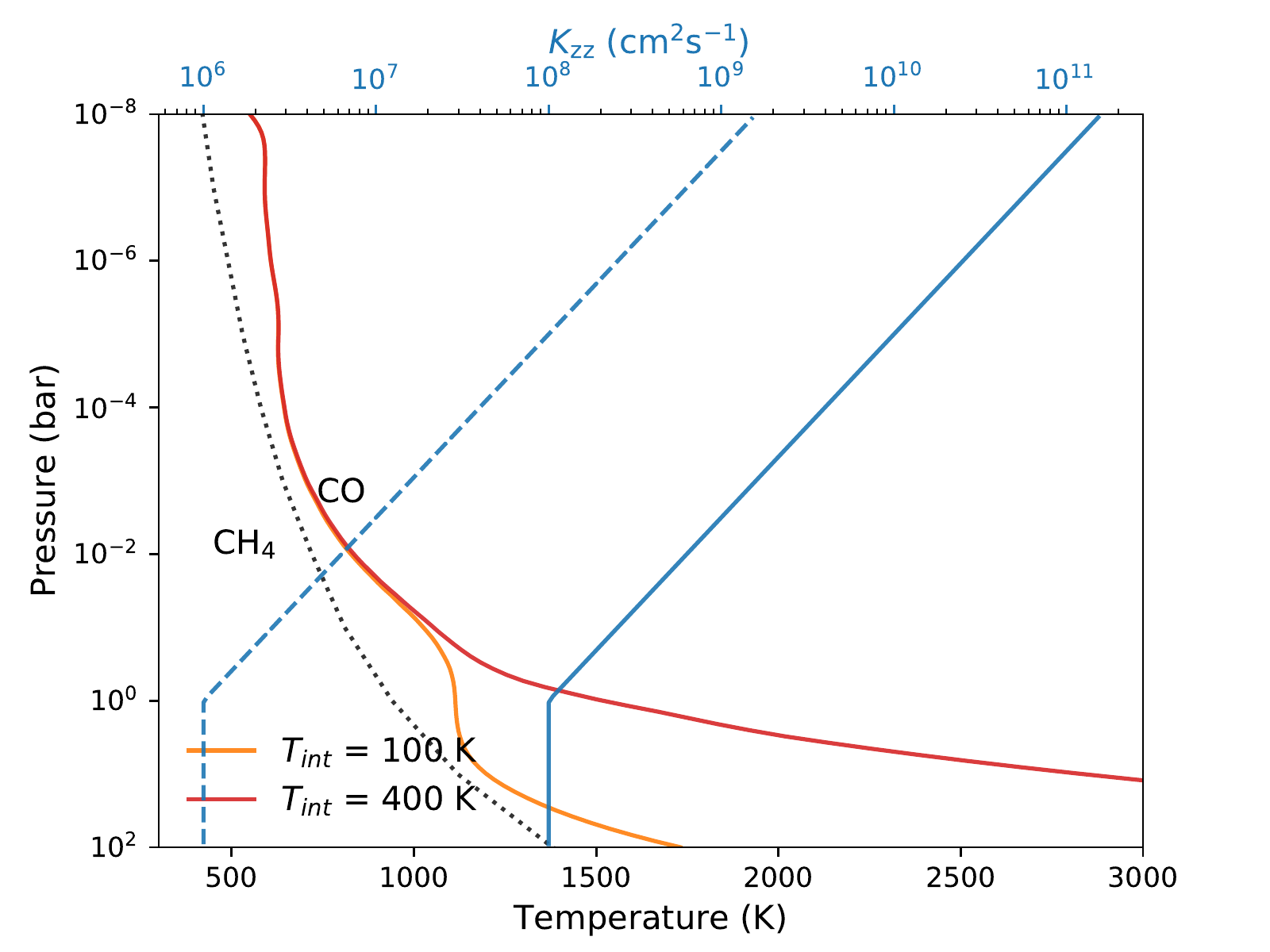}
\end{center}
\caption{The temperature-pressure and eddy diffusion ($K_\textrm{zz}$) profiles for GJ 436b, showing low (T$_{\textrm{int}}$ = 100 K) and high (T$_{\textrm{int}}$ = 400 K) internal heating and weak (dashed) and strong (solid) vertical mixing. The [\ce{CH4}]/[CO] = 1 equilibrium transition curve for 100 times solar metallicity is shown by the dotted curve.}
\label{fig:GJ436b-TPK} 
\end{figure}

\begin{figure*}[htp]
\begin{center}
\includegraphics[width=\columnwidth]{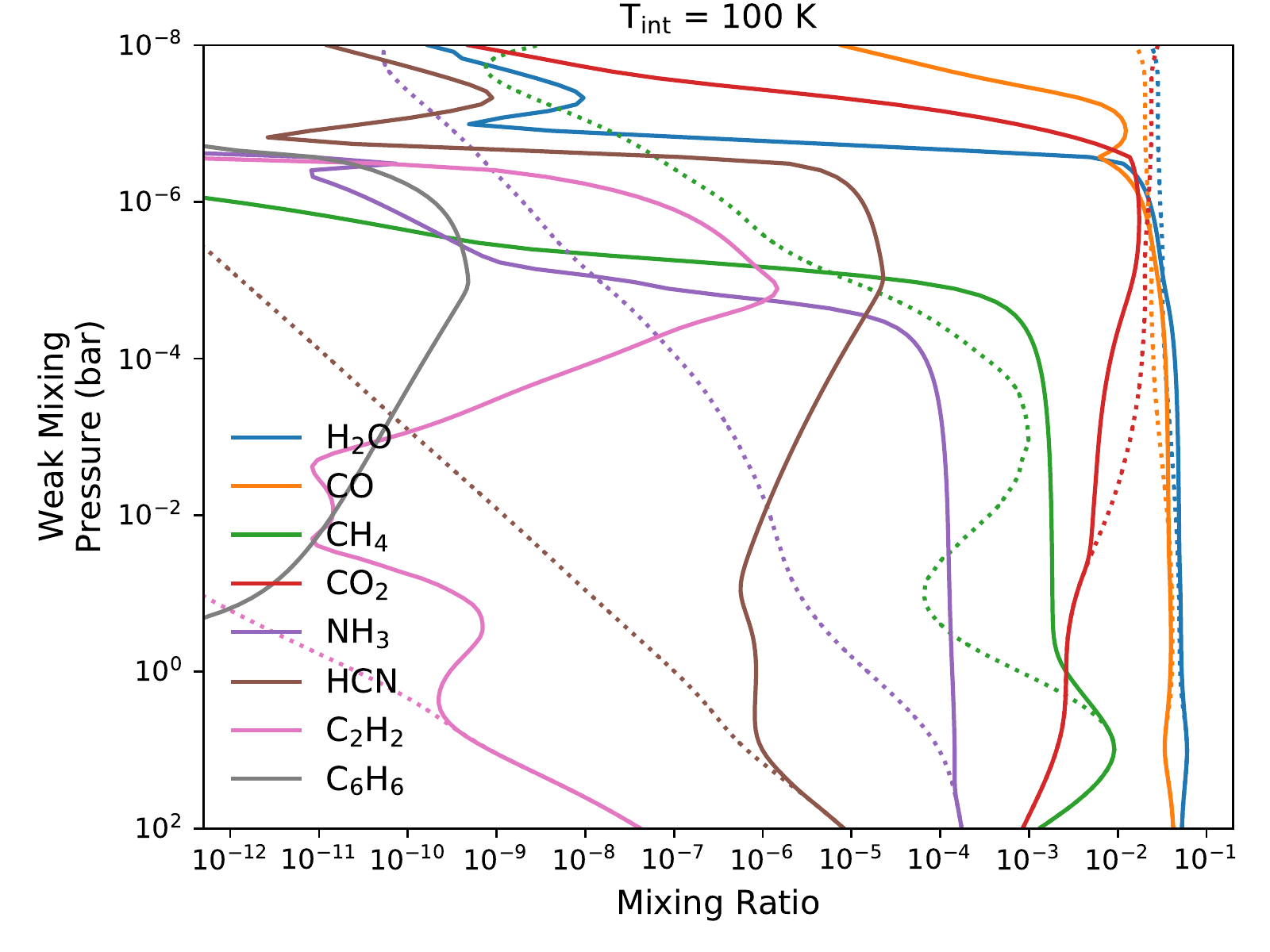}
\includegraphics[width=\columnwidth]{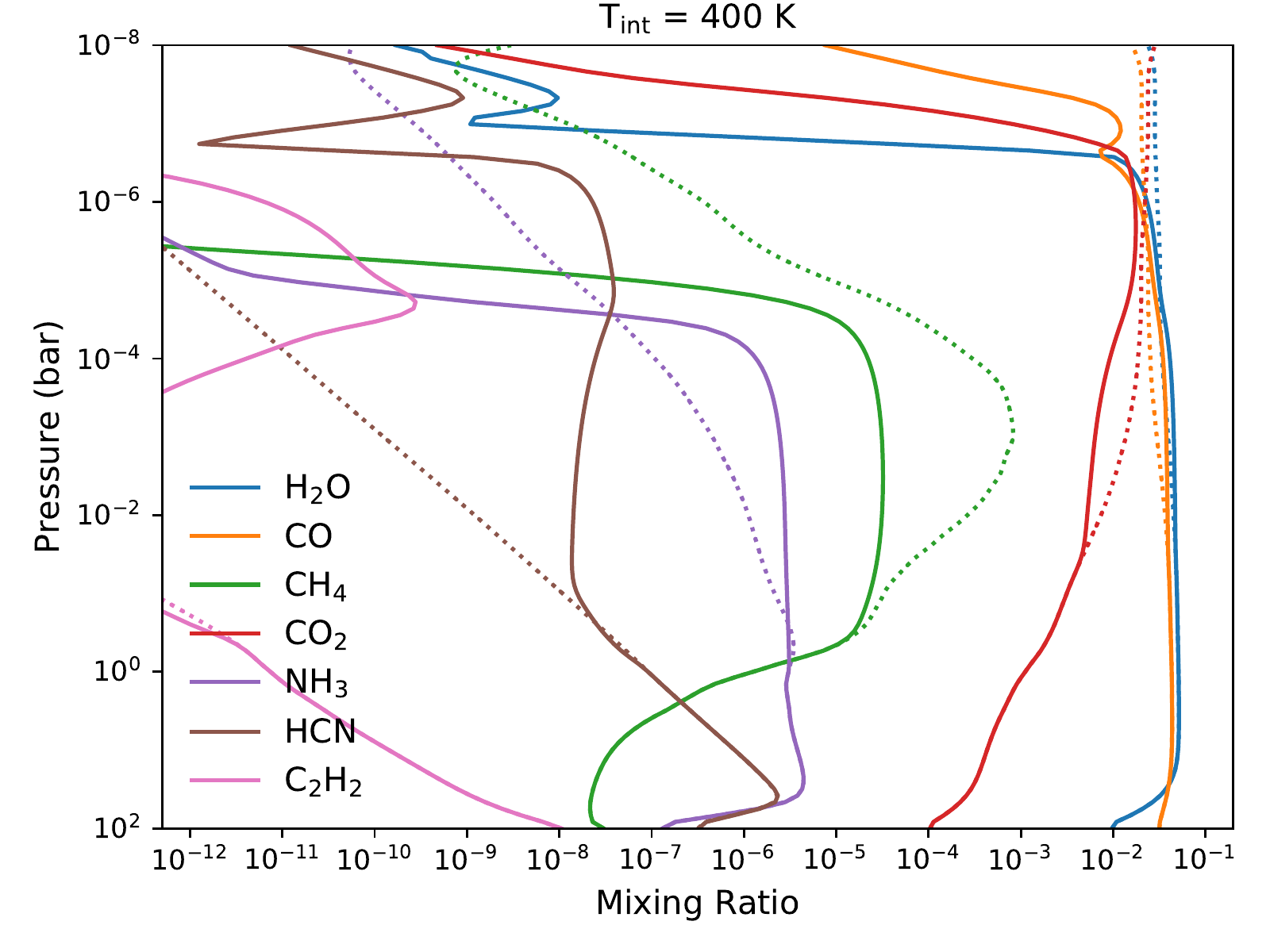}
\includegraphics[width=\columnwidth]{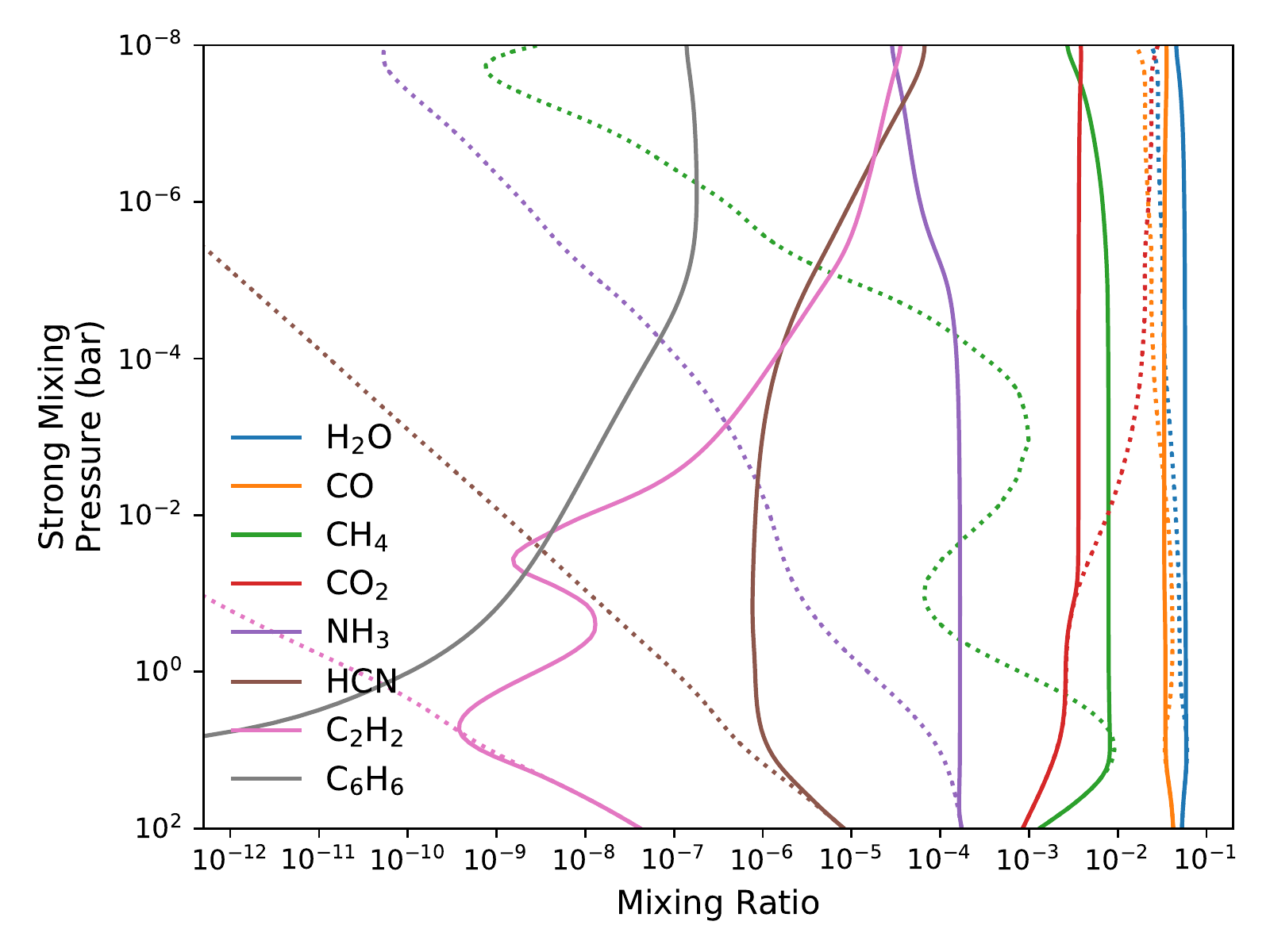}
\includegraphics[width=\columnwidth]{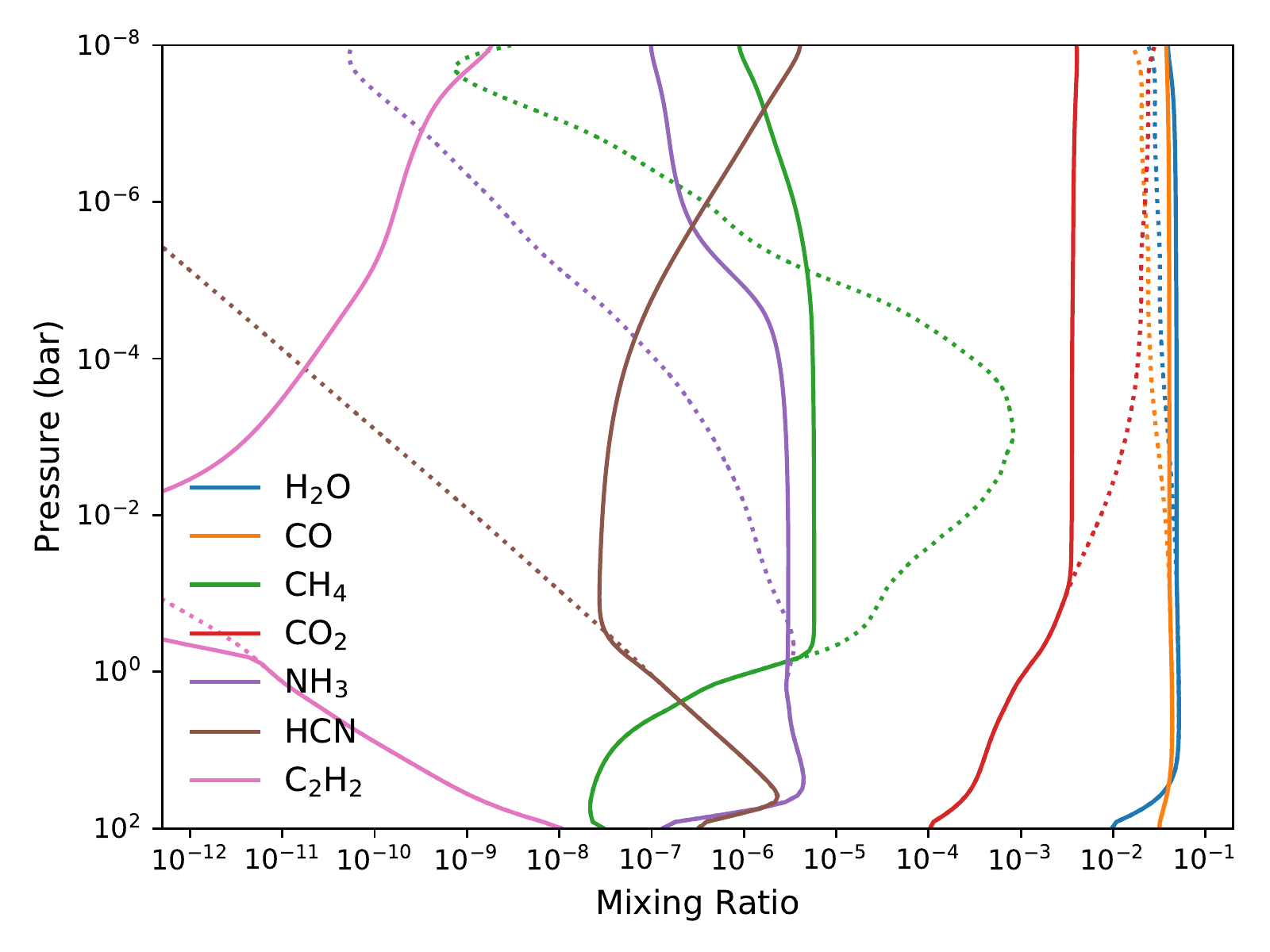}
\end{center}
\caption{The mixing ratio profiles (solid) along with equilibrium profiles (dotted) of several main species on GJ 436b for different assumption of internal temperature and vertical mixing. The left/right columns correspond to low/high ($T_{\textrm{int}}$ = 100 K/$T_{\textrm{int}}$ = 400 K) internal heating and the top/bottom rows correspond to weak/strong vertical mixing.}
\label{fig:GJ436b-mix}
\end{figure*}

\begin{figure*}[htp]
\begin{center}
\includegraphics[width=\columnwidth]{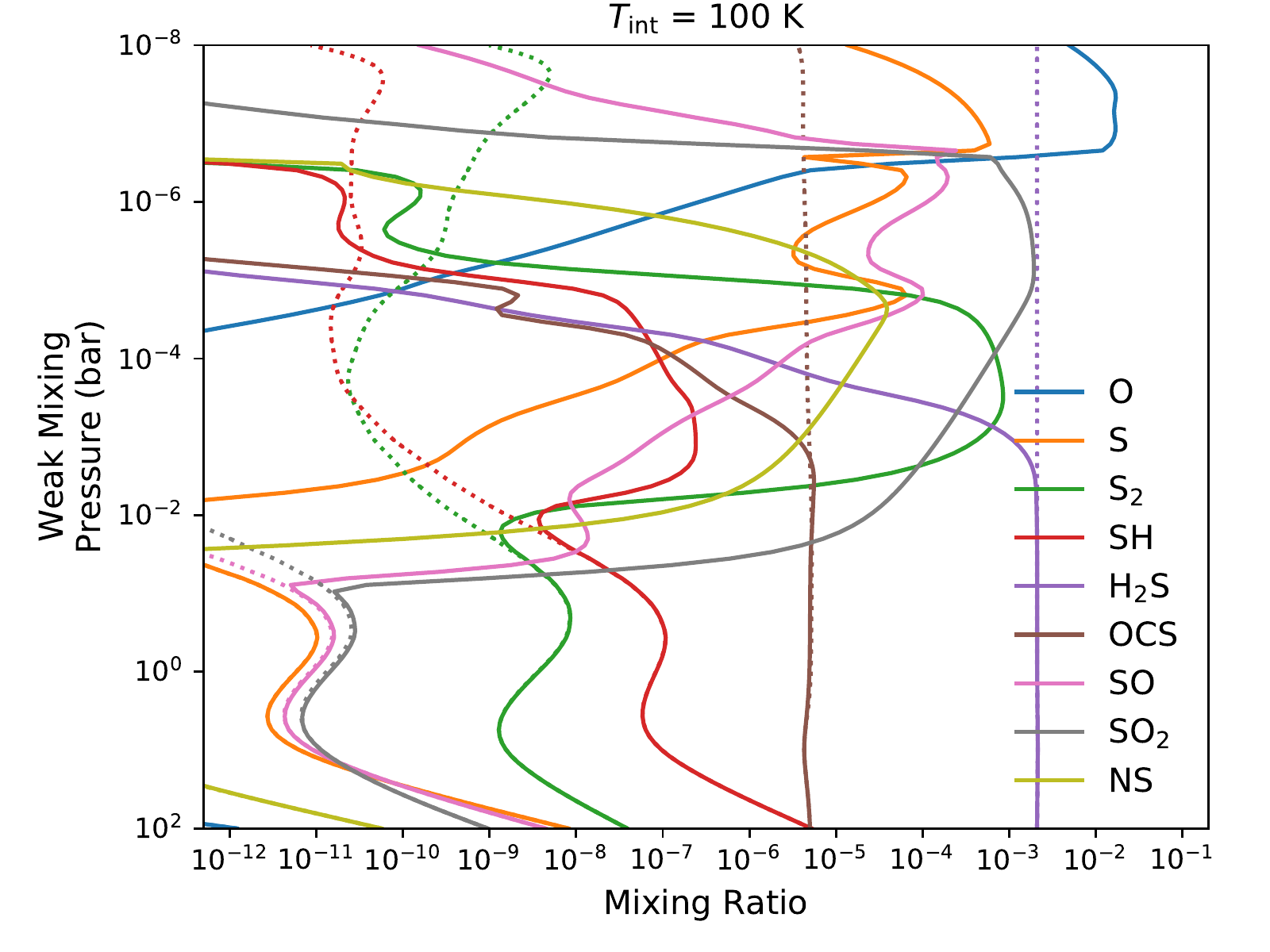}
\includegraphics[width=\columnwidth]{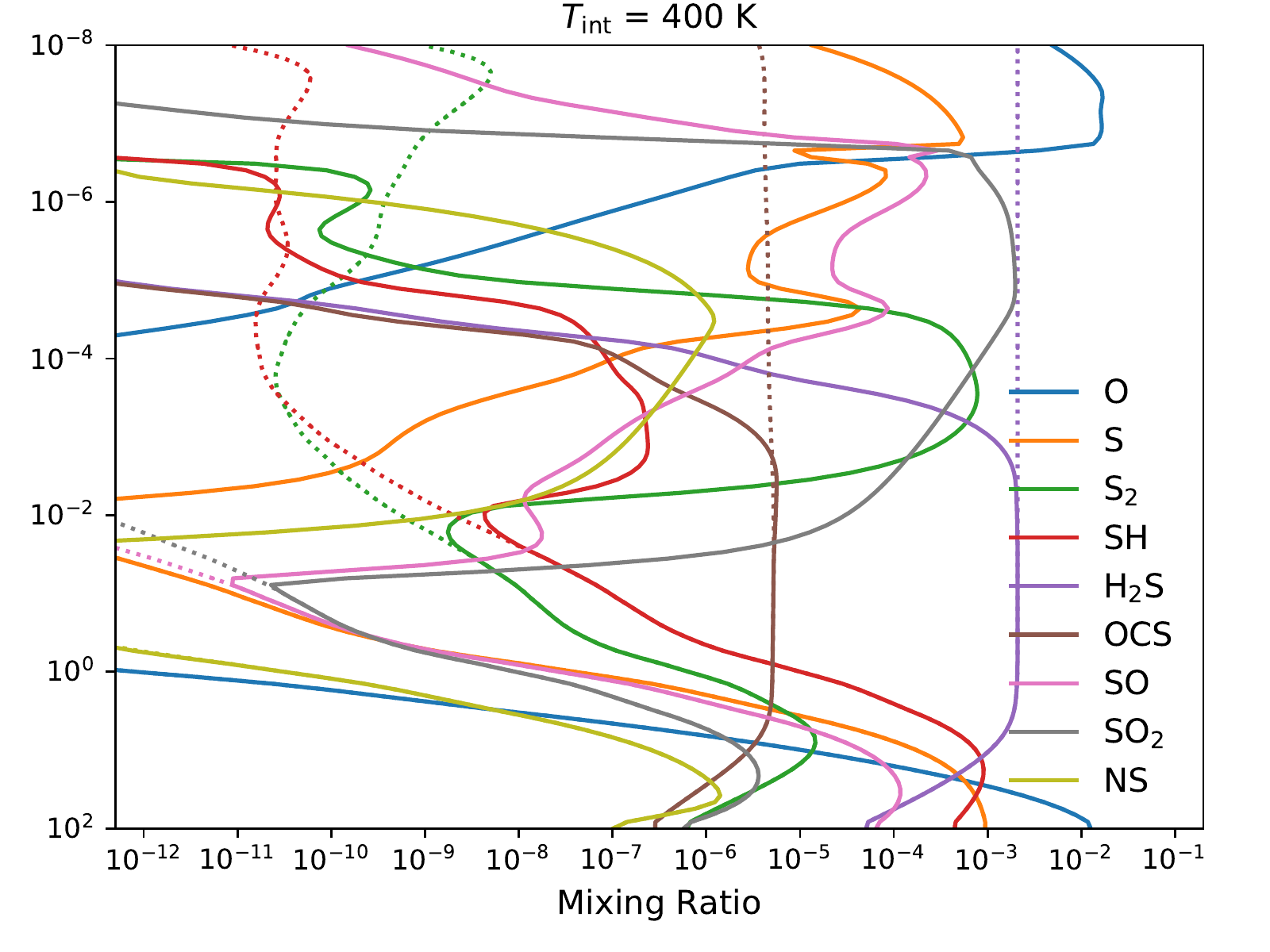}
\includegraphics[width=\columnwidth]{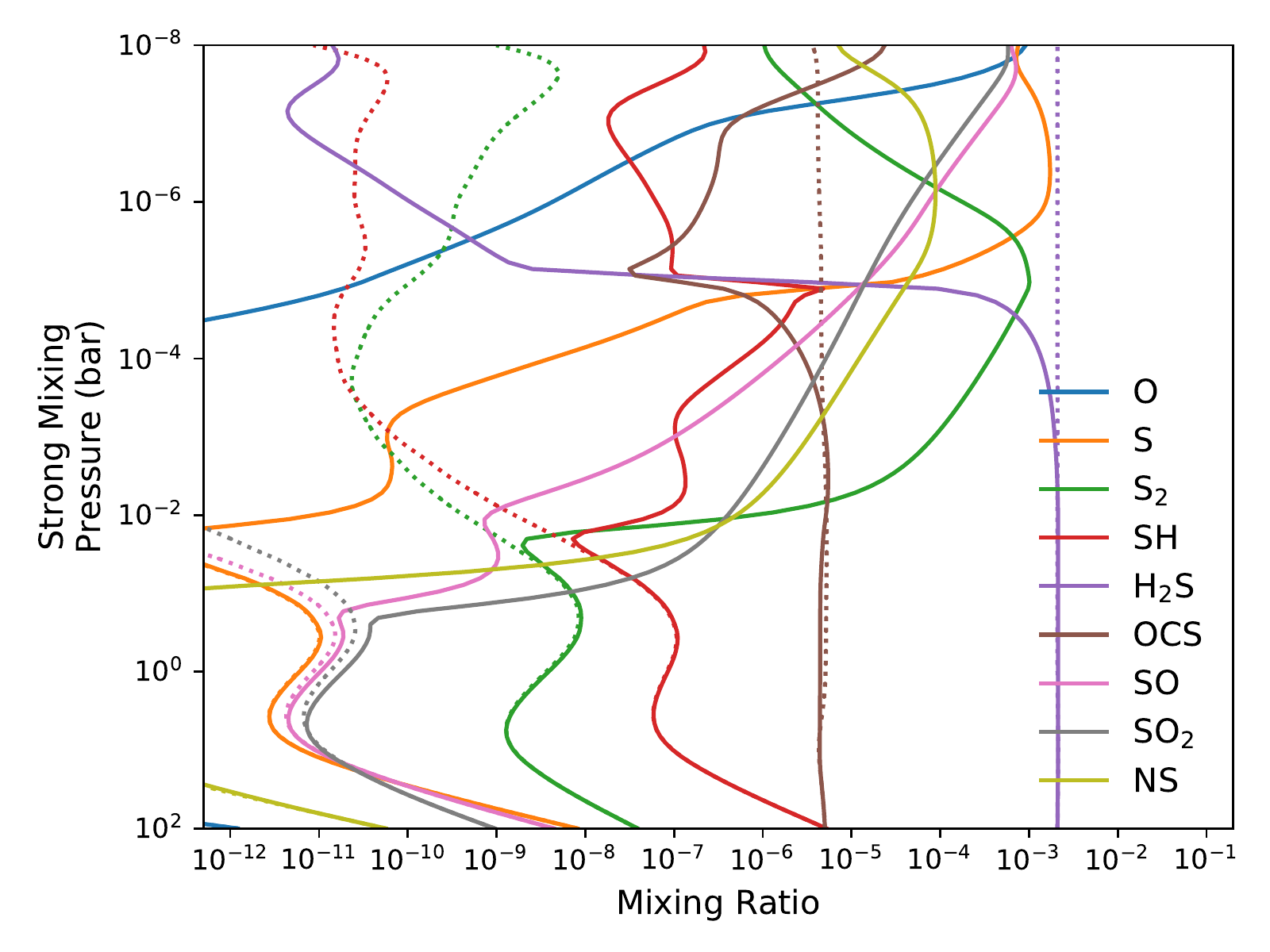}
\includegraphics[width=\columnwidth]{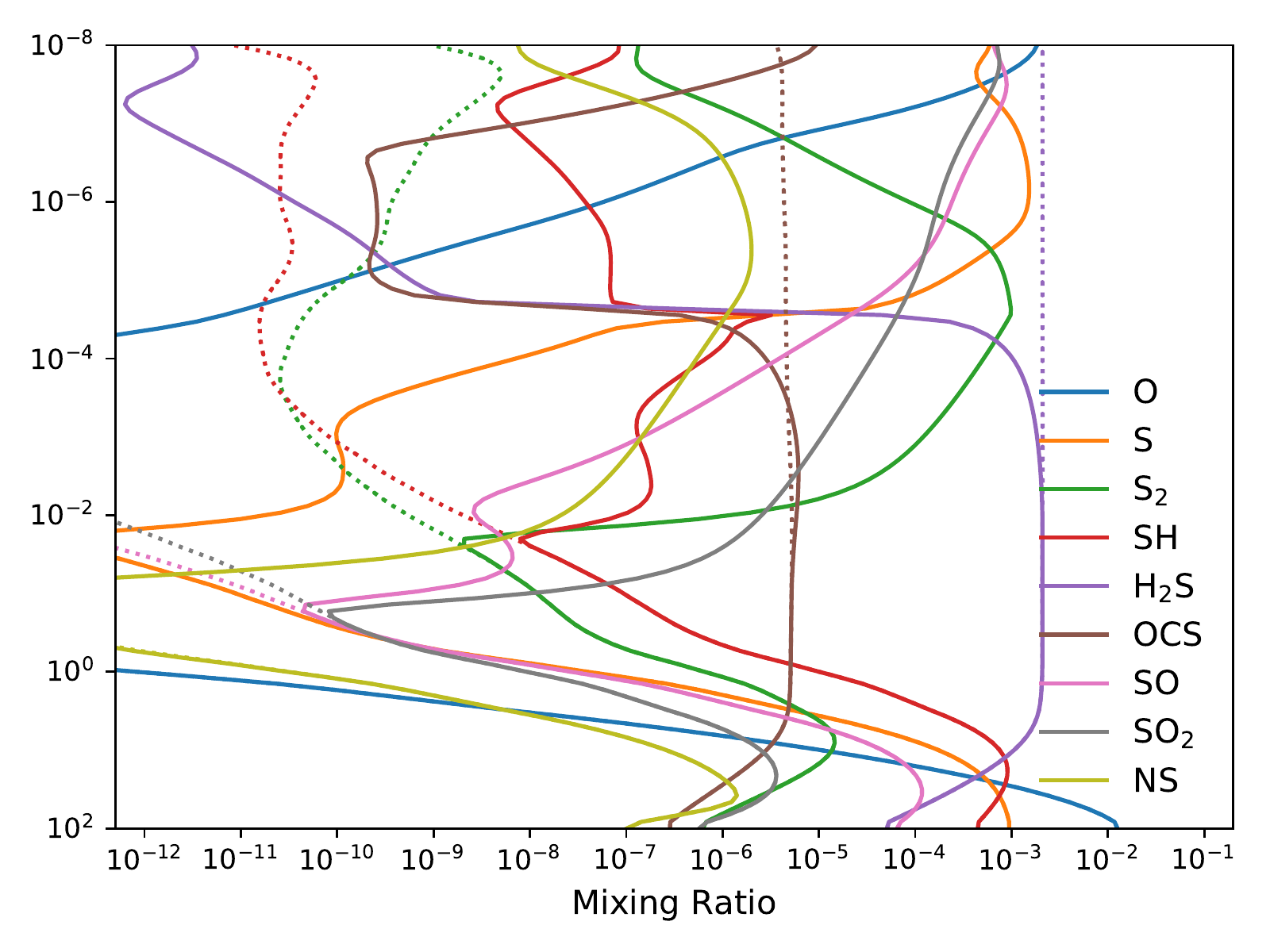}
\end{center}
\caption{Same as Figure \ref{fig:GJ436b-mix} but for atomic O and sulfur species.}
\label{fig:GJ436b-S}
\end{figure*}

\begin{figure*}[htp]
\begin{center}
\includegraphics[width=\columnwidth]{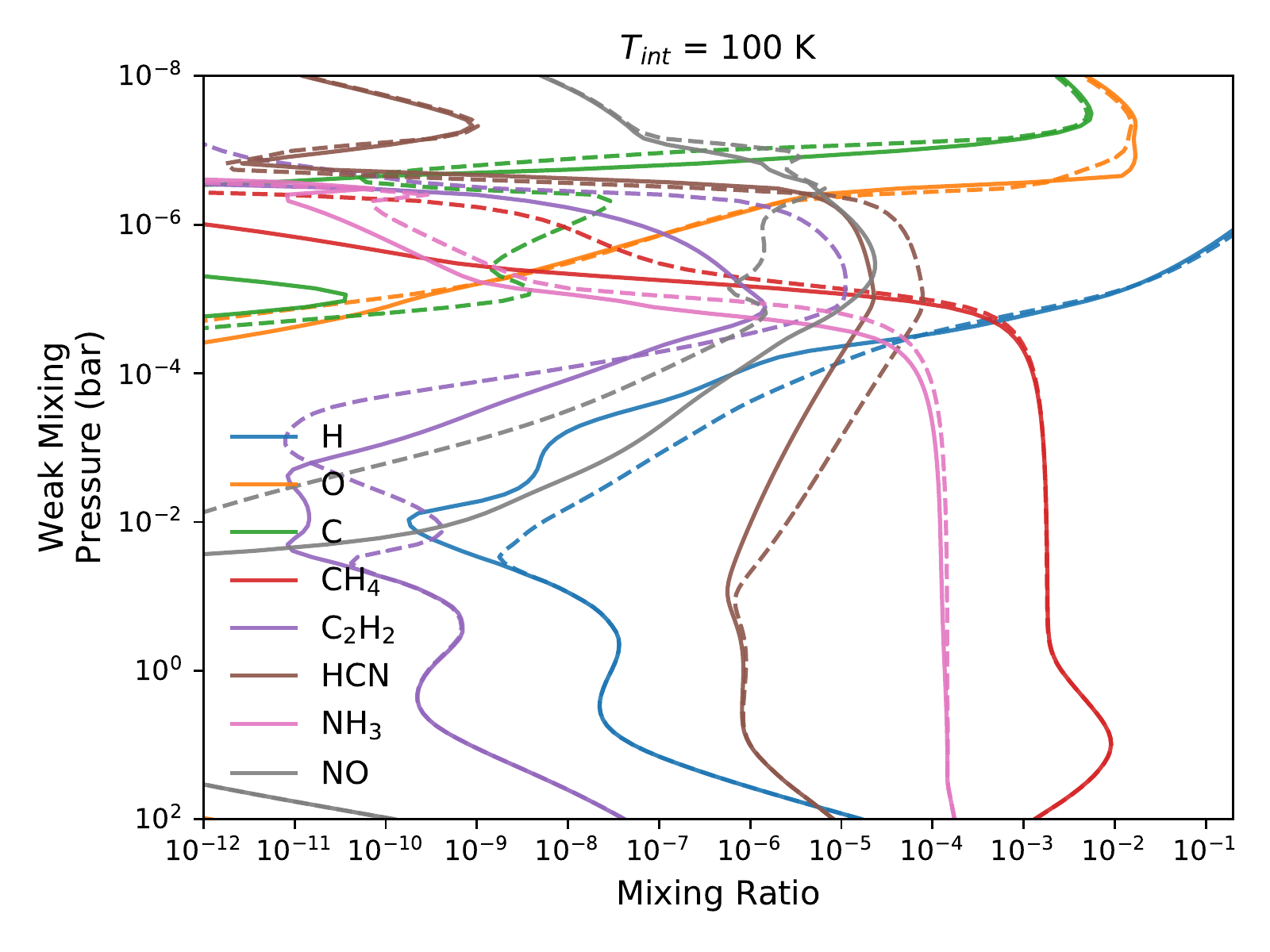}
\includegraphics[width=\columnwidth]{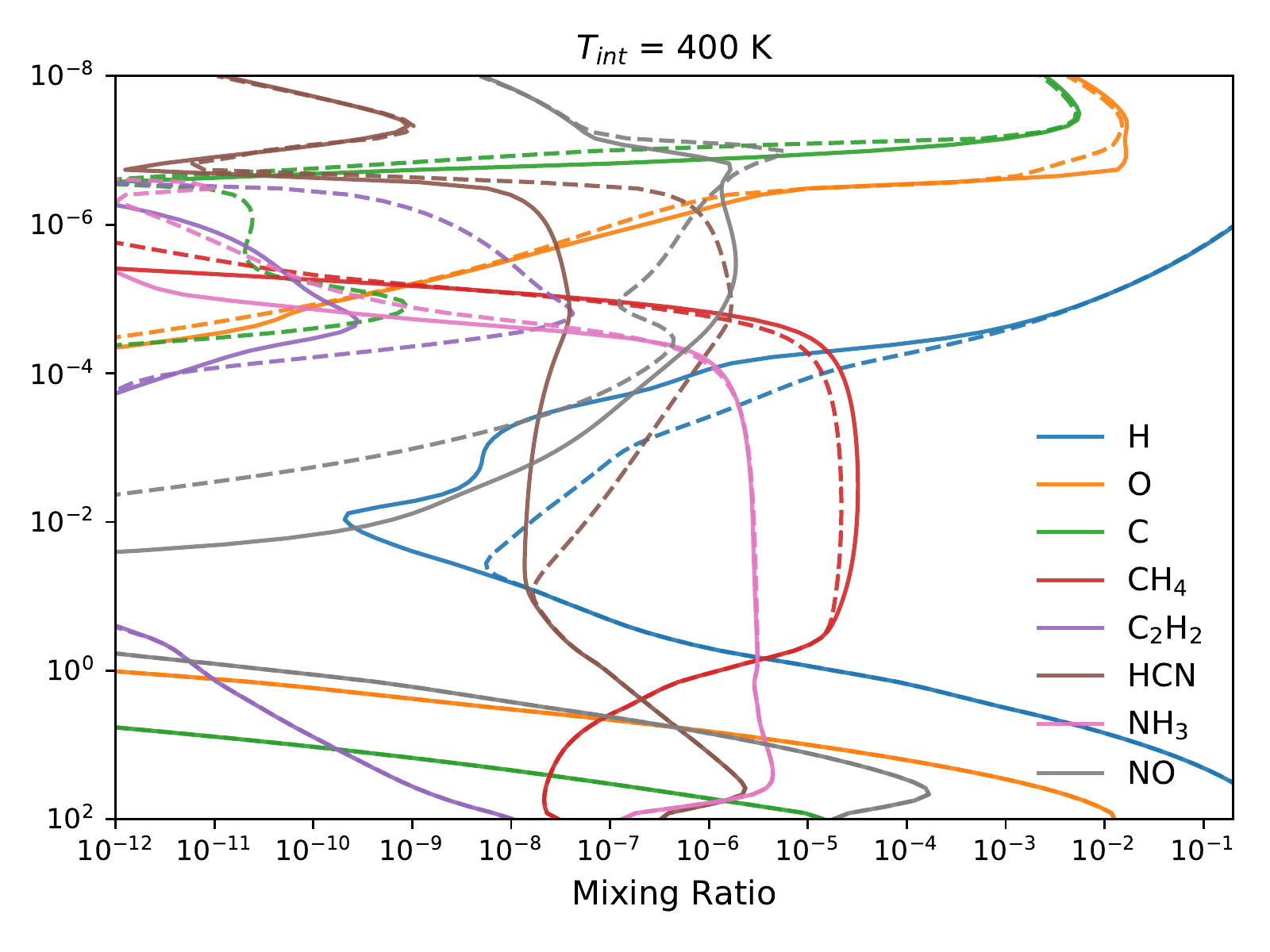}
\includegraphics[width=\columnwidth]{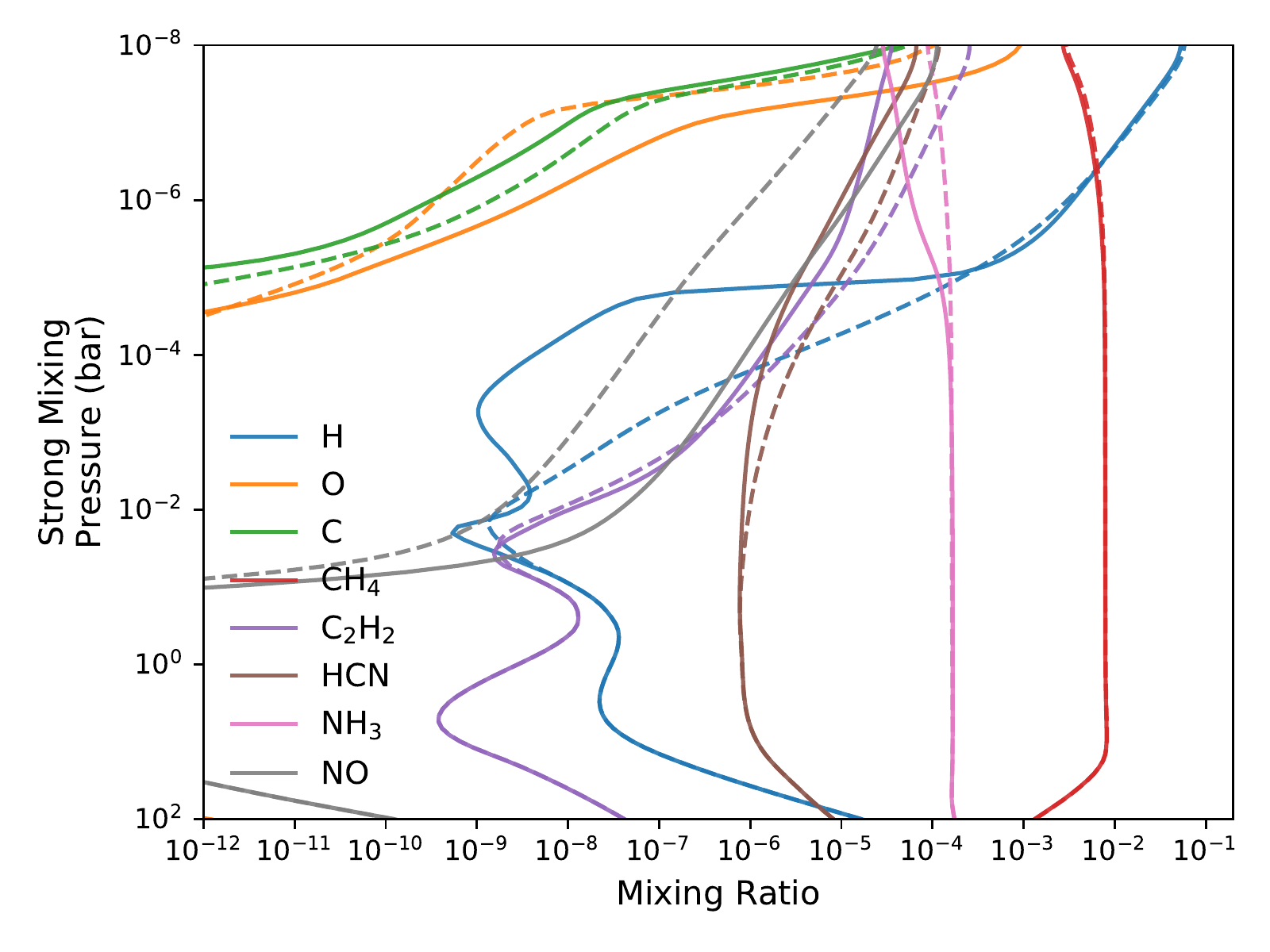}
\includegraphics[width=\columnwidth]{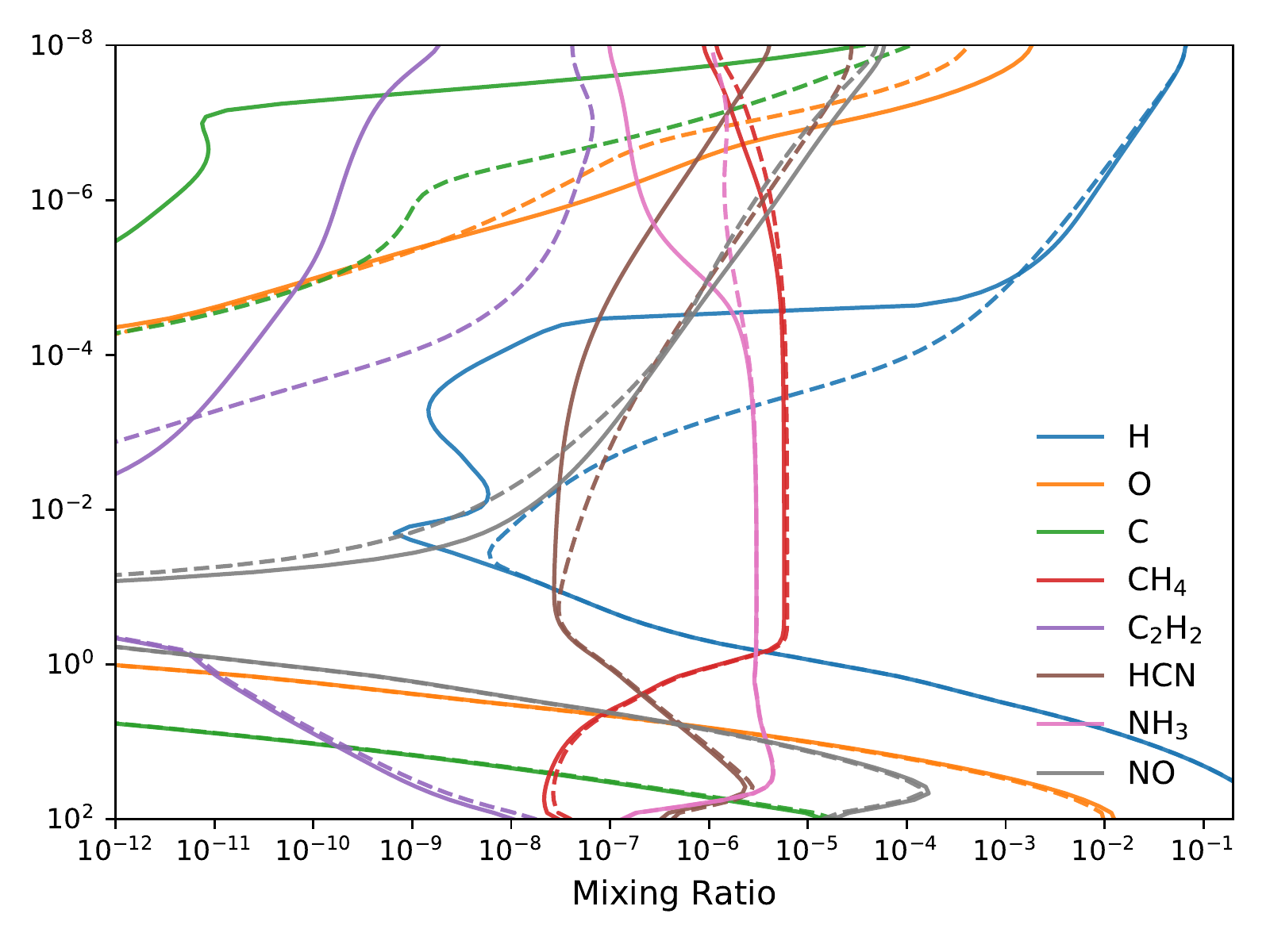}
\end{center}
\caption{The abundances of several main species that show differences from models including sulfur kinetics (solid) and without sulfur kinetics (dashed). Each panel corresponds to the same internal heating and vertical mixing as Figure \ref{fig:GJ436b-mix}.}
\label{fig:GJ436-noS}
\end{figure*}

\begin{figure}[htp]
\begin{center}
\includegraphics[width=\columnwidth]{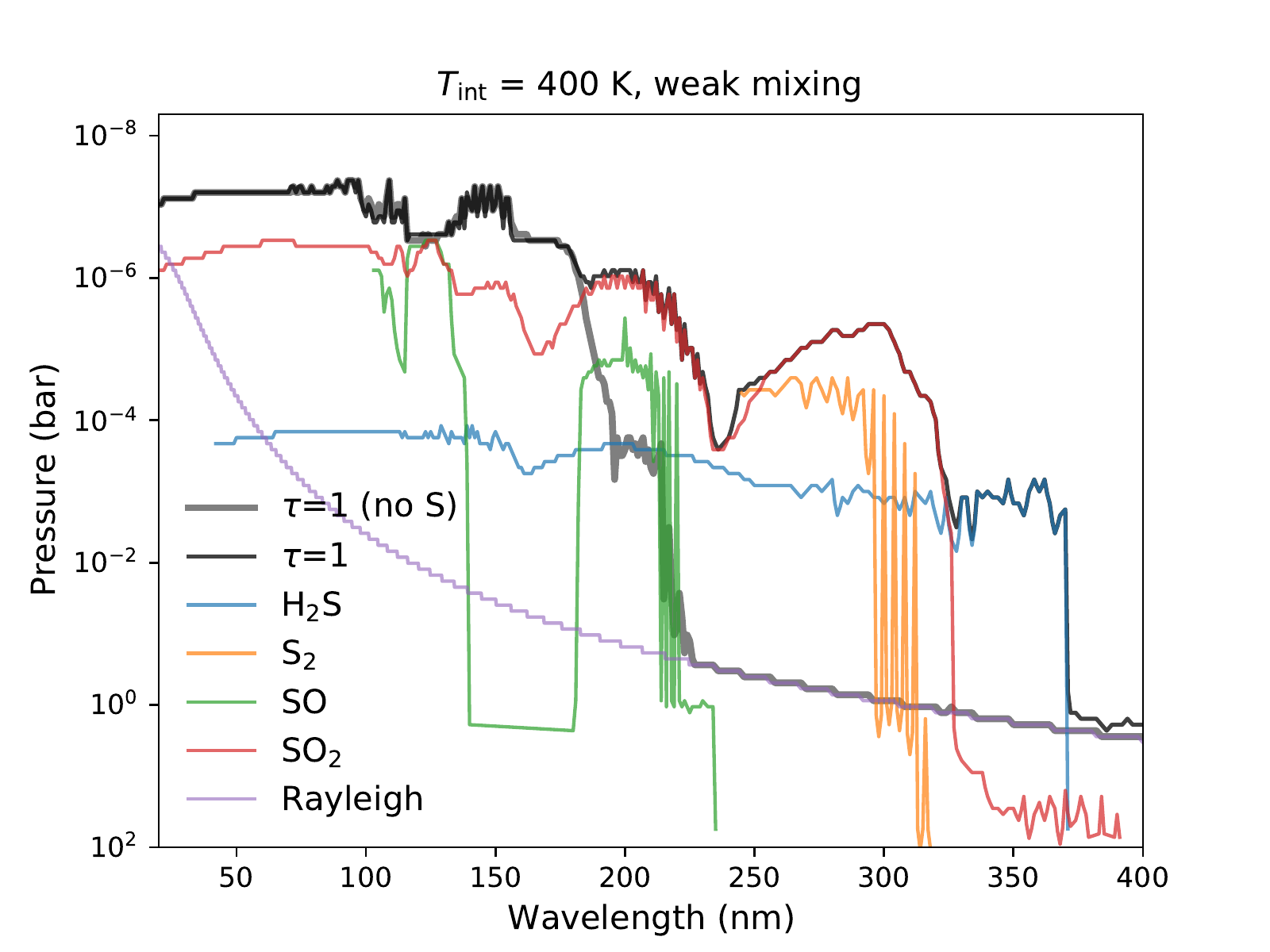}
\includegraphics[width=\columnwidth]{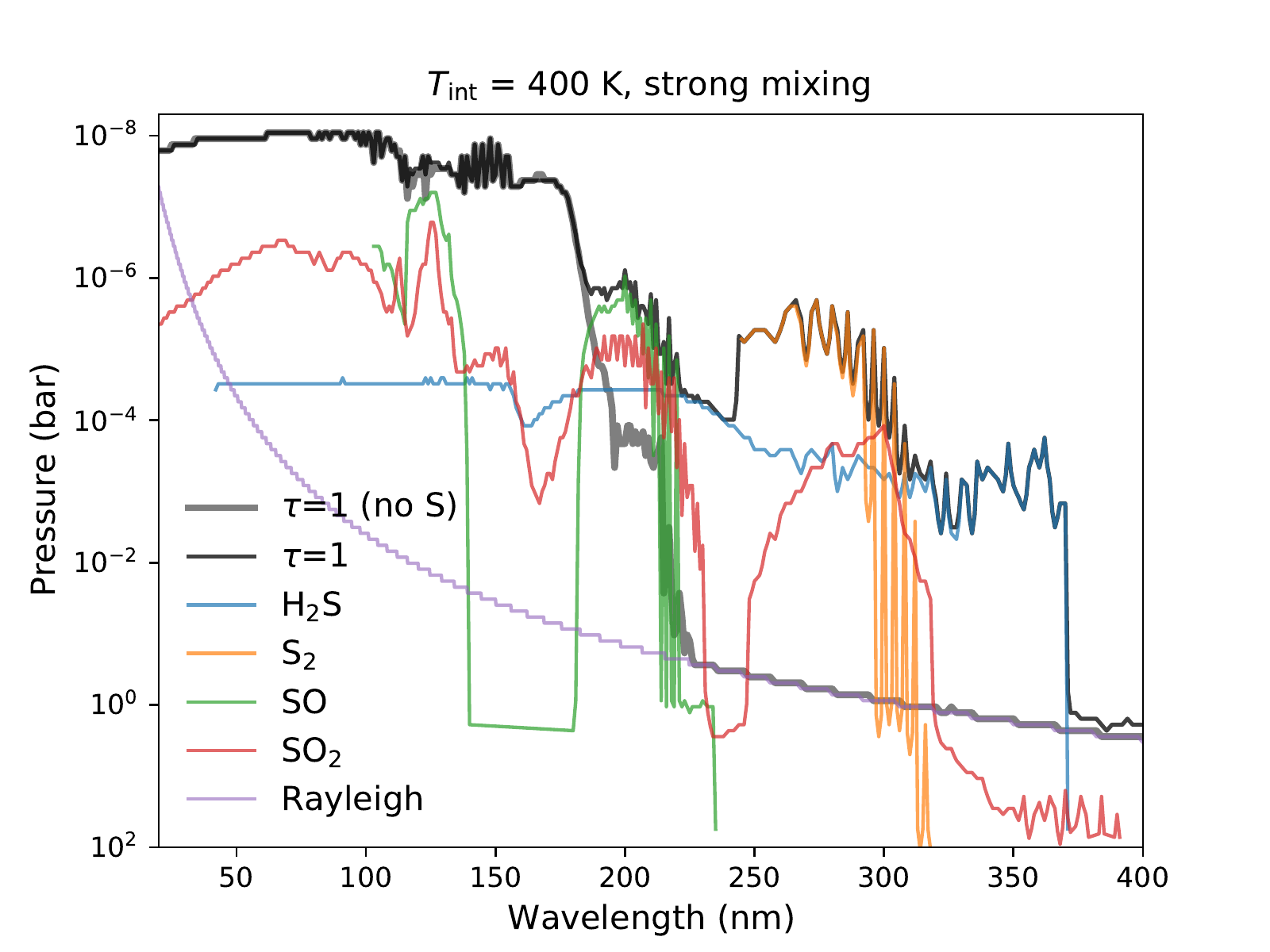}
\end{center}
\caption{The pressure level of optical depth $\tau$ = 1 for GJ 436b with high internal heating (T$_{\textrm{int}}$ = 400 K) while including (black) and excluding (grey) sulfur chemistry, along with the main contribution from sulfur species. The top and bottom panels are for weak and strong vertical mixing.}
\label{fig:GJ436b-photosphere}
\end{figure}

\subsubsection{Model Input}\label{sec:GJ436b_input}
Following the best-fit parameters in \cite{Morley2017}, we set up a low and a high internal heating scenario by running HELIOS assuming T$_{\textrm{int}}$ = 100 K and 400 K, respectively. The stellar UV flux of GJ 436 is adopted from the MUSCLES survey (version 2.2; \cite{France2016,Youngblood2016,Loyd2016}). The eddy diffusion profile is prescribed by (\ref{eq:Kzz}) with $P_{\textrm{tran}}$ = 1 bar, as where the radiative-convective transition is located in our radiative transfer calculation. We also explore the weak and strong vertical mixing scenarios, based on the GCM simulation by \cite{Lewis2010}. The average vertical wind from \cite{Lewis2010} translates to effective eddy diffusion coefficient from 10$^8$ cm$^2$ s$^{-1}$ at 100 bar to 10$^{11}$ cm$^2$ s$^{-1}$ at 0.1 mbar, assuming the mixing length to be the atmospheric scale height. Since this choice of mixing length generally overestimates the strength of eddy diffusion \citep{Smith1998,Vivien2013,Bordwell2018}, we consider it as the upper limit and set it for the strong vertical mixing scenario. We correspondingly have $K_{\textrm{deep}}$ = 10$^8$ cm$^2$ s$^{-1}$ for the strong mixing scenario and assume $K_{\textrm{deep}}$ = 10$^6$ cm$^2$ s$^{-1}$ for the weak mixing scenario.


\subsubsection{Effects of Vertical Mixing and Internal Heating}
Resolving the CO/\ce{CH4} abundance ratio is the leading question for the atmospheric compositions of GJ 436b. Since we did not vary the elemental abundance, the photospheric abundance of \ce{CH4} primarily depends on the quench level, which is controlled physically by the strength of vertical mixing and thermal structures. Figure \ref{fig:GJ436b-TPK} shows that the 100-times solar metallicity constrains both temperature profiles within the CO dominated region. As illustrated by the equilibrium profiles in Figure \ref{fig:GJ436b-mix}, for low internal heating ($T_{\textrm{int}}$ = 100 K), the temperature is close to the \ce{CH4}/CO transition and the equilibrium \ce{CH4} abundance oscillates below 10$^{-4}$ bar, whereas the equilibrium \ce{CH4} abundance decreases monotonically with increasing pressure from about 10$^{-4}$ bar for high internal heating ($T_{\textrm{int}}$ = 400 K). The twisting variation of \ce{CH4} with depth was pointed out by \cite{Karan2019}, suggesting it can lead to a non-monotonic correlation with increasing $K_{\textrm{zz}}$ for low internal heating. However, in the physically-motivated range of $K_{\textrm{zz}}$ we explored, \ce{CH4} is always quenched in the confined region below 1 bar where \ce{CH4} increases with depth, as shown in Figure \ref{fig:GJ436b-mix}. Therefore, stronger vertical mixing produces higher quenched \ce{CH4} abundance for low internal heating and conversely produces lower quenched \ce{CH4} abundance for high internal heating.

For low internal heating, \ce{CH4} remains in considerable amounts in both weak and strong mixing cases, with photospheric CO/\ce{CH4} ratio about 25 and 4, respectively. The amount of methane efficiently converts to other hydrocarbons (e.g., \ce{C2H2} and \ce{C6H6}) and HCN, especially in the photochemically active region above 1 mbar. For high internal heating, \ce{CH4} abundance is significantly reduced compared to that with low internal heating, regardless of vertical mixing. The photospheric CO/\ce{CH4} ratio for weak and strong mixing is about 2000 and 7000. \ce{C2H2} and HCN are also diminished with \ce{CH4}, except \ce{CH4} can still be transported to the upper atmosphere in the strong mixing scenario. 

For nitrogen species, \ce{N2} predominants under high metallicity and quenched \ce{NH3} exhibits greater abundances than equilibrium in all cases. The \ce{NH3}--\ce{N2} conversion is slower than \ce{CH4}--\ce{CO} and hence \ce{NH3} is quenched deeper than \ce{CH4}. Photochemically produced HCN can take over \ce{NH3} and \ce{CH4} in the upper atmosphere, but at higher altitude compared to HD 189733b due to the weaker UV flux of GJ 436. Interestingly, the quench level of \ce{NH3} appears to vary little with vertical mixing and mainly responds to the change of internal heating (see the pressure and temperature dependence of \ce{NH3}--\ce{N2} conversion in \cite{tsai18}). The insensitivity of \ce{NH3} to vertical mixing could provide additional constraints to the deep thermal property.

\subsubsection{Effects of Sulfur Species}\label{sec-GJ436-S}
Figure \ref{fig:GJ436b-S} shows the distribution of main sulfur species for each scenario. \ce{H2S} still remains the major sulfur-bearing molecule. The region where \ce{H2S} is stable extends to a lower pressure of about 0.1 mbar compared to HD 189733b because of less photochemically produced atomic H on GJ 436b. Above the \ce{H2S}-stable region, the sulfur species is more diverse than the S/\ce{H2S} dichotomy in a hot Jupiter. The stratospheric temperature of GJ 436b is too warm for elemental sulfur to grow into large allotropes but allows rich interaction of sulfur and oxygen species in the upper stratosphere. The distribution is sensitive to mixing processes: \ce{SO2} takes over \ce{H2S} for weak vertical mixing while S, \ce{S2}, and SO are in turn the leading sulfur species for strong vertical mixing. Since sulfur species {\it do not} quench in the deep region like \ce{CH4} and \ce{NH3}, they are not affected by internal heating. Instead, sulfur species are more sensitive to photochemical products transported from the upper atmosphere. 

\begin{figure}[htp]
\begin{center}
\includegraphics[width=\columnwidth]{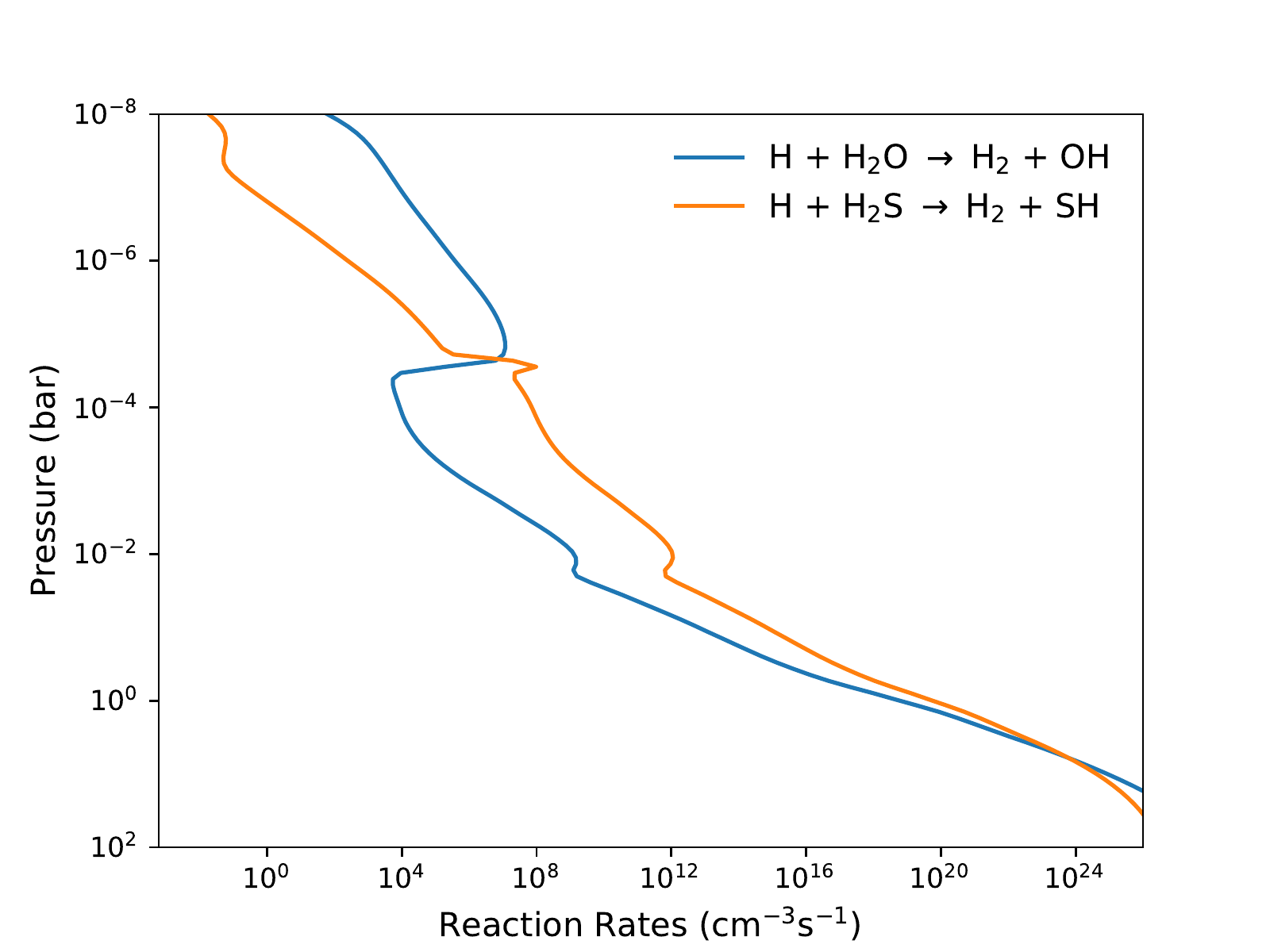}
\end{center}
\caption{The rates of reactions that are key to recycle H back to \ce{H2}, in the GJ 436b model including sulfur kinetics with $T_{\textrm{int}}$ = 400 K and weak vertical mixing (corresponding to the bottom right panel of Figure \ref{fig:GJ436-noS}).}
\label{fig:S-H2S-rate}
\end{figure}

Sulfur also impacts other non-sulfur species. Figure \ref{fig:GJ436-noS} compares our models of GJ 436b that include and exclude sulfur chemistry. H is considerably reduced between 0.1 and 10$^{-4}$ bar in the presence of sulfur, opposite to what is seen on HD 189733b. This is because  
the photolysis of SH which provides the catalytic H production on HD 189733b is absent as SH is less favored on GJ 436b. Instead, hydrogen is recycled faster by \ce{H + H2S -> H2 + SH} in the \ce{H2S}-stable region of GJ 436b, as indicated in Figure \ref{fig:S-H2S-rate}. The reduction of H subsequently slows down the production of \ce{C2H2} and HCN, even when \ce{CH4} abundance is almost unchanged. In the photochemically active region above $\sim$ 0.1 mbar, atomic C preferably combines with sulfur into OCS or CS, which further lowers \ce{C2H2} and HCN in the upper atmosphere. As \ce{NH3} being oxidized by atomic O into NO in this region, nitrogen sulfide (NS) accelerates \ce{NH3} oxidization while coupling to sulfur, analogous to the role of HCS for destroying \ce{CH4} on HD 189733b. The catalytic pathway for oxidizing \ce{NH3} is
\begin{eqnarray}
\begin{aligned} 
\ce{NH3 + H &-> NH2 + H2}\\
\ce{NH2 + S &-> NS + H2}\\
\ce{NS + O &-> NO + S}\\
\noalign{\vglue 5pt} 
\hline %
\noalign{\vglue 5pt} 
\mbox{net} : \ce{NH3 + H + O &-> NO + 2H2}.
\end{aligned}
\label{path:NH3-S}
\end{eqnarray}

Similar to HD 189733b, sulfur species raise the UV photosphere longward of 200 nm, as shown in Figure \ref{fig:GJ436b-photosphere}. The absorption feature around 200--300 nm reflects the aforementioned sensitivity to vertical mixing, with \ce{SO2} predominating in the weak mixing scenario and SO and \ce{S2} in the strong mixing scenario.

\subsubsection{Sensitivity to \ce{S_x} Polymerization}
Since the growth from \ce{S2} to larger sulfur allotropes is suppressed in our GJ 436b model, we perform a sensitivity test to see if \ce{S_x} beyond \ce{S2} can be produced with faster polysulfur recombination rates. The three-body recombination reactions that interconvert \ce{S2}--\ce{S4}--\ce{S8} are the main polymerization steps. We follow \cite{Zahnle2016} and raise the rate of \ce{S4} recombination by 10 and that of \ce{S8} recombination by 100 for a faster polysulfur forming test. We find the abundances of \ce{S4} and \ce{S8} remain low and almost unchanged. We confirm that the stratosphere of GJ 436b is too warm for elemental sulfur to grow beyond \ce{S2} into fair quantities, even after taken into account the uncertainties in the sulfur polymerization rates. 
   
\begin{figure}[htp]
\begin{center}
\includegraphics[width=\columnwidth]{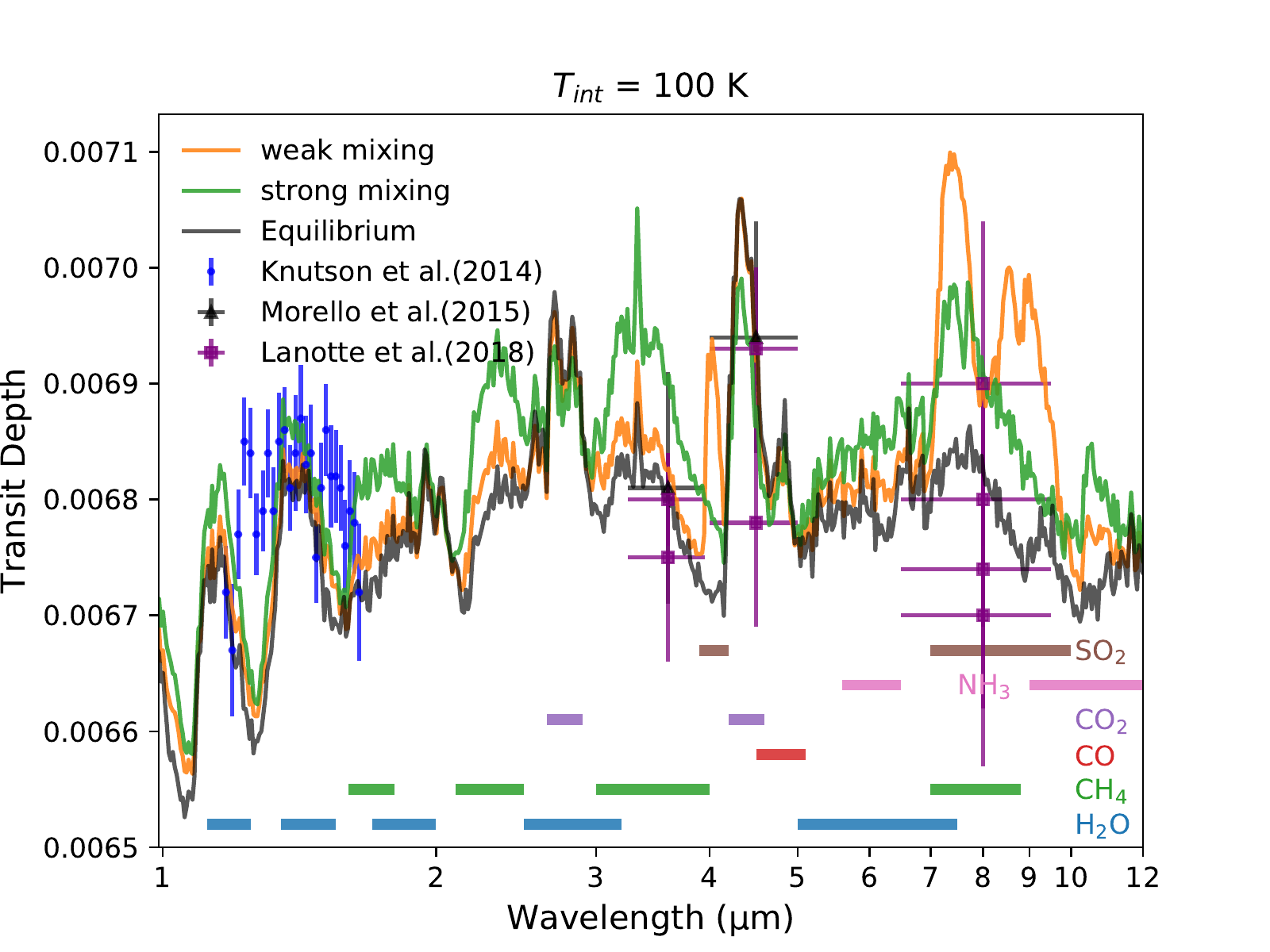}
\includegraphics[width=\columnwidth]{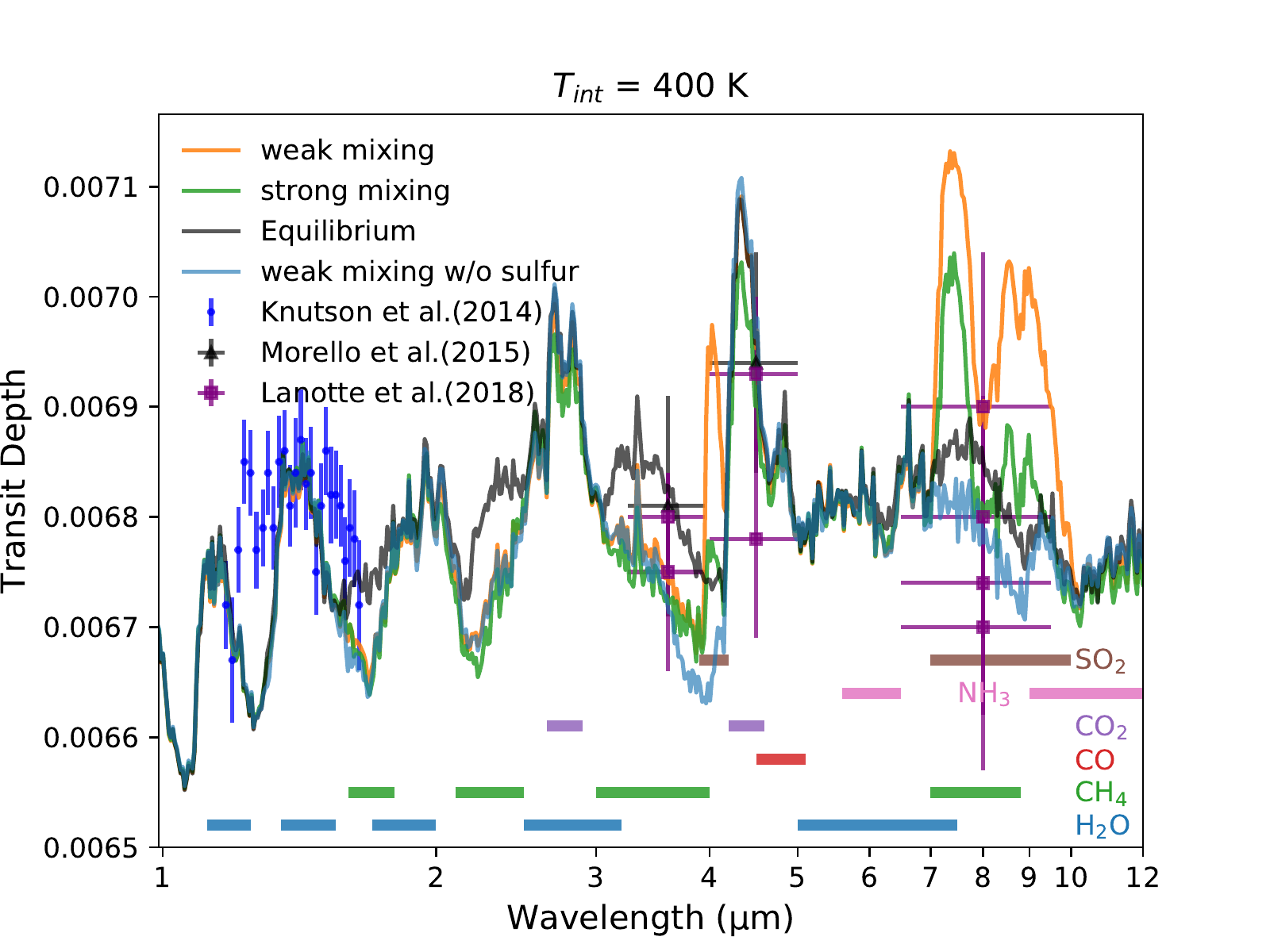}
\end{center}
\caption{Synthetic transmission spectra computed for our GJ 436b model assuming $T_{\textrm{int}}$ = 100 K (top) and 400 K (bottom) with weak and strong vertical mixing. The model without sulfur chemistry (for $T_{\textrm{int}}$ = 400 K and weak vertical mixing) is also plotted for comparison. The {\it HST}/WFC3 points from \cite{Knutson2014} have been shifted down by 200 ppm, following \cite{Lothringer2018}. The wavebands of main molecular absorption are indicated.}
\label{fig:GJ436-transit}
\end{figure}

\begin{figure}[htp]
\begin{center}
\includegraphics[width=\columnwidth]{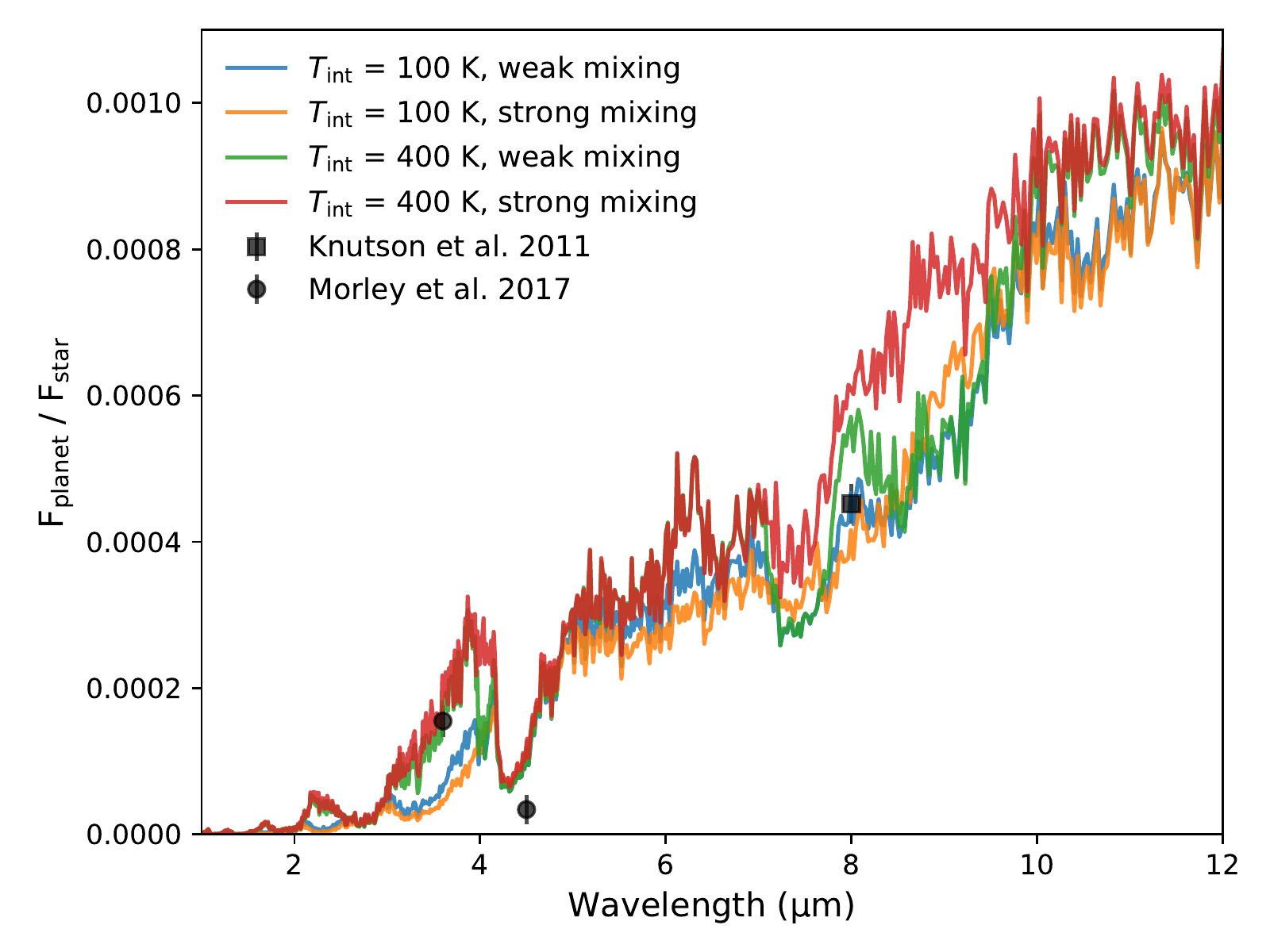}
\end{center}
\caption{Synthetic emission spectra computed for our GJ 436b models including sulfur chemistry, in comparison with {\it Spitzer} secondary-eclipse data.}
\label{fig:GJ436-emission}
\end{figure}


\subsubsection{Transmission and Emission Spectra}
The observational indication in transmission spectroscopy of varying vertical mixing and internal heating for GJ 436b is shown in Figure \ref{fig:GJ436-transit}. The early analysis of {\it Spitzer} data \citep{Knutson2011} has shown inter-epoch variability, which is reduced in the investigation of \cite{Lanotte2014,Morello2015}. Methane absorption at 2.1-2.5 and 3-4 $\mu$m and sulfur dioxide absorption at 7-10 $\mu$m are the most promising diagnostic features. For $T_{\textrm{int}}$ = 100 K, vertical mixing leads to higher \ce{CH4} abundance and the strong mixing scenario can be marginally ruled out by the {\it Spitzer} 3.6-$\mu$m data. For $T_{\textrm{int}}$ = 400 K, vertical mixing conversely reduces \ce{CH4} abundances, confirmed with previous work by \cite{Agundez2014} and \cite{Morley2017}. The 3.6 and 4.5 $\mu$m {\it Spitzer} data are consistent with our models under weak/strong vertical mixing or in chemical equilibrium. While \ce{CH4} is not sensitive to the strength of vertical mixing at high internal heating scenario, \ce{SO2} shows strong dependence on mixing processes and is favored in our weak mixing scenario, which can potentially be detectable by {\it JWST}/MIRI. In addition, \ce{S2} is more favored with strong vertical mixing and provides strong absorption features in the UV. 

Figure \ref{fig:GJ436-emission} shows the synthetic emission spectra for GJ 436b compared to {\it Spitzer} observations. While the 3.6 $\mu$m data prefers the $T_{\textrm{int}}$ = 400 K models, $T_{\textrm{int}}$ = 100 K models are favored by the 8 $\mu$m data. On the other hand, the already large column abundance of CO makes the thermal emission at 4.5 $\mu$m insensitive to internal heating or vertical mixing. The models somewhat overpredict the flux at 4.5 $\mu$m, as in the previous study \citep{Morley2017}

We conclude that our models demonstrate and confirm that the combination of moderately high ($\gtrsim$ 100 times) solar metallicity and internal heating can explain the low \ce{CH4}/CO ratio loosely constrained by the {\it Spitzer} 3.6 and 4.5 $\mu$m observations, regardless of the strength of mixing. Sulfur species do not quench in the deep region like \ce{CH4} or \ce{NH3} but closely associate with photolysis and mixing processes in the upper stratosphere. The independence from the thermal property of the interior makes sulfur chemistry a complementary avenue for characterizing GJ 436b, in conjunction with the long-standing quest for constraining \ce{CH4}/CO. 

\subsection{51 Eridani b}
51 Eridani b (51 Eri b) is a young Jupiter-mass giant around an F-type star at a wide orbit. Unlike irradiated hot Jupiters, the residual heat from the formation predominates over the stellar flux. In the discovery work, \cite{Macintosh2015} suggest 51 Eri b has an effective temperature 550 -- 750 K with vertically quenched \ce{CH4} and CO. Water vapor should not condense owing to its heat at depth, in contrast to our colder Jupiter. The combination of the hot-interior and photochemically-active stratosphere makes 51 Eri b a unique testbed for atmospheric chemistry, as  explored by \cite{Zahnle2016,Moses2016}. 

\begin{figure}[htp]
\begin{center}
\includegraphics[width=\columnwidth]{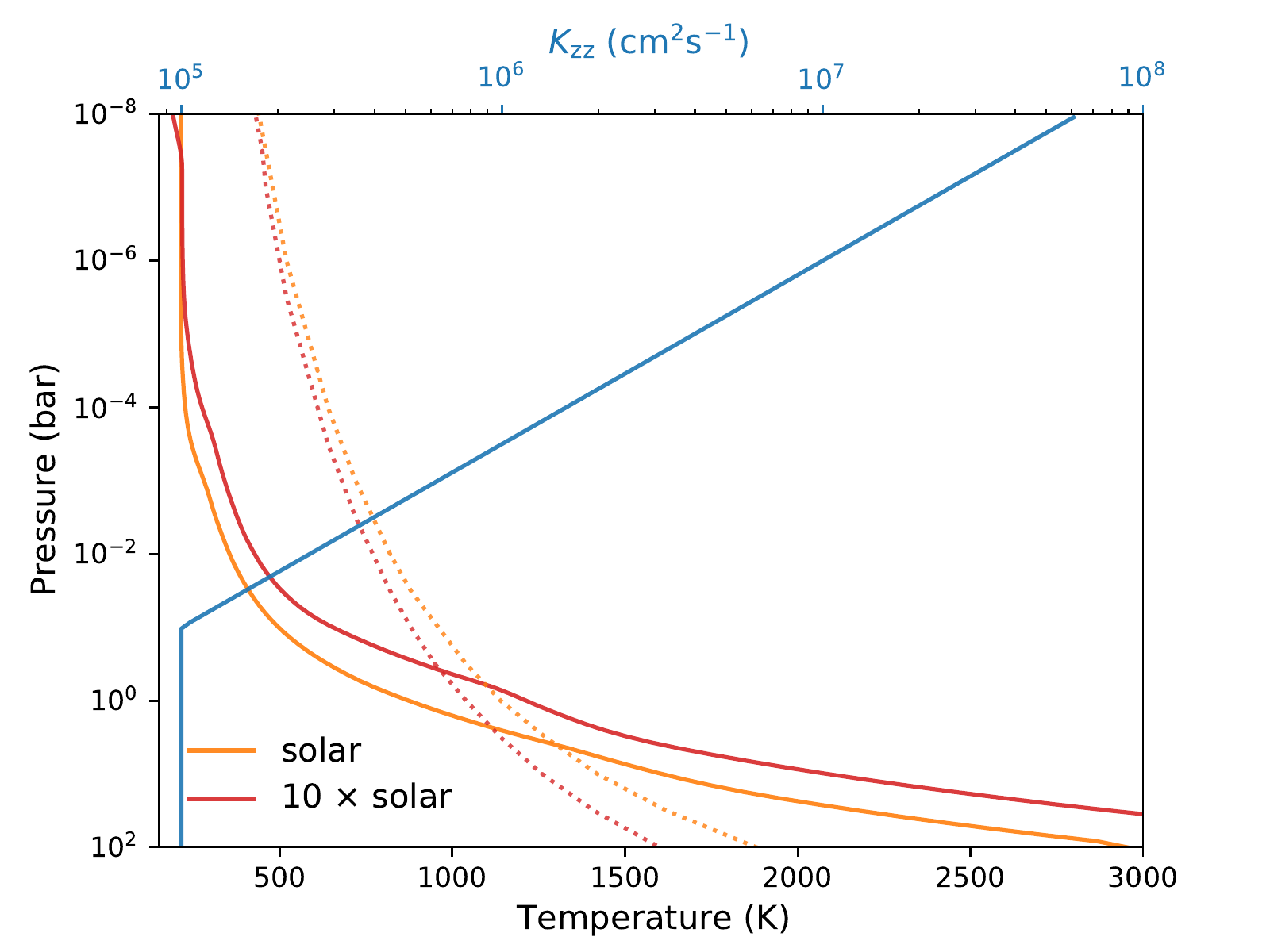}
\end{center}
\caption{The temperature-pressure and eddy diffusion ($K_\textrm{zz}$) profiles for 51 Eri b, assuming solar and 10 times solar metallicity. The [\ce{CH4}]/[CO] = 1 equilibrium transition curves corresponding to two metallicities are shown by the dotted curves.}
\label{fig:51Erib-TPK}
\end{figure}

\subsubsection{Model Setup}
We adopt $T_{\textrm{eff}}$ = 760 K as suggested by the retrieval work of \cite{Samland2017} for calculating the temperature profile of 51 Eri b (although we find little difference between setting $T_{\textrm{eff}}$ = 760 K and $T_{\textrm{eff}}$ = 700 K as assumed in previous work \citep{Moses2016,Zahnle2016}). \cite{Samland2017} also derive a 10 times super-solar metallicity based on the K-band emission. To explore the effects of metallicity, we construct one temperature profile with solar metallicity and one with 10 times solar metallicity. The resulting P-T profiles are shown in Figure \ref{fig:51Erib-TPK}. We did not include a thermosphere as \cite{Moses2016} have added an arbitrary 1000 K inversion layer above 1 $\mu$bar but found little effects on the chemistry. The eddy diffusion takes the same form by Equation (\ref{eq:Kzz}), with the radiative-convective transition P$_{\textrm{tran}}$ and $K_{\textrm{deep}}$ set to 0.1 bar and 10$^5$ cm$^2$s$^{-1}$, respectively.

 
The stellar UV flux of 51 Eridani is assembled from various sources. The observations from the International Ultraviolet Explorer (IUE)\footnote{\url{https://archive.stsci.edu/iue/}} covers the wavelength range between 115 and 198 nm. The EUV flux ($\lambda <$ 115 nm) is adopted from the synthetic spectrum of HR 8799 \citep{Sanz2011}\footnote{\url{http://sdc.cab.inta-csic.es/xexoplanets}}, following \cite{Moses2016}. For wavelengths greater than 198 nm, we use a theoretical stellar spectrum with a close temperature from ATLAS9 Model Atmosphere Database\footnote{\url{https://archive.stsci.edu/prepds/bosz/}} by BOSZ stellar atmosphere model \citep{BOSZ2017}, assuming $T_{\textrm{eff}}$ = 7250 K, log(g) = 4, log[Fe/H] = 0, radius = 1.45 M$_\odot$. The stellar irradiation received by the planet in our 51 Eridani model is stronger by about 50 $\%$ than previous work, since the orbit has been updated from 13-14 AU to 11.1 AU according to \cite{DeRosa2020}.	

\subsubsection{Disequilibrium Chemistry Compared with Previous Work}
\begin{figure*}[!ht]
\begin{center}
\includegraphics[width=\columnwidth]{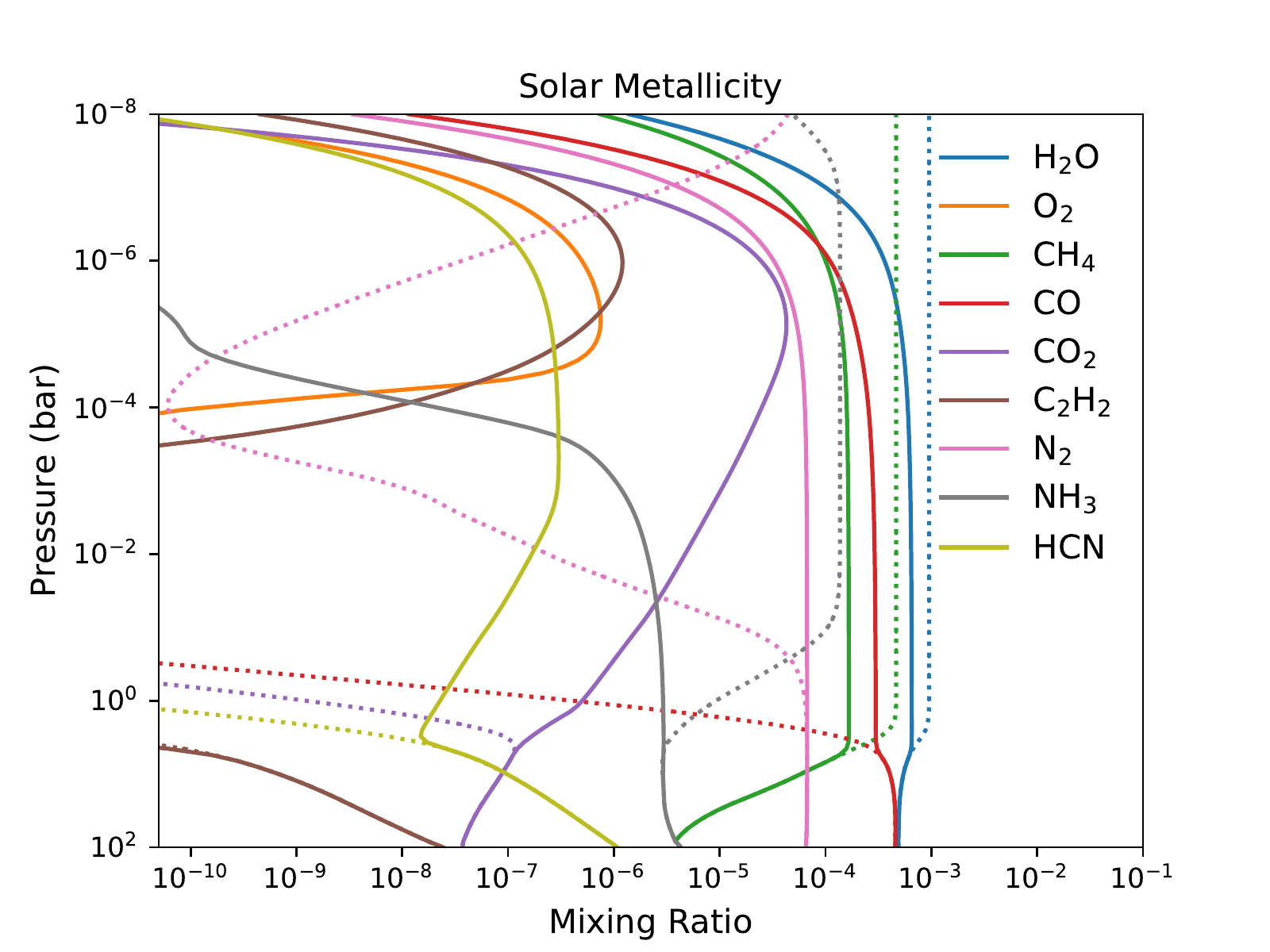}
\includegraphics[width=\columnwidth]{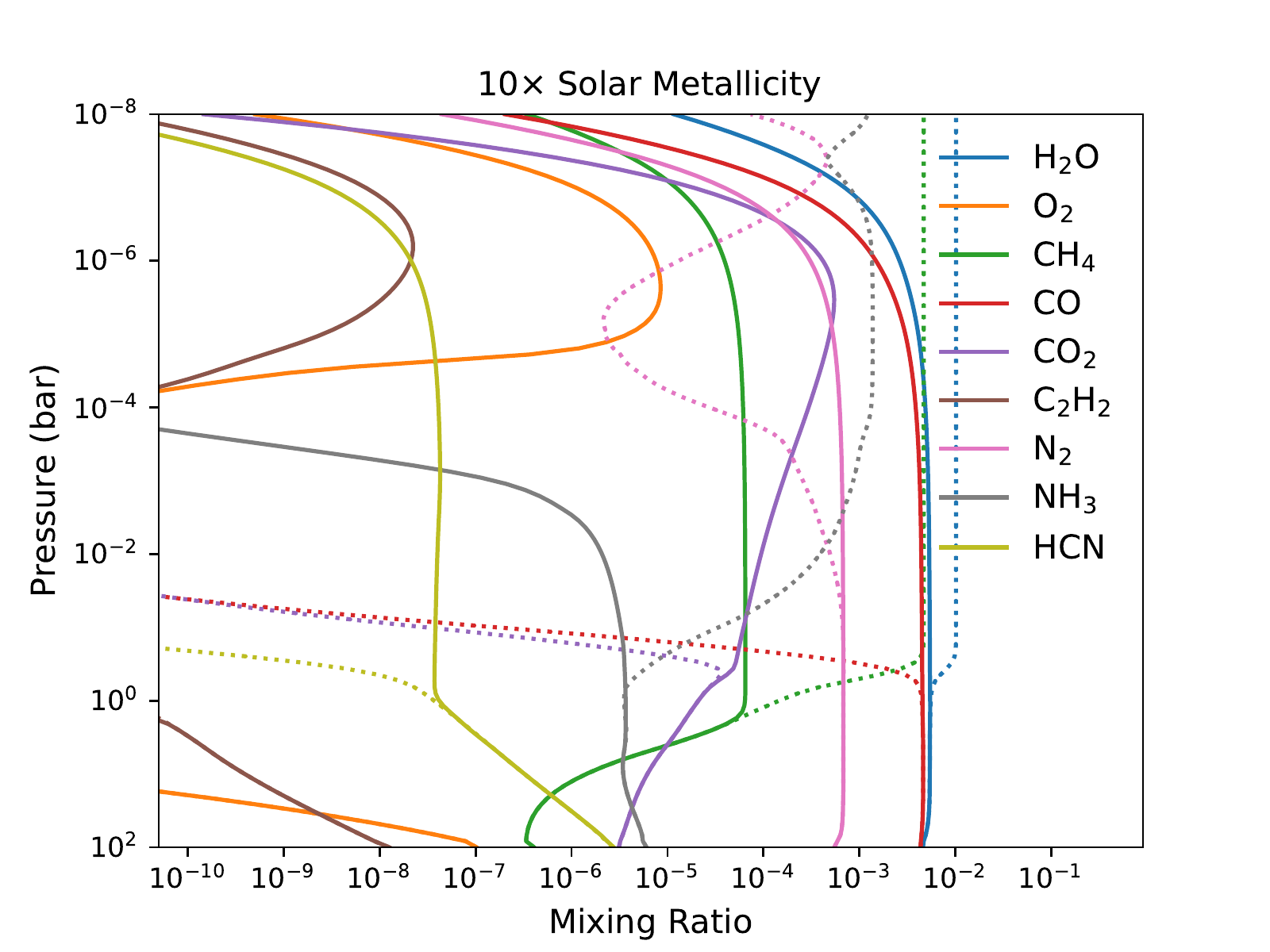}
\includegraphics[width=\columnwidth]{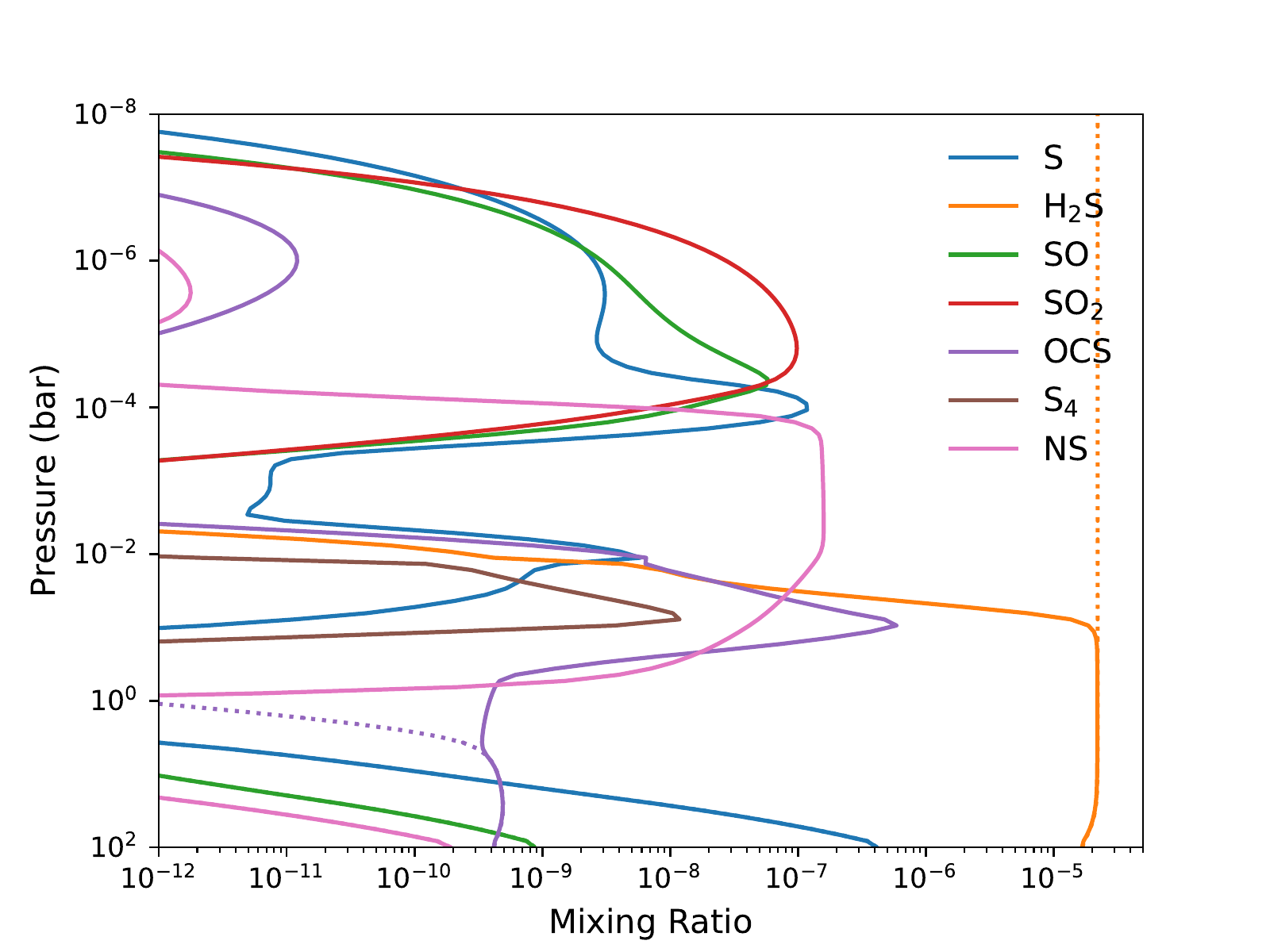}
\includegraphics[width=\columnwidth]{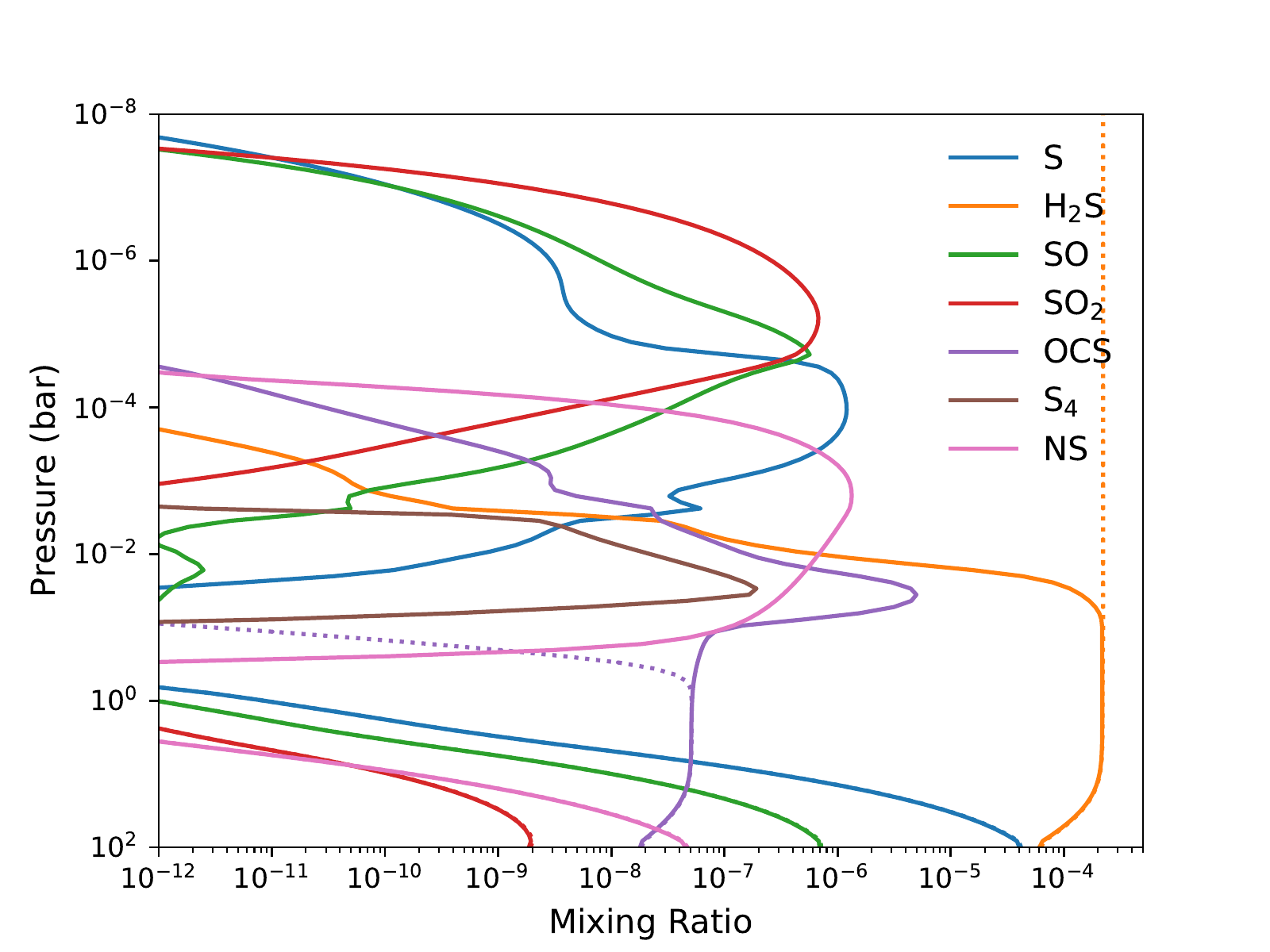}
\includegraphics[width=\columnwidth]{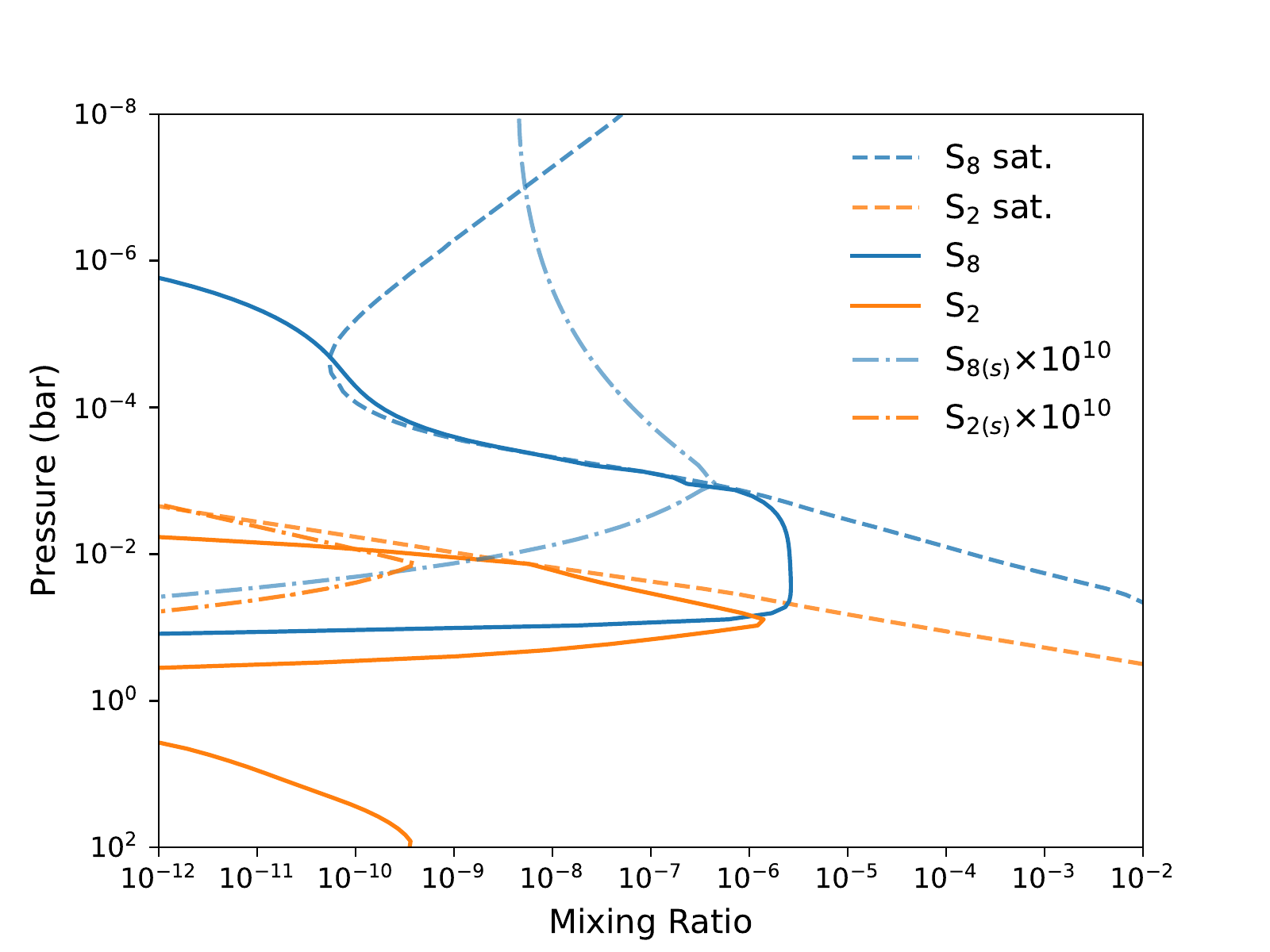}
\includegraphics[width=\columnwidth]{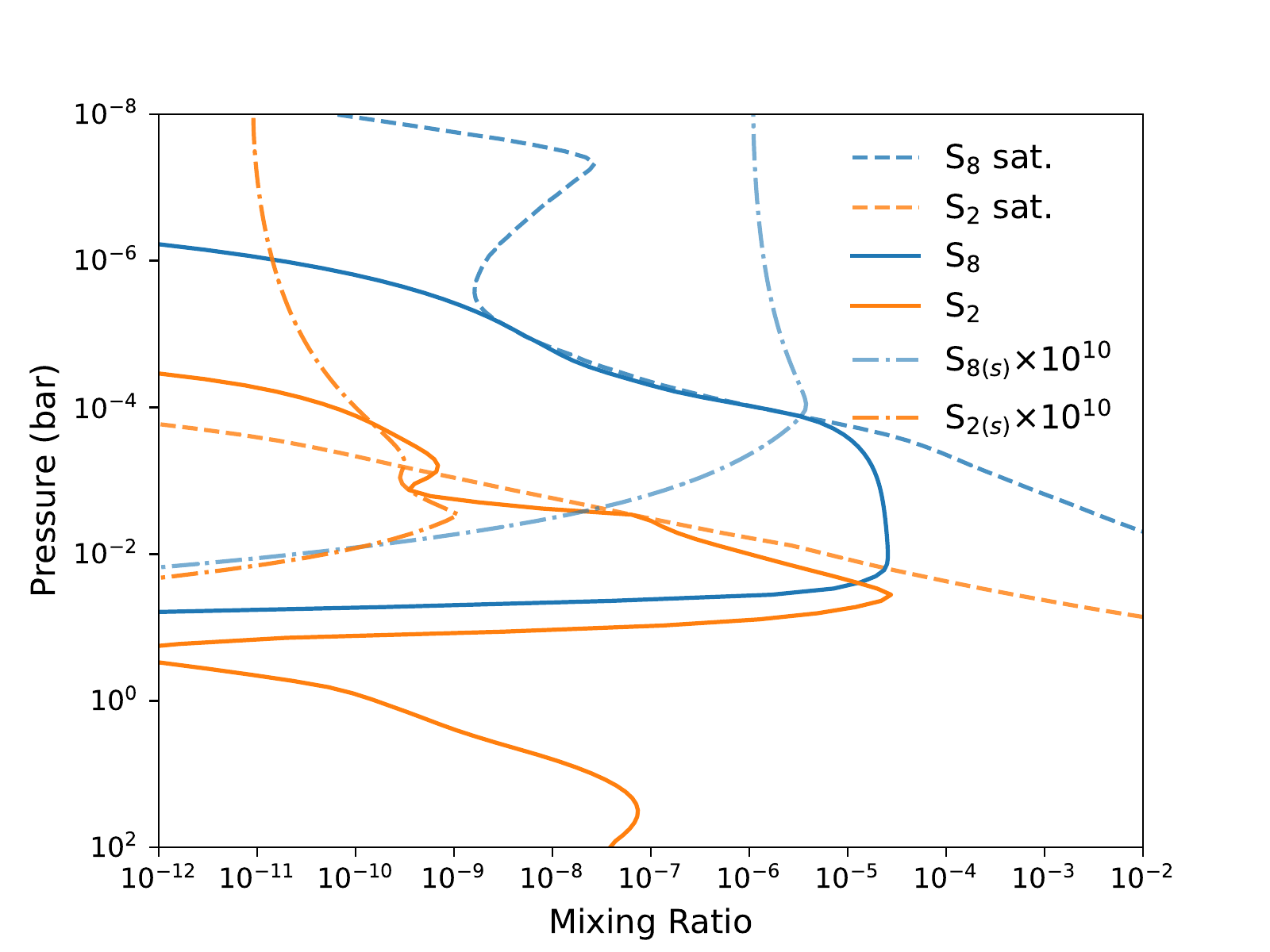}
\end{center}
\caption{The computed abundance profiles of 51 Eri b, assuming solar (left panels) and 10 times solar (right panels) metallicity. The top row presents the main species, with equilibrium profiles shown in dotted lines. The middle row shows the main sulfur species and the bottom has \ce{S2}/\ce{S8} vapor (solid), \ce{S2}/\ce{S8} condensate particles (dashed-dotted), and the saturation mixing ratios of \ce{S2}/\ce{S8} (dotted). The particles are plotted in the ratio of the number density of particles to the total number density of gas molecules and multiplied by 10$^{10}$.}
\label{fig:51Erib-mix}
\end{figure*}

\cite{Zahnle2016} investigate sulfur hazes in the atmosphere of 51 Eri b. \cite{Moses2016} use an extensive N-C-H-O chemical network ($\sim$ 1650 reactions), which include more complex hydrocarbons, to study the quenching and photochemical effects. The mixing ratios of the main species in our 51 Eri b model for solar and 10 times solar metallicity are displayed in the top left panel of Figure \ref{fig:51Erib-mix}. The main C, H, N, O chemical species in our model are overall consistent with both \cite{Zahnle2016} and \cite{Moses2016}, which we summarize in the following paragraph. 

The top row of Figure \ref{fig:51Erib-mix} shows how disequilibrium processes control the composition distribution. First, \ce{CH4}-CO conversion is quenched at about 1 bar thus CO predominates over \ce{CH4}. Likewise, \ce{N2} is the predominant nitrogen bearing species over \ce{NH3}. Second, without the fast recycling mechanism like that on a hot Jupiter, strong photolysis of water makes the upper atmosphere oxidizing and produces considerable \ce{CO2} and \ce{O2}. Third, \ce{C2H2} and \ce{HCN} are photochemically generated in the upper atmosphere, similar to hot Jupiters. While \ce{C6H6} is produced to about 10 ppb level in \cite{Moses2016}, \ce{C6H6} is greatly reduced in our model including sulfur, as seen for GJ 436b. 

The most outstanding difference between \cite{Zahnle2016} and our model in terms of sulfur chemistry is that \ce{S8} reaches condensable level in our atmosphere. Although produced at about the same level, \ce{S8} is close to saturation but does not condense in the nominal model of \cite{Zahnle2016}. Since we adopt the same saturation vapor pressure of sulfur allotropes \citep{Lyon2008} as \cite{Zahnle2016}, the different condensation behavior should be due to a warmer upper stratosphere in \cite{Zahnle2016}. \cite{Zahnle2016} indeed noted that \ce{S8} would condense if the temperature were just a few degrees lower.

\begin{figure*}[!ht]
\begin{center}
\includegraphics[width=\columnwidth]{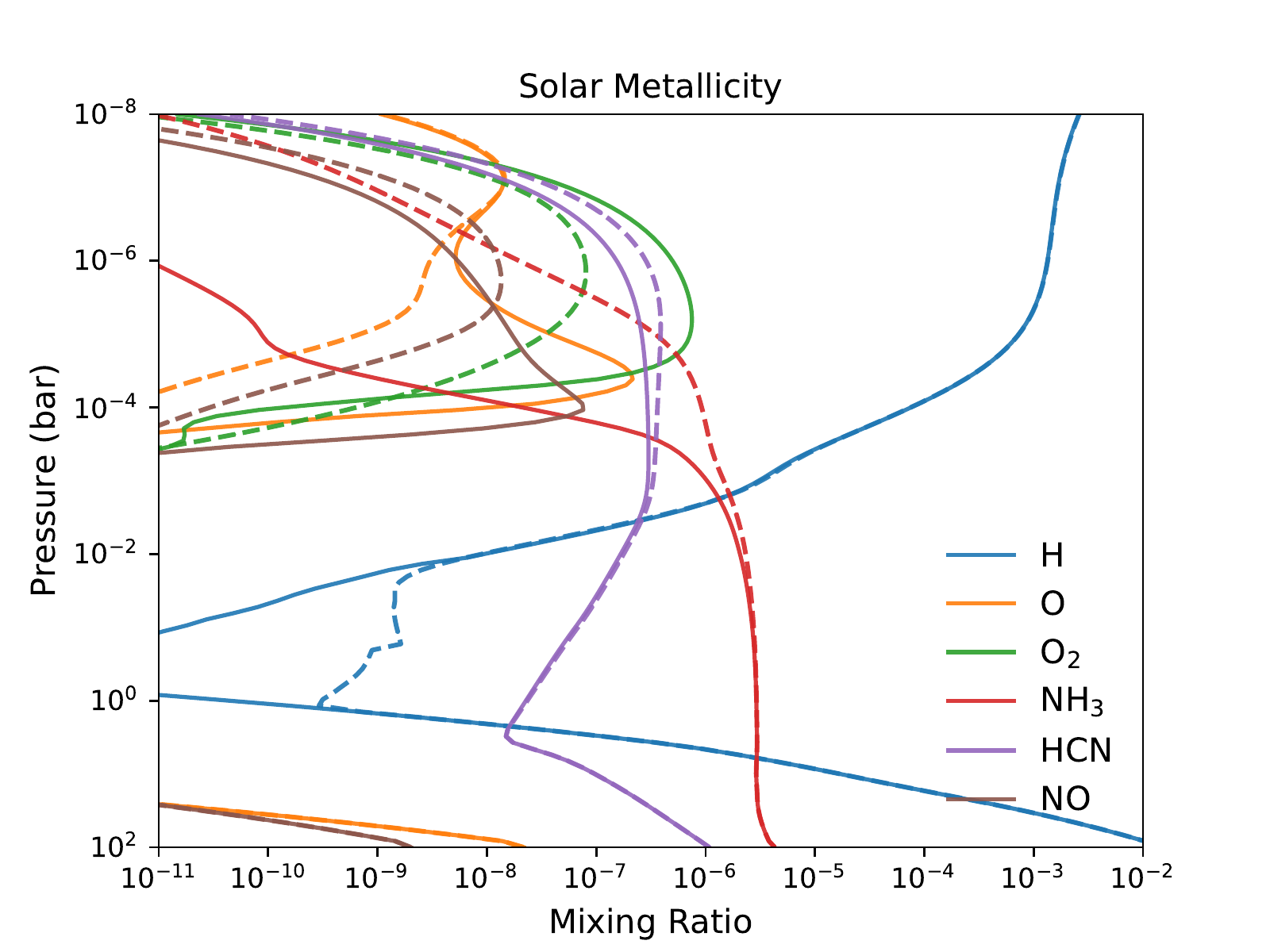}
\includegraphics[width=\columnwidth]{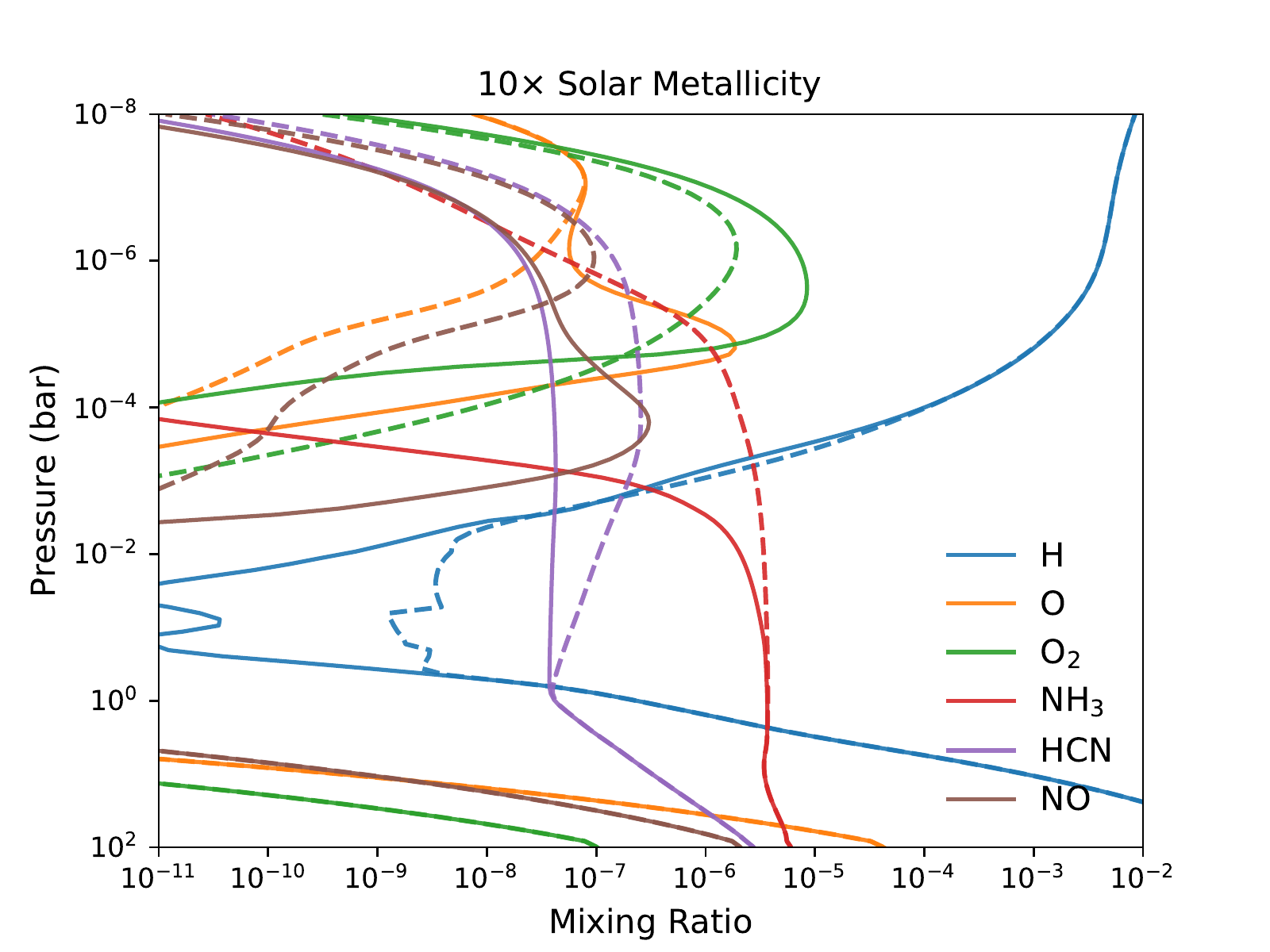}
\end{center}
\caption{The abundances of several main species that show differences from models including sulfur kinetics (solid) and without sulfur kinetics (dashed) for 51 Eri b.}
\label{fig:51Erib-noS}
\end{figure*}

\subsubsection{Effects of Super-Solar Metallicity}
The left and right columns of Figure \ref{fig:51Erib-mix} compare the results of solar and 10 times solar metallicity. The model with 10 times solar metallicity has slightly hotter troposphere (Figure \ref{fig:51Erib-TPK}) which favored CO over \ce{CH4}. Although the equilibrium abundance of \ce{CH4} in the stratosphere is increased in the 10 $\times$ solar metallicity model, \ce{CH4} is in fact decreased with a lower \ce{CH4}/CO ratio at the quench level. In the end, the 10 $\times$ solar metallicity model shows higher \ce{CO}, \ce{CO2}, \ce{H2O} and lower \ce{CH4} abundances, which in turn reduces other hydrocarbons as well. The mixing ratio of \ce{CO2} exceeds \ce{CH4} for 10 $\times$ solar metallicity and can reach $\sim$ 0.1$\%$ in the upper atmosphere. The production of \ce{O2} also raises with metallicity following the increase of water.

\subsubsection{Effects of Sulfur}\label{sec:51Eri-S}
Compared to HD 189733b and GJ 436b, \ce{H2S} can only remain stable against hydrogen abstraction (\ref{re:H2S}) at higher pressure about 0.05 bar. The reverse rate of (\ref{re:H2S}) significantly drops in the cooler stratosphere of 51 Eri b and prohibits the reformation of \ce{H2S}. 
The active SH radical produced by \ce{H2S} leads to a rich variety of sulfur species, as illustrated in the middle row of Figure \ref{fig:51Erib-mix}. Compared to \cite{Zahnle2016}, our model exhibits a more oxidized upper stratosphere due to stronger UV irradiaion from the closer orbit in our setting (11.1 AU compared to 13.2 AU). Nonetheless, both models predict efficient polymerization forming great abundance of \ce{S8}. Since \ce{S8} is the end pool of sulfur chain reactions, we find the condensation of \ce{S8} does not affect other sulfur species. Elemental S is still the leading sulfur species above the \ce{S8} condensing layers until being oxidized into SO and \ce{SO2} in the upper stratosphere.

The equilibrium abundance of \ce{H2S} scales with metallicity, which leads to more production of \ce{S8} vapor as metallicity increased. The 10 times solar metallicity model has slightly warmer temperature which allows higher saturation pressure of sulfur as well. In the end, both the gas and condensates of \ce{S2} and \ce{S8} increase with metallicity.  

The effects of coupling to sulfur on other species are highlighted in Figure \ref{fig:51Erib-noS}.  The most remarkable feature is the enhanced oxygen abundances in the upper atmosphere with sulfur. In the absence of sulfur, atomic O can be released from \ce{H2O} with the aid of \ce{CO2}: 
\begin{eqnarray}
\begin{aligned} 
\ce{H2O &->[h\nu] OH + H}\\
\ce{CO + OH &-> CO2 + H}\\
\ce{CO2 &->[h\nu] CO + O}\\
\noalign{\vglue 5pt} 
\hline %
\noalign{\vglue 5pt} 
\mbox{net} : \ce{H2O &->[2 h\nu] 2H + O}.
\end{aligned}
\label{re:path-ch4-co-S}
\end{eqnarray}
While sulfur is present, SO and \ce{SO2} dissociate more than \ce{CO2} around Ly-$\alpha$ and provide a faster channel to liberate O from \ce{H2O}: 
\begin{eqnarray}
\begin{aligned} 
\ce{H2O &->[h\nu] OH + H}\\
\ce{SO + OH &-> SO2 + H}\\
\ce{SO2 &->[h\nu] SO + O}\\
\noalign{\vglue 5pt} 
\hline %
\noalign{\vglue 5pt} 
\mbox{net} : \ce{H2O &->[2 h\nu] 2H + O}.
\end{aligned}
\end{eqnarray}
or
\begin{eqnarray}
\begin{aligned} 
\ce{H2O &->[h\nu] OH + H}\\
\ce{S + OH &-> SO + H}\\
\ce{SO &->[h\nu] S + O}\\
\noalign{\vglue 5pt} 
\hline %
\noalign{\vglue 5pt} 
\mbox{net} : \ce{H2O &->[2 h\nu] 2H + O}.
\end{aligned}
\end{eqnarray}
The excess atomic O readily reacts with OH to form \ce{O2}. This enhanced oxidized region along with NS accelerates the oxidization of \ce{NH3}, via the same pathway (\ref{path:NH3-S}) but more pronounced than that on GJ 436b. On the other hand, \ce{CH4} is unaffected because the intermediates HCS and CS are deficient in the colder stratosphere of 51 Eri b. Lastly, the coupling to \ce{H2S} also helps atomic H recycle back to \ce{H2} faster, as seen on GJ 436b. 

\begin{figure}[htp]
\begin{center}
\includegraphics[width=\columnwidth]{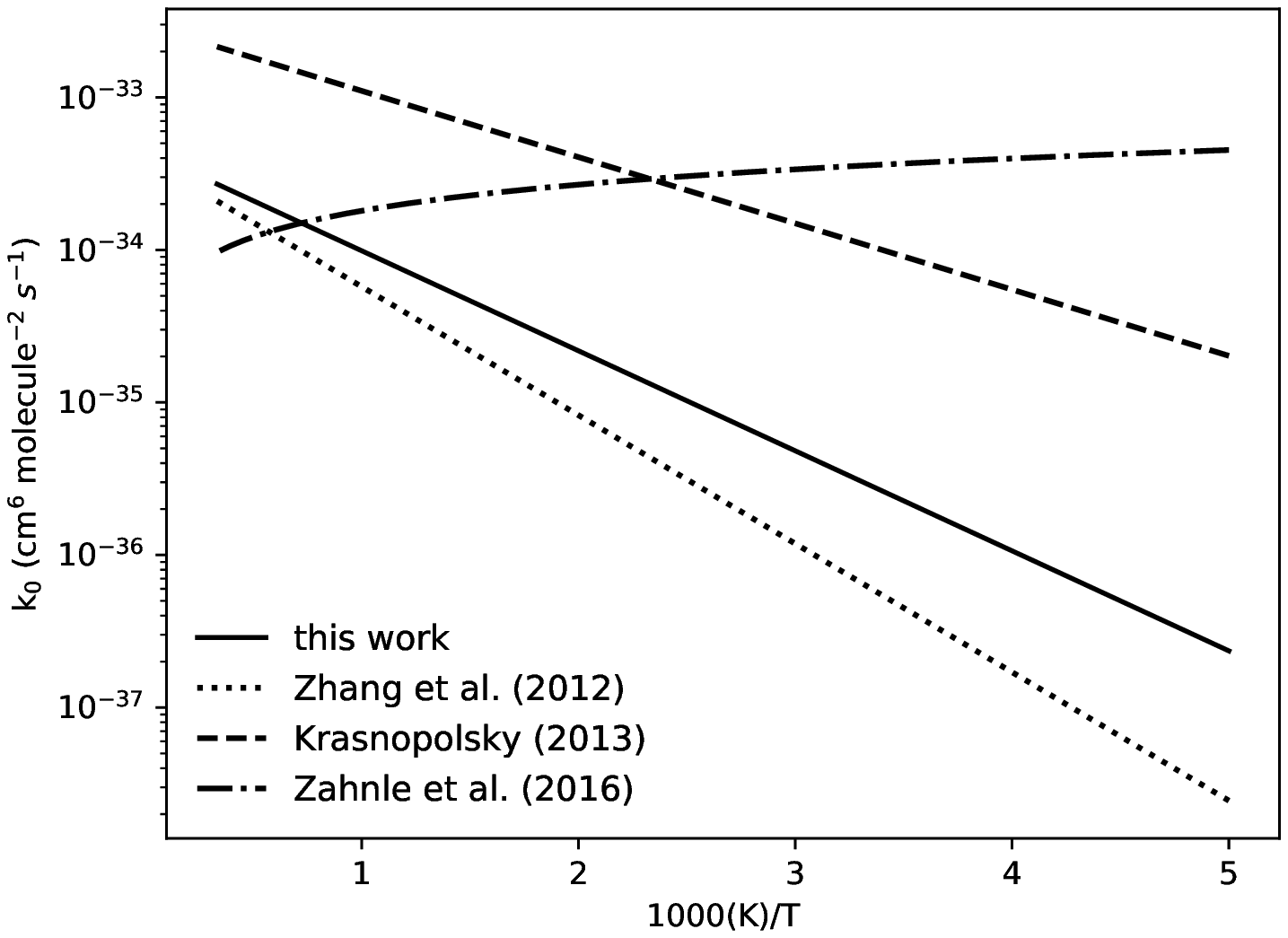}
\end{center}
\caption{The low-pressure limit rate coefficient of \ce{S + CO ->[\textrm{M}] OCS} estimated in this work (\ref{COS_rate}), compared to those in the literature.}
\label{fig:rate_COS}
\end{figure}

\begin{figure}[htp]
\includegraphics[width=\columnwidth]{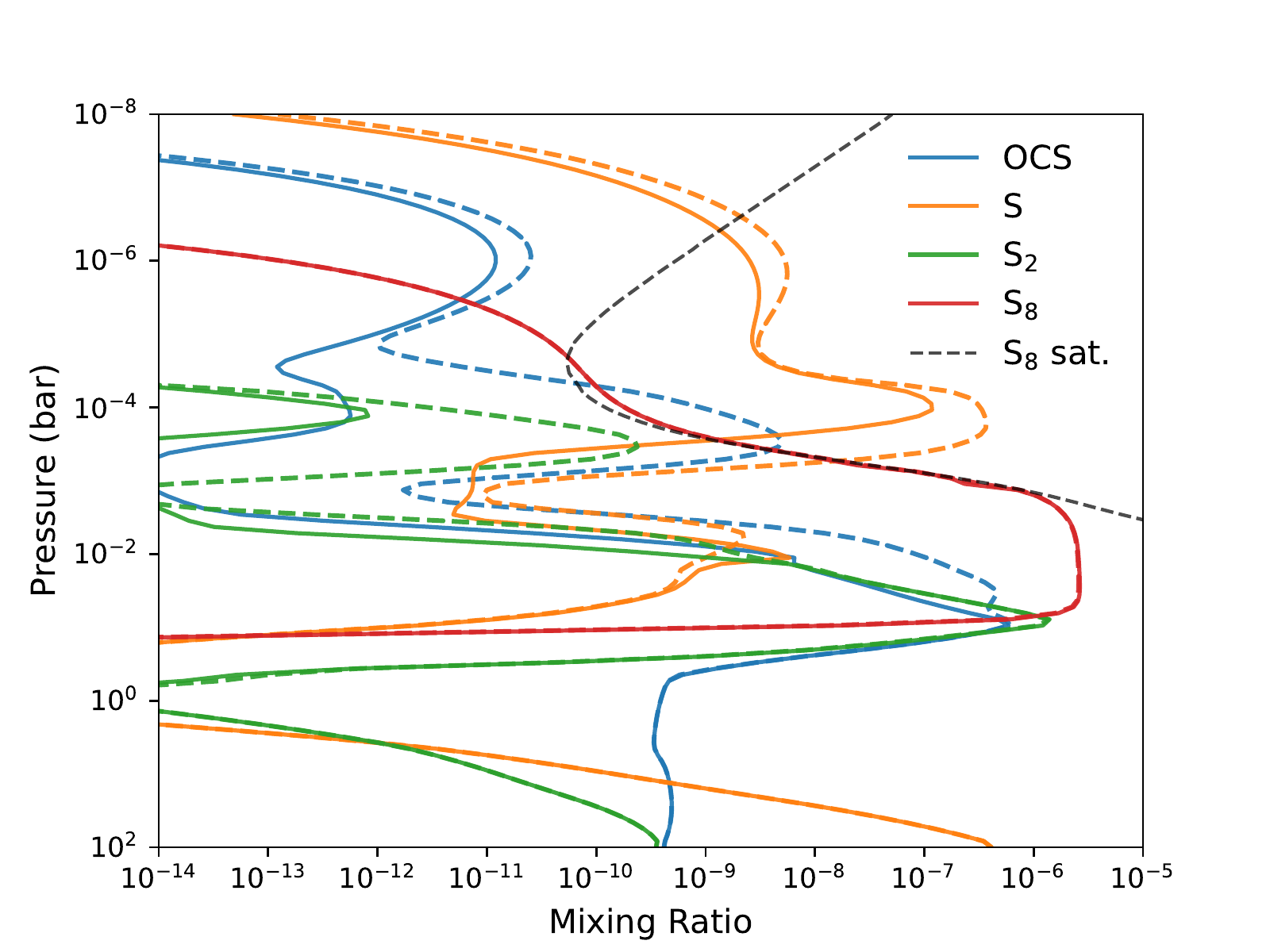}
\caption{Main sulfur species from our nominal model with solar metallicity (solid) compared to those adopting the faster rate of (\ref{Re_OCS}) from \cite{Zahnle2016} (dashed).}
\label{fig:sen-rev}
\end{figure}

\subsubsection{sensitivity to OCS recombination}
The fate of elemental S after being released from \ce{H2S} is critical in sulfur kinetics. Several reactions potentially control whether S proceeds to chain formation into larger polysulfur (\ce{S_x}), forming OCS, or being oxidized to SO, \ce{SO2}. To address the effects of kinetics uncertainties, \cite{Zahnle2016} explore the sensitivity to \ce{H2S} recombination and S$_x$ polymerization. The authors found a faster \ce{H2S} recombination counteracts the destruction of \ce{H2S} and reduces the production of \ce{S8}, while their results are not sensitive to the polymerizing rates of forming \ce{S4} and \ce{S8} within the tested ranges. We have tested the polymerization rates for GJ 436b and confirmed its general insensitivity to \ce{S8} formation. For 51 Eri b, we recognize that the recombination of OCS 
\begin{equation}
\ce{S + CO ->[\textrm{M}] OCS}
\label{Re_OCS}
\end{equation}
could be important in determining the oxidizing rate of sulfur. The rate coefficient of Reaction (\ref{Re_OCS}) has in fact not been measured. Only the reverse step of (\ref{Re_OCS}), the dissociation of OCS, has available data at high temperatures. Recently, \cite{Ranjan20} has also identified this reaction to modestly alter the CO abundance in a \ce{CO2}-\ce{N2} atmosphere and advocate laboratory investigation. Here, we will explain how the rate coefficient of Reaction (\ref{Re_OCS}) is estimated in our nominal model and then explore the sensitivity to the uncertainty for 51 Eri b.
 
Reaction (\ref{Re_OCS}) is a spin-forbidden reaction and usually many orders of magnitude slower than a typical three-body reaction. Since the measured high-temperature dissociation reaction has a high activation energy, extrapolating the dissociation reaction (the reversal of (\ref{Re_OCS})) to low temperatures will result in unrealistically rates. Instead, we estimate the activation energy from the well-studied analogous reaction, \ce{O + CO ->[\textrm{M}] CO2}. The pre-exponential factor is then determined by matching the reverse of dissociation reaction at 2000 K from \cite{Oya1994}. The low-pressure limit rate of (\ref{Re_OCS}) we estimate is
\begin{equation}\label{COS_rate}
k_\textrm{0} = 4.47 \times 10^{-34} \textrm{exp}(-1510/T). 
\end{equation}
We compare the rate coefficient (\ref{COS_rate}) with those assumed in \cite{Zahnle2016} and Venus literature \citep{Zhang2012,Krasnopolsky2013} in Figure \ref{fig:rate_COS}. The reaction rates show diverse values especially toward lower temperatures, the relevant temperature range for the stratosphere of 51 Eri b. Albeit the rate discrepancy in each model, the rate constants in \cite{Zhang2012}, \cite{Krasnopolsky2013} and this work exhibit consistent temperature dependence from 1000 K to 200 K, whereas that in \cite{Zahnle2016} has suprisingly almost no temperature dependence. Since rate constant of (\ref{Re_OCS}) from \cite{Zahnle2016} is the most different 
from the literature and yields fastest OCS forming rate, we will use the rate from \cite{Zahnle2016} as the upper limit to test the sensitivity.

We run our nominal model with solar metallicity but adopt the rate constant of (\ref{Re_OCS}) from \cite{Zahnle2016}. The effects of faster OCS recombination are illustrated in Figure \ref{fig:sen-rev}. With the OCS recombination rate from \cite{Zahnle2016}, OCS mixing ratio is significantly increased above 0.1 bar. \ce{S8} is slightly reduced but remains the major sulfur carrier between 10$^{-2}$ and 10$^{-4}$ bar, consistent with the model results in \cite{Zahnle2016}. The abundance of S and \ce{S2} are subsequently affected by more ample OCS photodissociation but that of \ce{S8} remains the same as set by condensation. Given these differences, we reiterate further investigation to pin down the reaction rate of OCS recombination. 



\begin{figure}[!ht]
\includegraphics[width=\columnwidth]{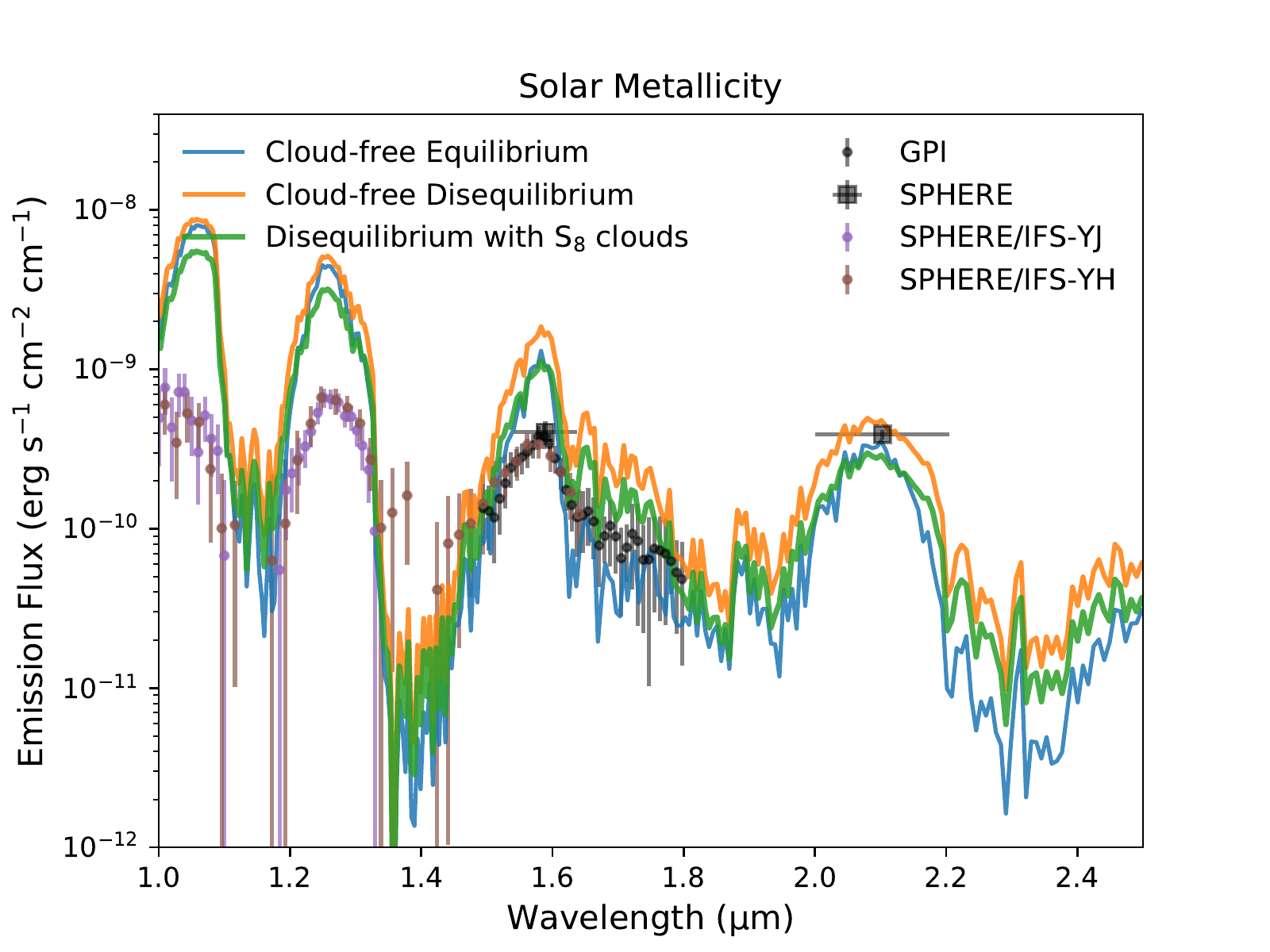}
\includegraphics[width=\columnwidth]{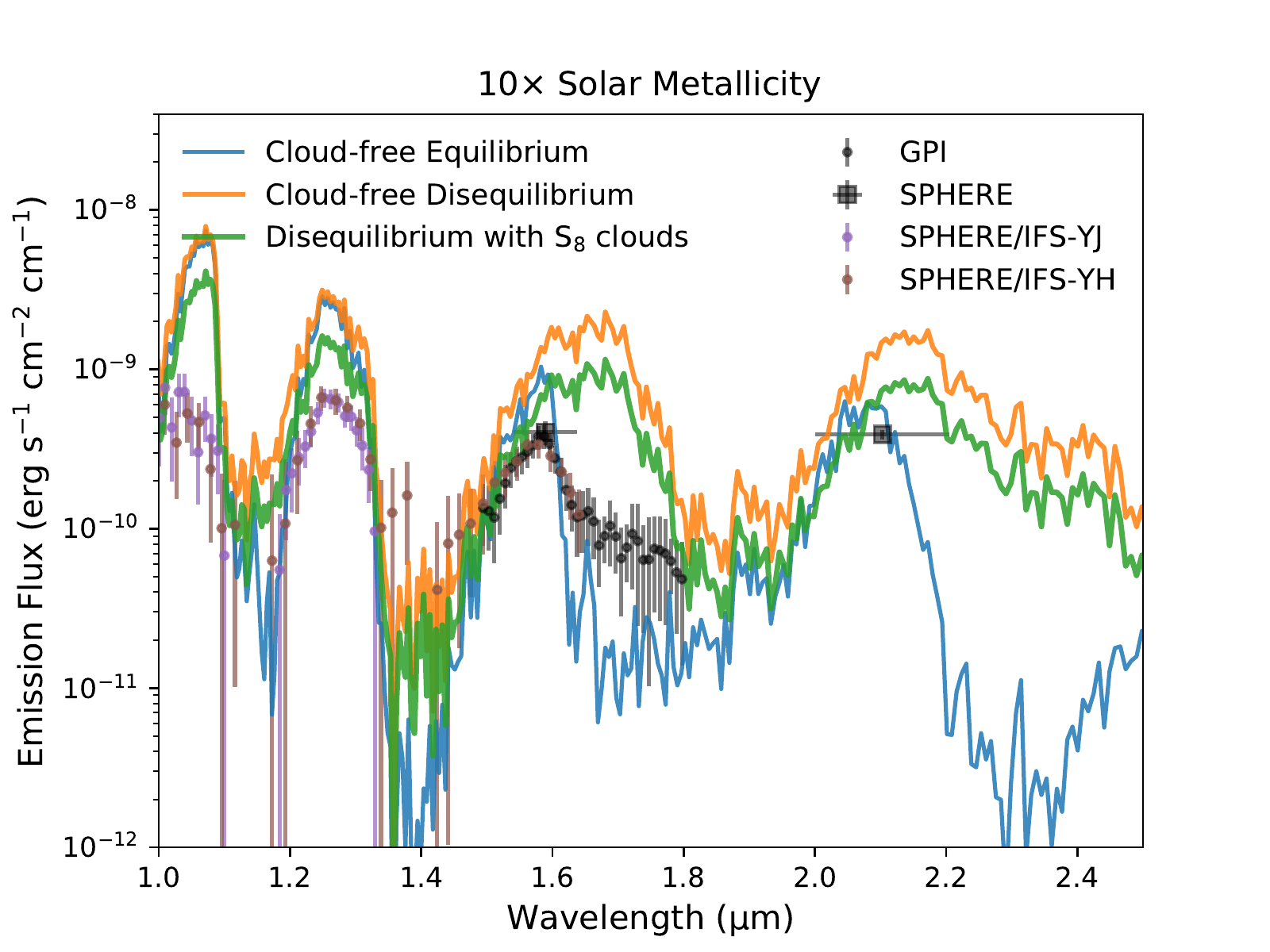}
\caption{Synthetic emission spectra of 51 Eri b produced from equilibrium abundances, disequilibrium abundances, and with \ce{S8} condensate layer. Data points show GPI observations from \cite{Macintosh2015} and SPHERE observations from \cite{Samland2017}.}
\label{fig:51Eri-emission}
\end{figure}

\begin{figure}[htp]
\includegraphics[width=\columnwidth]{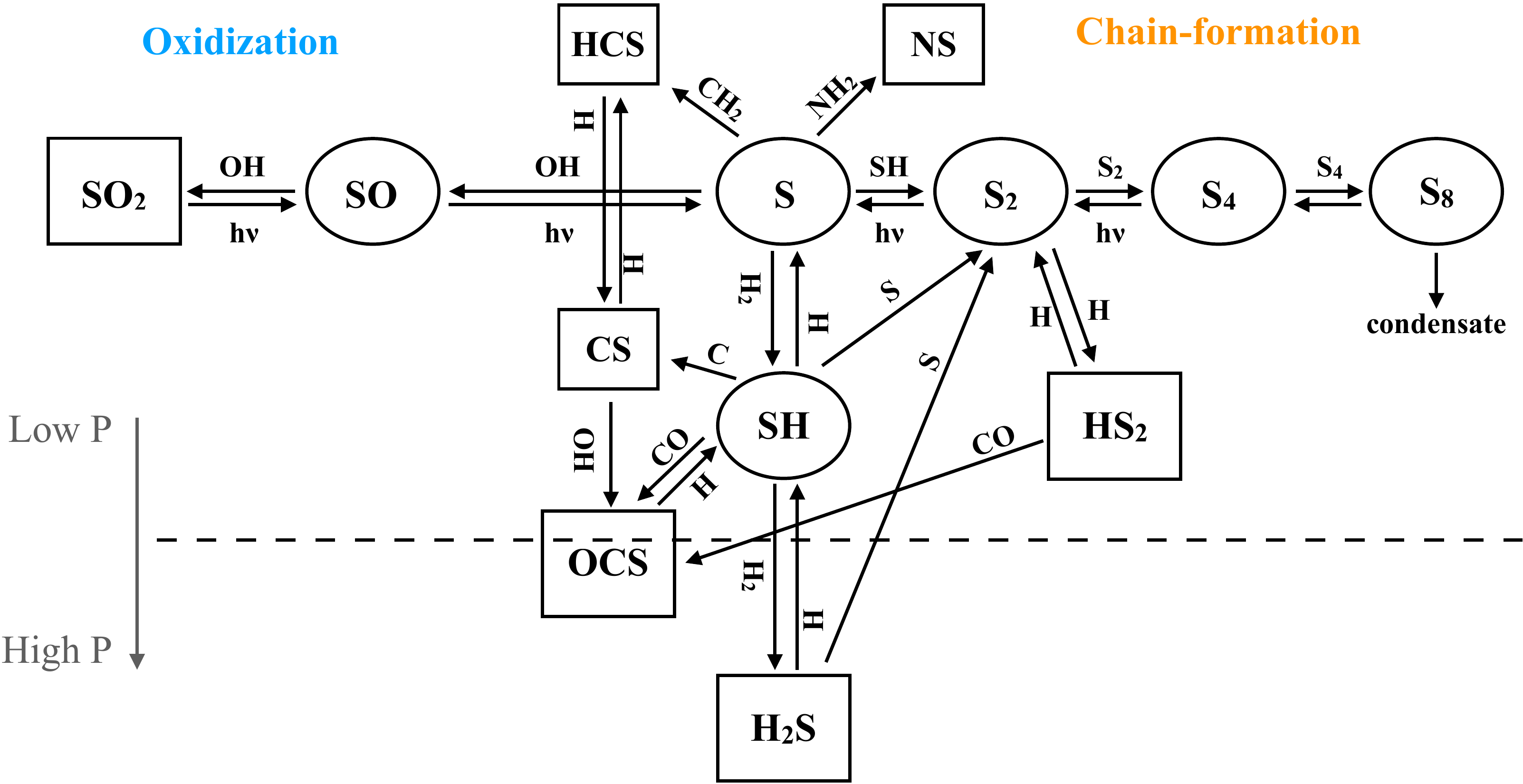}
\caption{A Schematic diagram illustrating the main pathways for sulfur kinetics in an \ce{H2}-dominated atmosphere. The dashed line represents the transition from the lower region where sulfur is predominantly locked in \ce{H2S} to the upper region where \ce{H2S} is subject to dissociation. Rectangles indicate stable species whereas ellipses indicate active radical or intermediate species.}
\label{fig:S-diagram}
\end{figure}

\subsubsection{Emission Spectra}
Figure \ref{fig:51Eri-emission} demonstrates the effects of disequilibrium chemistry and \ce{S8} clouds on the planetary emission spectra. For both metallicities, quenched \ce{CH4} and \ce{H2O} have lower abundances than equilibrium, leading to higher emission in the H and J bands from the deeper region. The 10 times solar metallicity further reduces \ce{CH4} and prompts the flux at 1.6 - 1.8 $\mu$m. We assume 1 $\mu$m particle size for \ce{S8} condensates, which scatter strongly and reduce the emission in this wavelength range. However, using the higher effective temperature and metallicity from \cite{Samland2017}, our models generate emission that are too high in the H and J bands and fail to reproduce the observed spectra. We conclude that either $T_{\textrm{eff}}$ is lower than that determined by \cite{Samland2017} and/or additional cloud layers \citep[e.g.][]{Moses2016} is required to match the lower observed emission.


\subsection{Sulfur Mechanism}
Figure \ref{fig:S-diagram} summerizes the important pathways for sulfur species in the irradiated \ce{H2}-dominated atmospheres we explored in this section. \ce{H2S} is the dominant molecule, which is thermochemically stable for a wide range of temperatures in the lower atmosphere followed by OCS. The photochemistry of sulfur is initiated from SH and S produced by \ce{H2S} dissociation, leading to multiple channels including chain formation and oxidization depending on the atmospheric condition. Sulfur chain formation is highly temperature sensitive where \ce{S2} is favored about 600 - 800 K and \ce{S8} can only form below $\sim$ 500 K (e.g. the stratosphere of 51 Eri b). When OH is sufficiently produced by \ce{H2O} photolysis, S will most likely be oxidized into SO and \ce{SO2} in the upper atmosphere. S also participates in accelerating \ce{CH4} and \ce{NH3} destruction via the coupling to \ce{CH2} or \ce{NH2}, as seen in our HD 189733b and GJ 436b models.

\begin{figure*}[htp]
\includegraphics[width=\columnwidth]{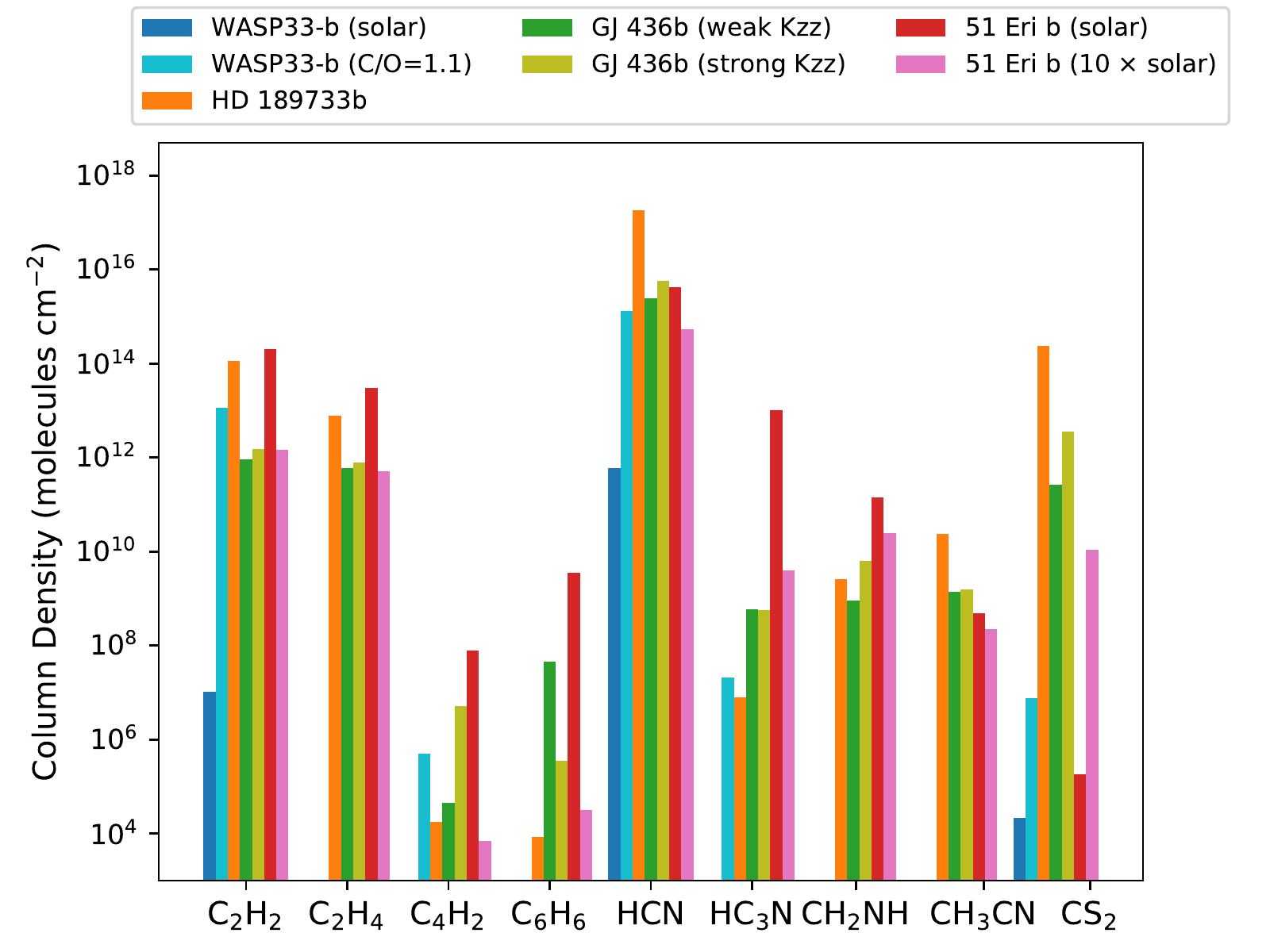}
\includegraphics[width=\columnwidth]{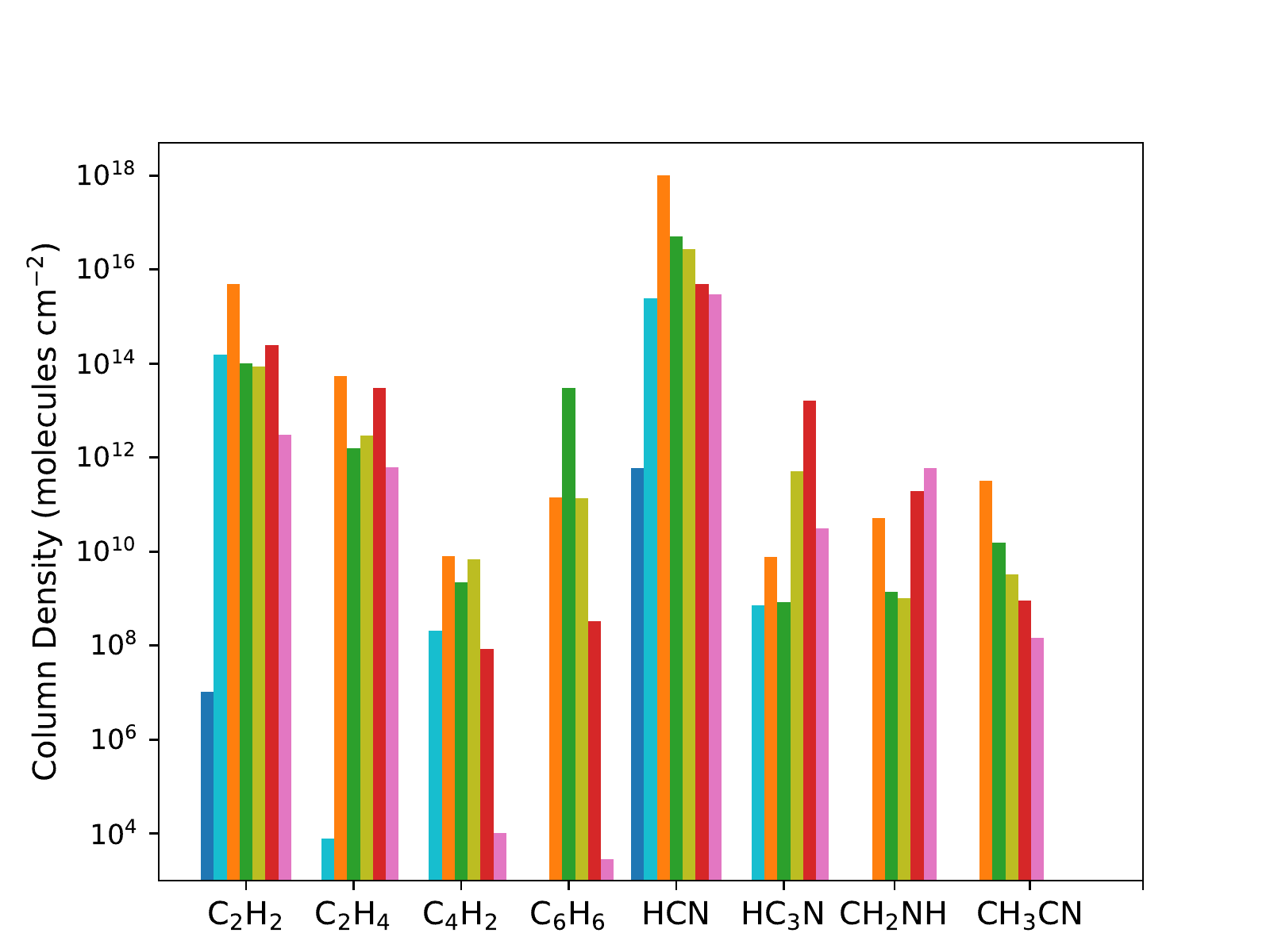}
\caption{The column number densities (molecules cm$^{-2}$) above 1 mbar of haze precursors for the simulated atmospheres in Section \ref{case}, including sulfur (left) and without sulfur (right). Some molecule abundances are negligible and not shown for WASP-33b. For GJ 436b, the models with T$_{\textrm{int}}$ = 400 K and weak/strong vertical mixing are used.}
\label{fig:haze-bar}
\end{figure*}


\subsection{Trends of Photochemical Hazy Precursors}
Figure \ref{fig:haze-bar} summarizes the column densities of haze precursors above 1 mbar for the simulated atmospheres in Section \ref{case}. Across the various irradiated \ce{H2}-atmospheres we explored, we find HCN consistently to be the most prevailing precursor. This is not surprising as HCN is a robust photochemical product of \ce{CH4} and \ce{NH3} and also recently been detected on HD 209458b \citep{Giacobbe2021}. Nonetheless, it does not necessarily imply HCN will lead to complex nitriles formation, since HCN is not the limiting factor as we discussed in Section \ref{sec:haze}. A more careful assessment at high temperatures is required before extrapolating the haze-forming mechanism below 200 K on Titan. We observe a general increasing trend with decreasing temperature for the more indicative nitrile precursors \ce{HC3N} but opposite for \ce{CH3CN}. The same trend is seen for the hydrocarbon precursors \ce{C4H2} and \ce{C6H6}. 

Only HCN and \ce{C2H2} can reach appreciable levels on WASP-33b, even as photochemical hazes are not expected on WASP-33b. For GJ 436b, most of the precursors are not too sensitive to eddy diffusion. For 51 Eri b, almost all precursors are reduced with increasing metallicity, except for \ce{CS2} since it contains no H. In fact, \ce{CS2} is most favored on HD 189733b, which suggests 
sulfur-containing hazes in the hot Jupiter condition as carbon and sulfur can couple closely. In addition to sulfur condensates, 51 Eri b might also be covered by nitrile-type hazes according to the precursor distribution.


 


\section{Discussion}\label{discussion}
\subsection{High-Temperature UV Cross Sections}
We have implemented layer-by-layer UV cross sections according to the temperature at each atmospheric level in VULCAN. Due to the sparsely available data, we did not perform systematic study in this work. Nonetheless, we have gained some insights through the case studies in Section \ref{case}.  

We found the effects of temperature dependence for \ce{H2O} is mostly negligible in a \ce{H2}-dominated atmosphere. However, this is solely based on the limited wavelength-range measurements we assembled. For the high-temperature ($T >$ 1000 K) cross sections, only wavelengths longer than about 190 nm are included (Figure \ref{fig:cross_T}). The high-temperature cross sections in the FUV could have larger effects.

We confirm the analysis in \cite{Venot2013a} that although the \ce{CO2} abundance is not directly influenced by the temperature dependence of \ce{CO2} photolysis, the shielding effects can impact other species. As \ce{CO2} absorb more strongly with increasing temperature, the UV photosphere is lifted to lower pressure. The production of radicals, such as H and OH, is reduced and subsequently alters other species. However, we also find that the shielding effects of \ce{CO2} are completely shadowed when sulfur species are included (e.g. see the right panel of Figure \ref{fig:HD189-S-noS}). The temperature dependence of \ce{CO2} photolysis should be more amplified in \ce{CO2}-dominated atmospheres.


\subsection{Implication of Ionization}
Ions are not included in this work. We are working on including ionchemistry in the next update of VULCAN. Photoionization is known to be critical in initiating the haze formation \citep{Wong2003,Krasnopolsky2009,Plessis2012,Lavvas2013}. Even thermoionization can be important for ultra hot Jupiters. In our study about WASP-33b, atomic Ti and V in the upper atmosphere are expected to be partly ionized and contribute to free electrons. Since Ti has an ionization threshold of 180 nm, compared to about 240 nm for Na, the effects of photoionization on Ti and V should be similar and probably smaller than those on the alkali atoms, as investigated in \cite{Lavvas2017}. An important outcome of photoionization is that the increased electrons can lead to more hydrogen anions (H-) than predicted by thermal equilibrium, which are found to be important opacity sources in some hot Jupiters atmospheres \citep{Lewis2020}. In terms of sulfur chemistry, several sulfur species have relatively lower-energy threshold of ionization and can be subject to photoionization. For example, atomic S starts to ionize from Lyman-$\alpha$. Since S is likely the dominant sulfur species in the hot Jupiter's stratosphere (Section \ref{sec:HD189-S}), S can be photoionized and ramify into various organic molecules through ion-exchange reactions.  

\subsection{More Intriguing Questions about Sulfur}
In Section \ref{case}, we find the coupling to sulfur chemistry impacts the core C-H-N-O kinetics in several ways for \ce{H2}-dominated atmospheres. The coupling effects essentially depend on if the sulfur-containing intermediates are active, which is not well-understood in general as it can vary with atmospheric conditions such as temperature and bulk compositions. \cite{Gersen2017} find that \ce{CH3S} and \ce{CH3SH} provide more efficient pathways for methane oxidization in the combustion (oxidizing) environment. \cite{He2020S} also observe the photochemical formation of \ce{CH3S} and \ce{CH3SH} in a \ce{CO2}-rich gas mixture in the experiments. Although we have included reactions involving \ce{CH3S} and \ce{CH3SH} in our sulfur mechanism, they are not identified to be important in the pathway analysis for all of the \ce{H2}-atmospheres we investigated. The chemical role of \ce{CH3SH} is worth further study in the broad context of biologically produced sulfur.   

The temperature profiles are fixed without considering the radiative feedback in this whole work. The radiative effects might be more prominent in the presence of sulfur, such as the absorption of SH and \ce{S2} in the optical and NUV. 51 Eri b or other directly imaged planets with a relatively cold stratosphere ($\lesssim$ 500 K) and under sufficient UV irradiation sit in the sweet spot for testing the radiative feedback on sulfur condensates.





\section{Summary}\label{sec:summary}
In this paper, we present a thorough theoretical framework of the updated photochemical code VULCAN. We validate our models for the atmospheres of hot Jupiters, Jupiter, and modern Earth and carry out comparative study on representative cases of extrasolar giant planets: WASP-33b, HD 189733b, GJ 436b, and 51 Eridani b. The highlights of our results are:

\renewcommand\labelitemi{\tiny$\bullet$}
\begin{itemize}
\item We have carefully validated the model of HD 189733b. The updated methanol scheme in \cite{Venot2020} is found to bring the quenching behavor of methane close to \cite{Moses11} and VULCAN. We pointed out the photochemical source plays a non trivial part in the model differences  
between \cite{Moses11}, \cite{Venot12}, and VULCAN.

\item We demonstrate advection transport in the downdraft can qualitatively explain the deep ammonia distribution in Jupiter, which can not be explained by eddy diffusion alone.
 
\item The implementation of surface boundary conditions and condensation in an oxygen-rich atmosphere is validated in the present-day Earth model. A general oxidation timescale analysis is provided for assessing the chemical lifetime of biosignature gases. 

\item The atmosphere of WASP-33b is not affected by vertical quenching but consisted of an upper photolytic region and a thermochemical equilibrium region below. For GJ 436b, we find \ce{NH3} insensitive to vertical mixing and the sulfur species governed by photolysis and mixing in the upper stratosphere are independent of the deep thermal structure, which can be complementary to the \ce{CH4}/CO metric for breaking degeneracies. The quenched CO always predominates over \ce{CH4} on 51 Eri b and sulfur aerosols (chiefly \ce{S8}) condense out in the stratosphere. 

\item We find the coupling to sulfur chemistry impact C-H-N-O kinetics. Sulfur can provide catalytic paths to destroy \ce{CH4} and \ce{NH3} and generally lower the hydrocarbon abundances. \ce{H2S} makes H recycled back to \ce{H2} faster on the cooler GJ 436b and 51 Eri b. The dissociation of SO and \ce{SO2} also make the upper atmosphere of 51 Eri b more oxidizing.

\item We suggest including several photochemical haze precursors such as \ce{C6H6} and \ce{HC3N}, which are more indicative than the commonly considered HCN and \ce{C2H2}. We observe a general increasing trend with decreasing temperature for \ce{C4H2}, \ce{C6H6}, and \ce{HC3N} but opposite for \ce{CH3CN}.


\end{itemize}


\section*{Model availability}
The results in this work are produced by version 2.0 of VULCAN (\url{https://github.com/exoclime/VULCAN/releases/tag/v2.0}). In addition to the public code, the configuration files used for the models in Section \ref{validation} are available on \url{https://github.com/exoclime/VULCAN} and the main model output in Section \ref{validation} and Section \ref{case} can be found in the supplementary material. 

Software: Python; Numpy \citep{numpy}; Scipy \citep{scipy}; Matplotlib \citep{matplotlib}

\acknowledgments
S.-M.T gratefully thanks M. Zhang for customizing PLATON to read non-equilibrium compositions. S.-M.T also thanks O. Venot and J. Moses for sharing the output of HD189733b for model comparison, C. Li for providing the retrieved ammonia results from Juno measurements, L.M. Lara for fruitful discussions about setting up photochemistry, P. Rimmer for the compiled observational data of Jupiter, and N. Wogan for pointing out a typo in Equation (14) in an earlier version of this paper. S.-M.T acknowledges support from PlanetS National Center of Competence in Research (NCCR) and University of Oxford. M.M. acknowledges support from NASA under the XRP grant No. 18-2XRP18\_2-0076. E.K.L. acknowledges support from the University of Oxford and CSH Bern through the Bernoulli fellowship. K.H. acknowledges support from the PlanetS National Center of Competence in Research (NCCR) of the Swiss National Science Foundation and the European Research Council Consolidator Grant EXOKLEIN (No. 771620). This work was supported by the European Research Council Advanced Grant EXOCONDENSE (\#740963; PI: R.T. Pierrehumbert).

\appendix
\section{Molecular diffusion and thermal diffusion factor}\label{app:Dzz}
For \ce{H2}-based atmospheres, we take 
\begin{equation}
D_{\ce{H2}-\ce{CH4}} = 2.2965 \times 10^{17} T^{0.765}/N
\end{equation}
from \cite{Marrero1972} as the reference and scale the molecular diffusion coefficient of \ce{H2} with other species according to Equation (\ref{eq:D_scale}). The thermal diffusion factor for H and He are approximately $\alpha_{\ce{H}}$ $\approx$ -0.1 and $\alpha_{\ce{He}}$ $\approx$ 0.145 \citep{moses2000a}. We assume $\alpha_i$ = -0.25 for the rest of species based on rigid sphere approximation \citep{Banks1973}. 

For \ce{N2}-based atmospheres, we take 
\begin{equation}
D_{\ce{N2}-\ce{CH4}} = 7.34 \times 10^{16} T^{0.75}/N
\end{equation}
from \cite{Banks1973} as the reference and scale the molecular diffusion coefficient of \ce{N2} with other species according to Equation (\ref{eq:D_scale}). The thermal diffusion factor of Ar is $\alpha_{\ce{Ar}}$ $\approx$ 0.17 and $\alpha_i$ = -0.25 for the rest species.

For \ce{CO2}-based atmospheres, we take 
\begin{equation}
D_{\ce{CO2}-\ce{H2}} = 7.51 \times 10^{16} T^{0.759}/N
\end{equation}
from \cite{Banks1973} as the reference and scale the molecular diffusion coefficient of \ce{CO2} with other species according to Equation (\ref{eq:D_scale}). The thermal diffusion factor of Ar is $\alpha_{\ce{Ar}}$ $\approx$ 0.17 and $\alpha_i$ = -0.25 for the rest species.

\begin{figure}[h]
\begin{center}
\includegraphics[width=0.4\columnwidth]{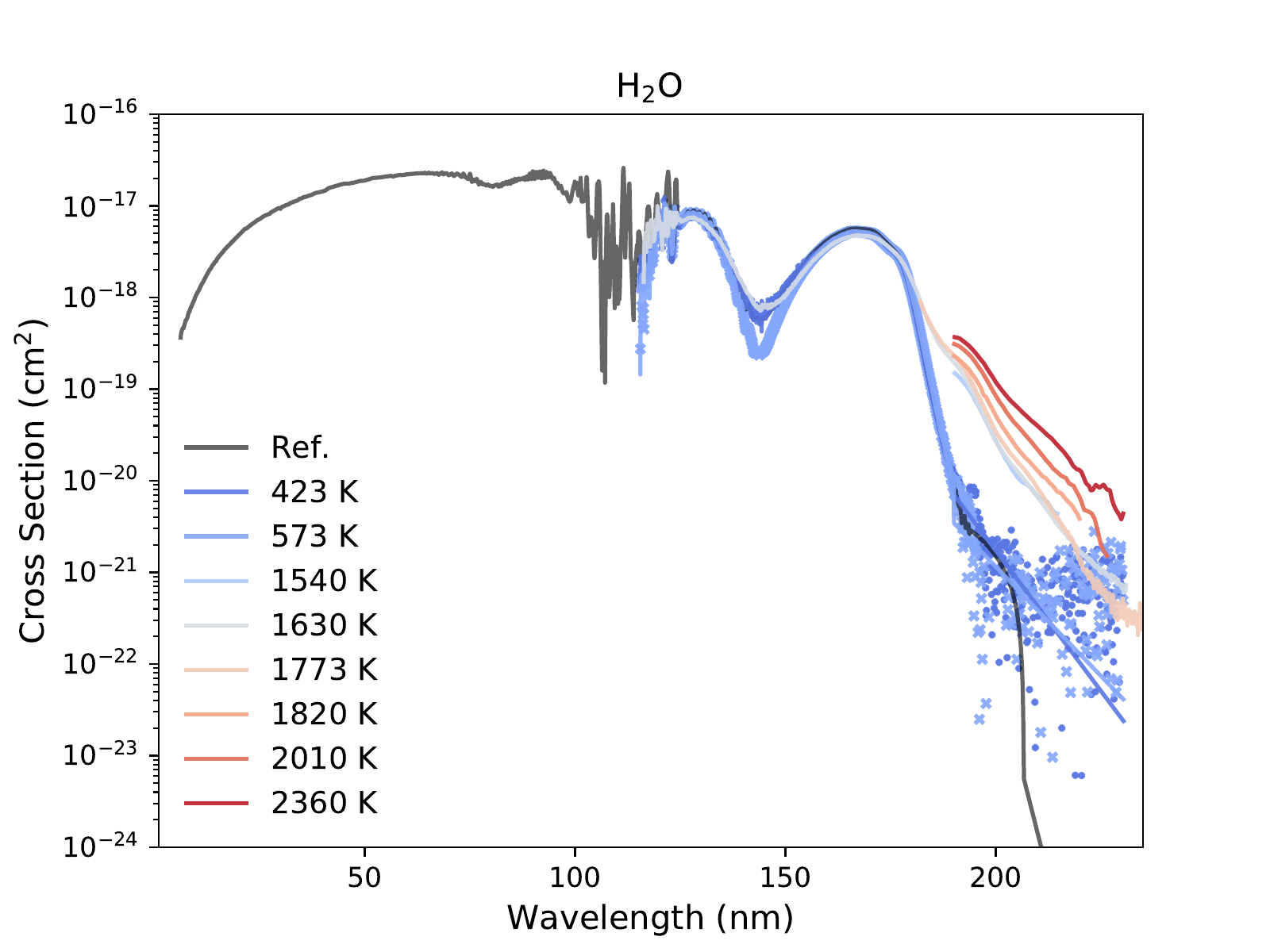}
\includegraphics[width=0.4\columnwidth]{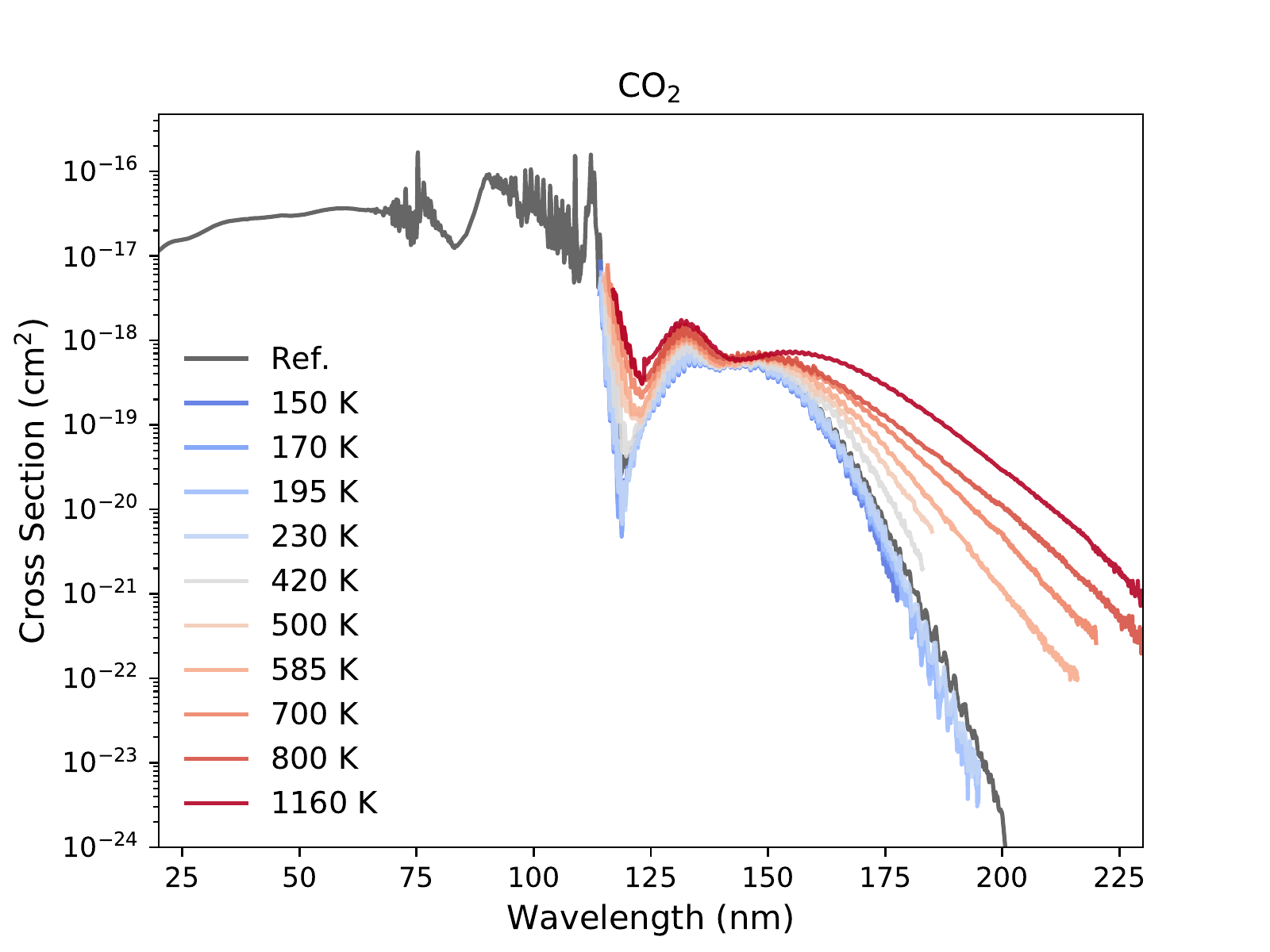}
\includegraphics[width=0.4\columnwidth]{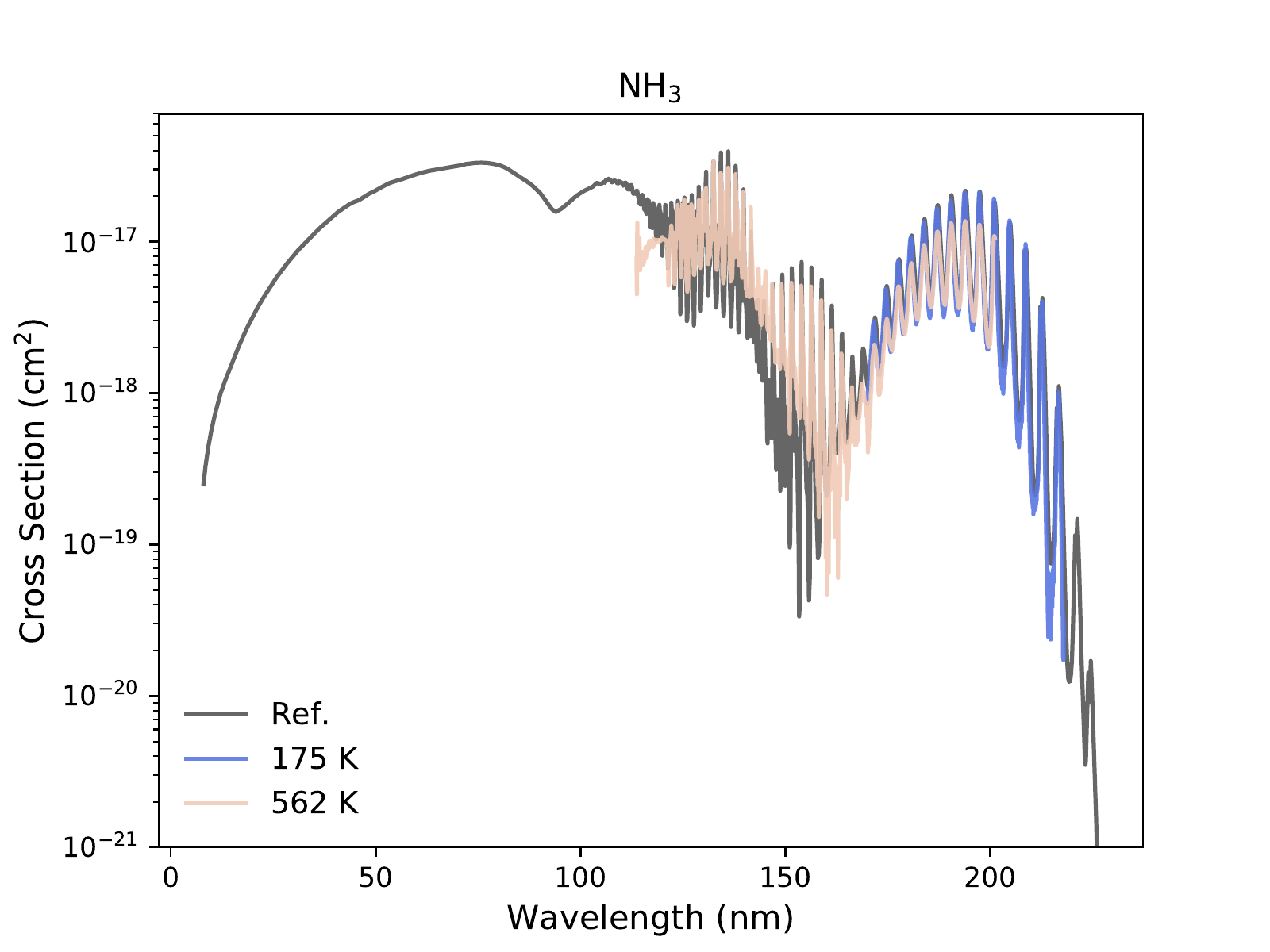}
\includegraphics[width=0.4\columnwidth]{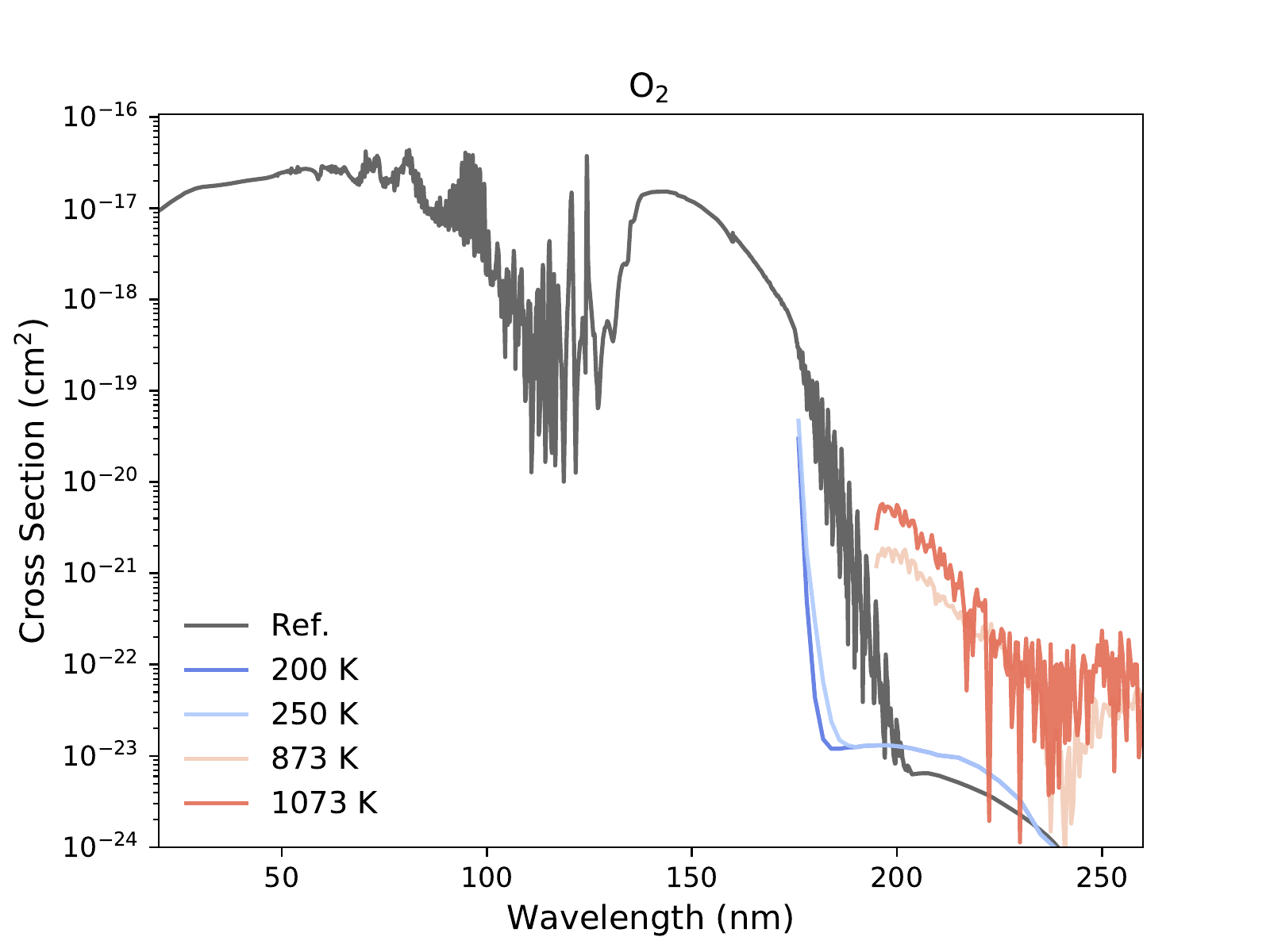}
\includegraphics[width=0.4\columnwidth]{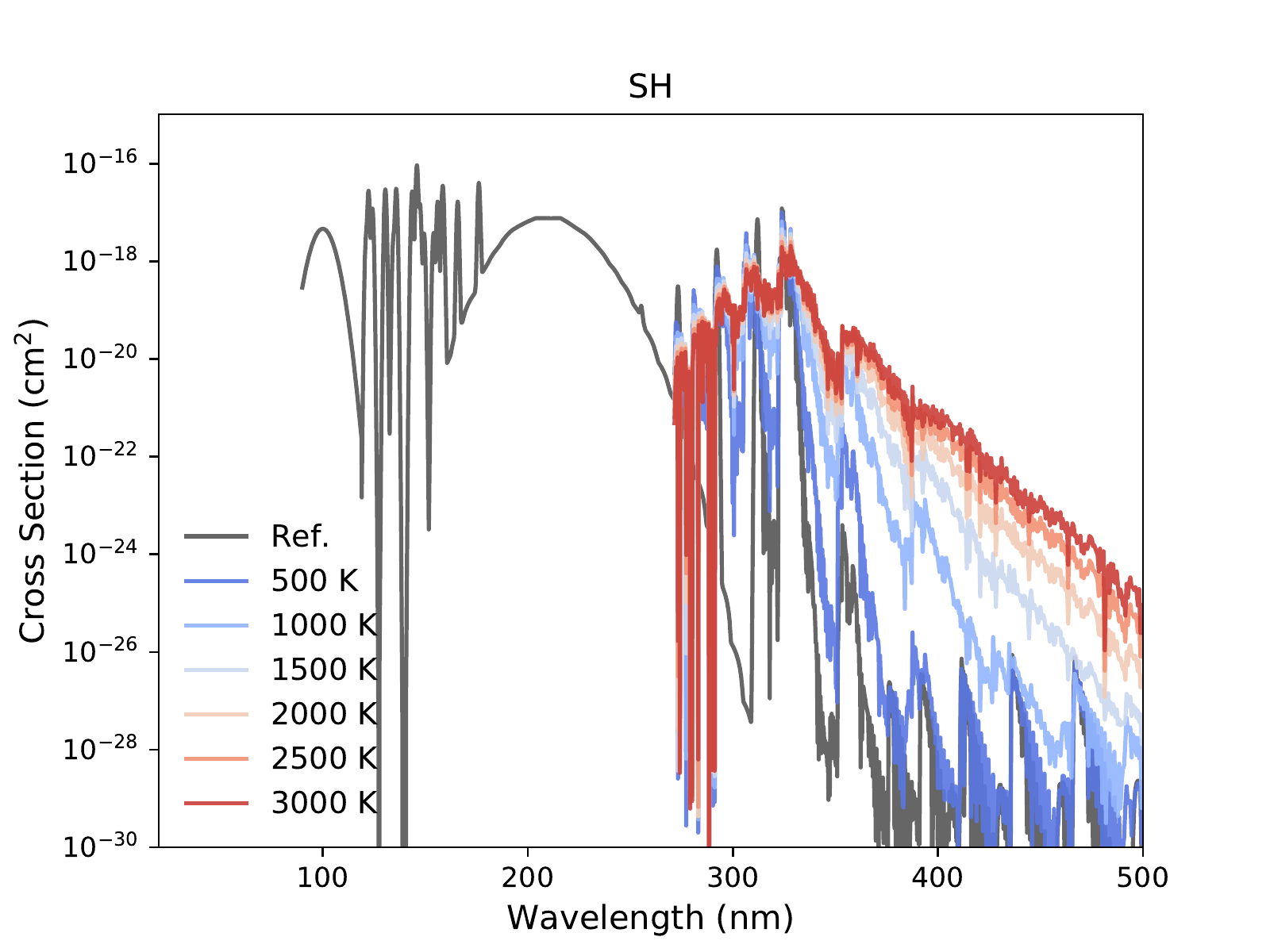}
\includegraphics[width=0.4\columnwidth]{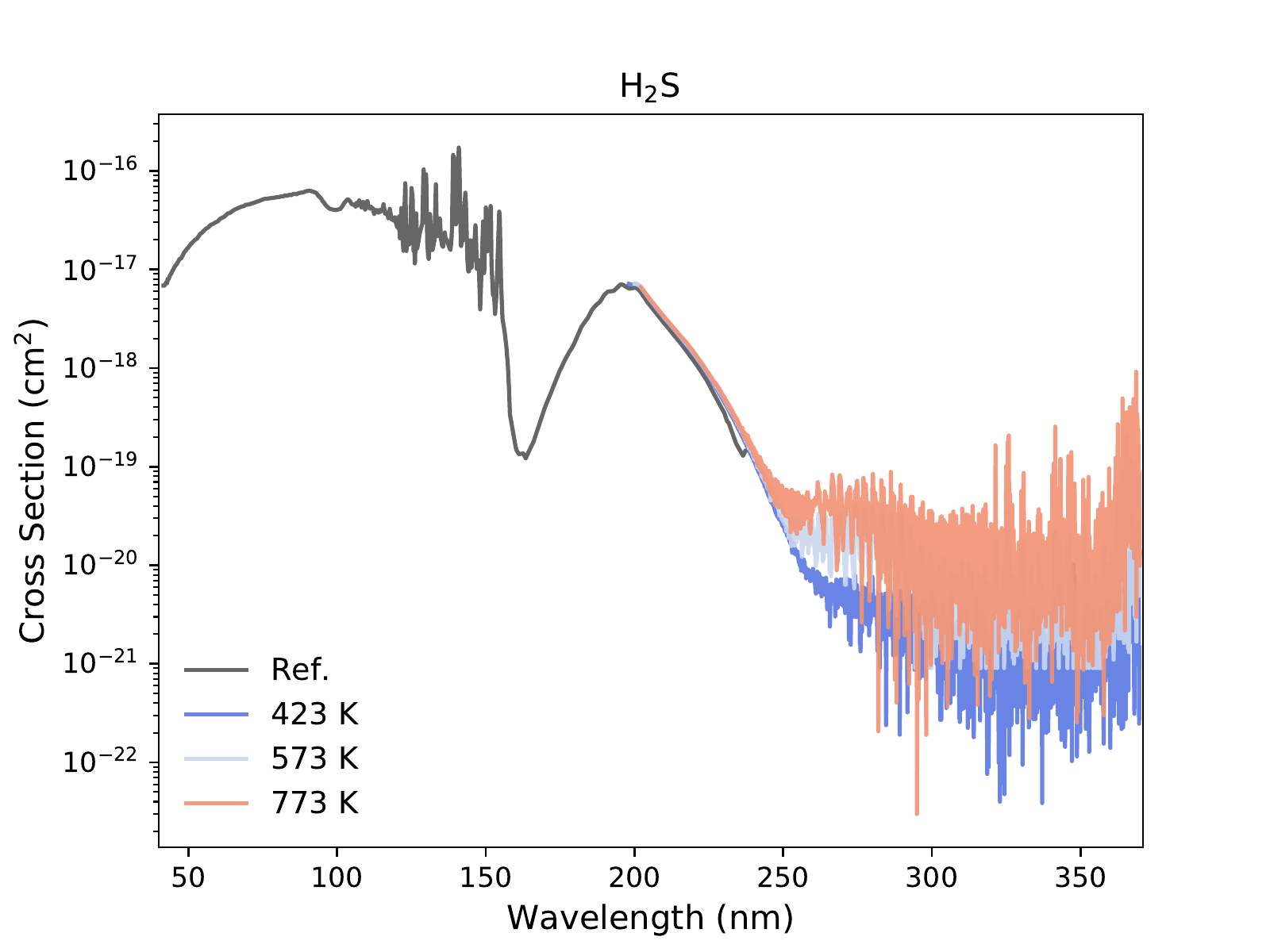}
\includegraphics[width=0.4\columnwidth]{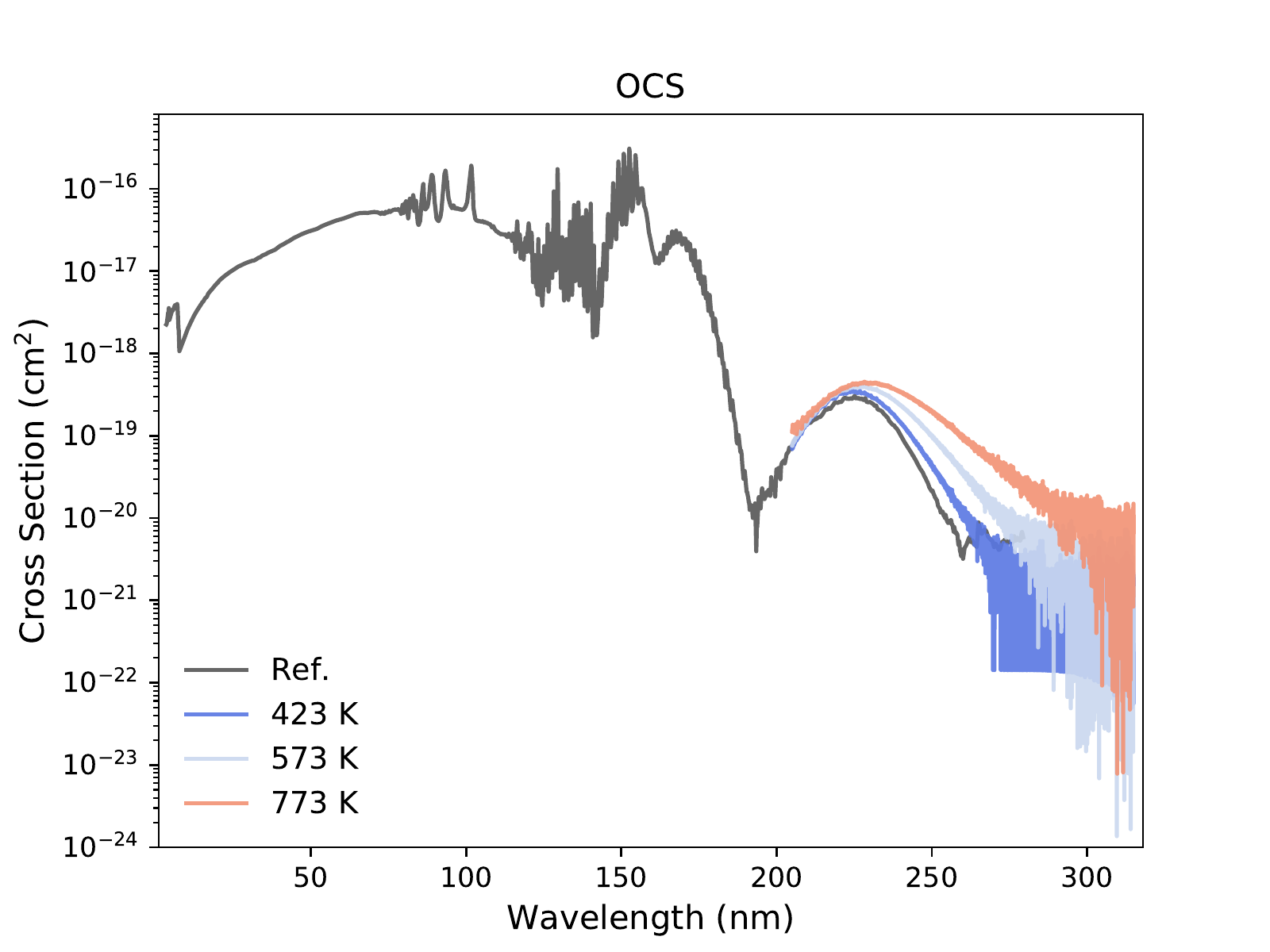}
\includegraphics[width=0.4\columnwidth]{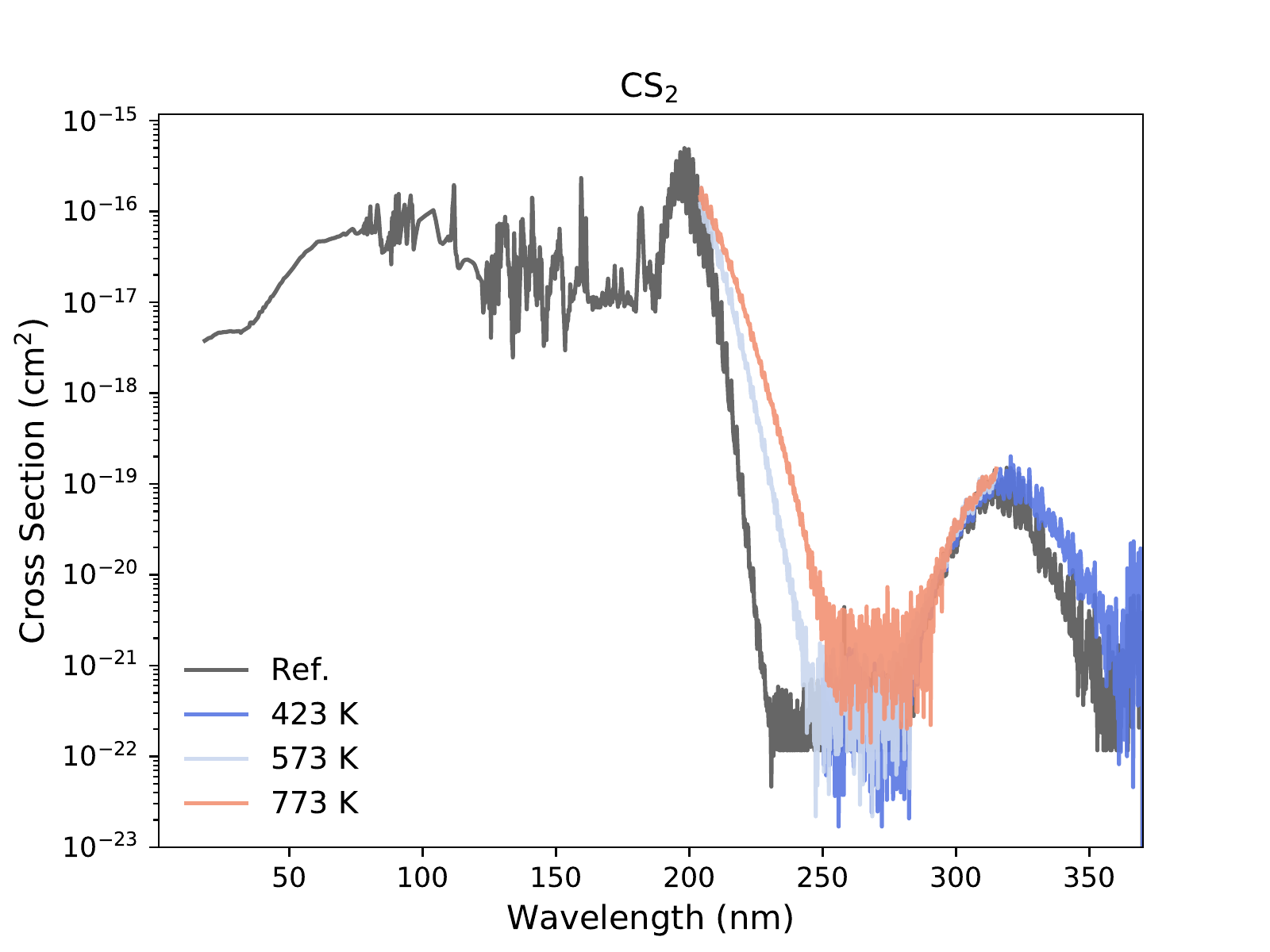}
\end{center}
\caption{Photoabsortion cross sections of \ce{H2O}, \ce{CO2}, and \ce{NH3} across various temperatures, with Ref. denoting the cross sections measured at room temperature. The measured cross sections of \ce{H2O} at 423 K (dot) and 573 K (cross) are noisy beyond 216 nm and we use linear fit for conservative estimate. The references of the cross sections are described in Section \ref{sec:Tcross}.}
\label{fig:cross_T}
\end{figure}

\section{Resolution Errors of Photolysis Rates}\label{app:resolution}

\begin{table}[htp]
\begin{center}
\caption{Errors from Low Spectral Resolution}\label{tab:res_error}
\begin{tabular}{l|c|c|c}
Bin (nm) & Stellar Flux (in $\%$)  & $\Delta$ J$_{\ce{H2}}$ \footnote{$\Delta$ J = $|$J - J$_0$$|$ / J$_0$ where J$_0$ is the reference photolysis rate calculated with constant 0.1 nm resolution}& $\Delta$ J$_{\ce{H2O}}$\\
\hline
0.2 & 0 & 0.0004 & 0.003\\
0.5 & 0.08 & 0.18 & 1.13 \\
1 & 0.08 & 0.44 & 0.87 \\ 
10 & 2.65 & 0.22 & 0.87 \\
\end{tabular}
\end{center}
\end{table}

Equation (\ref{eq:photo_rate}) is numerically computed in the form of finite sum. The wavelength grid in the code needs to properly resolve the line structures in the stellar flux and cross sections. This is especially important in the XUV where there are more fine structures from the band transition \citep{rimmer16}.

\begin{figure}[htp]
\begin{center}
\includegraphics[width=0.5\columnwidth]{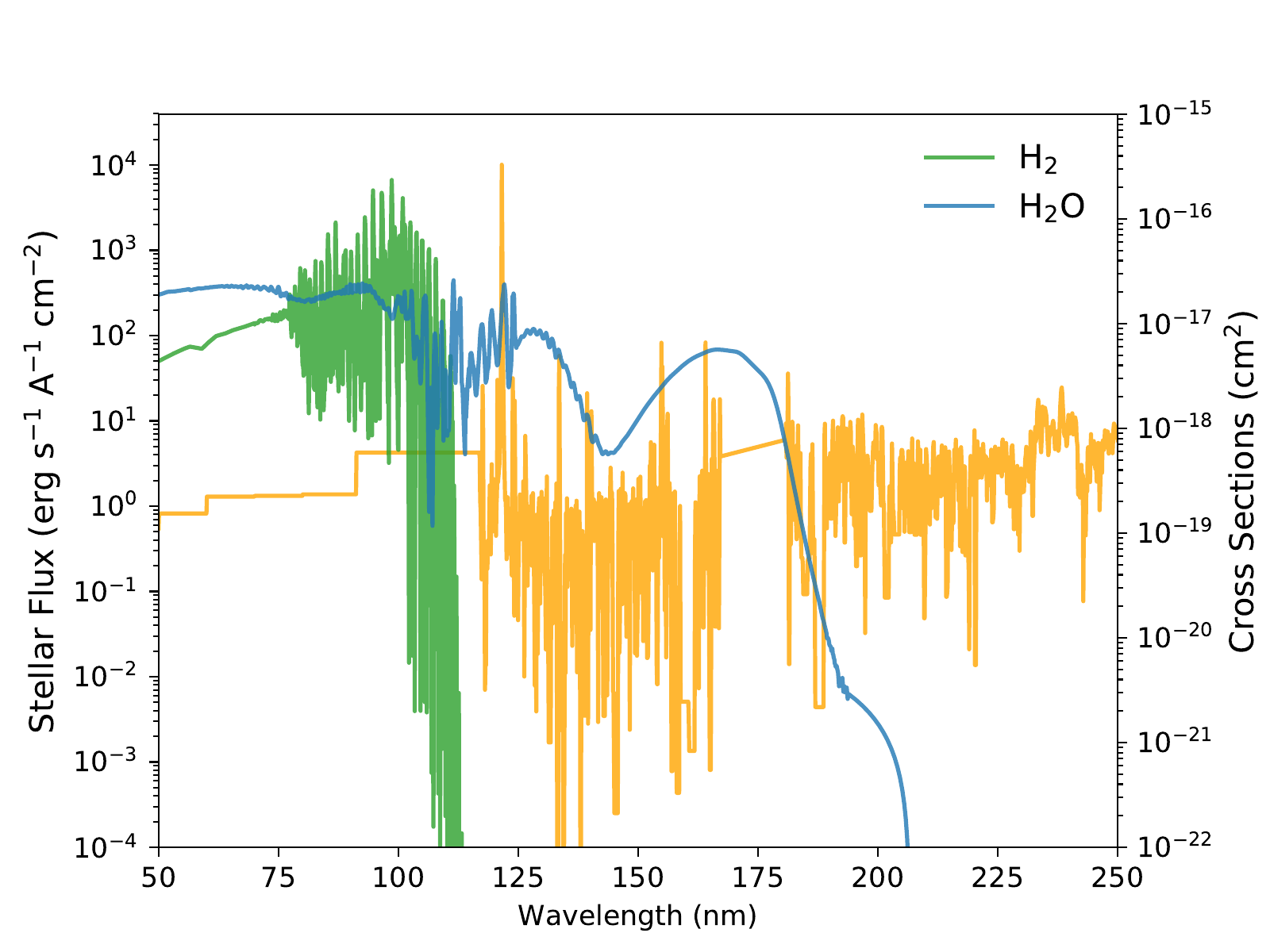}
\end{center}
\caption{The stellar UV flux of GJ 436 and cross sections of \ce{H2} and \ce{H2O}, showing the native resolution 0.1 nm adopted in the model.}
\label{fig:gj436-cross}
\end{figure}

We demonstrate the errors of computing Equation (\ref{eq:photo_rate}) with the GJ 436b model in Section \ref{sec:GJ436b}. The effects are emphasized with its host M-star showing more emission lines. Figure \ref{fig:gj436-cross} shows the stellar flux overplotted with the UV cross sections of \ce{H2} and \ce{H2O}. The stellar flux is adopted from the MUSCLES survey with a constant 0.1 nm resolution and the UV cross sections from the Lieden database have the same resolution of 0.1 nm. Therefore, we consider constant resolution of 0.1 nm as the reference for this test. The resolution for computing Equation (\ref{eq:photo_rate}) is varied from 0.2 nm to 10 nm, where the trapezoidal rule is applied for the integral. The errors with respect to summing-up the total stellar flux and the resultant photolysis rate of \ce{H2} and \ce{H2O} at the top of atmosphere (10$^{-8}$ bar) are summerized in Table \ref{tab:res_error}. Starting from the bin size of 0.5 nm, which is five times the native resolution of the flux and cross section data, the errors become comparable to the absolute value of the photolysis rate. The errors do not behave linearly with the bin size since the overestimate from the peak can offset the underestimate from the trough. The test shows the importance of using matching spectral resolution to attain accurate photolysis rates.

\begin{table*}[!h]
\begin{center}
\caption{List of forward reactions (backward reactions with even indexes are reversed numerically with thermodynamic data) and rate coefficients (cm$^3$ s$^{-1}$ for bimolecular reactions and s$^{-1}$ for k$_0$) of titanium and vanadium species. $E_a$ est. means the activation energy is estimated by the enthalpy difference as described in Section \ref{sec:Ti}.}
\label{tab:tio}
\begin{tabular}{|lll|}
\hline
Reaction & Rate Coefficient & Reference \\
\hline
\ce{TiO + O -> Ti + O2} & 5.42 $ \times 10^{-13}$ $\exp(-20794/T)$ &{\scriptsize $E_a$ est. with $A$\footnote{The pre-exponential factor.} from matching NIST 1994LIA/MIT (at room temperature)}\\
\ce{TiO + N ->  Ti + NO} & 4.76 $ \times 10^{-12}$ $\exp(-4772/T)$ &{\scriptsize $E_a$ est. with $A$ from matching NIST1993CLE/HON (at room temperature)}\\
\ce{TiO + N2 -> Ti + N2O} & 5.9 $\times 10^{-12}$ $\exp(-62193/T)$ &{\scriptsize $E_a$ est. with $A$ from matching NIST 1994LIA/MIT (at room temperature)}\\
\ce{TiO + H -> Ti + OH} & 2.0 $\times 10^{-10}$ $\exp(-28418/T)$ &{\scriptsize $E_a$ est. with $A$ based on \ce{FeO + H -> Fe + OH} from \cite{Rumminger1999}}\\
\ce{TiO + H2 -> Ti + H2O} & 2.0 $\times 10^{-11}$ $\exp(-21270/T)$ &{\scriptsize $E_a$ est. with $A$ based on \ce{FeO + H2 -> Fe + OH} from \cite{Decin2018}}\\
\ce{Ti + CO2 -> TiO + CO} & 7.01 $\times 10^{-11}$ $\exp(-1790/T)$ &{\scriptsize NIST 1993CAM/MCC7942-7946}\\
\ce{TiO2 + H -> TiO + OH} & $1.0 \times 10^{-10} \exp(-19500/T)$ &{\scriptsize $E_a$ est. with $A$ based on \ce{FeO2 + H -> FeO + OH} from \cite{Rumminger1999}}\\
\ce{TiO2 + O -> TiO + O2} & $1.0 \times 10^{-10} \exp(-11877/T)$ &{\scriptsize $E_a$ est.}\\
\ce{TiO2 + CO -> TiO + CO2} & $1.0 \times 10^{-10} \exp(-9160/T)$ &{\scriptsize $E_a$ est. with $A$ based on \ce{FeO2 + O -> FeO + O2} from \cite{Rumminger1999}}\\
\ce{TiH + H -> Ti + H2} & 8.3 $\times 10^{-11}$ &{\scriptsize Est. from \ce{FeH + H -> Fe + H2} \citep{Rumminger1999}}\\
\ce{TiH + O -> Ti + OH} & 1.66 $\times 10^{-10}$ &{\scriptsize Est. from \ce{FeH + O -> Fe + OH} \citep{Rumminger1999}}\\
\ce{TiH + CH3 -> Ti + CH4} & 1.0 $\times 10^{-10}$ &{\scriptsize Est. from \ce{FeH + CH3 -> Fe + CH4} \citep{Rumminger1999}}\\
\ce{TiC + H -> Ti + CH} & $1.0 \times 10^{-10} \exp(-20109/T)$ & $E_a$ est.\\
\ce{Ti + CO -> TiC + O} & $8.0 \times 10^{-10} \exp(-68842/T)$ & $E_a$ est.\\
\ce{Ti + CN -> TiC + N} & $5.0 \times 10^{-10} \exp(-29543/T)$ & $E_a$ est.\\
\ce{Ti + NO -> TiN + O} & $5.0 \times 10^{-11} \exp(-19706/T)$ &{\scriptsize $E_a$ est. with $A$ from matching \cite{campbell}}\\
\ce{TiN + H -> Ti + NH} & $10^{-10} \exp(-15655/T)$ & $E_a$ est.\\
\ce{VO + O -> V + O2} & $6.19 \times 10^{-12}$ $\exp(-14763/T)$&{\scriptsize $E_a$ est. with $A$ from matching NIST 1990RIT/WEI (at room temperature)}\\
\ce{V + N2O -> VO + N2} & $4.7 \times 10^{-11}$ $\exp(-1299/T)$&{\scriptsize NIST 2000CAM/KOL}\\
\ce{VO + H -> V + OH} & $1.66 \times 10^{-10}$ $\exp(-22386/T)$&{\scriptsize $E_a$ est. with $A$ based on \ce{FeO + H -> Fe + OH} from \cite{Rumminger1999}}\\
\ce{VO + H2 -> V + H2O} & $1.0 \times 10^{-11}$ $\exp(-15239/T)$&{\scriptsize $E_a$ est. with $A$ based on \ce{FeO + H2 -> Fe + H2O} from \cite{Decin2018}}\\
\ce{TiO2 ->[M] Ti + O2} &\hspace*{-0.8cm}\makecell[l]{$k_0$ = $1.38 \times T^{-1.8}$ $\exp(-94079/T)$\\$k_{\infty}$ = $2 \times 10^{17}$ $T^{-0.91}$ $\exp(-94079/T)$}&{\scriptsize $E_a$ est. $A$ based on \ce{FeO2 ->[M] Fe + O2} from \cite{Rumminger1999}}\\
\ce{TiO ->[\ce{M}] Ti + O} &\hspace*{-0.8cm}\makecell[l]{$k_0$ = $1.38 \times T^{-1.8}$ $\exp(-82202/T)$\\$k_{\infty}$ = $2 \times 10^{17}$ $T^{-0.91}$ $\exp(-76171/T)$}&{\scriptsize $E_a$ est. $A$ based on \ce{FeO ->[M] Fe + O} from \cite{Rumminger1999}}\\
\ce{VO ->[\ce{M}] V + O} &\hspace*{-0.8cm}\makecell[l]{$k_0$ = $1.38 \times T^{-1.8}$ $\exp(-76171/T)$\\$k_{\infty}$ = $2 \times 10^{17}$ $T^{-0.91}$ $\exp(-76171/T)$}&{\scriptsize $E_a$ est. $A$ based on \ce{FeO ->[M] Fe + O} from \cite{Rumminger1999}}\\
\hline
\end{tabular}
\end{center}
\end{table*}

\clearpage 

\begin{longtable}{|r p{.16\textwidth} | >{\centering\arraybackslash}p{.068\textwidth} | >{\centering\arraybackslash}p{.18\textwidth} | >{\centering\arraybackslash}p{.26\textwidth} |}
\caption{Full List of Photolysis Reactions in VULCAN} 
\label{tab:photo_rates}
\endfirsthead
\endhead
\hline
\mbox{  } Photolysis & \mbox{  } Reaction & Threshold (nm) & Temp. Dependence (K) & Reference \newline{\scriptsize Cross Sections / Quantum Yields ($\lambda$ in nm)}\\ 
\hline
\ce{H2O} &\ce{-> H + OH} & 207 & 300, 423, 573, 1230, 1540, 1630, 1820, 2010, 2360 &a, b, e / d, \cite{Stief1975,slanger82}\\
 	     &\ce{-> H2 + O(^1D)}  & & &\\
 	     &\ce{-> O + H + H}  & & &\\
\ce{CH4} & \ce{-> CH3 + H} & 145 & --- & a / d ($\lambda$ $<$ 97.7), \cite{gans11} ($\lambda$ $\geq$ 118.2)\\
 	     &\ce{-> ^1CH2 + H2}  &&&\\
 		 &\ce{-> ^1CH2 + H + H}  &&&\\
 		 &\ce{-> CH + H2 + H}  &&&\\
\ce{CH3} &\ce{-> CH + H2} & 220 & --- & a / \cite{lavvas2008,Kassner1994}\\             		             &\ce{-> CH2 + H} & & & \\
\ce{CH2} &  \ce{-> CH + H}& 275 & --- & a / a\\
 \ce{CO} &  \ce{-> C + O} & 166 & --- & a, b/ a\\
 \ce{H2} &  \ce{-> H + H} & 120 & --- & a / a\\
 \ce{C2H2} &  \ce{-> C2H + H} & 217 & --- & a / a, \cite{okabe83} \\
 \ce{CO2} &  \ce{-> CO + O} & 202 & 150, 170, 195, 230, 300, 420, 500, 585, 700, 800, 1160& a,  \cite{Venot2018} , e / b\\
		 	& \ce{-> CO + O(^1D)} & & &\\
\ce{C2H4} & \ce{-> C2H2 + H2} & 195 & --- & a / \cite{lavvas2008} and the references in\\
 		  &  \ce{-> C2H2 + H + H} & & &\\
 		  &  \ce{-> C2H3 + H} & & &\\
\ce{C2H6} &  \ce{-> C2H4 + H2} & 165 & --- & a / b, \cite{Lias1970} (104 $<$ $\lambda$ $<$ 105)\\
		  & \ce{-> C2H4 + H + H} & & &\\
		  & \ce{-> C2H2 + H2 + H2} & & &\\
		  & \ce{-> CH4 + ^1CH2} & & &\\
 		  &  \ce{-> CH3 + CH3} & & &\\
\ce{C4H2} & \ce{-> C2H2 + C2}& 197 & --- & a /  \cite{lavvas2008}\\
\ce{C6H6} & \ce{-> C6H5 + H}& 206 & --- & a / Est. from \cite{Kislov2004}\\
		  & \ce{-> C3H3 + C3H3} & & &\\
\ce{OH} & \ce{-> O + H}& 265 & --- & a / a, b  ($\lambda$ $<$ 91)\\
\ce{HCO} & \ce{-> H + CO} & 656 & --- & a / a\\
\ce{H2CO} & \ce{-> H2 + CO} & 360 & --- & a / b ($\lambda$ $<$ 250), d\\
		  &  \ce{-> H + HCO} & & &\\
\ce{CH3OH} &  \ce{-> CH3 + OH} & 220 & --- & a / b, \cite{Hagege1968}\\
		   & \ce{-> H2CO + H2} & & & \\
		   & \ce{-> CH3O + H} & & & \\
\ce{CH3CHO} &  \ce{-> CH4 + CO} & 350 & --- & a / b\\
			 &  \ce{-> CH3 + HCO} & & & \\
\ce{N2} &  \ce{-> N + N} & 150 & --- & a ($\lambda$ $<$ 100), c ($\lambda$ $>$ 120)  / a\\
\ce{NH3} &  \ce{-> NH2 + H} & 226 & 175, 300, 562 & a / b\\
         &  \ce{-> NH + H + H} & & &\\
\ce{HCN} &  \ce{-> H + CN} & 179  & --- & a / \cite{nuth82}\\
\ce{NO} &  \ce{-> N + O}& 202 & & a / a\\
\ce{NO2} & \ce{-> NO + O}& 398 & & b, \cite{voigt02} / b, d\\
\ce{NO3} & \ce{-> NO2 + O}& 703 & ---& b / b, \cite{sander}\\
         & \ce{-> NO + O2} &&&\\ 
\ce{N2O} & \ce{-> NO + O(^1D)}& 340 & 300, 714, 833, 1000, 1250, 1667, 2000 & a,c \citep{N2O}) / b\\
\ce{HNO2} &  \ce{-> NO + OH}& 591 & --- & b / b\\
\ce{HNO3} &  \ce{-> NO2 + OH}& 598 & --- & b / b\\
\ce{N2O5} &  \ce{-> NO3 + NO2}& 500 & --- & b / b  (from here)\\
          &  \ce{-> NO3 + NO + O}&  & &\\
\ce{N2H4} &  \ce{-> N2H3 + H}& 291 & --- &  c (BiehlStuhl(1991) and Vaghjiani(1993)	296K	191-291n) / \cite{lavvas2008}\\
\ce{O2} & \ce{-> O + O}& 240 & 200, 250, 300, 873, 1073 & a / b,d\\
        & \ce{-> O + O(^1D)} &&&\\
\ce{O3} & \ce{-> O2 + O}& 900 & --- & a / Matsumi(2002)\\
        & \ce{-> O2 + O(^1D)}  &&&\\
\ce{HO2} &  \ce{-> O + OH}& 275 & --- & a / a\\
\ce{H2O2} &  \ce{-> OH + OH}& 350 & --- & a / a\\
\ce{HNCO} &  \ce{-> NH + CO}& 354 & --- & b / b\\
          &  \ce{-> H + NCO}&  & &\\
\ce{SH} & \ce{-> S + H}& 345 & 300, 500, 750, 1000, 1250, 1500, 1750, 2000, 2250, 2500, 3000 & a, \cite{SH_cross} for $\lambda$ $\geq$ 314 nm / a\\
\ce{H2S} & \ce{-> SH + H}& 238 & 423, 573, 773 & a, e / a\\
\ce{SO} & \ce{-> S + O}& 235 & --- & a / a\\
\ce{SO2} & \ce{-> SO + O}& 220 & --- & a / a\\
         & \ce{-> S + O2}&     & --- & a / a\\
\ce{S2} &  \ce{-> S + S}& 283 & --- & a / a\\
\ce{S4} &  \ce{-> S + S}& 575 & --- & a / a\\
\ce{OCS} & \ce{-> CO + S}& 280 & --- & a / a\\
\ce{CS} &  \ce{-> C + S}& 160 & --- & a / a\\
\ce{CS2} & \ce{-> CS + S}& 278 & 300, 423, 573, 773 & a / a\\
\ce{CH3SH} & \ce{-> CH3S + H}& 310 & --- & a / b\\
		   & \ce{-> CH3 + SH} & & & \\

\hline
\end{longtable}

a: Leiden Observatory database\footnote{\url{http://home.strw.leidenuniv.nl/~ewine/photo}} \citep{Heays2017}
b: PHIDRATES database\footnote{\url{http://phidrates.space.swri.edu}} \citep{Huebner1992}
c: MPI-Mainz UV/VIS Spectral Atlas \footnote{\url{http://satellite.mpic.de/spectral_atlas/index.html}} \citep{mpi})
d: \cite{sander}
e: ExoMol database \footnote{\url{http://www.exomol.com/data/data-types/xsec_VUV}}

\section{The Choice of Zenith Angle}\label{app:mu}
A global or hemispheric average 1-D photochemical model requires specifying an effective zenith angle ($\overline{\theta}$) of the stellar beam to represent the planetary-mean actinic flux. The zenith angle of 48$^{\circ}$ -- 60$^{\circ}$ has been used for the hemispheric average in various photochemical models (e.g. \cite{Moses11,Hu2012}). For instance, a common choice in radiative transfer calculation is to take the flux-weighted cosine of the zenith angle \citep{Cronin2014}:
\begin{equation}
\overline{\mu_\textrm{I}} =  \frac{\int_0^{2\pi} \int_0^1 \mu F_0 \mu d\mu d\phi}{\int_0^{2\pi} \int_0^1 F_0 \mu d\mu d\phi}
\label{eq:mu1} 
\end{equation} 
where $\phi$ is the azimuth angle and $F_0$ is the stellar flux at the top of atmosphere. Equation (\ref{eq:mu1}) yields $\overline{\mu_\textrm{I}}$ = 2/3 or $\overline{\theta_\textrm{I}} \approx$ 48$^{\circ}$. For photochemistry calculation, actinic-flux-weighted cosine should be considered and Equation ({\ref{eq:mu1}}) becomes
\begin{equation}
\overline{\mu_\textrm{II}} =  \frac{\int_0^1 \mu J_0 d\mu}{\int_0^1 J_0 d \mu}
\label{eq:mu2} 
\end{equation} 
where $J_0$ is the actinic flux at the top of atmosphere (i.e. total intensity) and the azimuth term is dropped. Equation (\ref{eq:mu2}) now yields $\overline{\mu_\textrm{II}}$ = 1/2 or $\overline{\theta_\textrm{II}}$ = 60$^{\circ}$.

\cite{Hu2012} discuss this choice of mean zenith angle by further considering the optical depth of the level where the hemispheric average of the attenuated actinic flux is evaluated, e.g., the authors find a mean zenith angle of 57$^{\circ}$ and 48$^{\circ}$ correspond to optical depth 0.1 and 1, respectively. We will revisit the discussion of \cite{Hu2012} but with a different approach here. 

Instead of taking the average of $\mu$, we consider an effective zenith angle such that the resulting mean actinic flux matches the hemispheric-mean actinic flux, viz.

\begin{equation}
\begin{aligned}
J_0 \, \textrm{exp}(- \frac{\tau}{\overline{\mu}}) &= \frac{\int_0^{1} J_0 \, \textrm{exp} (- \frac{\tau}{\mu}) d\mu}{\int_{0}^1 d\mu}\\
&= J_0 \int_0^1 \textrm{exp} (- \frac{\tau}{\mu}) d\mu.
\end{aligned}
\label{eq:mu}
\end{equation}
 
\begin{figure}[htp]
\begin{center}
\includegraphics[width=0.5\columnwidth]{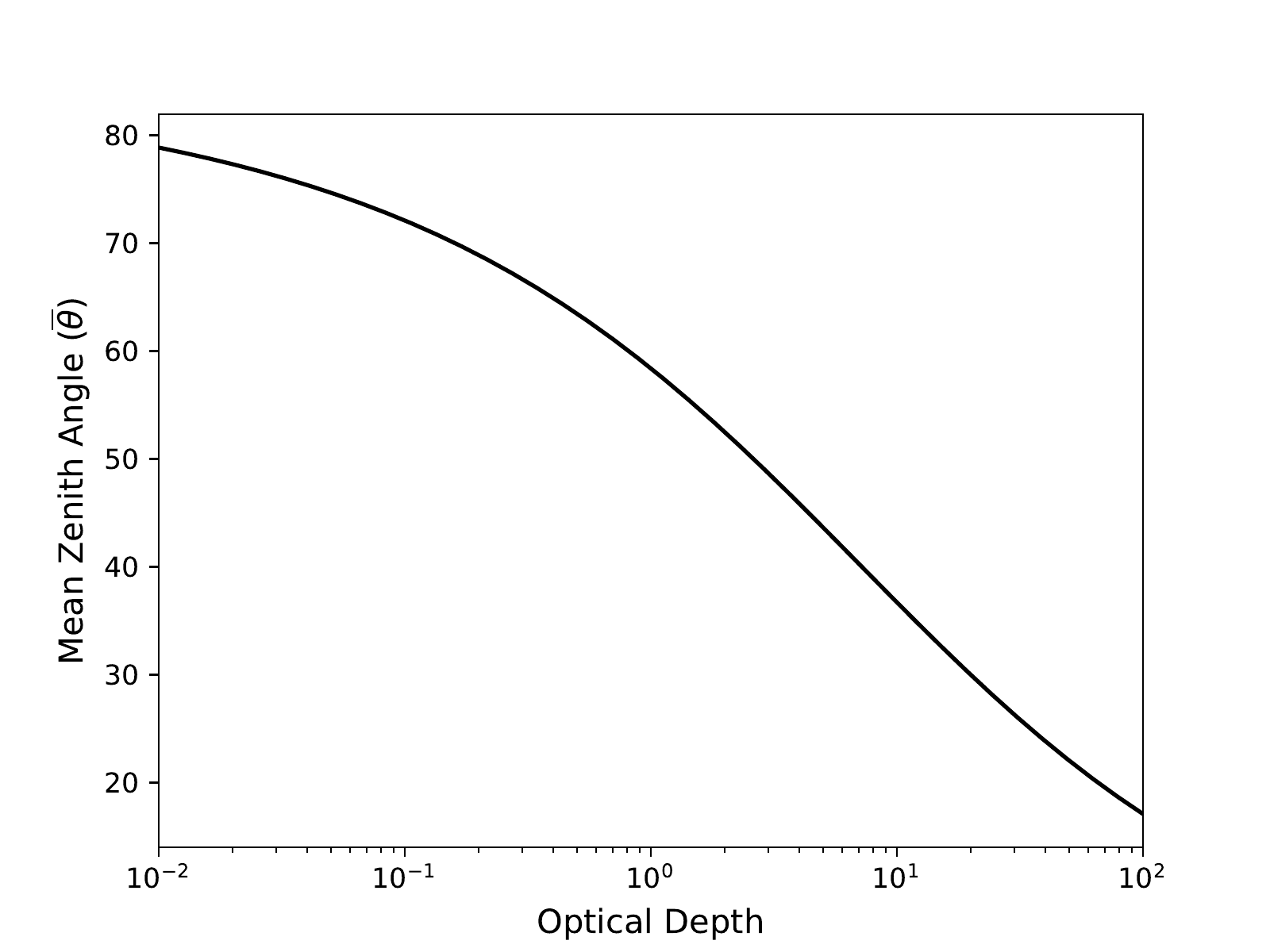}
\end{center}
\caption{Dayside mean zenith angle for photochemical calculation as a function of optical depth from Equation (\ref{eq:mu}).}
\label{fig:mu-tau}
\end{figure}

Equation (\ref{eq:mu}) can be evaluated numerically at the given optical depth, as illustrated in Figure \ref{fig:mu-tau}. We find that optical depth unity (i.e. where UV photons are mostly utilized for photochemistry) corresponds to $\overline{\theta}$ $\approx$ 58$^{\circ}$. Note that the evaluation in \cite{Hu2012} is similar to Equation (\ref{eq:mu}) but weighted by $\mu$ within the integral, which is effectively the mean stellar flux instead of the actinic flux. Evaluating Equation (\ref{eq:mu}) at $\tau = 1$, we find $\overline{\theta}$ $\approx$ 58$^{\circ}$ for a dayside-average model. For a terminator-average model, the denominator in Equation (\ref{eq:mu}) is integrated from -1 to 1 and we have $\overline{\theta}$ $\approx$ 67$^{\circ}$.  


\begin{figure}[htp]
\begin{center}
\includegraphics[width=0.5\columnwidth]{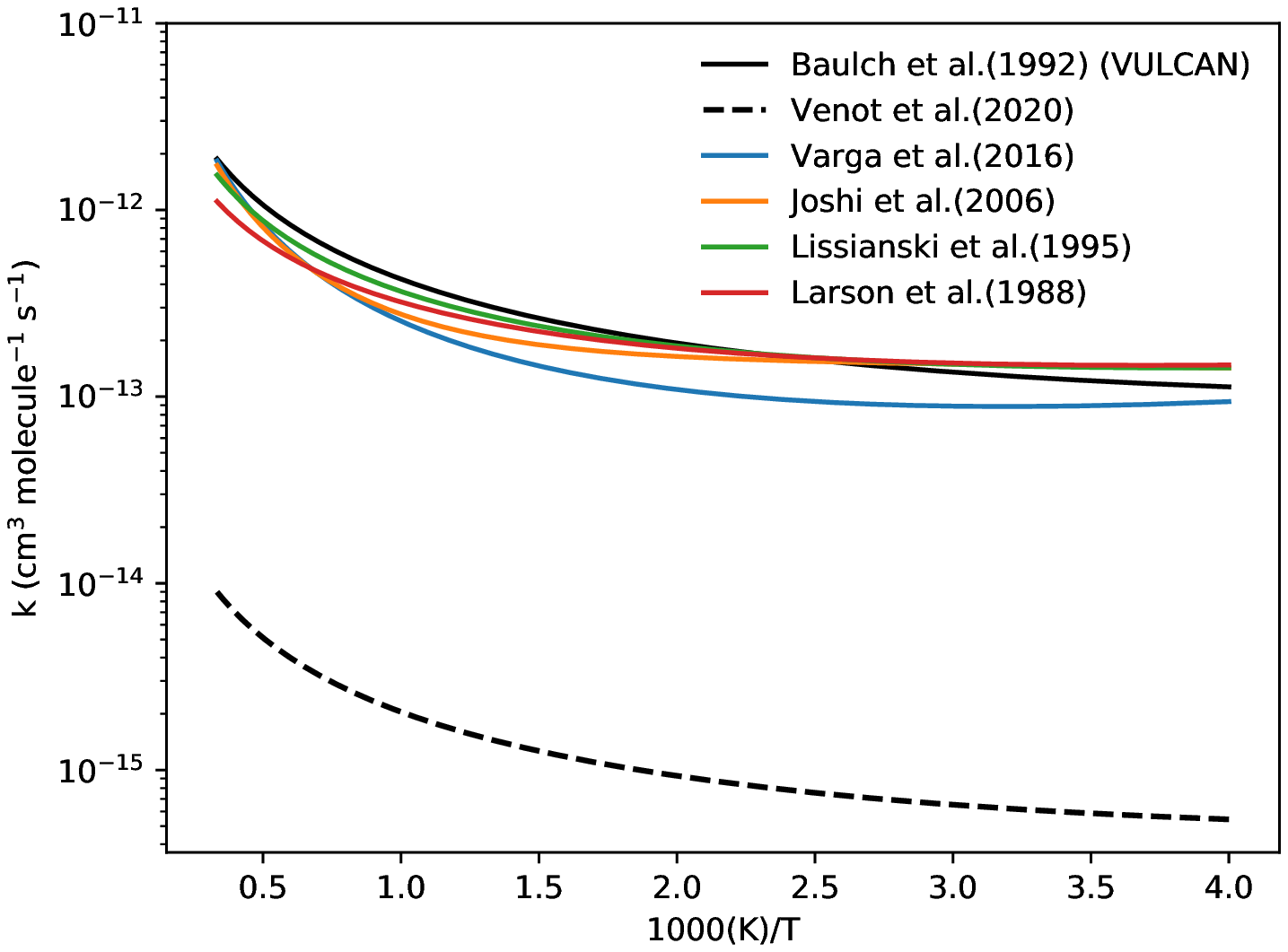}
\end{center}
\caption{Comparisons of selective rate constants of Reaction (\ref{re:CO}): \ce{CO + OH -> CO2 + H} with wide temperature ranges available on NIST.}
\label{fig:rate_CO2}
\end{figure}

\nocite{Barkley2008,Ehhalt1973,Funke2009,Hopfner2016,Inn1979,Georgii1980,Jaeschike1976}

\bibliographystyle{apj}

\bibliography{bib_vulcan2}

\label{lastpage}

\end{document}